\documentclass[12pt]{article}
\pdfoutput=1
\usepackage{jheppub}
\usepackage{epsfig}
\usepackage{amsmath}
\usepackage{amssymb}
\usepackage{amsfonts}
\usepackage{amsxtra}
\usepackage{amsthm}
\usepackage{mathrsfs}
\usepackage{mathdots}
\usepackage{makeidx}
\usepackage{graphics}
\usepackage{dsfont}
\usepackage{mathtools}
\usepackage{graphicx}
\usepackage{placeins}
\usepackage{bm}
\usepackage[capitalize]{cleveref}
\usepackage{empheq}
\usepackage{colortbl}
\usepackage{xcolor}
\usepackage{enumerate}
\usepackage{titlesec}
\usepackage{longtable}
\usepackage{hhline}
\usepackage{float}
\usepackage{color}
\usepackage{tikz}
\usepackage{xfrac}
\usepackage{footnote}
\usepackage{rotating}
\usepackage{lscape}
\usepackage{makecell}
\usepackage{environ}
\usepackage{tabularx}
\usepackage{subfiles}
\usepackage[export]{adjustbox}
\usepackage{ytableau}
\usepackage{tikz-3dplot}
\usepackage{slashed}
\usepackage{pifont}
\usepackage{multirow}
\usepackage{mdframed}
\usepackage{bbm}
\usepackage[normalem]{ulem}
\usepackage{arydshln}
\usepackage[inline]{enumitem} 
\usepackage{fancybox}
\usepackage[numbers,compress]{natbib}
\usepackage{subfiles}
\usepackage{crossreftools}
\usepackage{pdflscape}

\usetikzlibrary{positioning,trees,decorations.pathmorphing,decorations.markings,decorations.pathreplacing,calc,shapes,patterns,arrows,chains,arrows.meta,fit,fadings,decorations.markings,graphs,graphs.standard,quotes,plotmarks,calligraphy}

\DeclareMathOperator{\Res}{Res}
\DeclareMathOperator{\Disc}{Disc}
\DeclareMathOperator{\U}{U}
\DeclareMathOperator{\SU}{SU}
\DeclareMathOperator{\SO}{SO}
\DeclareMathOperator{\SL}{SL}
\DeclareMathOperator{\GL}{GL}
\DeclareMathOperator{\Sp}{Sp}
\DeclareMathOperator{\USp}{USp}
\DeclareMathOperator{\tr}{Tr}
\DeclareMathOperator{\str}{STr}
\DeclareMathOperator{\rank}{rank}
\DeclareMathOperator{\spa}{Span}
\DeclareMathOperator{\spec}{Spec}
\DeclareMathOperator{\sign}{sign}
\DeclareMathOperator{\perm}{perm}
\newcommand{\de}{\partial}
\newcommand{\Lag}{\mathcal{L}}
\newcommand{\Ham}{\mathcal{H}}
\newcommand{\cM}{\mathcal{M}}
\newcommand{\cN}{\mathcal{N}}
\newcommand{\tder}[2]{\frac{d#1}{d#2}}
\newcommand{\tderr}[1]{\frac{d}{d#1}}
\newcommand{\ntder}[3]{\frac{d^{#3}#1}{d#2^{#3}}}
\newcommand{\ntderr}[2]{\frac{d^{#2}}{d #1^{#2}}}
\newcommand{\pder}[2]{\frac{\de#1}{\de#2}}
\newcommand{\pderr}[1]{\frac{\de}{\de #1}}
\newcommand{\npder}[3]{\frac{\de^{#3}#1}{\de#2^{#3}}}
\newcommand{\npderr}[2]{\frac{\de^{#2}}{\de #1^{#2}}}
\newcommand{\lspinu}[2]{#1^{#2}}
\newcommand{\lspind}[2]{#1_{#2}}
\newcommand{\rspinu}[2]{\bar{#1}^{\dot{#2}}}
\newcommand{\rspind}[2]{\bar{#1}_{\dot{#2}}}
\newcommand{\ftheta}[2]{\vartheta\left[\begin{array}{c} 
		#1 \\ #2 
	\end{array} 
	\right]}
\newcommand{\coma}{\, , \quad}
\newcommand{\fstop}{\, .}
\newcommand{\hc}{\quad\text{ h.c.}}
\newcommand{\with}{\quad\text{with}\quad}
\newcommand{\aand}{\quad\text{and}\quad}
\newcommand{\AdS}{\text{AdS}}
\newcommand{\ID}{\mathds{1}}
\newcommand{\HB}{\text{HB}}

\def\CC{{\mathbb{C}}}
\def\QQ{{\mathbb{Q}}}
\def\RR{{\mathbb{R}}}
\def\ZZ{{\mathbb{Z}}}
\def\PP{{\mathbb{P}}}
\def\DD{{\mathbb{D}}}
\def\IC{{\bf{C}}}
\def\IH{{\bf{H}}}
\def\IS{{\bf {S}}}
\def\IR{{\bf {R}}}
\def\IZ{{\bf {Z}}}
\def\IX{{\bf {X}}}
\def\IY{{\bf {Y}}}
\def\IB{{\bf {B}}}
\def\IT{{\bf {T}}}
\def\IP{{\bf {P}}}
\def\IRP{{\bf {RP}}}
\renewcommand{\epsilon}{\varepsilon}

\theoremstyle{plain}

\theoremstyle{definition}

\let\oldFootnote\footnote
\newcommand\nextToken\relax

\renewcommand\footnote[1]{%
    \oldFootnote{#1}\futurelet\nextToken\isFootnote}

\newcommand\isFootnote{%
    \ifx\footnote\nextToken\textsuperscript{,}\fi}

\tikzset{cross/.style={cross out, draw=black, fill=none, minimum size=2*(#1-\pgflinewidth), inner sep=0pt, outer sep=0pt}, cross/.default={2pt}}

\newcommand{\Lpagenumber}{\ifdim\textwidth=\linewidth\else\bgroup
	\dimendef\margin=0 
	\ifodd\value{page}\margin=\oddsidemargin
	\else\margin=\evensidemargin
	\fi
	\raisebox{\dimexpr -3\topmargin-\headheight-\headsep-0.5\linewidth}[0pt][0pt]{%
		\rlap{\hspace{\dimexpr \margin+\textheight+2\footskip}%
			\llap{\rotatebox{90}{\hfill-- \thepage\ --\hfill}}}}%
	\egroup\fi}
\AddToHook{shipout/background}{\Lpagenumber}%

\newcommand*\widefbox[1]{\fbox{\hspace{2em}#1\hspace{2em}}}
\newcommand\superequiv{\mathrel{\rlap{\raisebox{\fontdimen22\textfont2}{$=$}}\raisebox{-0.5\fontdimen22\textfont2}{$ = $}}}

\usetikzlibrary{positioning}
\usetikzlibrary{chains}
\usetikzlibrary{arrows, arrows.meta ,fit,decorations.pathreplacing}
\tikzstyle{every picture}+=[remember picture]
\tikzstyle{na} = [baseline]
\tikzstyle{ligne}=[draw, thick]
\newcommand\addvmargin[1]{ \node[fit=(current bounding box),inner ysep=#1,inner xsep=0]{};}
\usetikzlibrary{arrows, decorations.markings, calc, fadings, decorations.pathreplacing, patterns, decorations.pathmorphing, positioning}
\tikzset{>={Latex[width=1.5mm,length=1.5mm]}}
\tikzset{bd/.style={circle, draw=black, inner sep=0pt, fill=black, minimum size=1.2mm}}
\tikzset{bld/.style={circle, draw=blue, inner sep=0pt, fill=blue, minimum size=1.2mm}}
\tikzset{wd/.style={circle, draw=black, inner sep=0pt, fill=white, minimum size=1.2mm}}
\tikzset{rd/.style={circle, draw=red, inner sep=0pt, fill=red, minimum size=.9mm}}
\tikzset{wrd/.style={circle, draw=red, inner sep=0pt, fill=white, minimum size=.9mm}}
\usetikzlibrary{graphs,graphs.standard,quotes}

\def\node#1#2{\overset{#1}{\underset{#2}{{\color{gray} \bullet}}}}
\def\Node#1#2{\overset{#1}{\underset{#2}{{ \bullet}}}}
\def\sqnode#1#2{\overset{#1}{\underset{#2}{{\color{gray} \blacksquare}}}}
\def\sqNode#1#2{\overset{#1}{\underset{#2}{{ \blacksquare}}}}
\def\sqwnode#1#2{\overset{#1}{\underset{#2}{{\square}}}}
\def\cver#1#2{\overset{{\llap{$\scriptstyle#1$}\displaystyle{\color{gray} \bullet}{\rlap{$\scriptstyle#2$}}}}{\scriptstyle\vert}}
\def\cvver#1#2{\overset{{\llap{$\scriptstyle#1$}\displaystyle{\color{gray} \bullet}{\rlap{$\scriptstyle#2$}}}}{\scriptstyle\uparrow\, \downarrow}}
\def\sqcver#1#2{\overset{{\llap{$\scriptstyle#1$}\displaystyle{\color{gray} \blacksquare}{\rlap{$\scriptstyle#2$}}}}{\scriptstyle\vert}}
\def\bnode#1#2{\overset{#1}{\underset{#2}{{\color{blue} \bullet}}}}
\def\pnode#1#2{\overset{#1}{\underset{#2}{{\color{violet} \bullet}}}}
\def\bsqnode#1#2{\overset{#1}{\underset{#2}{{\color{blue} \blacksquare}}}}
\def\bcver#1#2{\overset{{\llap{$\scriptstyle#1$}\displaystyle{\color{blue} \bullet}{\rlap{$\scriptstyle#2$}}}}{\scriptstyle\vert}}
\def\bsqcver#1#2{\overset{{\llap{$\scriptstyle#1$}\displaystyle{\color{blue} \blacksquare}{\rlap{$\scriptstyle#2$}}}}{\scriptstyle\vert}}

\def\node#1#2{\overset{#1}{\underset{#2}{\circ}}}
\def\sqnode#1#2{\overset{#1}{\underset{#2}{{\color{gray} \blacksquare}}}}
\def\rnode#1#2{\overset{#1}{\underset{#2}{{\color{red} \bullet}}}}
\def\bnode#1#2{\overset{#1}{\underset{#2}{{\color{blue} \bullet}}}}
\def\gnode#1#2{\overset{#1}{\underset{#2}{{\color{gray} \bullet}}}}
\def\sqgrnode#1#2{\overset{#1}{\underset{#2}{{\color{gray} \blacksquare}}}}
\def\sqbnode#1#2{\overset{#1}{\underset{#2}{{\color{blue} \blacksquare}}}}
\def\sqrnode#1#2{\overset{#1}{\underset{#2}{{\color{red} \blacksquare}}}}
\def\sqblnode#1#2{\overset{#1}{\underset{#2}{{ \blacksquare}}}}
\def\sqwnode#1#2{\overset{#1}{\underset{#2}{{ \square}}}}
\def\grcver#1#2{\overset{{\llap{$\scriptstyle#1$}\displaystyle{\color{gray} \bullet}{\rlap{$\scriptstyle#2$}}}}{\scriptstyle\vert}}
\def\rver#1#2{\overset{{\llap{$\scriptstyle#1$}\displaystyle{\color{red} \bullet}{\rlap{$\scriptstyle#2$}}}}{\scriptstyle\vert}}
\def\bver#1#2{\overset{{\llap{$\scriptstyle#1$}\displaystyle{\color{blue} \bullet}{\rlap{$\scriptstyle#2$}}}}{\scriptstyle\vert}}
\def\ver#1#2{\overset{{\llap{$\scriptstyle#1$}\displaystyle\circ{\rlap{$\scriptstyle#2$}}}}{\scriptstyle\vert}}
\def\wver#1#2{\overset{{\llap{$\scriptstyle#1$}\displaystyle{\square}{\rlap{$\scriptstyle#2$}}}}{\scriptstyle\vert}}
\def\bluesqver#1#2{\overset{{\llap{$\scriptstyle#1$}\displaystyle{\color{blue} \blacksquare}{\rlap{$\scriptstyle#2$}}}}{\scriptstyle\vert}}
\def\redsqver#1#2{\overset{{\llap{$\scriptstyle#1$}\displaystyle{\color{red} \blacksquare}{\rlap{$\scriptstyle#2$}}}}{\scriptstyle\vert}}
\def\grver#1#2{\overset{{\llap{$\scriptstyle#1$}\displaystyle{\color{gray} \blacksquare}{\rlap{$\scriptstyle#2$}}}}{\scriptstyle\vert}}
\def\grvcver#1#2{\overset{{\llap{$\scriptstyle#1$}\displaystyle{\color{gray} \bullet}{\rlap{$\scriptstyle#2$}}}}{\scriptstyle\uparrow\, \downarrow}}
\def\grvver#1#2{\overset{{\llap{$\scriptstyle#1$}\displaystyle{\color{gray} \blacksquare}{\rlap{$\scriptstyle#2$}}}}{\scriptstyle\uparrow\, \downarrow}}
\def\blver#1#2{\overset{{\llap{$\scriptstyle#1$}\displaystyle\blacksquare{\rlap{$\scriptstyle#2$}}}}{\scriptstyle\vert}}
\def\gruer#1#2{\underset{{\llap{$\scriptstyle#1$}\displaystyle{\color{gray} \bullet}{\rlap{$\scriptstyle#2$}}}}{\scriptstyle\vert}}
\def\grayr#1#2{\underset{{\llap{$\scriptstyle#1$}\displaystyle\bullet{\rlap{$\scriptstyle#2$}}}}{\scriptstyle\vert}}
\tikzstyle{every picture}+=[remember picture]
\tikzstyle{na} = [baseline=-.5ex]

\newcommand\longdownarrow{\rotatebox[origin=c]{-90}{\scalebox{1.5}{$\longrightarrow$}}} 
\newcommand{\re}{\,\mathbb{R}\mbox{e}\,}
\newcommand{\im}{\,\mathbb{I}\mbox{m}\,}
\newcommand{\trans}{\ensuremath{\mathsf T}}
\newcommand{\hyph}[1]{$#1$\nobreakdash-\hspace{0pt}}
\newcommand{\dvol}{d\mathrm{vol}}
\newcommand{\vol}{\mathrm{vol}}
\newcommand{\Vol}{\mathrm{Vol}}
\providecommand{\abs}[1]{\lvert#1\rvert}
\newcommand{\Nugual}[1]{\ensuremath{\mathcal{N}= #1}}
\newcommand{\ib}{\bar \imath}
\newcommand{\jb}{\bar \jmath}
\newcommand{\parfrac}[2]{\frac{\partial #1}{\partial #2}}
\newcommand{\delfrac}[2]{\frac{\delta #1}{\delta #2}}
\newcommand{\orders}[1]{\calO \Bigl( #1 \Bigr)}
\newcommand{\ud}[2]{^{#1}_{\phantom{#1}#2}}
\newcommand{\du}[2]{_{#1}^{\phantom{#1}#2}}
\newcommand{\ntitle}[1]{\begin{center} \LARGE \textbf{#1} \end{center}}
\newcommand{\subtitle}[1]{\par \ \par \textbf{#1} \par \ \par}
\newcommand{\mr}{\mathrm}
\newcommand{\rep}[1]{\ensuremath{\mathbf{#1}}}
\newcommand{\sub}[1]{{\lfloor #1 \rfloor}}

\newcommand{\eg}{e.g. }
\newcommand{\Eg}{E.g. }
\newcommand{\etal}{et al. }
\newcommand{\ie}{i.e. }
\newcommand{\ala}{\`a la }
\newcommand{\etc}{etc. }

\numberwithin{equation}{section}
\newcommand{\bes}[1]{\begin{equation} \begin{split} #1\end{split} \end{equation}}

\newcommand{\Dslash}{\slash\!\!\!\!D}
\newcommand{\mat}[1]{\begin{pmatrix} #1 \end{pmatrix}}
\newcommand{\smat}[1]{\big( \begin{smallmatrix} #1 \end{smallmatrix} \big)}
\newcommand{\be}{\begin{equation}} \newcommand{\ee}{\end{equation}}
\newcommand{\bea}{\begin{equation} \begin{aligned}} \newcommand{\eea}{\end{aligned} \end{equation}}
\newcommand{\tabs}{\rule[-1ex]{0pt}{3.5ex}}
\newcommand{\Tabs}{\rule[-2.2ex]{0pt}{6.4ex}}

\def\tilde{\widetilde}
\def\t{\tilde}
\def\hat{\widehat}
\def\h{\hat}
\def\bar{\overline}
\def\a{\alpha}
\def\b{\beta}
\def\g{\gamma}
\def\cm{\mathsf{m}}

\def\by{\times}
\def\rt2{\sqrt{2}}
\def\half {{1 \over 2}}
\def\Re{\mathop{\rm Re}}
\def\Im{\mathop{\rm Im}}
\def\d{\partial}
\def\dbar{\b{\d}}
\def\grad{\nabla}
\def\mod{{\rm mod}}
\def\det{\mathop{\rm det}}
\def\codim{{\mathop{\rm codim}}}
\def\coker{{\mathop {\rm coker}}}
\def\diff{{\rm diff}}
\def\Diff{{\rm Diff}}
\def\Tr{\mathop{\rm Tr}}
\def\tr{\mathop{\rm tr}}
\def\Res{\mathop{\rm Res}}
\def\abs#1{\left|#1\right|}
\def\norm#1{\left \|#1\right \|}
\def\vector#1{{\mbf \vec{#1}}}

\def\fa{\mathfrak{a}}
\def\fb{\mathfrak{b}}
\def\tfa{\tilde{\mathfrak{a}}}
\def\tfb{\tilde{\mathfrak{b}}}

\def\1{{\ds 1}}
\def\R{\hbox{$\bb R$}}
\def\C{\hbox{$\bb C$}}
\def\H{\hbox{$\bb H$}}
\def\Z{\hbox{$\bb Z$}}
\def\N{\hbox{$\bb N$}}
\def\P{\hbox{$\bb P$}}
\def\Sn{\hbox{$\bb S$}}

\def\tp{\thetb^+}
\def\ttp{\tilde \thetb^+}
\def\tm{\thetb^-}
\def\ttm{\tilde \thetb^-}
\def\tfour{\tp\tm \ttp\ttm}

\newcommand{\cA}{\mathcal{A}}
\newcommand{\cB}{\mathcal{B}}
\newcommand{\cC}{\mathcal{C}}
\newcommand{\cD}{\mathcal{D}}
\newcommand{\cE}{\mathcal{E}}
\newcommand{\cF}{\mathcal{F}}
\newcommand{\cG}{\mathcal{G}}
\newcommand{\cH}{\mathcal{H}}
\newcommand{\cI}{\mathcal{I}}
\newcommand{\cJ}{\mathcal{J}}
\newcommand{\cK}{\mathcal{K}}
\newcommand{\cL}{\mathcal{L}}
\newcommand{\cO}{\mathcal{O}}
\newcommand{\cP}{\mathcal{P}}
\newcommand{\cQ}{\mathcal{Q}}
\newcommand{\cR}{\mathcal{R}}
\newcommand{\cS}{\mathcal{S}}
\newcommand{\cT}{\mathcal{T}}
\newcommand{\cU}{\mathcal{U}}
\newcommand{\cV}{\mathcal{V}}
\newcommand{\cW}{\mathcal{W}}
\newcommand{\cX}{\mathcal{X}}
\newcommand{\cY}{\mathcal{Y}}
\newcommand{\cZ}{\mathcal{Z}}
\newcommand{\bB}{\mathbb{B}}
\newcommand{\bC}{\mathbb{C}}
\newcommand{\bF}{\mathbb{F}}
\newcommand{\bH}{\mathbb{H}}
\newcommand{\bI}{\mathbb{I}}
\newcommand{\bK}{\mathbb{K}}
\newcommand{\bN}{\mathbb{N}}
\newcommand{\bO}{\mathbb{O}}
\newcommand{\bP}{\mathbb{P}}
\newcommand{\bQ}{\mathbb{Q}}
\newcommand{\bR}{\mathbb{R}}
\newcommand{\bZ}{\mathbb{Z}}
\newcommand{\fg}{\mathfrak{g}}
\newcommand{\fm}{\mathfrak{m}}
\newcommand{\fn}{\mathfrak{n}}
\newcommand{\fu}{\mathfrak{u}}
\newcommand{\fz}{\mathfrak{z}}
\newcommand{\unit}{\mathbbm{1}}

\newcommand{\tj}{{\tilde j}}
\newcommand{\ti}{{\tilde i}}
\newcommand{\tl}{{\tilde l}}
\newcommand{\ta}{{\tilde a}}
\newcommand{\tM}{{\tilde M}}
\newcommand{\ttau}{{\tilde \tau}}
\newcommand{\ttauj}{{\tilde \tau}_{\tilde j}}
\newcommand{\ttaui}{{\tilde \tau}_{\tilde i}}
\newcommand{\br}{{\bar r}}
\newcommand{\bs}{{\bar s}}
\newcommand{\bt}{{\bar t}}
\newcommand{\bw}{{\overline w}}
\newcommand{\bz}{{\overline z}}
\newcommand{\bv}{{\bar v}}
\newcommand{\bpartial}{{\overline \partial}}
\newcommand{\e}{{\vec e}}

\def\SO{\mathrm{SO}}
\def\O{\mathrm{O}}
\def\SU{\mathrm{SU}}
\def\SL{\mathrm{SL}}
\def\GL{\mathrm{GL}}
\def\Sp{\mathrm{Sp}}
\def\Spin{\mathrm{Spin}}
\def\Usp{\mathrm{\USp}}
\def\Osp{\mathrm{Osp}}
\def\su{\mathfrak{su}}
\def\sl{\mathfrak{sl}}
\def\gl{\mathfrak{gl}}
\def\so{\mathfrak{so}}
\def\sp{\mathfrak{sp}}
\def\usp{\mathfrak{usp}}
\def\fp{\mathfrak{p}}
\def\fr{\mathfrak{r}}
\def\fh{\mathfrak{h}}
\def\fN{\mathfrak{N}}

\def\repf{\square}
\def\repfb{\overline{\square\mkern-1mu}}
\def\reps{\square\mkern-3mu\square}
\def\repsb{\overline{\square\mkern-3mu\square\mkern-1mu}}
\def\repa{\raise4pt\hbox{$\square$}\mkern-14mu\raise-4pt\hbox{$\square$}}
\def\repab{\overline{\raise4pt\hbox{$\square$}\mkern-14mu\raise-4pt\hbox{$\square$}\mkern-1mu}}

\def\smileface{\ensuremath{\hbox{\large$\bigcirc$}\mkern-15mu\raise-1pt\hbox{\scriptsize$\smallsmile$}%
\mkern-10mu\raise4pt\hbox{..}\mkern4mu}}
\def\frownface{\ensuremath{\hbox{\large$\bigcirc$}\mkern-15mu\raise-1pt\hbox{\scriptsize$\smallfrown$}%
\mkern-10mu\raise4pt\hbox{..}\mkern4mu}}

\DeclareMathOperator{\csch}{csch}

\newcommand{\ba}{\begin{array}}
\newcommand{\ea}{\end{array}}
\newcommand{\bi}{\begin{itemize}}
\newcommand{\ei}{\end{itemize}}
\def\vec#1{\bm{#1}}
\def\bea#1\eea{\allowdisplaybreaks \begin{align}#1\end{align}}
 \newcommand{\ben}{\begin{enumerate}}
\newcommand{\een}{\end{enumerate}}
\newcommand{\bean}{\begin{eqnarray*}}
\newcommand{\eean}{\end{eqnarray*}}
\newcommand{\eref}[1]{\eqref{#1}}
\newcommand{\sref}[1]{\S\ref{#1}}
\newcommand{\tref}[1]{Table~\ref{#1}}
\newcommand{\fref}[1]{Figure~\ref{#1}}
\newcommand{\pa}{\partial}
\newcommand{\oline}{\overline}
\newcommand{\PE}{\mathop{\rm PE}}
\newcommand{\PL}{\mathop{\rm PL}}
\newcommand{\VS}{\mathop{\rm VS}}
\newcommand{\res}{\mathop{\rm Res}}
\newcommand{\tQ}{\widetilde{Q}}
\newcommand{\tq}{\widetilde{q}}
\newcommand{\tA}{\widetilde{A}}
\newcommand{\tS}{\widetilde{S}}
\newcommand{\tX}{\widetilde{X}}
\newcommand{\tB}{\widetilde{B}}
\newcommand{\tw}{\widetilde{w}}
\newcommand{\hg}{\widehat{g}}
\newcommand{\ch}{\mathrm{ch}}
\newcommand{\sh}{\mathrm{sh}}

\newcommand{\BC}{\mathbb{C}}
\newcommand{\BR}{\mathbb{R}}
\newcommand{\BP}{\mathbb{P}}
\newcommand{\BT}{\mathbb{T}}
\newcommand{\BZ}{\mathbb{Z}}
\newcommand{\BF}{\mathbb{F}}
\newcommand{\BH}{\mathbb{H}}
\newcommand{\sC}{\mathscr{C}}
\newcommand{\sD}{\mathscr{D}}
\newcommand{\sH}{\mathscr{H}}
\newcommand{\sS}{\mathscr{S}}
\newcommand{\BU}{\mathbf{1}}
\newcommand{\W}{\mathcal{W}}

\newcommand{\diag}{\mathrm{diag}}
\newcommand{\End}{\mathrm{End}}
\newcommand{\tra}{\mathrm{Tr}_{\mathbf{adj}}}
\newcommand{\trR}{\mathrm{Tr}_{\mathbf{R}}}
\newcommand{\fund}{\mathbf{fund}}
\newcommand{\vect}{\mathrm{vector}}
\newcommand{\adj}{\mathbf{Adj}}

\newcommand{\fC}{\mathfrak{C}}
\newcommand{\frgl}{\mathfrak{gl}}
\newcommand{\frsl}{\mathfrak{sl}}
\newcommand{\frsu}{\mathfrak{su}}
\newcommand{\fru}{\mathfrak{u}}
\newcommand{\fro}{\mathfrak{o}}
\newcommand{\frso}{\mathfrak{so}}
\newcommand{\frg}{\mathfrak{g}}
\newcommand{\frsp}{\mathfrak{sp}}
\newcommand{\frW}{\mathfrak{W}}
\newcommand{\fun}{\mathbf{fund}}

\newcommand{\sgn}{\mathrm{sgn}}
\newcommand{\sd}{\mathrm{d}}
\newcommand{\tti}{\widetilde{t}}
\newcommand{\tU}{\widetilde{U}}
\newcommand{\tV}{\widetilde{V}}
\newcommand{\tH}{\widetilde{H}}
\newcommand{\tI}{\widetilde{I}}
\newcommand{\Sym}{\mathrm{Sym}}

\newcommand{\fflat}{\mathcal{F}^\flat}
\newcommand{\ff}{\mathcal{F}^\flat}

\definecolor{light-gray}{gray}{0.5}
\definecolor{new-green}{rgb}{0,0.7,0.3}

\newcommand{\purple}{\color{purple}}
\newcommand{\brown}{\color{brown}}
\newcommand{\blue}{\color{blue}}
\newcommand{\gray}{\color{light-gray}}
\newcommand{\red}{\color{red}}
\newcommand{\green}{\color{new-green}}

\def\skipper{\hskip.5em\relax}
\def\Esix#1#2#3#4#5#6{%
{\text{\small$ \left[ \begin{array}{c@{\skipper}c@{\skipper}c@{\skipper}c@{\skipper}c@{\skipper}c}
&&#2&& \\
#1 &  #3 &  #4 & #5 & #6
\end{array} \right]$}}}
\def\Eseven#1#2#3#4#5#6#7{%
{\text{\small$ \left[ \begin{array}{c@{\skipper}c@{\skipper}c@{\skipper}c@{\skipper}c@{\skipper}c}
&&#2&&& \\
#1 &  #3 &  #4 & #5 & #6&#7
\end{array} \right]$}}}
\def\Eeight#1#2#3#4#5#6#7#8{%
{\text{\small$\begin{array}{c@{\skipper}c@{\skipper}c@{\skipper}c@{\skipper}c@{\skipper}c@{\skipper}c@{\skipper}c}
&&#2&&&& \\
#1 &  #3 & #4  & #5 & #6 & #7 &  #8
\end{array}$}}}
\def\atwoEsix#1#2#3#4#5#6#7#8{%
{\text{\small$\left[#1,#2~;~\begin{array}{c@{\skipper}c@{\skipper}c@{\skipper}c@{\skipper}c@{\skipper}c}
&&#4&& \\
#3 &  #5 &  #6 & #7 & #8
\end{array}\right]$}}}
\def\aoneEseven#1#2#3#4#5#6#7#8{%
{\text{\small$\left[#1~;~\begin{array}{c@{\skipper}c@{\skipper}c@{\skipper}c@{\skipper}c@{\skipper}c}
&&#3&&& \\
#2 &  #4 &  #5 & #6 & #7 & #8
\end{array}\right]$}}}

\def\arr#1#2{\underset{#2}{\overset{#1}{\substack{\longrightarrow \\ \longleftarrow}}}}
\def\alr#1{\,\, \underset{\tilde{#1}}{\overset{#1}{\substack{\longrightarrow \\ \longleftarrow}}} \,\,\,}
\def\aup#1 {\overset{#1}{\uparrow} \, \overset{\tilde{#1}}{\downarrow}}

\newcommand{\Ot}{\mathrm{O3}}
\newcommand{\Ott}{\tilde{\mathrm{O3}}}
\newcommand{\Of}{\mathrm{O5}}
\newcommand{\ON}{\mathrm{ON}}
\tikzset{snake it/.style={decorate, decoration={snake, amplitude=.4mm, segment length=2mm,
                       post length=0mm,pre length=0mm}}}
                       
 \newcommand{\GCD}{\mathrm{GCD}}

\newcommand{\MA}{\mathds{A}}
\newcommand{\BM}{\mathds{B}}
\newcommand{\SM}{\mathds{S}}
\newcommand{\TM}{\mathds{T}}
\newcommand{\Bt}{\mathbf{t}}
\newcommand{\PM}{\mathds{P}}
\newcommand{\QM}{\mathds{Q}}
\newcommand{\ZM}{\mymathds{0}}
\newcommand{\BD}{\mathbb{D}}
\def\u{\mathfrak{u}}
\DeclareMathAlphabet{\mymathds}{U}{BOONDOX-ds}{m}{n}
\tikzstyle{double_border} = [draw, double, double distance=1pt]

\newcommand{\NM}[1]{{\color{red}{[\bf{NM}}: zzzz #1 ]}}
\newcommand{\AM}[1]{{\color{blue}{[\bf{AM}}: zzzz #1 ]}}
\newcommand{\SG}[1]{{\color{red}{[\bf{SG}}: zzzz #1 ]}}
\newcommand{\WH}[1]{{\color{orange}{[\bf{WH}}: zzzz #1 ]}}

\makeatletter
\newsavebox{\measure@tikzpicture}
\NewEnviron{scaletikzpicturetowidth}[1]{%
  \def\tikz@width{#1}%
  \def\tikzscale{1}\begin{lrbox}{\measure@tikzpicture}%
  \BODY
  \end{lrbox}%
  \pgfmathparse{#1/\wd\measure@tikzpicture}%
  \edef\tikzscale{\pgfmathresult}%
  \BODY
}
\makeatother

\newcommand*\circled[1]{\tikz[baseline=(char.base)]{
            \node[shape=circle,draw,inner sep=2pt] (char) {#1};}}

\makeatletter
\def\squarecorner#1{
    \pgf@x=\the\wd\pgfnodeparttextbox%
    \pgfmathsetlength\pgf@xc{\pgfkeysvalueof{/pgf/inner xsep}}%
    \advance\pgf@x by 2\pgf@xc%
    \pgfmathsetlength\pgf@xb{\pgfkeysvalueof{/pgf/minimum width}}%
    \ifdim\pgf@x<\pgf@xb%
        \pgf@x=\pgf@xb%
    \fi%
    \pgf@y=\ht\pgfnodeparttextbox%
    \advance\pgf@y by\dp\pgfnodeparttextbox%
    \pgfmathsetlength\pgf@yc{\pgfkeysvalueof{/pgf/inner ysep}}%
    \advance\pgf@y by 2\pgf@yc%
    \pgfmathsetlength\pgf@yb{\pgfkeysvalueof{/pgf/minimum height}}%
    \ifdim\pgf@y<\pgf@yb%
        \pgf@y=\pgf@yb%
    \fi%
    \ifdim\pgf@x<\pgf@y%
        \pgf@x=\pgf@y%
    \else
        \pgf@y=\pgf@x%
    \fi
    \pgf@x=#1.5\pgf@x%
    \advance\pgf@x by.5\wd\pgfnodeparttextbox%
    \pgfmathsetlength\pgf@xa{\pgfkeysvalueof{/pgf/outer xsep}}%
    \advance\pgf@x by#1\pgf@xa%
    \pgf@y=#1.5\pgf@y%
    \advance\pgf@y by-.5\dp\pgfnodeparttextbox%
    \advance\pgf@y by.5\ht\pgfnodeparttextbox%
    \pgfmathsetlength\pgf@ya{\pgfkeysvalueof{/pgf/outer ysep}}%
    \advance\pgf@y by#1\pgf@ya%
}
\makeatother

\pgfdeclareshape{square}{
    \savedanchor\northeast{\squarecorner{}}
    \savedanchor\southwest{\squarecorner{-}}

    \foreach \x in {east,west} \foreach \y in {north,mid,base,south} {
        \inheritanchor[from=rectangle]{\y\space\x}
    }
    \foreach \x in {east,west,north,mid,base,south,center,text} {
        \inheritanchor[from=rectangle]{\x}
    }
    \inheritanchorborder[from=rectangle]
    \inheritbackgroundpath[from=rectangle]
}

\tikzset{stretch/.initial=1}
\newcommand\drawloop[4][]%
   {\draw[shorten <=0pt, shorten >=0pt,#1]
      ($(#2)!\pgfkeysvalueof{/tikz/stretch}!(#2.#3)$)
      let \p1=($(#2.center)!\pgfkeysvalueof{/tikz/stretch}!(#2.north)-(#2)$),
          \n1= {veclen(\x1,\y1)*sin(0.5*(#4-#3))/sin(0.5*(180-#4+#3))}
      in arc [start angle={#3-90}, end angle={#4+90}, radius=\n1]%
   }

\definecolor{cE8}{HTML}{0000FF}
\definecolor{cE7}{HTML}{FF0000}
\definecolor{cE6}{HTML}{FFA500}

\tikzset{
	flavor/.style={rectangle, draw=black!100, thick, minimum size=2mm},
	gauge/.style={circle, thick, draw=black!100,fill=white!100,  minimum size=2mm, inner sep=0pt},
	bodyE6/.style={draw=cE6!100},
	bodyE7/.style={draw=cE7!100},
	bodyE8/.style={draw=cE8!100},
}

\crefname{table}{Table}{Tables}
\crefname{figure}{Figure}{Figures}
\crefname{section}{Section}{Sections}

\hypersetup{
	pdftitle={All Class S Theories of Type A Originate from Orbi-instantons},    
	pdfauthor={\textcopyright\ Simone Giacomelli, William Harding, Noppadol Mekareeya, Alessandro Mininno},     
	pdfsubject={HEP},   
	pdfcreator={pdfLaTex},   
	pdfproducer={LaTex}, 
	pdfkeywords={},
	colorlinks=true,
}

\preprint{ZMP-HH/24-25}

\title{All Class $\mathcal{S}$ Theories of Type-$A$ Originate from Orbi-instantons}
\author[a]{Simone Giacomelli,}
\author[a,b]{William Harding,}
\author[b,c]{Noppadol Mekareeya}
\author[d,*]{and Alessandro Mininno\note[*]{Part of this project was conducted when the author was affiliated to II. Institut f\"ur Theoretische Physik, Universit\"at Hamburg, Luruper Chaussee 149, 22607 Hamburg, Germany.}}
\affiliation[a]{Dipartimento di Fisica, Universit\`a di Milano-Bicocca, \\ Piazza della Scienza 3, I-20126 Milano, Italy}
\affiliation[b]{INFN, sezione di Milano-Bicocca, \\Piazza della Scienza 3,  I-20126 Milano, Italy}
\affiliation[c]{Department of Physics, Faculty of Science, Chulalongkorn University, \\ Phayathai Road,
Pathumwan, Bangkok 10330, Thailand}
\affiliation[d]{Department of Physics, University of Wisconsin-Madison,\\
	1150 University Avenue, Madison, WI 53706, USA}
\emailAdd{simone.giacomelli@unimib.it}
\emailAdd{w.harding@campus.unimib.it}
\emailAdd{n.mekareeya@gmail.com}
\emailAdd{mininno@physics.wisc.edu}

\abstract{A surprising relation between 4d $\mathcal{N}=2$ class $\mathcal{S}$ superconformal field theories of Type-$A$ and 6d $\mathcal{N}=(1,0)$ orbi-instanton theories is investigated. We find that all of the theories in the former class can be obtained by a series of deformations of the 4d theories arising from compactifying the latter on a torus. This is demonstrated by examining Fayet--Iliopoulos (FI) deformations of the $E_8$-shaped magnetic quivers of the orbi-instanton theories whose body fits into the affine $E_8$ Dynkin diagram with a tail attached. Turning on FI parameters at the appropriate gauge groups leads, in stages, to $E_7$-shaped, $E_6$-shaped, and general star-shaped quivers, where the latter are magnetic quivers for the class $\mathcal{S}$ theory of Type-$A$ on a sphere with punctures. Deforming a suitable star-shaped quiver, one obtains a magnetic quiver of the Type-$A$ class $\mathcal{S}$ theory with general genus and an arbitrary number of punctures. Given such a theory, we also propose the inverse algorithm, thereby determining a parent orbi-instanton theory. This is achieved by uplifting the corresponding magnetic quiver step by step to the star-shaped, $E_6$-shaped, $E_7$-shaped, and $E_8$-shaped quivers, where at each step all of the underbalanced nodes, possessing non-zero FI parameters, are dualized. The latter $E_8$-shaped quiver then characterizes the 6d orbi-instanton theory from which the class $\mathcal{S}$ theory in question originates.
}

\allowdisplaybreaks[1]

\begin{document} 

\maketitle

\section{Introduction and Summary}
\label{sec:intro}

Theories of class $\cS$ \cite{Gaiotto:2009we, Gaiotto:2009gz}, arising from compactification of the 6d $\cN=(2,0)$ theory on Riemann surfaces with punctures, constitute one of the largest and most interesting families of 4d $\cN=2$ superconformal field theories (SCFTs).\footnote{See, e.g., \cite{Akhond:2021xio,Argyres:2022mnu} for recent reviews.} They have been thoroughly studied over many years since their discovery. Another class of SCFTs that plays an important role in this paper consists of the so-called orbi-instanton theories. These are 6d $\cN=(1,0)$ theories realized on the worldvolume of M5-branes probing the M9-brane \cite{Horava:1996ma} on $\CC^2/\ZZ_k$. Many aspects of these theories, including the F-theory descriptions at a generic point on the tensor branch, compactification to lower spacetime dimensions, and the Higgs branch, were studied in \cite{Heckman:2013pva, DelZotto:2014hpa, Heckman:2015bfa, Zafrir:2015rga, Ohmori:2015tka, Hayashi:2015zka, Mekareeya:2017jgc, Frey:2018vpw, Cabrera:2019izd, Fazzi:2022hal, Fazzi:2022yca}. In particular, it was pointed out in \cite{Mekareeya:2017jgc} that, upon compactifying a 6d orbi-instanton theory on a torus, the resulting 4d theory belongs to class $\cS$ of Type-$A$ on a sphere with punctures bearing certain structures. Mass deformations of such 4d theories were studied, for example, in \cite{Giacomelli:2022drw, Giacomelli:2024dbd}, where it was observed that some other class $\cS$ theories can be obtained in this way. This leads to a natural question whether it is possible to reach {\it any} Type-$A$ class $\cS$ theory by a sequence of mass deformations of the 4d theory arising from a 6d orbi-instanton theory compactified on a two-torus. 

The main point of this paper is to give a positive answer to the above question, and to provide a systematic approach to do so. Our strategy is to work with the 3d mirror theory \cite{Intriligator:1996ex} of the 6d orbi-instanton theory compactified on a three-torus. In the modern terminology, this mirror theory is also known as the magnetic quiver for the 6d theory in question (see \eg~ \cite{Cabrera:2019izd}). Since the corresponding 4d theory admits a class $\cS$ description with a particular structure of the punctures, the associated magnetic quiver can be described as a star-shaped quiver \cite{Benini:2010uu} (see also \cite{Chacaltana:2010ks}) whose body fits into the affine $E_8$ Dynkin diagram and with a tail attached. We refer to this as an $E_8$-shaped quiver, or simply $E_8$ quiver for brevity. In fact, as pointed out in \cite{Mekareeya:2017jgc}, such an $E_8$-shaped quiver contains all data that characterize the 6d orbi-instanton theory. Note that the terminology ``$E_n$-shaped quiver'' can also be applied to any affine $E_n$ Dynkin diagram. Under mirror symmetry, mass deformations in the 4d theory correspond to Fayet--Iliopoulos (FI) deformations in the corresponding magnetic quiver. The latter have been extensively studied in \cite{vanBeest:2021xyt, Bourget:2023uhe}, using the quiver subtraction technique \cite{Hanany:2018uhm, Cabrera:2018ann} to determine the resulting theory after deformations.
We also generalize some of these results in this paper. We demonstrate that turning on appropriate FI parameters at various nodes in the $E_8$-shaped quiver, the theory flows to the $E_7$-shaped quiver. This method can be repeated to the $E_7$-shaped quivers, where the end results are the $E_6$-shaped quivers. This procedure can be implemented again on the $E_6$-shaped quivers, where, upon renormalization group (RG) flow, we land on general star-shaped quivers. The latter are indeed the magnetic quivers for general Type-$A$ class $\cS$ theories on a sphere. Moreover, it can be shown that, starting from an appropriate star-shaped quiver and turning on the FI parameters in a certain fashion, one can obtain a mirror theory of a class $\cS$ theory on a Riemann surface with general genus and an arbitrary number of punctures. It is worth emphasizing that, from the perspective of the 4d theory, this deformation is rather distinct from the aforementioned ones in the sense that it corresponds not only to mass deformations, but also to tuning to a specific singular locus on the Coulomb branch. 

Yet another crucial point of this paper is to determine a parent theory given a descendant that is an arbitrary Type-$A$ class $\cS$ theory on a sphere. By these terms, we mean that the descendant is obtained by a series of mass deformations from the parent theory arising directly from a 6d orbi-instanton theory compactified on a two-torus. Of course, for a given descendant, there can possibly be many parent theories that connect to it by various mass deformations. Our goal here is to determine one of them in terms of the $E_8$-shaped quiver which characterizes the corresponding 6d orbi-instanton theory. To achieve this, we uplift an arbitrary star-shaped quiver to an $E_6$-shaped candidate parent theory described in the previous paragraph. Such a candidate parent theory may contain an underbalanced node, rendering the whole quiver bad in the sense of \cite{Gaiotto:2008ak}. Nevertheless, we demonstrate that the FI parameter which deforms such an $E_6$-shaped quiver back into the star-shaped quiver can always be turned on at the underbalanced node. This allows us to implement a sequence of dualities proposed by Yaakov \cite{Yaakov:2013fza} to make the $E_6$-shaped quiver good. This process can be repeated algorithmically in the following sense. We can uplift the good $E_6$-shaped quiver to an $E_7$-shaped quiver which may contain an underbalanced node. We can dualize the latter until we obtain a good $E_7$-shaped quiver, which can then be uplifted to an $E_8$-shaped quiver. The final step is then to dualize all the underbalanced nodes until we arrive at a good $E_8$-shaped quiver. This then characterizes the 6d orbi-instanton parent theory. We dub this procedure the {\it inverse algorithm}. We test it against a plethora of star-shaped quivers and show that the end result is always a good $E_8$-shaped quiver. We also provide a systematic method to determine the F-theory description at a generic point on the tensor branch of the corresponding 6d orbi-instanton theory. In many cases, the direct mass deformations between the parent theory and its descendant can be seen clearly without going through the FI deformations of the magnetic quivers. Finally, we point out that the mirror theory of a class $\cS$ theory on a Riemann surface with an arbitrary genus and an arbitrary number of punctures can always be uplifted to a {\it good} star-shaped quiver, where dualization is not needed.

\subsubsection*{Structure of the Paper}

The paper is organized as follows. In Section \ref{sec:ReviewE8OrbiInst}, we review the orbi-instanton theories, corresponding to 6d $\mathcal{N}=(1,0)$ theories realized on M5-branes probing an M9-brane on $\CC^2/\ZZ_k$. In particular, in Section \ref{sec:AllE8OrbiInst}, we demonstrate that all $E_8$-shaped quivers come from orbi-instanton theories. In Section \ref{sec:MassDef}, we demonstrate how to determine the resulting theory arising from FI deformations of an $E_n$-shaped quiver. Although this section is significantly based on the analysis in \cite{vanBeest:2021xyt, Bourget:2023uhe}, we study new cases of FI deformations in Section \ref{example3}. The main results of the paper can be divided into two main parts. The first part is contained in Section \ref{sec:MassDefE8Orbi}, where we discuss FI deformations of $E_{n+1}$-shaped quivers to obtain $E_n$-shaped quivers with $n=6,\,7$, and how FI-deformed $E_6$-shaped quivers lead to general star-shaped quivers. The second part begins at Section \ref{sec:InvFormulas}, where we discuss the inverse algorithm. We uplift a general star-shaped quiver to an $E_6$-shaped quiver, and $E_n$-shaped quivers to $E_{n+1}$-shaped quivers with $n=6,\,7$. When the uplift leads to a bad candidate parent theory, there is an FI parameter turned on at the underbalanced node, and we can perform the duality at that node which leads to a good quiver. Finally, in Section \ref{sec:HigherGenusTheories} we explain how to obtain, via a series of FI deformations, the mirror theory of any Type-$A$ class $\mathcal{S}$ theory on a Riemann surface of arbitrary genus and arbitrary number of punctures starting from an appropriate star-shaped quiver. We also point out that the former can be uplifted to a good star-shaped quiver that does not need to be dualized. We briefly review the dualization proposed by Yaakov \cite{Yaakov:2013fza} in Appendix \ref{sec:Itamarduality}. In Appendix \ref{sec:MixDef2tail}, we demonstrate step by step the inversion algorithm and dualization for uplifting a general $E_7$-shaped quiver to the $E_8$-shaped quiver.

\subsubsection*{Conventions and Notation} 

\bi
\item In the following, we mostly consider unitary quivers, and we follow the usual language for which \tikz{\node[gauge] {};} represents a gauge node and its label is the rank of the unitary gauge group, while \tikz{\node[flavor] {};} denotes a flavor node and its label denotes the number of flavors. The line connecting two nodes represents a hypermultiplet in the bifundamental representation of the two groups connected by the line. If the line corresponds to multiple hypermultiplets, we will specify it. The adjoint representation is obtained by a line starting and ending on the same node.
\item Whenever there is no ambiguity, we will represent the quiver by drawing only the ranks of the gauge groups, implying that all the nodes are gauge nodes and bifundamental hypermultiplets connect them, e.g.,
\begin{equation}
\begin{tikzpicture}[baseline=0]
\node (quiv) {$\begin{array}{cccccccc}
		       &   &   &   &   & L &   &   \\
	          A & B & C & D & E & F & G & H   \\
	\end{array}$};
\node (mean) [right=2cm of quiv] {
 \begin{tikzpicture}[baseline=0,font=\footnotesize]
		\node[gauge, label=below:{$A$}] (A) {};
		\node[gauge, label=below:{$B$}] (B) [right=6mm of A] {};
		\node[gauge, label=below:{$C$}] (C) [right=6mm of B] {};
		\node[gauge, label=below:{$D$}] (D) [right=6mm of C] {};
		\node[gauge, label=below:{$E$}] (E) [right=6mm of D] {};
		\node[gauge, label=below:{$F$}] (F) [right=6mm of E] {};
		\node[gauge, label=below:{$G$}] (G) [right=6mm of F] {};
		\node[gauge, label=below:{$H$}] (H) [right=6mm of G] {};
		\node[gauge, label=above:{$L$}] (L) [above=4mm of F] {};
        \draw[thick] (A) -- (B) -- (C) -- (D) -- (E) -- (F) -- (G) -- (H);
        \draw[thick] (F) -- (L);
	\end{tikzpicture}
};
\draw[thick,<->] (quiv) -- node[above,midway,sloped] {meaning} (mean);
\end{tikzpicture}
\end{equation}
\item Except for Section \ref{sec:MassDef}, we will use colors in the quivers to distinguish the bodies of the affine $E$-type diagrams from the tails attached to them. In particular, the $E_8$ affine quiver will be drawn in {\color{cE8}{blue}}, while the $E_7$ will be {\color{cE7}{red}} and the $E_6$ {\color{cE6}{orange}}. In the $E_8$ case, we allow one tail attached to the body of the $E_8$ affine Dynkin diagram, whereas in the $E_7$ and $E_6$ cases, we allow at most two and three tails attached to the body, respectively.
\item We adopt the notions of excess number, balance, underbalance, and overbalance as in \cite{Gaiotto:2008ak}.  For a $\U(r)$ gauge group with $N_f$ hypermultiplets transforming in the fundamental representation, the excess number for such a gauge group is defined as
\begin{equation}
\texttt{e}_{\U(r)} = N_f - 2r~.
\end{equation}
If $\texttt{e}_{\U(r)} =0$, the $\U(r)$ gauge group is said to be balanced; if $\texttt{e}_{\U(r)} >0$, the gauge group is said to be overbalanced; and if $\texttt{e}_{\U(r)} <0$, the gauge group is said to be underbalanced.
\item For the $T_\rho[\SU(N)]$ theory, when no partition $\rho$ is specified, i.e. $T[\SU(N)]$, then $\rho=\left[1^N\right]$, and we drop $[\SU(N)]$ whenever there is no ambiguity and simply refer to it as $T_\rho$.
\ei

\section{Review of Orbi-instanton Theories}
\label{sec:ReviewE8OrbiInst}

The starting point of our analysis is the 6d $\mathcal{N}=(1,0)$ theories realized on M5-branes, probing an M9-brane \cite{Horava:1996ma} on $\BC^2/\BZ_k$ singularity. Due to the fact that the M5-brane is a codimension-$4$ object from the perspective of the M9-brane, these theories are also known as \textit{orbi-instantons}. Such theories are characterized by the M5-brane charge and the asymptotic holonomy, which is a homomorphism $\ZZ_k\rightarrow E_8$. The latter is encoded in the following set of non-negative integers $n_i$, known as the {\it Kac labels} \cite[\S8.6]{Kac_1990}:
\begin{equation}
\vec n= 		\begin{array}{cccccccc}
			&  &  &  &  & n_{3'} &  &  \\
			n_1 & n_2 & n_3 & n_4 & n_5 & n_6 & n_{4'} & n_{2'} 
		\end{array}
		\label{eq:KacE8}
	\end{equation}
satisfying the condition 
\bes{\label{sumdini}
\sum_{i} d_i n_i  = n_1+2(n_2+n_{2'})+3(n_3+n_{3'})+4(n_4+n_{4'})+5n_5+6n_6 =k~,} 
where $\vec d = \{d_i\}$ is defined as the set of the Dynkin labels of $\mathfrak{e}_8$:
\begin{equation}
\vec d= 		\begin{array}{cccccccc}
			&  &  &  &  & {3} &  &  \\
			1 & 2 & 3 & 4 & 5 & 6 & {4} & {2} 
		\end{array}
		\label{eq:dynE8}
	\end{equation}

It was pointed out in \cite{Mekareeya:2017jgc} that the Higgs branch of a 6d orbi-instanton theory can be realized from the Coulomb branch of the following 3d $\cN=4$ quiver theory:
\begin{equation} \label{magorbi}
    \begin{tikzpicture}[baseline=7,font=\footnotesize]
		\node (dots) {$\cdots$};
        \node[gauge, label=below:{$2$}] (t2) [left=3mm of dots] {};
        \node[gauge, label=below:{$1$}] (t1) [left=6mm of t2] {};
        \node[gauge, label=below:{$k-1$}] (tkm1) [right=3mm of dots] {};	
		\node[gauge, label=below:{$k$}] (k) [right=12mm of dots] {};
		\node[gauge, bodyE8, label=below:{$N_1$}] (A) [right=6mm of k] {};
		\node[gauge, bodyE8, label=below:{$N_2$}] (B) [right=6mm of A] {};
		\node[gauge, bodyE8, label=below:{$N_3$}] (C) [right=6mm of B] {};
		\node[gauge, bodyE8, label=below:{$N_4$}] (D) [right=6mm of C] {};
		\node[gauge, bodyE8, label=below:{$N_5$}] (E) [right=6mm of D] {};
		\node[gauge, bodyE8, label=below:{$N_6$}] (F) [right=6mm of E] {};
		\node[gauge, bodyE8, label=below:{$N_{4'}$}] (G) [right=6mm of F] {};
		\node[gauge, bodyE8, label=below:{$N_{2'}$}] (H) [right=6mm of G] {};
		\node[gauge, bodyE8, label=above:{$N_{3'}$}] (L) [above=4mm of F] {};
        \draw[thick] (t1) -- (t2) -- (dots) -- (tkm1) -- (k) -- (A);
        \draw[thick,bodyE8] (A) -- (B) -- (C) -- (D) -- (E) -- (F) -- (G) -- (H);
        \draw[thick,bodyE8] (F) -- (L);
	\end{tikzpicture}
\end{equation}
In contemporary terminology, the theory in \eref{magorbi} is also known as the {\it magnetic quiver} of the 6d orbi-instanton theory. For a given $k$, the integers $N_i$ are related to the Kac labels $n_i$ by the following relation (see \cite[(4.8)]{Mekareeya:2017jgc}):
\bes{ \label{CtimesN}
C\cdot (N_1, N_2, \ldots, N_6, N_{4'}, N_{2'}, N_{3'})^T = (k-n_1, -n_2, \ldots, -n_6, -n_{4'}, -n_{2'}, -n_{3'})^T~,
}
where $C$ is the affine Cartan matrix for the $E_8$ algebra given by
\begin{equation} \label{affineCartanE8}
C = \left(
\begin{array}{ccccccccc}
 2 & -1 & 0 & 0 & 0 & 0 & 0 & 0 & 0 \\
 -1 & 2 & -1 & 0 & 0 & 0 & 0 & 0 & 0 \\
 0 & -1 & 2 & -1 & 0 & 0 & 0 & 0 & 0 \\
 0 & 0 & -1 & 2 & -1 & 0 & 0 & 0 & 0 \\
 0 & 0 & 0 & -1 & 2 & -1 & 0 & 0 & 0 \\
 0 & 0 & 0 & 0 & -1 & 2 & -1 & 0 & -1 \\
 0 & 0 & 0 & 0 & 0 & -1 & 2 & -1 & 0 \\
 0 & 0 & 0 & 0 & 0 & 0 & -1 & 2 & 0 \\
 0 & 0 & 0 & 0 & 0 & -1 & 0 & 0 & 2 \\
\end{array}
\right)~.
\end{equation}
Note that $C$ has a one-dimensional kernel spanned by $\vec d$, and so $C\cdot \vec d=\vec 0$. Moreover, from \eref{CtimesN} and \eref{affineCartanE8}, we see that
\bes{
n_1 &= k+N_2-2N_1~,\\
n_j &= (N_{j-1} + N_{j+1})-2N_j~, \quad j=2, \ldots, 5~, \\
n_6 &= N_{3'}+N_{4'}+N_5-2N_6~, \\
n_{4'} &= N_{2'}+N_6-2N_{4'}~, \\
n_{3'} &= N_6-2N_{3'}~.
}
This means that the Kac label $n_i$ has an interpretation as the {\it excess number} for the node $N_i$. 

Let $\fm \geq 1$ be the number of tensor multiplets in the 6d theory.\footnote{The integer $\fm$ here was called $N_6$ in \cite{Mekareeya:2017jgc} and $N$ throughout \cite[Section 5]{Mekareeya:2017jgc}. This is equal to the number of M5-branes in question.} 
As pointed out in \cite[Section 5]{Mekareeya:2017jgc}, this is related to $N_i$ in the following way: $\fm$ is the largest non-negative integer such that each integer
\bes{ \label{defri}
\fr_i=N_i -\fm d_i
}
is non-negative. Since $\vec d$ is in the kernel of $C$, it follows that $\fr_i$ also satisfies \eref{CtimesN}. It is worth pointing out that the node $k$ in \eref{magorbi} is guaranteed to be balanced or overbalanced: $N_1 +(k-1) \geq 2k$, \ie~ $N_1 \geq k+1$, if
\bes{\label{n1cond}
\fm \geq k+1~.
}
This is always achievable since the number of M5-branes can be chosen at will. Note that the other nodes on the black tail of \eref{magorbi} are always balanced.

As emphasized in \cite{Mekareeya:2017jgc}, the Higgs branch of an orbi-instanton theory is {\it not} the moduli space of $E_8$ instantons on $\BC^2/\BZ_k$.  In fact, the magnetic quiver \eref{magorbi} can be considered as the coupled system
\bes{ \label{diaggauge}
\begin{tikzpicture}[baseline=0,font=\footnotesize]
		\node (dots) {$\cdots$};
        \node[gauge, label=below:{$2$}] (t2) [left=3mm of dots] {};
        \node[gauge, label=below:{$1$}] (t1) [left=6mm of t2] {};
	\node[gauge, label=below:{$k-1$}] (tkm1) [right=3mm of dots] {};	
        \node[flavor, label=below:{$k$}] (k) [right=12mm of dots] {};
        \node (plus) [right=3mm of k] {$+$};
        \node[flavor, label=below:{$k$}] (k2) [right=3mm of plus] {};
		\node[gauge, bodyE8, label=below:{$N_1$}] (A) [right=6mm of k2] {};
		\node[gauge, bodyE8, label=below:{$N_2$}] (B) [right=6mm of A] {};
		\node[gauge, bodyE8, label=below:{$N_3$}] (C) [right=6mm of B] {};
		\node[gauge, bodyE8, label=below:{$N_4$}] (D) [right=6mm of C] {};
		\node[gauge, bodyE8, label=below:{$N_5$}] (E) [right=6mm of D] {};
		\node[gauge, bodyE8, label=below:{$N_6$}] (F) [right=6mm of E] {};
		\node[gauge, bodyE8, label=below:{$N_{4'}$}] (G) [right=6mm of F] {};
		\node[gauge, bodyE8, label=below:{$N_{2'}$}] (H) [right=6mm of G] {};
		\node[gauge, bodyE8, label=above:{$N_{3'}$}] (L) [above=4mm of F] {};
        \draw[thick] (t1) -- (t2) -- (dots) -- (tkm1) -- (k);
        \draw[thick] (k2) -- (A);
        \draw[thick,bodyE8] (A) -- (B) -- (C) -- (D) -- (E) -- (F) -- (G) -- (H);
        \draw[thick,bodyE8] (F) -- (L);
	\end{tikzpicture}
}
where $+$ means that we gauge the $\SU(k)/\BZ_k$ diagonal subgroup of the $\SU(k) \times \SU(k)$ flavor symmetry denoted by a square node in each quiver. The left quiver flows to the $T[\SU(k)]$ theory, whose Coulomb and Higgs branches are isomorphic to the nilpotent cone of $\SU(k)$. The Coulomb branch of the right quiver is the moduli space of $E_8$ instantons on $\BC^2/\BZ_k$.\footnote{The Higgs branch of the right quiver in \eref{diaggauge} describes the moduli space of $\SU(k)$ instantons on $\BC^2/\hat{E}_8$ singularity.}

Given the Kac labels $n_i$ and the number of tensor multiplets $\fm$, the 6d F-theory quiver description  at a generic point on the tensor branch takes the form \cite{Heckman:2013pva, DelZotto:2014hpa, Heckman:2015bfa, Zafrir:2015rga, Ohmori:2015tka, Hayashi:2015zka, Mekareeya:2017jgc}
\begin{equation}
	 \overset{G_1}{1} \, \,  \overset{\mathfrak{su}(m_2)}{2}  \, \, \overset{\mathfrak{su}(m_3)}{2} \, \, \cdots \, \,   \overset{\mathfrak{su}(m_\fm)}{2} \, \, [\mathfrak{su}(k)]
\end{equation}
where $G_1$ and $m_i$ can be determined following the algorithm in \cite[Section 3.2]{Mekareeya:2017jgc}, and we have to add hypermultiplets appropriately for the gauge anomaly cancellation. Alternatively, and more conveniently, this 6d quiver can be obtained by performing the $E_8$ quiver subtraction \cite{Hanany:2018uhm, Cabrera:2018ann} $\fm$ times from quiver \eref{magorbi}, and applying 3d mirror symmetry \cite{Intriligator:1996ex} or reading off the magnetic objects from the corresponding brane system as in \cite{Cabrera:2019izd}. This procedure will be described in \cref{sec:procedure} and elucidated via numerous examples in \cref{sec:rank4onetail,sec:rank6onetail,sec:rank4twotails}.

We may turn on a nilpotent higgsing associated with a partition $\rho=\left[s_1^{r_1}, s_2^{r_2}, \cdots, s_l^{r_l}\right]$ of $k$, with $s_1 > s_2 > \cdots > s_l \geq 1$ and $\sum_{i=1}^l s_i r_i =k$. Upon doing so, the $\su(k)$ flavor symmetry is higgsed to $\mathfrak{s}(\oplus_i \u(r_i))$; see \cite[(2.7)]{Chacaltana:2012zy}. As pointed out in \cite[(5.77)]{Mekareeya:2017jgc}, in terms of the magnetic quiver, this amounts to replacing the left black quiver in \eref{diaggauge} by the one associated with $T_\rho[\SU(k)]$, namely
\bes{
\begin{tikzpicture}[baseline=0,font=\footnotesize]
            \node[gauge, label=below:{$s_l$}] (t1) {};
            \node[gauge, label=below:{$2s_l$}] (t2) [right=5mm of t1] {};
		\node (dots) [right=5mm of t2] {$\cdots$};
            \node[gauge, label=below:{$r_l s_l$}] (r1s1) [right=5mm of dots] {};
            \node (dots1) [right=5mm of r1s1] {$\cdots$};
            \node (xx) [gauge, label=below:{$k-r_1 s_1$}] [right=5mm of dots1] {};
            \node (dots2) [right=5mm of xx] {$\cdots$};
            \node (yy) [gauge, label=below:{$k-s_1$}] [right=5mm of dots2] {};
		\node[flavor, label=below:{$k$}] (k) [right=7mm of yy] {};
  \draw[thick] (t1) -- (t2) -- (dots) -- (r1s1) -- (dots1) -- (xx) -- (dots2) -- (yy) -- (k);
\end{tikzpicture}        
}
Upon gluing this to the right blue quiver in \eref{diaggauge}, we obtain
\bes{ \label{magorbi1}
    \begin{tikzpicture}[baseline=0,font=\footnotesize]
		\node (dots) {$\cdots$};
        \node[gauge, label=below:{$2s_l$}] (t2) [left=3mm of dots] {};
        \node[gauge, label=below:{$s_l$}] (t1) [left=6mm of t2] {};
        \node[gauge, label=below:{$k-s_1$}] (tkm1) [right=3mm of dots] {};	
		\node[gauge, label=below:{$k$}] (k) [right=12mm of dots] {};
		\node[gauge, bodyE8, label=below:{$N_1$}] (A) [right=6mm of k] {};
		\node[gauge, bodyE8, label=below:{$N_2$}] (B) [right=6mm of A] {};
		\node[gauge, bodyE8, label=below:{$N_3$}] (C) [right=6mm of B] {};
		\node[gauge, bodyE8, label=below:{$N_4$}] (D) [right=6mm of C] {};
		\node[gauge, bodyE8, label=below:{$N_5$}] (E) [right=6mm of D] {};
		\node[gauge, bodyE8, label=below:{$N_6$}] (F) [right=6mm of E] {};
		\node[gauge, bodyE8, label=below:{$N_{4'}$}] (G) [right=6mm of F] {};
		\node[gauge, bodyE8, label=below:{$N_{2'}$}] (H) [right=6mm of G] {};
		\node[gauge, bodyE8, label=above:{$N_{3'}$}] (L) [above=4mm of F] {};
        \draw[thick] (t1) -- (t2) -- (dots) -- (tkm1) -- (k) -- (A);
        \draw[thick,bodyE8] (A) -- (B) -- (C) -- (D) -- (E) -- (F) -- (G) -- (H);
        \draw[thick,bodyE8] (F) -- (L);
	\end{tikzpicture}
}
In order for the node $k$ to be balanced or overbalanced, we must have $N_1+(k-s_1) \geq 2k$, \ie~ $N_1 \geq k+s_1$. Again, from \eref{defri}, this is guaranteed provided that the number of tensor multiplets in the 6d theory satisfies
\bes{
\fm \geq k+s_1~,
}
which is always possible to achieve since the number of M5-branes can be chosen as one wishes.

\subsection{All \texorpdfstring{$E_8$}{E8}-Shaped Quivers from Orbi-instanton Theories} 
\label{sec:AllE8OrbiInst}

Upon torus compactification, the 6d orbi-instanton theories we have just described become 4d SCFTs with eight supercharges, which are known to have a class $\mathcal{S}$ description. In class $\mathcal{S}$ terminology, they correspond to spheres with three (untwisted) punctures of type $A_{\fm}$. The three punctures are not arbitrary and, as can be easily deduced from \eqref{magorbi1}, always satisfy the following constraints \cite[(4.1)]{Mekareeya:2017jgc}:
\begin{itemize}
\item One puncture is labeled by a partition with two elements;
\item Another puncture is labeled by a partition with three elements; 
\item The third puncture is unconstrained.
\end{itemize}
The corresponding 3d mirror indeed coincides with the magnetic quiver of the underlying orbi-instanton model \cite{Benini:2010uu} and is therefore given by \eqref{magorbi1}. We can immediately notice that, in all such quivers, we can fit the graph of the affine $E_8$ Dynkin diagram, and this is the only affine Dynkin diagram which fits in the quiver. We therefore refer to all the 3d mirrors of class $\mathcal{S}$ trinions satisfying the constraints listed above as $E_8$ quivers. 

The purpose of this section is to show that there is a one-to-one correspondence between $E_8$ quivers and orbi-instanton theories or, equivalently, all class $\mathcal{S}$ trinions satisfying the above constraints arise via torus compactification of orbi-instanton theories. In the rest of the paper we will argue that with suitable mass deformations we can generate from these all other class $\mathcal{S}$ theories on the sphere. 

The argument establishing the one-to-one correspondence is actually rather simple. Let us consider a generic $E_8$ quiver with the same structure as in \eqref{magorbi1}:
\bes{ \label{magorbi2}
    \begin{tikzpicture}[baseline=0,font=\footnotesize]
		\node (dots) {$\cdots$};
        \node[gauge, label=below:{$2s'_{l'}$}] (t2) [left=3mm of dots] {};
        \node[gauge, label=below:{$s'_{l'}$}] (t1) [left=6mm of t2] {};
        \node[gauge, label=below:{$\kappa-s'_1$}] (tkm1) [right=3mm of dots] {};	
		\node[gauge, label=below:{$\kappa$}] (k) [right=12mm of dots] {};
		\node[gauge, bodyE8, label=below:{$R_1$}] (A) [right=6mm of k] {};
		\node[gauge, bodyE8, label=below:{$R_2$}] (B) [right=6mm of A] {};
		\node[gauge, bodyE8, label=below:{$R_3$}] (C) [right=6mm of B] {};
		\node[gauge, bodyE8, label=below:{$R_4$}] (D) [right=6mm of C] {};
		\node[gauge, bodyE8, label=below:{$R_5$}] (E) [right=6mm of D] {};
		\node[gauge, bodyE8, label=below:{$R_6$}] (F) [right=6mm of E] {};
		\node[gauge, bodyE8, label=below:{$R_{4'}$}] (G) [right=6mm of F] {};
		\node[gauge, bodyE8, label=below:{$R_{2'}$}] (H) [right=6mm of G] {};
		\node[gauge, bodyE8, label=above:{$R_{3'}$}] (L) [above=4mm of F] {};
        \draw[thick] (t1) -- (t2) -- (dots) -- (tkm1) -- (k) -- (A);
        \draw[thick,bodyE8] (A) -- (B) -- (C) -- (D) -- (E) -- (F) -- (G) -- (H);
        \draw[thick,bodyE8] (F) -- (L);
	\end{tikzpicture}
} 
where the rank $R_i$ of the various nodes is arbitrary, apart from the fact that we require all nodes to be balanced or overbalanced. The shape of the quiver implies that, as for all orbi-instanton theories, the excess numbers $\texttt{e}_i$ of the nodes inside the $E_8$ affine Dynkin diagrams, colored in blue in \eqref{magorbi2}, satisfy the relation
\be\label{sumdini2} \texttt{e}_1+2(\texttt{e}_2+\texttt{e}_{2'})+3(\texttt{e}_3+\texttt{e}_{3'})+4(\texttt{e}_4+\texttt{e}_{4'})+5\texttt{e}_5+\texttt{e}_6= \kappa\fstop\ee

The key point is that, as we have already explained, for orbi-instanton theories, the excess number for the nodes fitting inside the $E_8$ Dynkin diagram in the 3d magnetic quiver can be identified with the Kac labels parametrizing the $E_8$ holonomy. The idea then is that, given any $E_8$ quiver as in \eqref{magorbi2}, we consider the excess number $\texttt{e}_i$ of the various nodes in the Dynkin diagram. These will be identified with the Kac labels $n_i$ of the parent 6d orbi-instanton theory:
\bes{
\texttt{e}_i = n_i\fstop
}
Due to \eqref{sumdini} and \eqref{sumdini2}, we find that the orbifold order $k$ for the orbi-instanton theory and the rank of the node $\kappa$ on the left of $R_1$ in \eqref{magorbi2} necessarily coincide:
\bes{
\kappa = k~.
}

Once the Kac labels are specified, the only further choice we have for the orbi-instanton theory is the number of the tensor multiplets in the 6d theory. This freedom can be exploited to tune the rank of the node $N_1$ in \eqref{magorbi}. As we have already seen, we can get any value of $N_1$ greater than $k$ by properly tuning the number $\fm$ of the tensor multiplets; see around \eqref{n1cond}. In particular, we can choose the number of the tensor multiplets in such a way that the rank $N_1$ coincides with the value of $R_1$ in \eqref{magorbi2}: 
\bes{
R_1 = N_1~.
}
Modulo a nilpotent higgsing of the orbi-instanton theory, we can further assume that the tail on the left of node $k$ in \eqref{magorbi1} coincides with the corresponding tail in \eqref{magorbi2}. As a result, we see that, starting from an arbitrary $E_8$ quiver \eqref{magorbi2}, we can find an orbi-instanton theory whose magnetic quiver \eqref{magorbi1} has the same rank for the node $N_1$ and all nodes on its left:
\bes{
\left[{s'}_1^{{r'}_1}, {s'}_2^{{r'}_2}, \cdots, {s'}_{l'}^{{r'}_{l'}}\right] = \left[s_1^{r_1}, s_2^{r_2}, \cdots, s_l^{r_l}\right]~.  
}
Furthermore, the two quivers have the same excess numbers for all the remaining nodes belonging to the $E_8$ Dynkin diagram. All that is left to do now is to check that this implies the two quivers actually coincide, namely that all nodes on the right of $N_1$ have the same rank, \ie $R_i = N_i$ for $i=2,\ldots, 6, 4', 2', 3'$.

We can now conclude the argument taking inspiration from \eqref{CtimesN}. Focusing on the nodes of the $E_8$ Dynkin diagram on the right of $N_1$ in \eqref{magorbi2}, we can relate the ranks and excess numbers via the equation 
\be\label{eqrank1} M{\bf R}=R_1v_1-{\bf n}\coma\ee 
where we have collected ranks and excess numbers (which are equal to the Kac labels) in vectors ${\bf R}$ and ${\bf n}$ respectively: 
\be {\bf R}^T=(R_2,R_3,R_4,R_5,R_6,R_{4'},R_{2'},R_{3'})\,;\quad {\bf n}^T=(n_2,n_3,n_4,n_5,n_6,n_{4'},n_{2'},n_{3'})\;,\ee 
and we set \be v_1^T=(1,0,0,0,0,0,0,0)\;.\ee 
Notice that Eq. \eqref{eqrank1} holds both for the given $E_8$ quiver \eqref{magorbi2} and the magnetic quiver of the corresponding orbi-instanton theory \eqref{magorbi1}, with the same vector on the right-hand side and the same matrix $M$. If we define,
\bes{
{\vec N}^T = (N_2,N_3,N_4,N_5,N_6,N_{4'},N_{2'},N_{3'})~,
}
it follows that the difference $\vec R- \vec N$ between the ranks of the nodes of the two quivers satisfies the equation 
\be\label{eqrank2} M (\vec R- \vec N)=0\;.\ee 
At this point, it suffices to notice that the matrix $M$ in \eqref{eqrank1} and \eqref{eqrank2} coincides with the {\it finite} $E_8$ Cartan matrix obtained by removing the first row and first column of the matrix $C$ defined in \eref{affineCartanE8}. Unlike $C$, this matrix is actually invertible, and therefore we have
\bes{
\vec R = \vec N~,
}
implying that the two quivers \eref{magorbi} and \eref{magorbi2} coincide and therefore describe the same class $\mathcal{S}$ theory.

	\section{Review of Rules of Fayet--Iliopoulos and Mass Deformations}
 \label{sec:MassDef}
	
	In \cite{vanBeest:2021xyt}, the authors considered circle compactifications of the orbi-instanton theories, studying the resulting 5d SCFTs. In particular, they wanted to understand how the geometry of the 5d Higgs branch changes when moving around the extended Coulomb branch. At the level of the 5d SCFT, this corresponds to turning on mass deformations, while the theory of the magnetic quiver is deformed by Fayet--Iliopoulos terms. In this section, we are going to review how FI deformations of unitary quivers work at the level of 3d $\mathcal{N}=4$ theories and how they modify the magnetic quiver. The content of \cref{introremarks,example1,example2} is largely a review based on \cite{vanBeest:2021xyt} and \cite{Bourget:2023uhe}, while Section \ref{example3} contains new examples of FI deformations.
	
\subsection{General Remarks about FI Deformations}
\label{introremarks}

By turning on a (say complex) FI parameter $\xi$ at a $\U(N)$ gauge node in a quiver
\begin{equation}
\begin{tikzpicture}[baseline,font=\footnotesize]
    \node[gauge,label=below:{$N$}] (Nu) {};
    \node[flavor,label=below:{$k$}] (ku) [right=12mm of Nu] {};
    \draw[thick] (Nu) to[out=135, in=215, looseness=12] node[left,pos=0.5] {$\Phi$} (Nu);
    \draw[thick] (Nu) -- node[above,pos=0.5] {$\widetilde{Q}_i,Q_i$} (ku);
\end{tikzpicture}
\end{equation}
the superpotential becomes 
\begin{equation} 
\mathcal{W}= \widetilde{Q}_i\Phi Q^i + \xi\Tr\Phi\fstop
\end{equation}
Consequently, we have to solve the F-term and D-term equations: 
\begin{equation}
\begin{split}
Q^i\widetilde{Q}_i &= \xi I_N\coma\\ 
Q^iQ_i^{\dagger}-\widetilde{Q}^{\dagger i} \widetilde{Q}_i &=0\coma  
\end{split}
\end{equation}
 where the summation over flavor indices indeed includes a summation over all the bifundamental hypermultiplets charged under the $\U(N)$ gauge node. 

From the F-terms, we deduce that $\Tr(Q^i\widetilde{Q}_i)=N\xi$ and, combining this with the identity $\Tr(Q^i\widetilde{Q}_i)=\Tr(\widetilde{Q}_iQ^i)$, we conclude that, for any quiver with unitary gauge groups and bifundamental hypermultiplets only (which is the only case in which we will be interested), the FI parameters must satisfy the relation 
\be\label{constr} \sum_i N_i\xi^i=0\coma\ee
where $N_i$ denotes the rank of the $i$-th node. We therefore see that we always need to turn on FI parameters at two nodes at least. We will now discuss several examples of FI deformations which arise frequently in the study of mass deformations of class $\mathcal{S}$ theories, as we will see in the next sections.

\subsection{FI Deformations at Nodes of Equal Rank}
\label{example1}

The easiest case to analyze is that of FI parameters turned on at two abelian nodes. Due to \eqref{constr}, the two FI parameters are $\xi$ and $-\xi$. In this case, the deformation induces a nontrivial expectation value for a chain of bifundamentals connecting the two nodes. These identify a subquiver that starts and ends at the nodes at which we have turned on the FI parameter. As a result of the nontrivial expectation value for the bifundamentals, all the unitary gauge groups along the subquiver are spontaneously broken as $\U(n_i)\rightarrow \U(n_i-1)$, whereas the other nodes in the quiver are unaffected. Among all the broken $\U(1)$ factors, the diagonal combination survives and gives rise to a new $\U(1)$ node which is coupled to all the nodes of the quiver connected to those of the subquiver.  Overall, this is equivalent to subtracting from the original quiver an abelian quiver which has the same shape as the subquiver. 

The case of FI parameters turned on at nodes of the same rank (say $k$) is not harder to analyze. All the gauge groups along the subquiver are broken as $\U(n_i)\rightarrow \U(n_i-k)$ and, finally, we need to add a $\U(k)$ node associated with the unbroken diagonal factor.  This operation corresponds to a modified quiver subtraction, in which we rebalance with a $\U(k)$ node. Of course, the quiver we subtract has $\U(k)$ nodes only.

For our analysis, we can also use a more elaborate variant of the above deformation which involves three nodes. Say we turn on FI parameters at the nodes $\U(k)$, $\U(n)$ and $\U(n+k)$. The FI parameters satisfy the relation \eqref{constr} and we further impose the constraint $\xi_k=\xi_n$, so that we still have only one independent parameter. The equations of motion can be solved as follows. We set the VEV of all the bifundamentals in the subquiver connecting the nodes $\U(k)$ and $\U(n+k)$ to be 
\be\label{solab1} \langle B_i\rangle=\sqrt{\xi_k}\delta_{ab}\,; \quad \langle \widetilde{B}_i\rangle=\langle B_i\rangle^T\,;\quad \text{for $a,b\leq k$ and 0 otherwise}, \ee 
while the VEV of the bifundamentals in the subquiver connecting the nodes $\U(n+k)$ and $\U(n)$ to be 
\be\label{solab2} \langle B_i\rangle=\sqrt{\xi_k}\delta_{ab}\,; \quad \langle \widetilde{B}_i\rangle=\langle B_i\rangle^T\,;\quad \text{for $a,b\leq n$ and 0 otherwise}. \ee 
Here, we are assuming that the two subquivers meet at the $\U(n+k)$ node only. 

Overall, the higgsing of the theory can be described in terms of a sequence of two modified quiver subtractions. We first subtract a quiver of $\U(n)$ nodes going from node $\U(n)$ to node $\U(n+k)$ and, as in the previous case, we rebalance with a $\U(n)$ node. Then, we subtract from the resulting quiver a quiver of $\U(k)$ nodes going from node $\U(k)$ to node $\U(n+k)$ and we rebalance with a $\U(k)$ node. A careful analysis of the higgsing reveals that the $\U(n)$ node we introduced at the first step should not be rebalanced when we perform the second subtraction. 

Let us illustrate the procedure for $k=1$ and $n=2$ in the case of the $E_8$ quiver: 
\be\label{E8MQ}
\begin{tikzpicture}[baseline=7,font=\footnotesize]
\node[gauge, label=below:{$1$},fill=red] (A)  {};
\node[gauge, label=below:{$2$}] (B) [right=6mm of A] {};
\node[gauge, label=below:{$3$},fill=red] (C) [right=6mm of B] {};
\node[gauge, label=below:{$4$}] (D) [right=6mm of C] {};
\node[gauge, label=below:{$5$}] (E) [right=6mm of D] {};
\node[gauge, label=below:{$6$}] (F) [right=6mm of E] {};
\node[gauge, label=below:{$4$}] (G) [right=6mm of F] {};
\node[gauge, label=below:{$2$},fill=red] (H) [right=6mm of G] {};
\node[gauge, label=above:{$3$}] (I) [above=4mm of F] {};
\draw[thick] (A) -- (B) -- (C) -- (D) -- (E) -- (F) -- (G) -- (H);
\draw[thick] (F) -- (I);
\end{tikzpicture} 
\ee 
The quiver is depicted in \eqref{E8MQ}, where we turn on FI parameters at the nodes in {\color{red}{red}}. We first subtract a (non-affine) $A_6$ quiver with $\U(2)$ nodes such that the leftmost node is aligned with the {\color{red}{red}} node $3$ in \eqref{E8MQ} and the rightmost node is aligned with the {\color{red}{red}} node $2$ in \eqref{E8MQ}, getting 
\be\label{E8MQ2}
\begin{tikzpicture}[baseline=0]
\node (Q1) {
\begin{tikzpicture}[baseline=0,font=\footnotesize]
\node[gauge, label=below:{$1$},fill=red] (A)  {};
\node[gauge, label=below:{$2$},fill=green] (B) [right=6mm of A] {};
\node[gauge, label=below:{$3$},fill=red] (C) [right=6mm of B] {};
\node[gauge, label=below:{$4$}] (D) [right=6mm of C] {};
\node[gauge, label=below:{$5$}] (E) [right=6mm of D] {};
\node[gauge, label=below:{$6$}] (F) [right=6mm of E] {};
\node[gauge, label=below:{$4$}] (G) [right=6mm of F] {};
\node[gauge, label=below:{$2$},fill=red] (H) [right=6mm of G] {};
\node[gauge, label=above:{$3$},fill=green] (I) [above=4mm of F] {};
\draw[thick] (A) -- (B) -- (C) -- (D) -- (E) -- (F) -- (G) -- (H);
\draw[thick] (F) -- (I);
\node[gauge, label=below:{$2$}] (J) [below=12mm of C] {};
\node[gauge, label=below:{$2$}] (K) [right=6mm of J] {};
\node[gauge, label=below:{$2$}] (L) [right=6mm of K] {};
\node[gauge, label=below:{$2$}] (M) [right=6mm of L] {};
\node[gauge, label=below:{$2$}] (N) [right=6mm of M] {};
\node[gauge, label=below:{$2$}] (O) [right=6mm of N] {};
\draw[thick] (J) -- (K) -- (L) -- (M) -- (N) -- (O);
\node [left=4mm of J] {$-$};
\end{tikzpicture}};
\node (Q2) [right=12mm of Q1] {
\begin{tikzpicture}[baseline=0,font=\footnotesize]
\node[gauge, label=below:{$1$},fill=red] (A)  {};
\node[gauge, label=below:{$2$},fill=green] (B) [right=6mm of A] {};
\node[gauge, label=below:{$1$},fill=red] (C) [right=6mm of B] {};
\node[gauge, label=below:{$2$}] (D) [right=6mm of C] {};
\node[gauge, label=below:{$3$}] (E) [right=6mm of D] {};
\node[gauge, label=below:{$4$}] (F) [right=6mm of E] {};
\node[gauge, label=below:{$2$}] (G) [right=6mm of F] {};
\node[gauge, label=above:{$2$},fill=blue] (H) [above=4mm of B] {};
\node[gauge, label=above:{$3$},fill=green] (I) [above=4mm of F] {};
\draw[thick] (A) -- (B) -- (C) -- (D) -- (E) -- (F) -- (G);
\draw[thick] (F) -- (I);
\draw[thick] (H) -- (B);
\draw[thick] (H) -- (I); 
\end{tikzpicture}};
\draw[->,thick] (Q1) -- (Q2);
\end{tikzpicture} 
\ee 
The $\U(2)$ node in {\color{blue}{blue}} in \eqref{E8MQ2} arises after subtraction to rebalance the $\U(2)$ and $\U(3)$ nodes in {\green green}. Then, we subtract a (non-affine) $A_3$ abelian quiver such that the leftmost node is aligned with the left {\color{red}{red}} node $1$ in \eqref{E8MQ2} and the rightmost node is aligned with the right {\color{red}{red}} node $1$ in \eqref{E8MQ2}, obtaining 
\be\label{E8MQ3}
\begin{tikzpicture}[baseline=0]
\node (Q1) {
\begin{tikzpicture}[baseline=0,font=\footnotesize]
\node[gauge, label=below:{$1$},fill=red] (A)  {};
\node[gauge, label=below:{$2$}] (B) [right=6mm of A] {};
\node[gauge, label=below:{$1$},fill=red] (C) [right=6mm of B] {};
\node[gauge, label=below:{$2$},fill=green] (D) [right=6mm of C] {};
\node[gauge, label=below:{$3$}] (E) [right=6mm of D] {};
\node[gauge, label=below:{$4$}] (F) [right=6mm of E] {};
\node[gauge, label=below:{$2$}] (G) [right=6mm of F] {};
\node[gauge, label=above:{$2$}] (H) [above=4mm of B] {};
\node[gauge, label=above:{$3$}] (I) [above=4mm of F] {};
\draw[thick] (A) -- (B) -- (C) -- (D) -- (E) -- (F) -- (G);
\draw[thick] (F) -- (I);
\draw[thick] (H) -- (B);
\draw[thick] (H) -- (I); 
\node[gauge, label=below:{$1$}] (J) [below=12mm of A] {};
\node[gauge, label=below:{$1$}] (K) [right=6mm of J] {};
\node[gauge, label=below:{$1$}] (L) [right=6mm of K] {};
\draw[thick] (J) -- (K) -- (L);
\node [left=4mm of J] {$-$};
\end{tikzpicture}};
\node (Q2) [right=12mm of Q1] {
\begin{tikzpicture}[baseline=0,font=\footnotesize]
\node[gauge, label=below:{$1$},fill=blue] (A)  {};
\node[gauge, label=below:{$2$},fill=green] (B) [right=6mm of A] {};
\node[gauge, label=below:{$3$}] (C) [right=6mm of B] {};
\node[gauge, label=below:{$4$}] (D) [right=6mm of C] {};
\node[gauge, label=below:{$3$}] (E) [right=6mm of D] {};
\node[gauge, label=below:{$2$}] (F) [right=6mm of E] {};
\node[gauge, label=below:{$1$}] (G) [right=6mm of F] {};
\node[gauge, label=above:{$2$}] (H) [above=4mm of D] {};
\draw[thick] (A) -- (B) -- (C) -- (D) -- (E) -- (F) -- (G);
\draw[thick] (H) -- (D);
\end{tikzpicture}};
\draw[->,thick] (Q1) -- (Q2);
\end{tikzpicture}
\ee 
The $\U(1)$ node in {\color{blue}{blue}} is introduced again to rebalance and, as we have explained before, is connected only to the $\U(2)$ node in {\green green}, and not to the other $\U(2)$ nodes. In general, this FI deformation implements the mass deformation from the $E_8$ to the $E_7$ rank-1 theories.

\subsection{Turning on FI Deformations at Nodes of Rank \texorpdfstring{$n$}{n} and \texorpdfstring{$2n$}{2n}}
\label{example2}

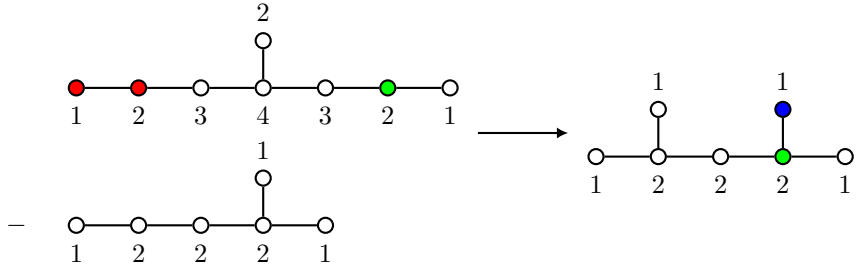
\begin{figure}[!htp]
\begin{center}
\begin{tikzpicture}[baseline=0]
\node (Q1) {
\begin{tikzpicture}[baseline=0,font=\footnotesize]
\node[gauge, label=below:{$1$},fill=red] (A)  {};
\node[gauge, label=below:{$2$},fill=red] (B) [right=6mm of A] {};
\node[gauge, label=below:{$3$}] (C) [right=6mm of B] {};
\node[gauge, label=below:{$4$}] (D) [right=6mm of C] {};
\node[gauge, label=below:{$3$}] (E) [right=6mm of D] {};
\node[gauge, label=below:{$2$},fill=green] (F) [right=6mm of E] {};
\node[gauge, label=below:{$1$}] (G) [right=6mm of F] {};
\node[gauge, label=above:{$2$}] (H) [above=4mm of D] {};
\draw[thick] (A) -- (B) -- (C) -- (D) -- (E) -- (F) -- (G);
\draw[thick] (H) -- (D);
\node[gauge, label=below:{$1$}] (I) [below=16mm of A] {};
\node[gauge, label=below:{$2$}] (J) [right=6mm of I] {};
\node[gauge, label=below:{$2$}] (K) [right=6mm of J] {};
\node[gauge, label=below:{$2$}] (L) [right=6mm of K] {};
\node[gauge, label=below:{$1$}] (M) [right=6mm of L] {};
\node[gauge, label=above:{$1$}] (N) [above=4mm of L] {};
\draw[thick] (I) -- (J) -- (K) -- (L) -- (M);
\draw[thick] (L) -- (N);
\node [left=4mm of I] {$-$};
\end{tikzpicture}};
\node (Q2) [right=12mm of Q1] {
\begin{tikzpicture}[baseline=0,font=\footnotesize]
\node[gauge, label=below:{$1$}] (A)  {};
\node[gauge, label=below:{$2$}] (B) [right=6mm of A] {};
\node[gauge, label=below:{$2$}] (C) [right=6mm of B] {};
\node[gauge, label=below:{$2$},fill=green] (D) [right=6mm of C] {};
\node[gauge, label=below:{$1$}] (E) [right=6mm of D] {};
\node[gauge, label=above:{$1$}] (F) [above=4mm of B] {};
\node[gauge, label=above:{$1$},fill=blue] (G) [above=4mm of D] {};
\draw[thick] (A) -- (B) -- (C) -- (D) -- (E);
\draw[thick] (F) -- (B);
\draw[thick] (G) -- (D);
\end{tikzpicture}};
\draw[->,thick] (Q1) -- (Q2);
\end{tikzpicture} 
\end{center}
\caption{The $\SO(12)$-preserving FI deformation of the rank-1 $E_7$ theory. We turn on FI parameters at the {\color{red}{red}} nodes, and we indicate in {\color{blue}{blue}} the $\U(1)$ node for rebalancing, attached to the $\U(2)$ node in {\green green}.}
\label{e7gaugedef2}
\end{figure}

Let us now consider the following case: the quiver contains a tail of the form $\U(1)-U(2)-\dots$ and we turn on FI parameters at the $\U(1)$ and $\U(2)$ nodes only (as before, the generalization to the case $\U(k)-U(2k)-\dots$ is obvious). Because of \eqref{constr}, we set $\xi_1=-2\xi_2=2\xi$. If we denote the $\U(1)\times \U(2)$ bifundamentals as $\widetilde{Q}$, $Q$ and the other $\U(2)$ fundamentals as $\widetilde{P}_i$, $P^i$, the relevant F-terms are 
\be \widetilde{Q}Q=2\xi\,;\quad \widetilde{P}_iP^i-Q\widetilde{Q}=\xi I_2\coma\ee 
which are solved by 
\be\label{u1u2}\widetilde{Q}=\sqrt{\xi}(1,1)\,;\; Q=\widetilde{Q}^T\,;\; \widetilde{P}=\sqrt{\xi}\left(\begin{array}{cccc}1 & 0 & 0 & \dots \\ 0 & 1 & 0 & \dots\\\end{array}\right)\,;\; P^T=\sqrt{\xi}\left(\begin{array}{cccc}0 & 1 & 0 & \dots \\ 1 & 0 & 0 & \dots\\\end{array}\right)\coma\ee 
where the index $i$ labels the columns of $\widetilde{P}$ and the rows of $P$. The VEV described above breaks spontaneously $\U(1)\times \U(2)$ to a diagonal $\U(1)$ subgroup. It is easy to check that D-terms are satisfied as well. The VEV for $\widetilde{P}_iP^i$ has the form $\left(\begin{array}{cc}0 & \xi \\ \xi & 0 \\\end{array}\right)$. This propagates along the quiver breaking all the groups as $\U(m)\rightarrow \U(m-2)$ until we find a junction, where we can ``decompose'' the VEV as 
\be \left(\begin{array}{cc}0 & \xi \\ \xi & 0 \\\end{array}\right)=\left(\begin{array}{cc}0 & \xi \\ 0 & 0 \\\end{array}\right)+\left(\begin{array}{cc}0 & 0 \\ \xi & 0 \\\end{array}\right)\fstop\ee 
Two nodes connected to the junction are higgsed as $\U(m)\rightarrow \U(m-1)$ and the VEV does not propagate any further. All the nodes connected to the subquiver of nodes which are (partially) higgsed are now coupled to a new $\U(1)$ node, which is left unbroken by the VEV. 
We give an example of this process in Figure \ref{e7gaugedef2}, which provides a way to understand, at the level of the 3d mirror theory, the flow from the $E_7$ theory to $\SU(2)$ SQCD with 6 flavors. As is clear from the picture, this is equivalent to subtracting a D-shaped quiver with three abelian nodes (while all the other nodes have rank 2). The number of nodes in the D-shaped quiver is dictated by the length of the tail in the original quiver.

\begin{figure}[!htp]
\begin{center}
\begin{tikzpicture}[baseline=0]
\node (Q1) {
\begin{tikzpicture}[baseline=0,font=\footnotesize]
\node[gauge, label=below:{$1$}] (A)  {};
\node[gauge, label=below:{$2$}] (B) [right=6mm of A] {};
\node[gauge, label=below:{$3$}] (C) [right=6mm of B] {};
\node[gauge, label=below:{$4$},fill=green] (D) [right=6mm of C] {};
\node[gauge, label=below:{$5$}] (E) [right=6mm of D] {};
\node[gauge, label=below:{$6$}] (F) [right=6mm of E] {};
\node[gauge, label=below:{$4$},fill=red] (G) [right=6mm of F] {};
\node[gauge, label=below:{$2$},fill=red] (H) [right=6mm of G] {};
\node[gauge, label=below:{$1$},fill=cyan] (I) [right=6mm of H] {};
\node[gauge, label=above:{$3$}] (J) [above=4mm of F] {};
\draw[thick] (A) -- (B) -- (C) -- (D) -- (E) -- (F) -- (G) -- (H) -- (I);
\draw[thick] (F) -- (J);
\node[gauge, label=below:{$2$}] (K) [below=16mm of E] {};
\node[gauge, label=below:{$4$}] (L) [right=6mm of K] {};
\node[gauge, label=below:{$4$}] (M) [right=6mm of L] {};
\node[gauge, label=below:{$2$}] (N) [right=6mm of M] {};
\node[gauge, label=above:{$2$}] (O) [above=4mm of L] {};
\draw[thick] (K) -- (L) -- (M) -- (N) ;
\draw[thick] (L) -- (O);
\node [left=4mm of K] {$-$};
\end{tikzpicture}};
\node (Q2) [right=12mm of Q1] {
\begin{tikzpicture}[baseline=0,font=\footnotesize]
\node[gauge, label=below:{$1$}] (A)  {};
\node[gauge, label=below:{$2$}] (B) [right=6mm of A] {};
\node[gauge, label=below:{$3$}] (C) [right=6mm of B] {};
\node[gauge, label=below:{$4$},fill=green] (D) [right=6mm of C] {};
\node[gauge, label=below:{$3$}] (E) [right=6mm of D] {};
\node[gauge, label=below:{$2$}] (F) [right=6mm of E] {};
\node[gauge, label=below:{$1$}] (G) [right=6mm of F] {};
\node[gauge, label=left:{$2$},fill=blue] (H) [above=4mm of D] {};
\node[gauge, label=left:{$1$},fill=cyan] (I) [above=4mm of H] {};
\draw[thick] (A) -- (B) -- (C) -- (D) -- (E) -- (F) -- (G);
\draw[thick] (D) -- (H) -- (I);
\end{tikzpicture}};
\draw[->,thick] (Q1) -- (Q2);
\end{tikzpicture} 
\end{center}
\caption{The $\SU(8)\times \SU(2)$ preserving mass deformation of the rank-2 $\SO(16)\times \SU(2)$ theory. We turn on FI parameters at {\color{red}{red}} nodes. As a result of the higgsing, a new {\color{blue}{blue}} $\U(2)$ node appears for rebalancing the $\U(4)$ node in {\green green}. This $\U(2)$ node then connects back with the $\U(1)$ node in {\color{cyan}{cyan}}.}\label{su3massdef}
\end{figure}
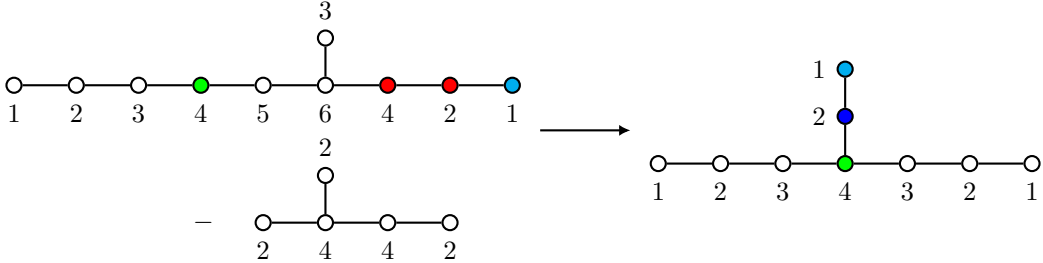

Let us now consider a more complicated example involving a class $\mathcal{S}$ trinion. The SCFT has global symmetry $\SO(16)\times \SU(2)$ and its Coulomb branch is two-dimensional. We wish to turn on a mass deformation which breaks the symmetry to $\SU(8)\times \SU(2)$, leading to another trinion theory. In order to preserve an $A_7\times A_1$ balanced subquiver, we turn on FI parameters at $\U(2)$ and $\U(4)$ nodes as in Figure \ref{su3massdef}. 

\noindent The F-terms are solved as explained above. The only difference is that we have to tensor all the matrices in \eqref{u1u2} by the $2\times 2$ identity matrix. The central node is higgsed to $\U(2)$ and the neighboring nodes are higgsed as $\U(n)\rightarrow \U(n-2)$. We further need to add a $\U(2)$ node attached to the unhiggsed $\U(1)$ and $\U(4)$ nodes.

\subsection{FI Deformations at Several Nodes of Different Rank}\label{example3}

By combining the techniques discussed in the previous sections, we can analyze more complicated deformations involving several nodes in the quiver. In this section, we will discuss in detail the case, most relevant for our analysis, in which we turn on the FI deformation at a balanced node in the quiver and at all nodes connected to it (which may or may not be balanced, it does not matter for our argument). In order to solve Eq. \eqref{constr}, we will set the value of the FI parameter at the balanced node to $-2\lambda$ and to $\lambda$ at all nodes connected to it. The balancing condition indeed guarantees that \eqref{constr} is automatically solved. We will focus for simplicity on star-shaped quivers with three tails, which is the only relevant case for us. 

Let us start by discussing the case in which the balanced node belongs to one of the tails. By construction, we have only two nodes connected to it and, therefore, the relevant part of the quiver is of the form 
\be\label{quiverpart}
\begin{tikzpicture}[baseline=0,font=\footnotesize]
\node (dots) {$\cdots$};
\node[gauge, label=below:{$A$},label=above:{\purple\scriptsize$\lambda$}] (A) [right=3mm of dots] {};
\node[gauge, label=below:{$B$},label=above:{\purple\scriptsize$-2\lambda$}] (B) [right=6mm of A] {};
\node[gauge, label=below:{$C$},label=above:{\purple\scriptsize$\lambda$}] (C) [right=6mm of B] {};
\node (dots2) [right=3mm of C] {$\cdots$};
\draw[thick] (dots) -- (A) -- (B) -- (C) -- (dots2);
\end{tikzpicture}
\ee
where we have added above each node the value of the corresponding FI parameter and below the rank of the node. Indeed, the balancing condition tells us that $2B=A+C$. We can assume without loss of generality that $A<C$, which implies that the node $C$ is closer to the central node.

In this situation, the solution of the equations of motion is obtained by combining the solutions discussed in the previous two examples. As in Section \ref{example1}, we have VEVs for the bifundamentals between nodes $A$, $B$ and $B$, $C$ which higgs the three gauge groups reducing their rank by $A$. At the level of the quiver, this is implemented by subtracting a linear quiver with three nodes all of rank $A$ and rebalancing with a $\U(A)$ node. As a result of the subtraction, the node $A$ disappears, the node $B$ is higgsed to $\frac{C-A}{2}$, while node $C$ is higgsed down to $C-A$, which is twice the new rank of the $B$ node. 

At this stage, the solution to the equations of motion can be found by combining the above with a further VEV for the bifundamental between $B$ and $C$ which is analogous to the one described in Section \ref{example2}. The VEV propagates all the way to the central node and the effect of the higgsing can be implemented by subtracting a D-shaped quiver: all nodes of the D-shaped quiver have rank $C-A$ except three, whose rank is $\frac{C-A}{2}$. We finally rebalance with a $\U(\frac{C-A}{2})$ node. Notice that the $\U(A)$ rebalancing node we have introduced earlier should not be rebalanced at this stage. Overall, we can handle this case by combining the two basic moves we have described earlier.  

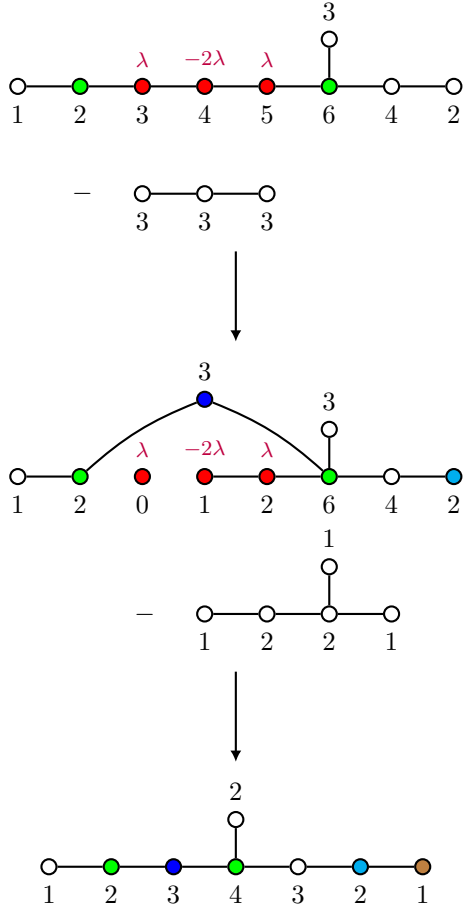
\begin{figure}
    \centering
    \begin{tikzpicture}[baseline=-4cm]
\node (Q1) {
\begin{tikzpicture}[baseline=0,font=\footnotesize]
\node[gauge, label=below:{$1$}] (A)  {};
\node[gauge, label=below:{$2$},fill=green] (B) [right=6mm of A] {};
\node[gauge, label=below:{$3$},fill=red,label=above:{\purple\scriptsize$\lambda$}] (C) [right=6mm of B] {};
\node[gauge, label=below:{$4$},fill=red,label=above:{\purple\scriptsize$-2\lambda$}] (D) [right=6mm of C] {};
\node[gauge, label=below:{$5$},fill=red,label=above:{\purple\scriptsize$\lambda$}] (E) [right=6mm of D] {};
\node[gauge, label=below:{$6$},fill=green] (F) [right=6mm of E] {};
\node[gauge, label=below:{$4$}] (G) [right=6mm of F] {};
\node[gauge, label=below:{$2$}] (H) [right=6mm of G] {};
\node[gauge, label=above:{$3$}] (I) [above=4mm of F] {};
\draw[thick] (A) -- (B) -- (C) -- (D) -- (E) -- (F) -- (G) -- (H);
\draw[thick] (F) -- (I);
\node[gauge, label=below:{$3$}] (J) [below=12mm of C] {};
\node[gauge, label=below:{$3$}] (K) [right=6mm of J] {};
\node[gauge, label=below:{$3$}] (M) [right=6mm of K] {};
\draw[thick] (J) -- (K) -- (M);
\node [left=4mm of J] {$-$};
\end{tikzpicture}
};
\node (Q2) [below=12mm of Q1] {
\begin{tikzpicture}[baseline=0,font=\footnotesize]
\node[gauge, label=below:{$1$}] (A)  {};
\node[gauge, label=below:{$2$},fill=green] (B) [right=6mm of A] {};
\node[gauge, label=below:{$0$},fill=red,label=above:{\purple\scriptsize$\lambda$}] (C) [right=6mm of B] {};
\node[gauge, label=below:{$1$},fill=red,label=above:{\purple\scriptsize$-2\lambda$}] (D) [right=6mm of C] {};
\node[gauge, label=below:{$2$},fill=red,label=above:{\purple\scriptsize$\lambda$}] (E) [right=6mm of D] {};
\node[gauge, label=below:{$6$},fill=green] (F) [right=6mm of E] {};
\node[gauge, label=below:{$4$}] (G) [right=6mm of F] {};
\node[gauge, label=below:{$2$},fill=cyan] (H) [right=6mm of G] {};
\node[gauge, label=above:{$3$}] (I) [above=4mm of F] {};
\node[gauge, label=above:{$3$},fill=blue] (J) [above=8mm of D] {};
\draw[thick] (A) -- (B);
\draw[thick] (D) -- (E) -- (F) -- (G) -- (H);
\draw[thick] (F) -- (I);
\draw[thick] (B) to[bend left=10] (J);
\draw[thick] (J) to[bend left=10] (F);
\node[gauge, label=below:{$1$}] (K) [below=16mm of D] {};
\node[gauge, label=below:{$2$}] (L) [right=6mm of K] {};
\node[gauge, label=below:{$2$}] (M) [right=6mm of L] {};
\node[gauge, label=below:{$1$}] (N) [right=6mm of M] {};
\node[gauge, label=above:{$1$}] (O) [above=4mm of M] {};
\draw[thick] (K) -- (L) -- (M) -- (N);
\draw[thick] (M) -- (O);
\node [left=4mm of K] {$-$};
\end{tikzpicture}
};
\node (Q3) [below=12mm of Q2] {
\begin{tikzpicture}[baseline=0,font=\footnotesize]
\node[gauge, label=below:{$1$}] (A)  {};
\node[gauge, label=below:{$2$},fill=green] (B) [right=6mm of A] {};
\node[gauge, label=below:{$3$},fill=blue] (C) [right=6mm of B] {};
\node[gauge, label=below:{$4$},fill=green] (D) [right=6mm of C] {};
\node[gauge, label=below:{$3$}] (E) [right=6mm of D] {};
\node[gauge, label=below:{$2$},fill=cyan] (F) [right=6mm of E] {};
\node[gauge, label=below:{$1$},fill=brown] (G) [right=6mm of F] {};
\node[gauge, label=above:{$2$}] (H) [above=4mm of D] {};
\draw[thick] (A) -- (B) -- (C) -- (D) -- (E) -- (F) -- (G);
\draw[thick] (H) -- (D);
\end{tikzpicture}};
\draw[->,thick] (Q1) -- (Q2);
\draw[->,thick] (Q2) -- (Q3);
\end{tikzpicture} 
    \caption{Example of FI deformations at several nodes of different rank.}
    \label{fig:E8Example2}
\end{figure}

Let us illustrate the procedure we have just described with a concrete example in Figure \ref{fig:E8Example2}, taking again the $E_8$ quiver as a starting point. 
We select as node $B$ in \eqref{quiverpart} the node $\U(4)$ in the long tail. In the case at hand, in Figure \ref{fig:E8Example2}, we have $A=3$ and $C=5$. In the first subtraction, the $\U(3)$ node in {\blue blue} arises to rebalance the $\U(2)$ and the $\U(6)$ nodes in {\green green} to which it connects. In the second subtraction, this $\U(3)$ node in {\blue blue} is left untouched, and the $\U(1)$ node in {\brown brown} arises to rebalance the $\U(2)$ node in {\color{cyan}{cyan}} to which it connects.
As we can see, the result is again the $E_7$ quiver.  

Let us conclude this section by discussing the case in which the balanced node coincides with the central node in the quiver. In this case, the balanced node is connected to three nodes and the relevant part of the quiver is 
\be\label{quiverpart2}
\begin{tikzpicture}[baseline=7,font=\footnotesize]
\node (dots) {$\cdots$};
\node[gauge, label=below:{$A$}] (A) [right=3mm of dots] {};
\node[gauge, label=below:{$B$}] (B) [right=6mm of A] {};
\node[gauge, label=below:{$C$}] (C) [right=6mm of B] {};
\node[gauge, label=left:{$D$}] (D) [above=4mm of B] {};
\node (dots2) [right=3mm of C] {$\cdots$};
\node (dots3) [above=2mm of D] {$\vdots$};
\draw[thick] (dots) -- (A) -- (B) -- (C) -- (dots2);
\draw[thick] (B) -- (D) -- (dots3);
\end{tikzpicture}
\ee
Indeed, in \eqref{quiverpart2}, the central node $B$ is the one with the largest rank, and due to its balancing, we have the relation $D=2B-A-C$. As we have explained before, the FI parameter at the central node is equal to $-2\lambda$, while the other three nodes in \eqref{quiverpart2} have the FI parameter equal to $\lambda$. 

It turns out that, in this case, we can solve the equations of motion only by using the basic move discussed in Section \ref{example1}, more precisely we need to perform consecutively three quiver subtractions. It is perhaps easier in this case to provide the VEV for the three mesons in the adjoint of $\U(B)$ built out of the bifundamental fields which solve the equations of motion. We denote them as $M_{AB}$, $M_{BC}$ and $M_{DB}$. The VEV for the elementary fields can then be easily extracted. 
We present the VEV for the mesons (which are $B\times B$ matrices) in block form: 
\be\label{blockform}
\begin{array}{r} B-C \,\{\\ \\ B-A \,\{\\\end{array}
\left(\begin{array}{c|c|c}
\hspace*{10pt} &\hspace*{10pt}&\hspace*{10pt} \\
\hline
 & & \\
\hline
 & & \\
\end{array}\right) \begin{array}{l}\\ \}\, A+C-B\\ \\\end{array}
\ee
The three diagonal blocks in \eqref{blockform} are square matrices of size $B-C$, $A+C-B$ and $B-A$ respectively. Each of the three mesons has two diagonal blocks proportional to the identity, while all other blocks are zero. Explicitly 
\be\label{solFI}
M_{AB}= \left(\begin{array}{c|c|c}
I &&\\
\hline
& I &\\
\hline
\hspace*{10pt}&\hspace*{10pt}& \hspace*{10pt}\\
\end{array}\right);\quad M_{BC}= \left(\begin{array}{c|c|c}
\hspace*{10pt} &\hspace*{10pt}&\hspace*{10pt}\\
\hline
& I &\\
\hline
&& I\\
\end{array}\right);\quad M_{DB}= \left(\begin{array}{c|c|c}
I &&\\
\hline
\hspace*{10pt}& \hspace*{10pt} &\hspace*{10pt}\\
\hline
&& I\\
\end{array}\right)\; .
\ee  
Notice that the rank of $M_{DB}$ is $2B-A-C$, which is equal to $D$ due to the balancing of the central node, as we have explained. 

The higgsing induced by the VEVs \eqref{blockform} can be described as a sequence of three quiver subtractions, where each time we subtract a linear quiver with three nodes of the same rank (say $n$) and, as explained in Section \ref{example1}, we rebalance with a node of rank $n$. The first subtraction involves the nodes $A$, $B$ and $C$ and we set $n=A+C-B$. The second subtraction involves the nodes $D$, $B$ and $C$ and we have $n=B-A$. Finally, the third subtraction involves the nodes $A$, $B$ and $D$ and we set $n=B-C$. As in the previous cases, we should not rebalance the balancing nodes introduced in previous steps. 

\begin{figure}[!htp]
    \centering
    \begin{tikzpicture}[baseline=-4cm]
\node (Q1) {
\begin{tikzpicture}[baseline=0,font=\footnotesize]
\node[gauge, label=below:{$1$}] (00) {};
\node[gauge, label=below:{$2$}] (0) [right=6mm of 00] {};
\node[gauge, label=below:{$3$}] (A) [right=6mm of 0] {};
\node[gauge, label=below:{$4$}] (B) [right=6mm of A] {};
\node[gauge, label=below:{$5$}] (C) [right=6mm of B] {};
\node[gauge, label=below:{$6$},fill=green] (D) [right=6mm of C] {};
\node[gauge, label=below:{$7$},fill=red,label=above:{\purple\scriptsize$\lambda$}] (E) [right=6mm of D] {};
\node[gauge, label=below:{$8$},fill=red,label={[label distance=-1.5mm]60:{\purple\scriptsize$-2\lambda$}}] (F) [right=6mm of E] {};
\node[gauge, label=below:{$5$},fill=red,label=above:{\purple\scriptsize$\lambda$}] (G) [right=6mm of F] {};
\node[gauge, label=below:{$2$},fill=green] (H) [right=6mm of G] {};
\node[gauge, label=right:{$4$},fill=red,label=above:{\purple\scriptsize$\lambda$}] (I) [above=6mm of F] {};
\draw[thick] (00) -- (0) -- (A) -- (B) -- (C) -- (D) -- (E) -- (F) -- (G) -- (H);
\draw[thick] (F) -- (I);
\node[gauge, label=below:{$4$}] (J) [below=12mm of E] {};
\node[gauge, label=below:{$4$}] (K) [right=6mm of J] {};
\node[gauge, label=below:{$4$}] (M) [right=6mm of K] {};
\draw[thick] (J) -- (K) -- (M);
\node [left=4mm of J] {$-$};
\end{tikzpicture}
};
\node (Q2) [below=10mm of Q1] {
\begin{tikzpicture}[baseline=0,font=\footnotesize]
\node[gauge, label=below:{$1$}] (00) {};
\node[gauge, label=below:{$2$}] (0) [right=6mm of 00] {};
\node[gauge, label=below:{$3$}] (A) [right=6mm of 0] {};
\node[gauge, label=below:{$4$}] (B) [right=6mm of A] {};
\node[gauge, label=below:{$5$}] (C) [right=6mm of B] {};
\node[gauge, label=below:{$6$},fill=green] (D) [right=6mm of C] {};
\node[gauge, label=below:{$3$},fill=red,label=above:{\purple\scriptsize$\lambda$}] (E) [right=6mm of D] {};
\node[gauge, label=below:{$4$},fill=red,label={[label distance=-1.5mm]60:{\purple\scriptsize$-2\lambda$}}] (F) [right=6mm of E] {};
\node[gauge, label=below:{$1$},fill=red,label=above:{\purple\scriptsize$\lambda$}] (G) [right=6mm of F] {};
\node[gauge, label=below:{$2$},fill=green] (H) [right=6mm of G] {};
\node[gauge, label=right:{$4$},fill=red,label=above:{\purple\scriptsize$\lambda$}] (I) [above=6mm of F] {};
\node[gauge, label=above:{$4$},fill=blue] (J) [above=10mm of E] {};
\draw[thick] (00) -- (0) -- (A) -- (B) -- (C) -- (D) -- (E) -- (F) -- (G) -- (H);
\draw[thick] (F) -- (I);
\draw[thick] (D) to[bend left=20] (J);
\draw[thick] (I) -- (J);
\draw[thick] (J) to[bend left=50] (H);
\node[gauge, label=below:{$1$}] (K) [below=16mm of F] {};
\node[gauge, label=below:{$1$}] (N) [right=6mm of K] {};
\node[gauge, label=above:{$1$}] (O) [above=4mm of K] {};
\draw[thick] (K) -- (N);
\draw[thick] (K) -- (O);
\node [left=4mm of K] {$-$};
\end{tikzpicture}
};
\node (Q3) [below=10mm of Q2] {
\begin{tikzpicture}[baseline=0,font=\footnotesize]
\node[gauge, label=below:{$1$}] (00) {};
\node[gauge, label=below:{$2$}] (0) [right=6mm of 00] {};
\node[gauge, label=below:{$3$}] (A) [right=6mm of 0] {};
\node[gauge, label=below:{$4$}] (B) [right=6mm of A] {};
\node[gauge, label=below:{$5$}] (C) [right=6mm of B] {};
\node[gauge, label=below:{$6$},fill=green] (D) [right=6mm of C] {};
\node[gauge, label=below:{$3$},fill=red,label=above:{\purple\scriptsize$\lambda$}] (E) [right=6mm of D] {};
\node[gauge, label=below:{$3$},fill=red,label={[label distance=-1.5mm]60:{\purple\scriptsize$-2\lambda$}}] (F) [right=6mm of E] {};
\node[gauge, label=below:{$0$},fill=red,label=above:{\purple\scriptsize$\lambda$}] (G) [right=6mm of F] {};
\node[gauge, label=below:{$2$},fill=green] (H) [right=6mm of G] {};
\node[gauge, label=right:{$3$},fill=red,label=above:{\purple\scriptsize$\lambda$}] (I) [above=6mm of F] {};
\node[gauge, label=above:{$4$},fill=blue] (J) [above=10mm of E] {};
\node[gauge, label=below:{$1$},fill=cyan] (L) [below=8mm of G] {};
\draw[thick] (00) -- (0) -- (A) -- (B) -- (C) -- (D) -- (E) -- (F);
\draw[thick] (F) -- (I);
\draw[thick] (D) to[bend left=20] (J);
\draw[thick] (I) -- (J);
\draw[thick] (J) to[bend left=50] (H);
\draw[thick] (L) to[bend left=5] (E);
\draw[thick] (H) -- (L);
\node[gauge, label=below:{$3$}] (K) [below=18mm of E] {};
\node[gauge, label=below:{$3$}] (N) [right=6mm of K] {};
\node[gauge, label=left:{$3$}] (O) [above=4mm of N] {};
\draw[thick] (K) -- (N) -- (O);
\node [left=4mm of K] {$-$};
\end{tikzpicture}
};
\node (Q4) [below=10mm of Q3] {
\begin{tikzpicture}[baseline=0,font=\footnotesize]
\node[gauge, label=below:{$1$}] (00) {};
\node[gauge, label=below:{$2$}] (0) [right=6mm of 00] {};
\node[gauge, label=below:{$3$}] (A) [right=6mm of 0] {};
\node[gauge, label=below:{$4$}] (B) [right=6mm of A] {};
\node[gauge, label=below:{$5$}] (C) [right=6mm of B] {};
\node[gauge, label=below:{$6$},fill=green] (D) [right=6mm of C] {};
\node[gauge, label=below:{$4$},fill=blue] (E) [right=6mm of D] {};
\node[gauge, label=below:{$2$},fill=green] (F) [right=6mm of E] {};
\node[gauge, label=below:{$1$},fill=cyan] (G) [right=6mm of F] {};
\node[gauge, label=above:{$3$},fill=brown] (H) [above=4mm of D] {};
\draw[thick] (00) -- (0) -- (A) -- (B) -- (C) -- (D) -- (E) -- (F) -- (G);
\draw[thick] (H) -- (D);
\end{tikzpicture}};
\draw[->,thick] (Q1) -- (Q2);
\draw[->,thick] (Q2) -- (Q3);
\draw[->,thick] (Q3) -- (Q4);
\end{tikzpicture} 
    \caption{Example of FI deformation involving a balanced central node in a star-shaped quiver.}
    \label{fig:CentralNodeExample}
\end{figure}
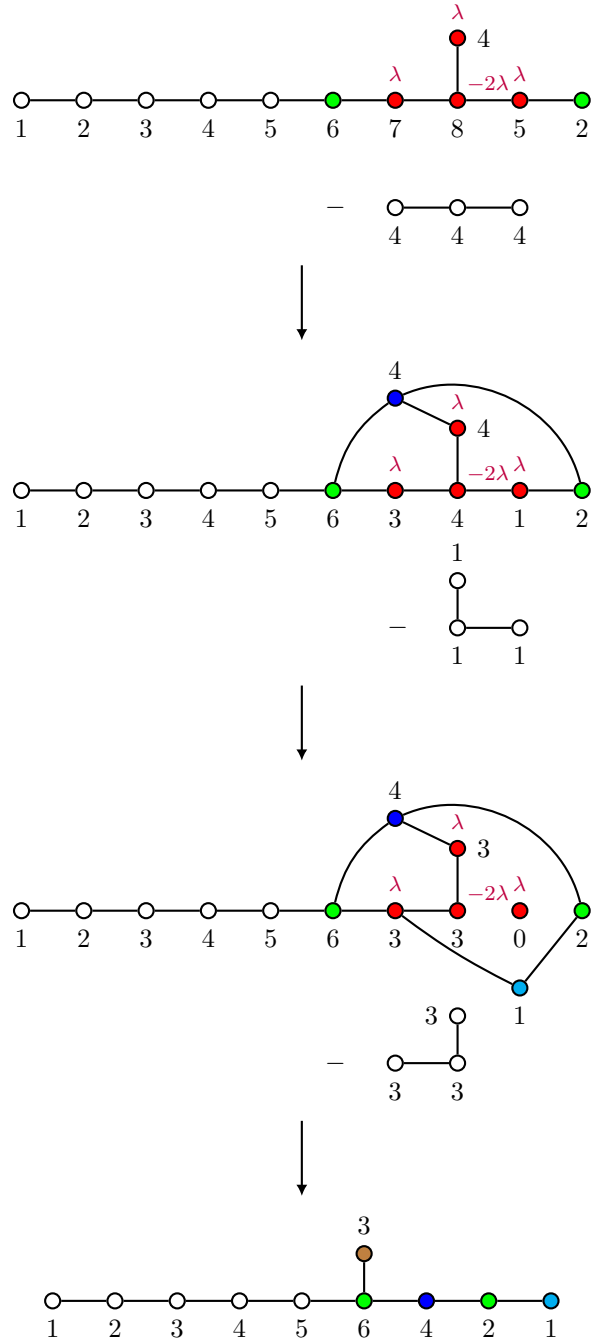

Let us illustrate the above procedure in the case of the quiver 
\bes{ \label{quiv1ex}
\begin{tikzpicture}[baseline=0,font=\footnotesize]
\node[gauge, label=below:{$1$}] (00)  {};
\node[gauge, label=below:{$2$}] (0) [right=6mm of 00] {};
\node[gauge, label=below:{$3$}] (A) [right=6mm of 0] {};
\node[gauge, label=below:{$4$}] (B) [right=6mm of A] {};
\node[gauge, label=below:{$5$}] (C) [right=6mm of B] {};
\node[gauge, label=below:{$6$}] (D) [right=6mm of C] {};
\node[gauge, label=below:{$7$}] (E) [right=6mm of D] {};
\node[gauge, label=below:{$8$}] (F) [right=6mm of E] {};
\node[gauge, label=below:{$5$}] (G) [right=6mm of F] {};
\node[gauge, label=below:{$2$}] (H) [right=6mm of G] {};
\node[gauge, label=above:{$4$}] (I) [above=4mm of F] {};
\draw[thick] (00) -- (0) -- (A);
\draw[thick] (A) -- (B) -- (C) -- (D) -- (E) -- (F) -- (G) -- (H) (F) -- (I);
\end{tikzpicture}
} 
Comparing \eqref{quiv1ex} and \eqref{quiverpart2} we see that $A+C-B=4$, $B-A=1$ and $B-C=3$. The sequence of three subtractions is then implemented as in Figure \ref{fig:CentralNodeExample}. In the first subtraction, the $\U(4)$ node in {\blue blue} arises to rebalance the $\U(6)$ and the $\U(2)$, that we depict in {\green green}, to which it connects. The second subtraction leaves the {\green green} $\U(6)$ untouched, but a {\color{cyan}{cyan}} $\U(1)$ arises to rebalance the {\red red} $\U(3)$ and the {\green green} $\U(2)$. Finally, in the last subtraction, we add a {\color{brown}{brown}} $\U(3)$ to rebalance the $\U(6)$. The last quiver is drawn so that the $E_7$-shaped quiver is easily recognizable, and the colors help in visualizing which node is which.

\section{Mass Deformations of Orbi-instanton Theories}
\label{sec:MassDefE8Orbi}

In this section, we will start exploring RG flows between class $\mathcal{S}$ theories triggered by mass deformations. Our main tool is the analysis of FI deformations for the corresponding magnetic quivers, which are reviewed in Section \ref{sec:MassDef}. We start from $E_8$ trinions which correspond to orbi-instanton theories reduced to 4d, as we have explained before, and discuss in detail various types of deformations which lead to other families of star-shaped quivers. As we will see, we can (often in several ways) deform every $E_8$ quiver to an $E_7$ quiver (in which we can fit the graph of the $E_7$ affine Dynkin diagram); this can then be deformed to an $E_6$ quiver which in turn can lead to more general star-shaped quivers with more than three tails. Our goal is not to provide a full catalog of possible RG flows, but rather to describe explicitly the families of deformations we will need in Section \ref{sec:InvFormulas} to argue that every star-shaped quiver can be realized as the deformation of an $E_8$ quiver. 
 
\subsection{Deformations from \texorpdfstring{$E_8$}{E8} Quivers To \texorpdfstring{$E_7$}{E7} Quivers}
\label{sec:MassDefE8E7}
In this section, we explain how to obtain via FI deformations $E_7$-shaped quivers from $E_8$-shaped quivers that are magnetic quivers for orbi-instantons.

\subsubsection{Deformations Involving \texorpdfstring{$E_8$}{E8} Nodes}
\label{sec:MassDef-Body}
Let us start by considering RG flows which do not break (the subgroup of) the $\SU(k)$ global symmetry of the parent 6d theory. This corresponds to activating the FI deformation only at nodes fitting inside the $E_8$ affine Dynkin diagram. We start by noticing that every time a node is balanced, i.e. the corresponding Kac label vanishes, it is possible to turn on an FI parameter at that node and the surrounding ones, leading to an $E_7$ quiver. 

Let us consider a generic $E_8$ magnetic quiver for an orbi-instanton theory, i.e. 
\begin{equation}\label{eq:generalE8quiv}
	\begin{tikzpicture}[baseline=7,font=\footnotesize]
		\node (dots) {$\cdots$};
		\node[gauge, label=below:{$k$}] (k) [right=6mm of dots] {};
		\node[gauge, bodyE8, label=below:{$A$}] (A) [right=6mm of k] {};
		\node[gauge, bodyE8, label=below:{$B$}] (B) [right=6mm of A] {};
		\node[gauge, bodyE8, label=below:{$C$}] (C) [right=6mm of B] {};
		\node[gauge, bodyE8, label=below:{$D$}] (D) [right=6mm of C] {};
		\node[gauge, bodyE8, label=below:{$E$}] (E) [right=6mm of D] {};
		\node[gauge, bodyE8, label=below:{$F$}] (F) [right=6mm of E] {};
		\node[gauge, bodyE8, label=below:{$G$}] (G) [right=6mm of F] {};
		\node[gauge, bodyE8, label=below:{$H$}] (H) [right=6mm of G] {};
		\node[gauge, bodyE8, label=above:{$L$}] (L) [above=4mm of F] {};
		\draw[thick] (dots.east) -- (k.west);
		\draw[thick] (k.east) -- (A.west);
		\draw[thick,bodyE8] (A.east) -- (B.west);
		\draw[thick,bodyE8] (B.east) -- (C.west);
		\draw[thick,bodyE8] (C.east) -- (D.west);
		\draw[thick,bodyE8] (D.east) -- (E.west);
		\draw[thick,bodyE8] (E.east) -- (F.west);
		\draw[thick,bodyE8] (F.east) -- (G.west);
		\draw[thick,bodyE8] (G.east) -- (H.west);
		\draw[thick,bodyE8] (F.north) -- (L.south);
	\end{tikzpicture}
\end{equation}
where we have {\color{cE8}{colored}} the part of the quiver associated to the $E_8$ affine Dynkin diagram to distinguish it from the tail, and, for example, we assume that $n_6$ is balanced, i.e. $n_6 = 0$. This imposes that 
\begin{equation}\label{eq:n6rankconstraint}
	2F = E+G+L\fstop
\end{equation}
We can now turn FI deformations at node $F$ and the nodes surrounding it, i.e. $E$, $G$ and $L$, remembering that their ranks are constrained by \eqref{eq:n6rankconstraint}. The resulting quiver can be obtained by implementing the procedure described at the end of Section \ref{example3} and reads
\begin{equation}\label{eq:E7quiverFromn6=0}
	\begin{tikzpicture}[baseline=7,font=\footnotesize]
		\node (dots) {$\cdots$};
		\node[gauge, label=below:{$k$}] (k) [right=6mm of dots] {};
		\node[gauge, bodyE7, label=below:{$A$}] (A) [right=6mm of k] {};
		\node[gauge, bodyE7, label=below:{$B$}] (B) [right=6mm of A] {};
		\node[gauge, bodyE7, label=below:{$C$}] (C) [right=6mm of B] {};
		\node[gauge, bodyE7, label=below:{$D$}] (D) [right=6mm of C] {};
		\node[gauge, bodyE7, label=above:{$G+E-F$}] (E) [right=9mm of D] {};
		\node[gauge, bodyE7, label=below:{$H$}] (F) [right=9mm of E] {};
		\node[gauge, bodyE7, label=above:{$F-E$}] (G) [right=6mm of F] {};
		\node[gauge, bodyE7, label=above:{$F-G$}] (FG) [above=4mm of D] {};
		\draw[thick] (dots.east) -- (k.west);
		\draw[thick] (k.east) -- (A.west);
		\draw[thick,bodyE7] (A.east) -- (B.west);
		\draw[thick,bodyE7] (B.east) -- (C.west);
		\draw[thick,bodyE7] (C.east) -- (D.west);
		\draw[thick,bodyE7] (D.east) -- (E.west);
		\draw[thick,bodyE7] (E.east) -- (F.west);
		\draw[thick,bodyE7] (F.east) -- (G.west);
		\draw[thick,bodyE7] (D.north) -- (FG.south);
	\end{tikzpicture}
\end{equation}
which, as we have made clear by the choice of {\color{cE7}{colors}}, is a $E_7$ quiver, with one tail. The tail is simply inherited from the parent quiver.

This is not the only possible FI deformation that one can turn on in \eqref{eq:generalE8quiv}. Provided that at least one of the Kac labels is zero, it is always possible to activate FI deformations involving only nodes inside the $E_8$ affine Dynkin diagram, which lead to an $E_7$ quiver with one tail. In this case, the tail is identical to that of the parent $E_8$ quiver. The deformation can be implemented simply following the rules discussed in Section \ref{sec:MassDef}. The resulting quivers are listed in Table \ref{tab:E8toE7-singletail-body}. 

Possible FI deformations involving nodes in the tail are explained in Section \ref{sec:E8toE7-TailDeform}, while some of the deformations that lead to $E_7$ quivers with two tails are discussed in Section \ref{sec:E8toE7-MixDef2tail} and are listed in Table \ref{tab:E8toE7-2tail-body}. 

\begin{table}[!htp]
\begin{tabular}{c|c|c}
	Balanced Node & Constraint & $E_7$ Quiver\\
	\hline
	$n_{6} = 0$ & $2F = E+G+L$ & 
	\begin{tikzpicture}[baseline=0,font=\footnotesize]
		\node (dots) {$\cdots$};
		\node[gauge, label=below:{$k$}] (k) [right=6mm of dots] {};
		\node[gauge, bodyE7, label=below:{$A$}] (A) [right=6mm of k] {};
		\node[gauge, bodyE7, label=below:{$B$}] (B) [right=6mm of A] {};
		\node[gauge, bodyE7, label=below:{$C$}] (C) [right=6mm of B] {};
		\node[gauge, bodyE7, label=below:{$D$}] (D) [right=6mm of C] {};
		\node[gauge, bodyE7, label=above:{$G+E-F$}] (E) [right=9mm of D] {};
		\node[gauge, bodyE7, label=below:{$H$}] (F) [right=9mm of E] {};
		\node[gauge, bodyE7, label=above:{$F-E$}] (G) [right=6mm of F] {};
		\node[gauge, bodyE7, label=above:{$F-G$}] (FG) [above=4mm of D] {};
		\draw[thick] (dots.east) -- (k.west);
		\draw[thick] (k.east) -- (A.west);
		\draw[thick,bodyE7] (A.east) -- (B.west);
		\draw[thick,bodyE7] (B.east) -- (C.west);
		\draw[thick,bodyE7] (C.east) -- (D.west);
		\draw[thick,bodyE7] (D.east) -- (E.west);
		\draw[thick,bodyE7] (E.east) -- (F.west);
		\draw[thick,bodyE7] (F.east) -- (G.west);
		\draw[thick,bodyE7] (D.north) -- (FG.south);
	\end{tikzpicture} \\ \hline
	$n_{3'} = 0$ & $2L=F$ & 
	\begin{tikzpicture}[baseline=0,font=\footnotesize]
		\node (dots) {$\cdots$};
		\node[gauge, label=below:{$k$}] (k) [right=6mm of dots] {};
		\node[gauge, bodyE7, label=below:{$A$}] (A) [right=6mm of k] {};
		\node[gauge, bodyE7, label=below:{$B$}] (B) [right=6mm of A] {};
		\node[gauge, bodyE7, label=below:{$C$}] (C) [right=6mm of B] {};
		\node[gauge, bodyE7, label=below:{$D$}] (D) [right=6mm of C] {};
		\node[gauge, bodyE7, label=below:{$L$}] (E) [right=6mm of D] {};
		\node[gauge, bodyE7, label=below:{$H$}] (F) [right=6mm of E] {};
		\node[gauge, bodyE7, label=above:{$G-L$}] (G) [right=6mm of F] {};
		\node[gauge, bodyE7, label=above:{$E-L$}] (FG) [above=4mm of D] {};
		\draw[thick] (dots.east) -- (k.west);
		\draw[thick] (k.east) -- (A.west);
		\draw[thick,bodyE7] (A.east) -- (B.west);
		\draw[thick,bodyE7] (B.east) -- (C.west);
		\draw[thick,bodyE7] (C.east) -- (D.west);
		\draw[thick,bodyE7] (D.east) -- (E.west);
		\draw[thick,bodyE7] (E.east) -- (F.west);
		\draw[thick,bodyE7] (F.east) -- (G.west);
		\draw[thick,bodyE7] (D.north) -- (FG.south);
	\end{tikzpicture} \\ \hline
	$n_{2'} = 0$ & $2H=G$ & 
	\begin{tikzpicture}[baseline=0,font=\footnotesize]
		\node (dots) {$\cdots$};
		\node[gauge, label=below:{$k$}] (k) [right=6mm of dots] {};
		\node[gauge, bodyE7, label=below:{$A$}] (A) [right=6mm of k] {};
		\node[gauge, bodyE7, label=below:{$B$}] (B) [right=6mm of A] {};
		\node[gauge, bodyE7, label=below:{$C$}] (C) [right=6mm of B] {};
		\node[gauge, bodyE7, label=below:{$D$}] (D) [right=6mm of C] {};
		\node[gauge, bodyE7, label=above:{$E-H$}] (E) [right=9mm of D] {};
		\node[gauge, bodyE7, label=below:{$F-2H$}] (F) [right=9mm of E] {};
		\node[gauge, bodyE7, label=above:{$L-H$}] (G) [right=9mm of F] {};
		\node[gauge, bodyE7, label=above:{$H$}] (FG) [above=4mm of D] {};
		\draw[thick] (dots.east) -- (k.west);
		\draw[thick] (k.east) -- (A.west);
		\draw[thick,bodyE7] (A.east) -- (B.west);
		\draw[thick,bodyE7] (B.east) -- (C.west);
		\draw[thick,bodyE7] (C.east) -- (D.west);
		\draw[thick,bodyE7] (D.east) -- (E.west);
		\draw[thick,bodyE7] (E.east) -- (F.west);
		\draw[thick,bodyE7] (F.east) -- (G.west);
		\draw[thick,bodyE7] (D.north) -- (FG.south);
	\end{tikzpicture} \\ \hline
	$n_{1} = 0$ & $2A=k+B$ & 
	\begin{tikzpicture}[baseline=0,font=\footnotesize]
		\node (dots) {$\cdots$};
		\node[gauge, label=below:{$k$}] (k) [right=6mm of dots] {};
		\node[gauge, bodyE7, label=above:{$C+2k-2A$}] (A) [right=9mm of k] {};
		\node[gauge, bodyE7, label=below:{$D+2k-2A$}] (B) [right=9mm of A] {};
		\node[gauge, bodyE7, label=above:{$E+2k-2A$}] (C) [right=9mm of B] {};
		\node[gauge, bodyE7, label=below:{$F+2k-2A$}] (D) [right=9mm of C] {};
		\node[gauge, bodyE7, label=above:{$G+k-A$}] (E) [right=9mm of D] {};
		\node[gauge, bodyE7, label=below:{$H$}] (F) [right=9mm of E] {};
		\node[gauge, bodyE7, label=above:{$A-k$}] (G) [right=9mm of F] {};
		\node[gauge, bodyE7, label=above:{$L+k-A$}] (H) [above=4mm of D] {};
		\draw[thick] (dots.east) -- (k.west);
		\draw[thick] (k.east) -- (A.west);
		\draw[thick,bodyE7] (A.east) -- (B.west);
		\draw[thick,bodyE7] (B.east) -- (C.west);
		\draw[thick,bodyE7] (C.east) -- (D.west);
		\draw[thick,bodyE7] (D.east) -- (E.west);
		\draw[thick,bodyE7] (E.east) -- (F.west);
		\draw[thick,bodyE7] (F.east) -- (G.west);
		\draw[thick,bodyE7] (D.north) -- (H.south);
	\end{tikzpicture} \\ \hline
	$n_{2} = 0$ & $2B=A+C$ & 
	\begin{tikzpicture}[baseline=0,font=\footnotesize]
		\node (dots) {$\cdots$};
		\node[gauge, label=below:{$k$}] (k) [right=6mm of dots] {};
		\node[gauge, bodyE7, label=above:{$A$}] (A) [right=6mm of k] {};
		\node[gauge, bodyE7, label=below:{$D+2A-2B$}] (B) [right=9mm of A] {};
		\node[gauge, bodyE7, label=above:{$E+2A-2B$}] (C) [right=9mm of B] {};
		\node[gauge, bodyE7, label=below:{$F+2A-2B$}] (D) [right=9mm of C] {};
		\node[gauge, bodyE7, label=above:{$G+A-B$}] (E) [right=9mm of D] {};
		\node[gauge, bodyE7, label=below:{$H$}] (F) [right=9mm of E] {};
		\node[gauge, bodyE7, label=above:{$B-A$}] (G) [right=9mm of F] {};
		\node[gauge, bodyE7, label=above:{$L+A-B$}] (H) [above=4mm of D] {};
		\draw[thick] (dots.east) -- (k.west);
		\draw[thick] (k.east) -- (A.west);
		\draw[thick,bodyE7] (A.east) -- (B.west);
		\draw[thick,bodyE7] (B.east) -- (C.west);
		\draw[thick,bodyE7] (C.east) -- (D.west);
		\draw[thick,bodyE7] (D.east) -- (E.west);
		\draw[thick,bodyE7] (E.east) -- (F.west);
		\draw[thick,bodyE7] (F.east) -- (G.west);
		\draw[thick,bodyE7] (D.north) -- (H.south);
	\end{tikzpicture} \\ \hline
	$n_{3} = 0$ & $2C=B+D$ & 
		\begin{tikzpicture}[baseline=0,font=\footnotesize]
		\node (dots) {$\cdots$};
		\node[gauge, label=below:{$k$}] (k) [right=6mm of dots] {};
		\node[gauge, bodyE7, label=above:{$A$}] (A) [right=6mm of k] {};
		\node[gauge, bodyE7, label=below:{$B$}] (B) [right=6mm of A] {};
		\node[gauge, bodyE7, label=above:{$E+2B-2C$}] (C) [right=9mm of B] {};
		\node[gauge, bodyE7, label=below:{$F+2B-2C$}] (D) [right=9mm of C] {};
		\node[gauge, bodyE7, label=above:{$G+B-C$}] (E) [right=9mm of D] {};
		\node[gauge, bodyE7, label=below:{$H$}] (F) [right=9mm of E] {};
		\node[gauge, bodyE7, label=above:{$C-B$}] (G) [right=9mm of F] {};
		\node[gauge, bodyE7, label=above:{$L+B-C$}] (H) [above=4mm of D] {};
		\draw[thick] (dots.east) -- (k.west);
		\draw[thick] (k.east) -- (A.west);
		\draw[thick,bodyE7] (A.east) -- (B.west);
		\draw[thick,bodyE7] (B.east) -- (C.west);
		\draw[thick,bodyE7] (C.east) -- (D.west);
		\draw[thick,bodyE7] (D.east) -- (E.west);
		\draw[thick,bodyE7] (E.east) -- (F.west);
		\draw[thick,bodyE7] (F.east) -- (G.west);
		\draw[thick,bodyE7] (D.north) -- (H.south);
	\end{tikzpicture} \\ \hline
	$n_{4} = 0$ & $2D=C+E$& 
	\begin{tikzpicture}[baseline=0,font=\footnotesize]
		\node (dots) {$\cdots$};
		\node[gauge, label=below:{$k$}] (k) [right=6mm of dots] {};
		\node[gauge, bodyE7, label=below:{$A$}] (A) [right=6mm of k] {};
		\node[gauge, bodyE7, label=below:{$B$}] (B) [right=6mm of A] {};
		\node[gauge, bodyE7, label=above:{$C$}] (C) [right=6mm of B] {};
		\node[gauge, bodyE7, label=below:{$F+2C-2D$}] (D) [right=9mm of C] {};
		\node[gauge, bodyE7, label=above:{$G+C-D$}] (E) [right=9mm of D] {};
		\node[gauge, bodyE7, label=below:{$H$}] (F) [right=9mm of E] {};
		\node[gauge, bodyE7, label=above:{$D-C$}] (G) [right=9mm of F] {};
		\node[gauge, bodyE7, label=above:{$L+C-D$}] (H) [above=4mm of D] {};
		\draw[thick] (dots.east) -- (k.west);
		\draw[thick] (k.east) -- (A.west);
		\draw[thick,bodyE7] (A.east) -- (B.west);
		\draw[thick,bodyE7] (B.east) -- (C.west);
		\draw[thick,bodyE7] (C.east) -- (D.west);
		\draw[thick,bodyE7] (D.east) -- (E.west);
		\draw[thick,bodyE7] (E.east) -- (F.west);
		\draw[thick,bodyE7] (F.east) -- (G.west);
		\draw[thick,bodyE7] (D.north) -- (H.south);
	\end{tikzpicture} \\ \hline
	$n_{5} = 0$ & $2E=D+F$& 
	\begin{tikzpicture}[baseline=0,font=\footnotesize]
		\node (dots) {$\cdots$};
		\node[gauge, label=below:{$k$}] (k) [right=6mm of dots] {};
		\node[gauge, bodyE7, label=below:{$A$}] (A) [right=6mm of k] {};
		\node[gauge, bodyE7, label=below:{$B$}] (B) [right=6mm of A] {};
		\node[gauge, bodyE7, label=below:{$C$}] (C) [right=6mm of B] {};
		\node[gauge, bodyE7, label=below:{$D$}] (D) [right=6mm of C] {};
		\node[gauge, bodyE7, label=above:{$G+D-E$}] (E) [right=9mm of D] {};
		\node[gauge, bodyE7, label=below:{$H$}] (F) [right=9mm of E] {};
		\node[gauge, bodyE7, label=above:{$E-D$}] (G) [right=9mm of F] {};
		\node[gauge, bodyE7, label=above:{$L+D-E$}] (H) [above=4mm of D] {};
		\draw[thick] (dots.east) -- (k.west);
		\draw[thick] (k.east) -- (A.west);
		\draw[thick,bodyE7] (A.east) -- (B.west);
		\draw[thick,bodyE7] (B.east) -- (C.west);
		\draw[thick,bodyE7] (C.east) -- (D.west);
		\draw[thick,bodyE7] (D.east) -- (E.west);
		\draw[thick,bodyE7] (E.east) -- (F.west);
		\draw[thick,bodyE7] (F.east) -- (G.west);
		\draw[thick,bodyE7] (D.north) -- (H.south);
	\end{tikzpicture} \\ \hline
	$n_{4'} = 0$ & $2G=F+H$& 
	\begin{tikzpicture}[baseline=0,font=\footnotesize]
		\node (dots) {$\cdots$};
		\node[gauge, label=below:{$k$}] (k) [right=6mm of dots] {};
		\node[gauge, bodyE7, label=below:{$A$}] (A) [right=6mm of k] {};
		\node[gauge, bodyE7, label=below:{$B$}] (B) [right=6mm of A] {};
		\node[gauge, bodyE7, label=below:{$C$}] (C) [right=6mm of B] {};
		\node[gauge, bodyE7, label=below:{$D$}] (D) [right=6mm of C] {};
		\node[gauge, bodyE7, label=above:{$E+H-G$}] (E) [right=9mm of D] {};
		\node[gauge, bodyE7, label=below:{$H$}] (F) [right=9mm of E] {};
		\node[gauge, bodyE7, label=above:{$L+H-G$}] (G) [right=9mm of F] {};
		\node[gauge, bodyE7, label=above:{$G-H$}] (H) [above=4mm of D] {};
		\draw[thick] (dots.east) -- (k.west);
		\draw[thick] (k.east) -- (A.west);
		\draw[thick,bodyE7] (A.east) -- (B.west);
		\draw[thick,bodyE7] (B.east) -- (C.west);
		\draw[thick,bodyE7] (C.east) -- (D.west);
		\draw[thick,bodyE7] (D.east) -- (E.west);
		\draw[thick,bodyE7] (E.east) -- (F.west);
		\draw[thick,bodyE7] (F.east) -- (G.west);
		\draw[thick,bodyE7] (D.north) -- (H.south);
	\end{tikzpicture} 
\end{tabular}
\caption{Some examples of FI deformations involving the balanced nodes of \eqref{eq:generalE8quiv}. We list the balanced node, the consequent constraint on the ranks of the balanced node and the surrounding ones and the corresponding resulting $E_7$-shaped quiver with one tail, assuming of turning on FI deformations at the balanced node and all the surrounding nodes.}
\label{tab:E8toE7-singletail-body}
\end{table}

\subsubsection{Deformations Involving the Tail}
\label{sec:E8toE7-TailDeform}
Hitherto, we have considered the deformations corresponding to turning on an FI parameter at the node inside the $E_8$ affine Dynkin diagram with the zero Kac label. In this section, we consider deformations corresponding to turning on the FI parameter at a node on the quiver tail. 
 This procedure can be applied regardless of the values of the Kac labels. A particularly interesting case is when none of the Kac labels in the $E_8$ affine Dynkin diagram are zero. An example of which is as follows:
\begin{equation}
	\begin{tikzpicture}[baseline=7,font=\footnotesize]
		\node (dots) {$\cdots$};
		\node[gauge, label=below:{$30$}] (k) [right=6mm of dots] {};
		\node[gauge, bodyE8, label=below:{$31$}] (A) [right=6mm of k] {};
		\node[gauge, bodyE8, label=below:{$33$}] (B) [right=6mm of A] {};
		\node[gauge, bodyE8, label=below:{$36$}] (C) [right=6mm of B] {};
		\node[gauge, bodyE8, label=below:{$40$}] (D) [right=6mm of C] {};
		\node[gauge, bodyE8, label=below:{$45$}] (E) [right=6mm of D] {};
		\node[gauge, bodyE8, label=below:{$51$}] (F) [right=6mm of E] {};
		\node[gauge, bodyE8, label=below:{$33$}] (G) [right=6mm of F] {};
		\node[gauge, bodyE8, label=below:{$16$}] (H) [right=6mm of G] {};
		\node[gauge, bodyE8, label=above:{$25$}] (L) [above=4mm of F] {};
		\draw[thick] (dots.east) -- (k.west);
		\draw[thick] (k.east) -- (A.west);
		\draw[thick,bodyE8] (A.east) -- (B.west);
		\draw[thick,bodyE8] (B.east) -- (C.west);
		\draw[thick,bodyE8] (C.east) -- (D.west);
		\draw[thick,bodyE8] (D.east) -- (E.west);
		\draw[thick,bodyE8] (E.east) -- (F.west);
		\draw[thick,bodyE8] (F.east) -- (G.west);
		\draw[thick,bodyE8] (G.east) -- (H.west);
		\draw[thick,bodyE8] (F.north) -- (L.south);
	\end{tikzpicture}
\end{equation}

Let us consider the following quiver and turn on the FI parameters $2\lambda$ and $-\lambda$ at the nodes $\U(1)$ and $\U(2)$ on the tail, respectively:
\begin{equation}
	\begin{tikzpicture}[baseline=7,font=\footnotesize]
		\node (dots) {$\cdots$};
		\node [gauge, label=below:{$2$}] (n2) [left=6mm of dots] {};
		\node[gauge, label=below:{$1$}] (n1) [left=6mm of n2] {};
		\node[gauge, label=below:{$k$}] (k) [right=6mm of dots] {};
		\node[gauge, bodyE8, label=below:{$A$}] (A) [right=6mm of k] {};
		\node[gauge, bodyE8, label=below:{$B$}] (B) [right=6mm of A] {};
		\node[gauge, bodyE8, label=below:{$C$}] (C) [right=6mm of B] {};
		\node[gauge, bodyE8, label=below:{$D$}] (D) [right=6mm of C] {};
		\node[gauge, bodyE8, label=below:{$E$}] (E) [right=6mm of D] {};
		\node[gauge, bodyE8, label=below:{$F$}] (F) [right=6mm of E] {};
		\node[gauge, bodyE8, label=below:{$G$}] (G) [right=6mm of F] {};
		\node[gauge, bodyE8, label=below:{$H$}] (H) [right=6mm of G] {};
		\node[gauge, bodyE8, label=above:{$L$}] (L) [above=4mm of F] {};
		\draw[thick] (n1.east) -- (n2.west);
		\draw[thick] (n2.east) -- (dots.west);
		\draw[thick] (dots.east) -- (k.west);
		\draw[thick] (k.east) -- (A.west);
		\draw[thick,bodyE8] (A.east) -- (B.west);
		\draw[thick,bodyE8] (B.east) -- (C.west);
		\draw[thick,bodyE8] (C.east) -- (D.west);
		\draw[thick,bodyE8] (D.east) -- (E.west);
		\draw[thick,bodyE8] (E.east) -- (F.west);
		\draw[thick,bodyE8] (F.east) -- (G.west);
		\draw[thick,bodyE8] (G.east) -- (H.west);
		\draw[thick,bodyE8] (F.north) -- (L.south);
	\end{tikzpicture}
\end{equation}
Using the argument provided in Figure \ref{e7gaugedef2}, we see that the resulting $E_7$ quiver is 
\begin{equation}
	\begin{tikzpicture}[baseline=7,font=\footnotesize]
		\node (dots) {$\cdots$};
		\node [gauge, label=below:{$2$}] (n2) [left=6mm of dots] {};
		\node[gauge, label=below:{$1$}] (n1) [left=6mm of n2] {};
		\node[gauge, label=below:{$k-2$}] (k) [right=9mm of dots] {};
		\node[gauge, label=above:{$A-2$}] (A) [right=9mm of k] {};
		\node[gauge, label=below:{$B-2$}] (B) [right=9mm of A] {};
		\node[gauge, bodyE7, label=above:{$C-2$}] (C) [right=9mm of B] {};
		\node[gauge, bodyE7, label=below:{$D-2$}] (D) [right=9mm of C] {};
		\node[gauge, bodyE7, label=above:{$E-2$}] (E) [right=9mm of D] {};
		\node[gauge, bodyE7, label=below:{$F-2$}] (F) [right=9mm of E] {};
		\node[gauge, bodyE7, label=above:{$G-1$}] (G) [right=9mm of F] {};
		\node[gauge, bodyE7, label=below:{$H$}] (H) [right=6mm of G] {};
		\node[gauge, bodyE7, label=below:{$1$}] (I) [right=6mm of H] {};
		\node[gauge, bodyE7, label=above:{$L-1$}] (L) [above=4mm of F] {};
		\draw[thick] (n1.east) -- (n2.west);
		\draw[thick] (n2.east) -- (dots.west);
		\draw[thick] (dots.east) -- (k.west);
		\draw[thick] (k.east) -- (A.west);
		\draw[thick] (A.east) -- (B.west);
		\draw[thick] (B.east) -- (C.west);
		\draw[thick,bodyE7] (C.east) -- (D.west);
		\draw[thick,bodyE7] (D.east) -- (E.west);
		\draw[thick,bodyE7] (E.east) -- (F.west);
		\draw[thick,bodyE7] (F.east) -- (G.west);
		\draw[thick,bodyE7] (G.east) -- (H.west);
		\draw[thick,bodyE7] (H.east) -- (I.west);
		\draw[thick,bodyE7] (F.north) -- (L.south);
	\end{tikzpicture}
\end{equation}

Another option is to apply the method of Section \ref{example3} which always holds whenever one of the nodes in the tail is balanced. The corresponding FI deformation involves only the balanced node and the two nodes connected to it. In a special case where the node $k$ is balanced, we can obtain in this way an $E_7$ quiver with a single tail: starting from a quiver of the form
\begin{equation}
	\begin{tikzpicture}[baseline=7,font=\footnotesize]
		\node (dots) {$\cdots$};
		\node [gauge, label=below:{$2$}] (n2) [left=6mm of dots] {};
		\node[gauge, label=below:{$1$}] (n1) [left=6mm of n2] {};
		\node[gauge, label=above:{$k-a$}] (k2) [right=9mm of dots] {};
		\node[gauge, label=below:{$k$}] (k) [right=9mm of k2] {};
		\node[gauge, bodyE8, label=above:{$k+a$}] (A) [right=9mm of k] {};
		\node[gauge, bodyE8, label=below:{$B$}] (B) [right=6mm of A] {};
		\node[gauge, bodyE8, label=below:{$C$}] (C) [right=6mm of B] {};
		\node[gauge, bodyE8, label=below:{$D$}] (D) [right=6mm of C] {};
		\node[gauge, bodyE8, label=below:{$E$}] (E) [right=6mm of D] {};
		\node[gauge, bodyE8, label=below:{$F$}] (F) [right=6mm of E] {};
		\node[gauge, bodyE8, label=below:{$G$}] (G) [right=6mm of F] {};
		\node[gauge, bodyE8, label=below:{$H$}] (H) [right=6mm of G] {};
		\node[gauge, bodyE8, label=above:{$L$}] (L) [above=4mm of F] {};
		\draw[thick] (n1.east) -- (n2.west);
		\draw[thick] (n2.east) -- (dots.west);
		\draw[thick] (dots.east) -- (k2.west);
		\draw[thick] (k2.east) -- (k.west);
		\draw[thick] (k.east) -- (A.west);
		\draw[thick,bodyE8] (A.east) -- (B.west);
		\draw[thick,bodyE8] (B.east) -- (C.west);
		\draw[thick,bodyE8] (C.east) -- (D.west);
		\draw[thick,bodyE8] (D.east) -- (E.west);
		\draw[thick,bodyE8] (E.east) -- (F.west);
		\draw[thick,bodyE8] (F.east) -- (G.west);
		\draw[thick,bodyE8] (G.east) -- (H.west);
		\draw[thick,bodyE8] (F.north) -- (L.south);
	\end{tikzpicture}
\end{equation}
we land on
\begin{equation}
	\begin{tikzpicture}[baseline=7,font=\footnotesize]
		\node (dots) {$\cdots$};
		\node [gauge, label=below:{$2$}] (n2) [left=6mm of dots] {};
		\node[gauge, label=below:{$1$}] (n1) [left=6mm of n2] {};
		\node[gauge, label=above:{$k-a$}] (k) [right=9mm of dots] {};
		\node[gauge, label=below:{$B-2a$}] (A) [right=9mm of k] {};
		\node[gauge, bodyE7, label=above:{$C-2a$}] (B) [right=9mm of A] {};
		\node[gauge, bodyE7, label=below:{$D-2a$}] (C) [right=9mm of B] {};
		\node[gauge, bodyE7, label=above:{$E-2a$}] (D) [right=9mm of C] {};
		\node[gauge, bodyE7, label=below:{$F-2a$}] (E) [right=9mm of D] {};
		\node[gauge, bodyE7, label=above:{$G-a$}] (F) [right=9mm of E] {};
		\node[gauge, bodyE7, label=below:{$H$}] (G) [right=9mm of F] {};
		\node[gauge, bodyE7, label=below:{$a$}] (H) [right=6mm of G] {};
		\node[gauge, bodyE7, label=above:{$L-a$}] (L) [above=4mm of E] {};
		\draw[thick] (n1.east) -- (n2.west);
		\draw[thick] (n2.east) -- (dots.west);
		\draw[thick] (dots.east) -- (k.west);
		\draw[thick] (k.east) -- (A.west);
		\draw[thick] (A.east) -- (B.west);
		\draw[thick,bodyE7] (B.east) -- (C.west);
		\draw[thick,bodyE7] (C.east) -- (D.west);
		\draw[thick,bodyE7] (D.east) -- (E.west);
		\draw[thick,bodyE7] (E.east) -- (F.west);
		\draw[thick,bodyE7] (F.east) -- (G.west);
		\draw[thick,bodyE7] (G.east) -- (H.west);
		\draw[thick,bodyE7] (E.north) -- (L.south);
	\end{tikzpicture}
\end{equation}
Note, however, that, in general, unless the chosen balanced node is the first or last in the tail, the resulting $E_7$ quiver will have two tails.

\subsubsection{Mixed Deformations: \texorpdfstring{$E_7$}{E7} Quivers with Two Tails}
\label{sec:E8toE7-MixDef2tail}
So far, we have considered deformations that lead to an $E_7$ quiver with a single tail attached. However, it is possible to generate $E_7$-shaped quivers with two tails attached, illustrated below:
\begin{equation}
\label{eq:generalE7quiv-2tails1}
    \begin{tikzpicture}[baseline=7,font=\footnotesize]
			\node (dots) {$\cdots$};
			\node[gauge, label=below:{$k_1$}] (k1) [right=3mm of dots] {};
			\node[gauge, bodyE7, label=below:{$\alpha_1$}] (A) [right=6mm of k1] {};
			\node[gauge, bodyE7, label=below:{$\alpha_2$}] (B) [right=6mm of A] {};
			\node[gauge, bodyE7, label=below:{$\alpha_3$}] (C) [right=6mm of B] {};
			\node[gauge, bodyE7, label=below:{$\alpha_4$}] (D) [right=6mm of C] {};
			\node[gauge, bodyE7, label=below:{$\alpha_{3'}$}] (E) [right=6mm of D] {};
			\node[gauge, bodyE7, label=below:{$\alpha_{2'}$}] (F) [right=6mm of E] {};
			\node[gauge, bodyE7, label=below:{$\alpha_{1'}$}] (G) [right=6mm of F] {};
			\node[gauge, bodyE7, label=above:{$\alpha_{2''}$}] (H) [above=4mm of D] {};
			\node[gauge, label=below:{$k_2$}] (k2) [right=6mm of G] {};
            \node (dots2) [right=3mm of k2] {$\cdots$};
			\draw[thick] (dots.east) -- (k1.west);
			\draw[thick] (k1.east) -- (A.west);
			\draw[thick,bodyE7] (A.east) -- (B.west);
			\draw[thick,bodyE7] (B.east) -- (C.west);
			\draw[thick,bodyE7] (C.east) -- (D.west);
			\draw[thick,bodyE7] (D.east) -- (E.west);
			\draw[thick,bodyE7] (E.east) -- (F.west);
			\draw[thick,bodyE7] (F.east) -- (G.west);
			\draw[thick,bodyE7] (D.north) -- (H.south);
            \draw[thick] (G.east) -- (k2.west);
            \draw[thick] (k2.east) -- (dots2.west);
		\end{tikzpicture}
\end{equation}
Generally this involves turning on FI parameters at certain nodes in the $E_8$ affine Dynkin diagram, and at a node on the tail of the $E_8$-shaped quiver.  This is thus referred to as a {\it mixed deformation}. Let us emphasize again that we are not aiming at classifying all possible deformations; rather we would like to list some way to obtain an $E_7$-shaped quiver with two tails starting from an $E_8$ orbi-instanton theory and a choice of mass deformation. 

We first give an example and list other possibilities in Table \ref{tab:E8toE7-2tail-body}. Let us consider again a generic $E_8$ quiver as in \eqref{eq:generalE8quiv} and turn on FI deformations at nodes $2$ / $2'$ (with equal and opposite value for the FI parameters) of the $E_8$ quiver. We can introduce a parameter $\tilde{k}$, defined as
\begin{equation}
	\tilde{k} = B-H\coma
\end{equation}
or, analogously, it can be expressed purely in terms of the excess numbers of the $E_8$ quiver, i.e.
	\begin{equation}
	\tilde{k} = n_3+2n_4+3n_5+4n_6+3n_{4'}+2n_{2'}+2n_{3'}\fstop
\end{equation}
Indeed, in order to satisfy \eqref{constr}, we should also turn on the deformation at some other nodes and we choose the node with rank $\tilde{k}$ in the tail (assuming it exists).
Following the rules discussed in Section \ref{example1}, we find that this deformation leads to the following $E_7$-shaped quiver with two tails:
\begin{equation}
	\begin{tikzpicture}[baseline=7,font=\footnotesize]
		\node (dots) {$\cdots$};
		\node[gauge, label=below:{$k-\tilde{k}$}] (tk) [right=6mm of dots] {};
		\node[gauge, bodyE7, label=above:{$A-\tilde{k}$}] (A) [right=9mm of tk] {};
		\node[gauge, bodyE7, label=below:{$H$}] (B) [right=9mm of A] {};
		\node[gauge, bodyE7, label=below:{$L$}] (C) [right=9mm of B] {};
		\node[gauge, bodyE7, label=below:{$F-H$}] (D) [right=9mm of C] {};
		\node[gauge, bodyE7, label=above:{$E-H$}] (E) [right=9mm of D] {};
		\node[gauge, bodyE7, label=below:{$D-H$}] (F) [right=9mm of E] {};
		\node[gauge, bodyE7, label=above:{$C-H$}] (G) [right=9mm of F] {};
		\node[gauge, bodyE7, label=above:{$G-H$}] (H) [above=4mm of D] {};
		\node[gauge, label=below:{$\tilde{k}$}] (k2) [right=9mm of G] {};
		\node (dots2) [right=6mm of k2] {$\cdots$};
		\draw[thick] (dots.east) -- (tk.west);
		\draw[thick] (tk.east) -- (A.west);
		\draw[thick,bodyE7] (A.east) -- (B.west);
		\draw[thick,bodyE7] (B.east) -- (C.west);
		\draw[thick,bodyE7] (C.east) -- (D.west);
		\draw[thick,bodyE7] (D.east) -- (E.west);
		\draw[thick,bodyE7] (E.east) -- (F.west);
		\draw[thick,bodyE7] (F.east) -- (G.west);
		\draw[thick,bodyE7] (D.north) -- (H.south);
		\draw[thick] (G.east) -- (k2.west);
		\draw[thick] (k2.east) -- (dots2.west);
	\end{tikzpicture}
\end{equation}
We list other possible FI deformations that similarly lead to an $E_7$-shaped quiver in Table \ref{tab:E8toE7-2tail-body}. All the deformations provided in the table involve one node in the tail of rank $\tilde{k}$, besides the nodes explicitly reported in the first column. 

	\begin{landscape}
		\pagestyle{empty}
			\renewcommand{\arraystretch}{1.1}
            \setlength\LTleft{-0.79cm}
			\setlength{\LTcapwidth}{1.2\textwidth}
			\begin{longtable}{c|c|c}
   \caption{Possible FI deformations of the $E_8$ orbi-instanton theory \eqref{eq:generalE8quiv} that lead to an $E_7$-shaped quiver with two tails. We list the nodes where the FI deformations are turned on, the value of $\tilde{k}$ and the corresponding $E_7$ quiver.}  \label{tab:E8toE7-2tail-body}\\
		Nodes & $\tilde{k}$ & $E_7$ Quiver \\
		\hline
  \endfirsthead  
  Nodes & $\tilde{k}$ & $E_7$ Quiver \\
		\hline
  \endhead
		$2$ / $2'$ & $\begin{aligned} &B-H=\\&n_3+2n_4+3n_5+4n_6+3n_{4'}+2n_{2'}+2n_{3'}\end{aligned}$& 
	\begin{tikzpicture}[baseline=0,font=\footnotesize]
		\node (dots) {$\cdots$};
		\node[gauge, label=below:{$k-\tilde{k}$}] (tk) [right=6mm of dots] {};
		\node[gauge, bodyE7, label=above:{$A-\tilde{k}$}] (A) [right=9mm of tk] {};
		\node[gauge, bodyE7, label=below:{$H$}] (B) [right=9mm of A] {};
		\node[gauge, bodyE7, label=below:{$L$}] (C) [right=9mm of B] {};
		\node[gauge, bodyE7, label=below:{$F-H$}] (D) [right=9mm of C] {};
		\node[gauge, bodyE7, label=above:{$E-H$}] (E) [right=9mm of D] {};
		\node[gauge, bodyE7, label=below:{$D-H$}] (F) [right=9mm of E] {};
		\node[gauge, bodyE7, label=above:{$C-H$}] (G) [right=9mm of F] {};
		\node[gauge, bodyE7, label=above:{$G-H$}] (H) [above=4mm of D] {};
		\node[gauge, label=below:{$\tilde{k}$}] (k2) [right=9mm of G] {};
		\node (dots2) [right=6mm of k2] {$\cdots$};
		\draw[thick] (dots.east) -- (tk.west);
		\draw[thick] (tk.east) -- (A.west);
		\draw[thick,bodyE7] (A.east) -- (B.west);
		\draw[thick,bodyE7] (B.east) -- (C.west);
		\draw[thick,bodyE7] (C.east) -- (D.west);
		\draw[thick,bodyE7] (D.east) -- (E.west);
		\draw[thick,bodyE7] (E.east) -- (F.west);
		\draw[thick,bodyE7] (F.east) -- (G.west);
		\draw[thick,bodyE7] (D.north) -- (H.south);
		\draw[thick] (G.east) -- (k2.west);
		\draw[thick] (k2.east) -- (dots2.west);
	\end{tikzpicture} \\ \hline
		$3$ / $3'$ & $\begin{aligned} &C-L=\\&n_4+2n_5+3n_6+2n_{4'}+n_{2'}+2n_{3'}\end{aligned}$& 
			\begin{tikzpicture}[baseline=0,font=\footnotesize]
			\node (dots) {$\cdots$};
			\node[gauge, label=below:{$k-\tilde{k}$}] (tk) [right=3mm of dots] {};
			\node[gauge, bodyE7, label=above:{$A-\tilde{k}$}] (A) [right=9mm of tk] {};
			\node[gauge, bodyE7, label=below:{$B-\tilde{k}$}] (B) [right=9mm of A] {};
			\node[gauge, bodyE7, label=above:{$L$}] (C) [right=9mm of B] {};
			\node[gauge, bodyE7, label=below:{$G$}] (D) [right=9mm of C] {};
			\node[gauge, bodyE7, label=above:{$F-L$}] (E) [right=9mm of D] {};
			\node[gauge, bodyE7, label=below:{$E-L$}] (F) [right=9mm of E] {};
			\node[gauge, bodyE7, label=above:{$D-L$}] (G) [right=9mm of F] {};
			\node[gauge, bodyE7, label=above:{$H$}] (H) [above=4mm of D] {};
			\node[gauge, label=below:{$\tilde{k}$}] (k2) [right=9mm of G] {};
			\node (dots2) [right=3mm of k2] {$\cdots$};
			\draw[thick] (dots.east) -- (tk.west);
			\draw[thick] (tk.east) -- (A.west);
			\draw[thick,bodyE7] (A.east) -- (B.west);
			\draw[thick,bodyE7] (B.east) -- (C.west);
			\draw[thick,bodyE7] (C.east) -- (D.west);
			\draw[thick,bodyE7] (D.east) -- (E.west);
			\draw[thick,bodyE7] (E.east) -- (F.west);
			\draw[thick,bodyE7] (F.east) -- (G.west);
			\draw[thick,bodyE7] (D.north) -- (H.south);
			\draw[thick] (G.east) -- (k2.west);
			\draw[thick] (k2.east) -- (dots2.west);
		\end{tikzpicture} \\ \hline
		$4$ / $4'$ & $\begin{aligned} &D-G=\\ &n_5+2n_6+2n_{4'}+n_{2'}+n_{3'}\end{aligned}$& 
		\begin{tikzpicture}[baseline=0,font=\footnotesize]
			\node (dots) {$\cdots$};
			\node[gauge, label=below:{$k-\tilde{k}$}] (tk) [right=3mm of dots] {};
			\node[gauge, bodyE7, label=above:{$A-\tilde{k}$}] (A) [right=9mm of tk] {};
			\node[gauge, bodyE7, label=below:{$B-\tilde{k}$}] (B) [right=9mm of A] {};
			\node[gauge, bodyE7, label=above:{$C-\tilde{k}$}] (C) [right=9mm of B] {};
			\node[gauge, bodyE7, label=below:{$G$}] (D) [right=9mm of C] {};
			\node[gauge, bodyE7, label=above:{$L$}] (E) [right=9mm of D] {};
			\node[gauge, bodyE7, label=below:{$F-G$}] (F) [right=9mm of E] {};
			\node[gauge, bodyE7, label=above:{$E-G$}] (G) [right=9mm of F] {};
			\node[gauge, bodyE7, label=above:{$H$}] (H) [above=4mm of D] {};
			\node[gauge, label=below:{$\tilde{k}$}] (k2) [right=9mm of G] {};
			\node (dots2) [right=3mm of k2] {$\cdots$};
			\draw[thick] (dots.east) -- (tk.west);
			\draw[thick] (tk.east) -- (A.west);
			\draw[thick,bodyE7] (A.east) -- (B.west);
			\draw[thick,bodyE7] (B.east) -- (C.west);
			\draw[thick,bodyE7] (C.east) -- (D.west);
			\draw[thick,bodyE7] (D.east) -- (E.west);
			\draw[thick,bodyE7] (E.east) -- (F.west);
			\draw[thick,bodyE7] (F.east) -- (G.west);
			\draw[thick,bodyE7] (D.north) -- (H.south);
			\draw[thick] (G.east) -- (k2.west);
			\draw[thick] (k2.east) -- (dots2.west);
		\end{tikzpicture} \\ \hline
		$5$ / $3'+2'$ & $\begin{aligned} &E-L-H=\\ &n_6+n_{4'}+n_{2'}+n_{3'}\end{aligned}$& 
			\begin{tikzpicture}[baseline=0,font=\footnotesize]
			\node (dots) {$\cdots$};
			\node[gauge, label=below:{$k-\tilde{k}$}] (tk) [right=3mm of dots] {};
			\node[gauge, bodyE7, label=above:{$A-\tilde{k}$}] (A) [right=9mm of tk] {};
			\node[gauge, bodyE7, label=below:{$B-\tilde{k}$}] (B) [right=9mm of A] {};
			\node[gauge, bodyE7, label=above:{$C-\tilde{k}$}] (C) [right=9mm of B] {};
			\node[gauge, bodyE7, label=below:{$D-\tilde{k}$}] (D) [right=9mm of C] {};
			\node[gauge, bodyE7, label=above:{$L$}] (E) [right=9mm of D] {};
			\node[gauge, bodyE7, label=below:{$G-H$}] (F) [right=9mm of E] {};
			\node[gauge, bodyE7, label=above:{$F-L-H$}] (G) [right=9mm of F] {};
			\node[gauge, bodyE7, label=above:{$H$}] (H) [above=4mm of D] {};
			\node[gauge, label=below:{$\tilde{k}$}] (k2) [right=9mm of G] {};
			\node (dots2) [right=3mm of k2] {$\cdots$};
			\draw[thick] (dots.east) -- (tk.west);
			\draw[thick] (tk.east) -- (A.west);
			\draw[thick,bodyE7] (A.east) -- (B.west);
			\draw[thick,bodyE7] (B.east) -- (C.west);
			\draw[thick,bodyE7] (C.east) -- (D.west);
			\draw[thick,bodyE7] (D.east) -- (E.west);
			\draw[thick,bodyE7] (E.east) -- (F.west);
			\draw[thick,bodyE7] (F.east) -- (G.west);
			\draw[thick,bodyE7] (D.north) -- (H.south);
			\draw[thick] (G.east) -- (k2.west);
			\draw[thick] (k2.east) -- (dots2.west);
		\end{tikzpicture} \\ \hline
		$6$ / $3'+2'+1$ & $\begin{aligned} &A+L+H-F=\\ &n_2+2n_3+3n_4+4n_5+5n_6+3n_{4'}+n_{2'}+2n_{3'}\end{aligned}$& 
		\begin{tikzpicture}[baseline=0,font=\footnotesize]
			\node (dots) {$\cdots$};
			\node[gauge, label=below:{$k-\tilde{k}$}] (tk) [right=3mm of dots] {};
			\node[gauge, bodyE7, label=above:{$A-\tilde{k}$}] (A) [right=9mm of tk] {};
			\node[gauge, bodyE7, label=below:{$G-H$}] (B) [right=9mm of A] {};
			\node[gauge, bodyE7, label=above:{$L$}] (C) [right=9mm of B] {};
			\node[gauge, bodyE7, label=below:{$E+\tilde{k}-A$}] (D) [right=9mm of C] {};
			\node[gauge, bodyE7, label=above:{$D+\tilde{k}-A$}] (E) [right=9mm of D] {};
			\node[gauge, bodyE7, label=below:{$C+\tilde{k}-A$}] (F) [right=9mm of E] {};
			\node[gauge, bodyE7, label=above:{$B+\tilde{k}-A$}] (G) [right=9mm of F] {};
			\node[gauge, bodyE7, label=above:{$H$}] (H) [above=4mm of D] {};
			\node[gauge, label=below:{$\tilde{k}$}] (k2) [right=9mm of G] {};
			\node (dots2) [right=3mm of k2] {$\cdots$};
			\draw[thick] (dots.east) -- (tk.west);
			\draw[thick] (tk.east) -- (A.west);
			\draw[thick,bodyE7] (A.east) -- (B.west);
			\draw[thick,bodyE7] (B.east) -- (C.west);
			\draw[thick,bodyE7] (C.east) -- (D.west);
			\draw[thick,bodyE7] (D.east) -- (E.west);
			\draw[thick,bodyE7] (E.east) -- (F.west);
			\draw[thick,bodyE7] (F.east) -- (G.west);
			\draw[thick,bodyE7] (D.north) -- (H.south);
			\draw[thick] (G.east) -- (k2.west);
			\draw[thick] (k2.east) -- (dots2.west);
		\end{tikzpicture}
	\end{longtable}
	\end{landscape}

\subsection{Deformations from \texorpdfstring{$E_7$}{E7} Quivers To \texorpdfstring{$E_6$}{E6} Quivers}
\label{sec:MassDefE7E6}

In this section, we are going to describe in detail several FI deformations which lead to $E_6$-shaped quivers starting from $E_7$ quivers.

\subsubsection{Deformations Starting from \texorpdfstring{$E_7$}{E7} Quivers with One Tail}

Let us start by considering the following generic $E_7$-shaped quiver with one tail:
\begin{equation}
\label{eq:generalE7quiv-1tail}
    \begin{tikzpicture}[baseline=7,font=\footnotesize]
			\node (dots) {$\cdots$};
			\node[gauge, label=below:{$k$}] (k) [right=3mm of dots] {};
			\node[gauge, bodyE7, label=below:{$A$}] (A) [right=6mm of k] {};
			\node[gauge, bodyE7, label=below:{$B$}] (B) [right=6mm of A] {};
			\node[gauge, bodyE7, label=below:{$C$}] (C) [right=6mm of B] {};
			\node[gauge, bodyE7, label=below:{$D$}] (D) [right=6mm of C] {};
			\node[gauge, bodyE7, label=below:{$E$}] (E) [right=6mm of D] {};
			\node[gauge, bodyE7, label=below:{$F$}] (F) [right=6mm of E] {};
			\node[gauge, bodyE7, label=below:{$G$}] (G) [right=6mm of F] {};
			\node[gauge, bodyE7, label=above:{$H$}] (H) [above=4mm of D] {};
			\draw[thick] (dots.east) -- (k.west);
			\draw[thick] (k.east) -- (A.west);
			\draw[thick,bodyE7] (A.east) -- (B.west);
			\draw[thick,bodyE7] (B.east) -- (C.west);
			\draw[thick,bodyE7] (C.east) -- (D.west);
			\draw[thick,bodyE7] (D.east) -- (E.west);
			\draw[thick,bodyE7] (E.east) -- (F.west);
			\draw[thick,bodyE7] (F.east) -- (G.west);
			\draw[thick,bodyE7] (D.north) -- (H.south);
		\end{tikzpicture}
\end{equation}
and we assume $B\geq H$. In this case, it is possible to turn on the FI parameters at nodes $B$ and $H$, and assuming that there exists a node in the tail with rank $\tilde{k}=B-H$, we must also turn on the FI parameter at that node. By following the rules reviewed in Section \ref{sec:MassDef}, the resulting quiver is 
\begin{equation}
    \begin{tikzpicture}[baseline=7,font=\footnotesize]
			\node (dots) {$\cdots$};
			\node[gauge, label=below:{$k-\tilde{k}$}] (k) [right=3mm of dots] {};
			\node[gauge, bodyE6, label=above:{$A-\tilde{k}$}] (A) [right=9mm of k] {};
			\node[gauge, bodyE6, label=below:{$H$}] (B) [right=9mm of A] {};
			\node[gauge, bodyE6, label=below:{$E$}] (C) [right=9mm of B] {};
			\node[gauge, bodyE6, label=above:{$D-H$}] (D) [right=9mm of C] {};
			\node[gauge, bodyE6, label=below:{$C-H$}] (E) [right=9mm of D] {};
			\node[gauge, bodyE6, label=left:{$F$}] (F) [above=4mm of C] {};
			\node[gauge, bodyE6, label=above:{$G$}] (G) [above=4mm of F] {};
			\node[gauge, label=below:{$\tilde{k}$}] (tk) [right=9mm of E] {};
            \node (dots2) [right=3mm of tk] {$\cdots$};
			\draw[thick] (dots.east) -- (k.west);
			\draw[thick] (k.east) -- (A.west);
			\draw[thick,bodyE6] (A.east) -- (B.west);
			\draw[thick,bodyE6] (B.east) -- (C.west);
			\draw[thick,bodyE6] (C.east) -- (D.west);
			\draw[thick,bodyE6] (D.east) -- (E.west);
			\draw[thick,bodyE6] (C.north) -- (F.south);
			\draw[thick,bodyE6] (F.north) -- (G.south);
			\draw[thick] (E.east) -- (tk.west);
            \draw[thick] (tk.east) -- (dots2.west);
		\end{tikzpicture}
  \label{eq:BHdef-E71tailtoE62tail}
\end{equation}
This means that we have generated an $E_6$ quiver, marked in {\color{cE6}{orange}}, with two tails, where the tail ending with the node $k-\tilde{k}$ originates from the tail of the $E_7$ quiver where all ranks are reduced by $\tilde{k}$. The other tail ends with a node of rank $\tilde{k}$. This is not the only way to obtain the $E_6$ quiver, and we list several other similar options in Table \ref{tab:E7toE6-singletail-body}. 

\begin{table}
	\begin{center}
	\begin{tabular}{c|c|c}
		Nodes & Condition & $E_6$ Quiver \\
		\hline
		$\begin{aligned}&B,\,H,\\ \tilde{k}&=B-H\end{aligned}$ & $B\geq H$ & 
	\begin{tikzpicture}[baseline=0,font=\footnotesize]
			\node (dots) {$\cdots$};
			\node[gauge, label=below:{$k-\tilde{k}$}] (k) [right=3mm of dots] {};
			\node[gauge, bodyE6, label=above:{$A-\tilde{k}$}] (A) [right=9mm of k] {};
			\node[gauge, bodyE6, label=below:{$H$}] (B) [right=9mm of A] {};
			\node[gauge, bodyE6, label=below:{$E$}] (C) [right=9mm of B] {};
			\node[gauge, bodyE6, label=above:{$D-H$}] (D) [right=9mm of C] {};
			\node[gauge, bodyE6, label=below:{$C-H$}] (E) [right=9mm of D] {};
			\node[gauge, bodyE6, label=left:{$F$}] (F) [above=4mm of C] {};
			\node[gauge, bodyE6, label=above:{$G$}] (G) [above=4mm of F] {};
			\node[gauge, label=below:{$\tilde{k}$}] (tk) [right=9mm of E] {};
            \node (dots2) [right=3mm of tk] {$\cdots$};
			\draw[thick] (dots.east) -- (k.west);
			\draw[thick] (k.east) -- (A.west);
			\draw[thick,bodyE6] (A.east) -- (B.west);
			\draw[thick,bodyE6] (B.east) -- (C.west);
			\draw[thick,bodyE6] (C.east) -- (D.west);
			\draw[thick,bodyE6] (D.east) -- (E.west);
			\draw[thick,bodyE6] (C.north) -- (F.south);
			\draw[thick,bodyE6] (F.north) -- (G.south);
			\draw[thick] (E.east) -- (tk.west);
            \draw[thick] (tk.east) -- (dots2.west);
		\end{tikzpicture} \\ \hline
		$\begin{aligned}&A,\,G,\\ \tilde{k}&=A-G\end{aligned}$ & $A\geq G$& 
			\begin{tikzpicture}[baseline=0,font=\footnotesize]
			\node (dots) {$\cdots$};
			\node[gauge, label=below:{$k-\tilde{k}$}] (k) [right=3mm of dots] {};
			\node[gauge, bodyE6, label=above:{$G$}] (A) [right=9mm of k] {};
			\node[gauge, bodyE6, label=below:{$H$}] (B) [right=9mm of A] {};
			\node[gauge, bodyE6, label=below:{$D-G$}] (C) [right=9mm of B] {};
			\node[gauge, bodyE6, label=above:{$C-G$}] (D) [right=9mm of C] {};
			\node[gauge, bodyE6, label=below:{$B-G$}] (E) [right=9mm of D] {};
			\node[gauge, bodyE6, label=left:{$E-G$}] (F) [above=4mm of C] {};
			\node[gauge, bodyE6, label=above:{$F-G$}] (G) [above=4mm of F] {};
			\node[gauge, label=below:{$\tilde{k}$}] (tk) [right=9mm of E] {};
            \node (dots2) [right=3mm of tk] {$\cdots$};
			\draw[thick] (dots.east) -- (k.west);
			\draw[thick] (k.east) -- (A.west);
			\draw[thick,bodyE6] (A.east) -- (B.west);
			\draw[thick,bodyE6] (B.east) -- (C.west);
			\draw[thick,bodyE6] (C.east) -- (D.west);
			\draw[thick,bodyE6] (D.east) -- (E.west);
			\draw[thick,bodyE6] (C.north) -- (F.south);
			\draw[thick,bodyE6] (F.north) -- (G.south);
			\draw[thick] (E.east) -- (tk.west);
            \draw[thick] (tk.east) -- (dots2.west);
		\end{tikzpicture} \\ \hline
		$\begin{aligned}&C,\,E,\\ \tilde{k}&=C-E\end{aligned}$ & $C\geq E$& 
		\begin{tikzpicture}[baseline=0,font=\footnotesize]
			\node (dots) {$\cdots$};
			\node[gauge, label=below:{$k-\tilde{k}$}] (k) [right=3mm of dots] {};
			\node[gauge, bodyE6, label=above:{$A-\tilde{k}$}] (A) [right=9mm of k] {};
			\node[gauge, bodyE6, label=below:{$B-\tilde{k}$}] (B) [right=9mm of A] {};
			\node[gauge, bodyE6, label=below:{$E$}] (C) [right=9mm of B] {};
			\node[gauge, bodyE6, label=above:{$H$}] (D) [right=9mm of C] {};
			\node[gauge, bodyE6, label=below:{$D-E$}] (E) [right=9mm of D] {};
			\node[gauge, bodyE6, label=left:{$F$}] (F) [above=4mm of C] {};
			\node[gauge, bodyE6, label=above:{$G$}] (G) [above=4mm of F] {};
			\node[gauge, label=below:{$\tilde{k}$}] (tk) [right=9mm of E] {};
            \node (dots2) [right=3mm of tk] {$\cdots$};
			\draw[thick] (dots.east) -- (k.west);
			\draw[thick] (k.east) -- (A.west);
			\draw[thick,bodyE6] (A.east) -- (B.west);
			\draw[thick,bodyE6] (B.east) -- (C.west);
			\draw[thick,bodyE6] (C.east) -- (D.west);
			\draw[thick,bodyE6] (D.east) -- (E.west);
			\draw[thick,bodyE6] (C.north) -- (F.south);
			\draw[thick,bodyE6] (F.north) -- (G.south);
			\draw[thick] (E.east) -- (tk.west);
            \draw[thick] (tk.east) -- (dots2.west);
		\end{tikzpicture} \\ \hline
		$\begin{aligned}&H,\,F,\\ \tilde{k}&=H-F\end{aligned}$ & $H>F$& 
			\begin{tikzpicture}[baseline=0,font=\footnotesize]
			\node (dots) {$\cdots$};
			\node[gauge, label=above:{$k-\tilde{k}$}] (k) [right=3mm of dots] {};
			\node[gauge, bodyE6, label=below:{$A-\tilde{k}$}] (A) [right=9mm of k] {};
			\node[gauge, bodyE6, label=above:{$B-\tilde{k}$}] (B) [right=9mm of A] {};
			\node[gauge, bodyE6, label=below:{$C-\tilde{k}$}] (C) [right=9mm of B] {};
			\node[gauge, bodyE6, label=above:{$D-F-\tilde{k}$}] (D) [right=9mm of C] {};
			\node[gauge, bodyE6, label=below:{$E-F$}] (E) [right=9mm of D] {};
			\node[gauge, bodyE6, label=left:{$F$}] (F) [above=6mm of C] {};
			\node[gauge, bodyE6, label=above:{$G$}] (G) [above=6mm of F] {};
			\node[gauge, label=below:{$\tilde{k}$}] (tk) [right=9mm of E] {};
            \node (dots2) [right=3mm of tk] {$\cdots$};
			\draw[thick] (dots.east) -- (k.west);
			\draw[thick] (k.east) -- (A.west);
			\draw[thick,bodyE6] (A.east) -- (B.west);
			\draw[thick,bodyE6] (B.east) -- (C.west);
			\draw[thick,bodyE6] (C.east) -- (D.west);
			\draw[thick,bodyE6] (D.east) -- (E.west);
			\draw[thick,bodyE6] (C.north) -- (F.south);
			\draw[thick,bodyE6] (F.north) -- (G.south);
			\draw[thick] (E.east) -- (tk.west);
            \draw[thick] (tk.east) -- (dots2.west);
		\end{tikzpicture} \\ \hline
		$\begin{aligned}&H,\,F,\\ \tilde{k}&=F-H\end{aligned}$ & $H<F$& 
		\begin{tikzpicture}[baseline=0,font=\footnotesize]
			\node (dots) {$\cdots$};
			\node[gauge, label=above:{$k-\tilde{k}$}] (k) [right=3mm of dots] {};
			\node[gauge, bodyE6, label=below:{$A-\tilde{k}$}] (A) [right=9mm of k] {};
			\node[gauge, bodyE6, label=above:{$B-\tilde{k}$}] (B) [right=9mm of A] {};
			\node[gauge, bodyE6, label=below:{$C-\tilde{k}$}] (C) [right=9mm of B] {};
			\node[gauge, bodyE6, label=below:{$H$}] (D) [right=9mm of C] {};
			\node[gauge, bodyE6, label=below:{$G$}] (E) [right=9mm of D] {};
			\node[gauge, bodyE6, label=right:{$D-H-\tilde{k}$}] (F) [above=6mm of C] {};
			\node[gauge, bodyE6, label=above:{$E-H-\tilde{k}$}] (G) [above=6mm of F] {};
			\node[gauge, label=below:{$\tilde{k}$}] (tk) [right=9mm of E] {};
            \node (dots2) [right=3mm of tk] {$\cdots$};
			\draw[thick] (dots.east) -- (k.west);
			\draw[thick] (k.east) -- (A.west);
			\draw[thick,bodyE6] (A.east) -- (B.west);
			\draw[thick,bodyE6] (B.east) -- (C.west);
			\draw[thick,bodyE6] (C.east) -- (D.west);
			\draw[thick,bodyE6] (D.east) -- (E.west);
			\draw[thick,bodyE6] (C.north) -- (F.south);
			\draw[thick,bodyE6] (F.north) -- (G.south);
			\draw[thick] (E.east) -- (tk.west);
            \draw[thick] (tk.east) -- (dots2.west);
		\end{tikzpicture} 
	\end{tabular}
\end{center}
	\caption{Possible FI deformations of the $E_7$ quiver \eqref{eq:generalE7quiv-1tail} that lead to an $E_6$-shaped quiver with two tails. We list the nodes where the FI deformations are turned on, the conditions for which the deformation is possible and the corresponding $E_6$ quiver.}
	\label{tab:E7toE6-singletail-body}
\end{table}

\subsubsection{Deformations Starting from \texorpdfstring{$E_7$}{E7} Quiver with Two Tails}

It is possible to generalize the deformations seen in the previous section to $E_7$ quivers that have two tails, i.e.
\begin{equation}
\label{eq:generalE7quiv-2tails}
    \begin{tikzpicture}[baseline=7,font=\footnotesize]
			\node (dots) {$\cdots$};
			\node[gauge, label=below:{$k_1$}] (k1) [right=3mm of dots] {};
			\node[gauge, bodyE7, label=below:{$A$}] (A) [right=6mm of k1] {};
			\node[gauge, bodyE7, label=below:{$B$}] (B) [right=6mm of A] {};
			\node[gauge, bodyE7, label=below:{$C$}] (C) [right=6mm of B] {};
			\node[gauge, bodyE7, label=below:{$D$}] (D) [right=6mm of C] {};
			\node[gauge, bodyE7, label=below:{$E$}] (E) [right=6mm of D] {};
			\node[gauge, bodyE7, label=below:{$F$}] (F) [right=6mm of E] {};
			\node[gauge, bodyE7, label=below:{$G$}] (G) [right=6mm of F] {};
			\node[gauge, bodyE7, label=above:{$H$}] (H) [above=4mm of D] {};
			\node[gauge, label=below:{$k_2$}] (k2) [right=6mm of G] {};
            \node (dots2) [right=3mm of k2] {$\cdots$};
			\draw[thick] (dots.east) -- (k1.west);
			\draw[thick] (k1.east) -- (A.west);
			\draw[thick,bodyE7] (A.east) -- (B.west);
			\draw[thick,bodyE7] (B.east) -- (C.west);
			\draw[thick,bodyE7] (C.east) -- (D.west);
			\draw[thick,bodyE7] (D.east) -- (E.west);
			\draw[thick,bodyE7] (E.east) -- (F.west);
			\draw[thick,bodyE7] (F.east) -- (G.west);
			\draw[thick,bodyE7] (D.north) -- (H.south);
            \draw[thick] (G.east) -- (k2.west);
            \draw[thick] (k2.east) -- (dots2.west);
		\end{tikzpicture}
\end{equation}
We are not going to list all the quivers that result for each of the deformations in Table \ref{tab:E7toE6-singletail-body}, since the analysis is very similar, and we focus for definiteness on the deformation involving nodes $B$ and $H$. As before, if $B>H$, we need also to turn on an FI deformation at a node with rank $\tilde{k}=B-H$ in the tail that ends with $k_1$. The resulting quiver is similar to \eqref{eq:BHdef-E71tailtoE62tail}, but with three tails, i.e.
\begin{equation}
    \begin{tikzpicture}[baseline=30,font=\footnotesize]
			\node (dots) {$\cdots$};
			\node[gauge, label=below:{$k_1-\tilde{k}$}] (k1) [right=3mm of dots] {};
			\node[gauge, bodyE6, label=above:{$A-\tilde{k}$}] (A) [right=9mm of k1] {};
			\node[gauge, bodyE6, label=below:{$H$}] (B) [right=9mm of A] {};
			\node[gauge, bodyE6, label=below:{$E$}] (C) [right=9mm of B] {};
			\node[gauge, bodyE6, label=above:{$D-H$}] (D) [right=9mm of C] {};
			\node[gauge, bodyE6, label=below:{$C-H$}] (E) [right=9mm of D] {};
			\node[gauge, bodyE6, label=left:{$F$}] (F) [above=4mm of C] {};
			\node[gauge, bodyE6, label=left:{$G$}] (G) [above=4mm of F] {};
			\node[gauge, label=left:{$k_2$}] (k2) [above=4mm of G] {};
            \node (dots2) [above=3mm of k2] {$\vdots$};
            \node[gauge, label=above:{$\tilde{k}$}] (tk) [right=6mm of E] {};
            \node (dots3) [right=3mm of tk] {$\cdots$};
			\draw[thick] (dots.east) -- (k1.west);
			\draw[thick] (k1.east) -- (A.west);
			\draw[thick,bodyE6] (A.east) -- (B.west);
			\draw[thick,bodyE6] (B.east) -- (C.west);
			\draw[thick,bodyE6] (C.east) -- (D.west);
			\draw[thick,bodyE6] (D.east) -- (E.west);
			\draw[thick,bodyE6] (C.north) -- (F.south);
			\draw[thick,bodyE6] (F.north) -- (G.south);
			\draw[thick] (E.east) -- (tk.west);
            \draw[thick] (tk.east) -- (dots3.west);
            \draw[thick] (G.north) -- (k2.south);
            \draw[thick] (k2.north) -- (dots2.south);
		\end{tikzpicture}
  \label{eq:BHdef-E71tailtoE63tail}
\end{equation}
If, instead, $H\geq B$, the FI deformation must be turned on at a node with rank $\tilde{k}=H-B$ in the tail that ends with $k_2$ in \eqref{eq:generalE7quiv-2tails}. The resulting quiver is
\begin{equation}
    \begin{tikzpicture}[baseline=30,font=\footnotesize]
			\node (dots) {$\cdots$};
			\node[gauge, label=below:{$k_1$}] (k1) [right=3mm of dots] {};
			\node[gauge, bodyE6, label=above:{$A$}] (A) [right=6mm of k1] {};
			\node[gauge, bodyE6, label=below:{$B$}] (B) [right=6mm of A] {};
			\node[gauge, bodyE6, label=below:{$E-\tilde{k}$}] (C) [right=9mm of B] {};
			\node[gauge, bodyE6, label=above:{$D-B-\tilde{k}$}] (D) [right=9mm of C] {};
			\node[gauge, bodyE6, label=below:{$C-B$}] (E) [right=9mm of D] {};
			\node[gauge, bodyE6, label=left:{$F-\tilde{k}$}] (F) [above=4mm of C] {};
			\node[gauge, bodyE6, label=left:{$G-\tilde{k}$}] (G) [above=4mm of F] {};
			\node[gauge, label=left:{$k_2-\tilde{k}$}] (k2) [above=4mm of G] {};
            \node (dots2) [above=3mm of k2] {$\vdots$};
            \node[gauge, label=above:{$\tilde{k}$}] (tk) [right=6mm of E] {};
            \node (dots3) [right=3mm of tk] {$\cdots$};
			\draw[thick] (dots.east) -- (k1.west);
			\draw[thick] (k1.east) -- (A.west);
			\draw[thick,bodyE6] (A.east) -- (B.west);
			\draw[thick,bodyE6] (B.east) -- (C.west);
			\draw[thick,bodyE6] (C.east) -- (D.west);
			\draw[thick,bodyE6] (D.east) -- (E.west);
			\draw[thick,bodyE6] (C.north) -- (F.south);
			\draw[thick,bodyE6] (F.north) -- (G.south);
			\draw[thick] (E.east) -- (tk.west);
            \draw[thick] (tk.east) -- (dots3.west);
            \draw[thick] (G.north) -- (k2.south);
            \draw[thick] (k2.north) -- (dots2.south);
		\end{tikzpicture}
  \label{eq:BHdef-E71tailtoE63tail-2}
\end{equation}

Notice that the vast majority of star-shaped quivers with three tails, including the 3d mirror of $T_N$ theory, are of the form \eqref{eq:BHdef-E71tailtoE63tail-2}. In Section \ref{sec:TN}, we will see that all of them can be realized as the deformation of an $E_7$ quiver.

The above analysis can be repeated verbatim for the other deformations listed in Table \ref{tab:E7toE6-singletail-body}. Interestingly, we observe that the deformation involving the $A$ and $G$ nodes does not generically lead to a star-shaped quiver, except for very special cases. Nevertheless, all other deformations in Table \ref{tab:E7toE6-singletail-body} can be generalized to the two tail case and lead to $E_6$-shaped quivers with three tails, as in \eqref{eq:BHdef-E71tailtoE63tail} and \eqref{eq:BHdef-E71tailtoE63tail-2}.

\subsection{From \texorpdfstring{$E_6$}{E6} Quivers to Generic Star-shaped Quivers}
\label{sec:E6toStarShaped}

We are now ready to discuss how it is possible to obtain any generic star-shaped quiver starting from FI deformations on an $E_6$-shaped quiver.  For pedagogical reasons, let us start by discussing deformations of the $T_{N+1}$ theory and generalize it further.

\subsubsection{Example: the Trinion Theory} 

We first consider a simple example of the $T_{N+1}$ theory, whose mirror theory is
\bes{\label{eq:mirrorTN+1}
    \begin{tikzpicture}[baseline=0,font=\footnotesize]
        \node[gauge,bodyE6,label=below:$N+1$] (C) {};
        \node[gauge,bodyE6,label=right:$N$] (t11) [above=4mm of C] {};
        \node[gauge,bodyE6,label=right:$N-1$] (t12) [above=4mm of t11] {};
        \node (t13) [above=3mm of t12] {$\vdots$};
        \node[gauge,label=right:$1$] (t14) [above=1.5mm of t13] {};
        \node[gauge,bodyE6,label=below:$N$] (t21) [right=9mm of C] {};
        \node[gauge,bodyE6,label=below:$N-1$] (t22) [right=9mm of t21] {};
        \node  (t23) [right=3mm of t22] {$\cdots$};
        \node[gauge,label=below:$1$] (t24) [right=3mm of t23] {};
        \node[gauge,bodyE6,label=below:$N$] (t31) [left=9mm of C] {};
        \node[gauge,bodyE6,label=below:$N-1$] (t32) [left=9mm of t31] {};
        \node  (t33) [left=3mm of t32] {$\cdots$};
        \node[gauge,label=below:$1$] (t34) [left=3mm of t33] {};
        \draw[thick] (t12)--(t13) -- (t14);
        \draw[thick] (t22)--(t23) -- 
        (t24);
        \draw[thick] (t32)--(t33) -- (t34);
        \draw[thick,bodyE6] (C) -- (t11) -- (t12);
        \draw[thick,bodyE6] (C) -- (t21) -- (t22);
        \draw[thick,bodyE6] (C) -- (t31) -- (t32);
    \end{tikzpicture}
}
Turning on the FI parameters with opposite signs, say at the leftmost $\U(1)$ and the rightmost $\U(1)$ nodes, we obtain
\bes{
    \begin{tikzpicture}[baseline=0,font=\footnotesize]
        \node[gauge,bodyE6,label=below:$N$] (C) {};
        \node[gauge,bodyE6,label=right:$N$,fill=green] (t11) [above=4mm of C] {};
        \node[gauge,bodyE6,label=right:$N-1$] (t12) [above=4mm of t11] {};
        \node (t13) [above=3mm of t12] {$\vdots$};
        \node[gauge,label=right:$1$] (t14) [above=1.5mm of t13] {};
        \node[gauge,bodyE6,label=below:$N-1$] (t21) [right=9mm of C] {};
        \node[gauge,bodyE6,label=below:{$N-2$}] (t22) [right=9mm of t21] {};
        \node  (t23) [right=3mm of t22] {$\cdots$};
        \node[gauge,label=below:$1$] (t24) [right=3mm of t23] {};
        \node[gauge,bodyE6,label=below:$N-1$] (t31) [left=9mm of C] {};
        \node[gauge,bodyE6,label=below:{$N-2$}] (t32) [left=9mm of t31] {};
        \node  (t33) [left=3mm of t32] {$\cdots$};
        \node[gauge,label=below:$1$] (t34) [left=3mm of t33] {};
        \node[gauge,label=left:$1$] (t11x) [above=4mm of t31] {};
        \draw[thick] (t11)--(t11x);
        \draw[thick] (t12)--(t13) -- (t14);
        \draw[thick] (t22)--(t23) -- 
        (t24);
        \draw[thick] (t32)--(t33) -- (t34);
        \draw[thick,bodyE6] (C) -- (t11) -- (t12);
        \draw[thick,bodyE6] (C) -- (t21) -- (t22);
        \draw[thick,bodyE6] (C) -- (t31) -- (t32);
    \end{tikzpicture}
}
Upon reiterating this process, namely turning on the FI parameters with opposite signs at the leftmost and rightmost $\U(1)$ nodes, the left and right tails shrink, and there are more and more $\U(1)$ nodes attaching to the green node forming a bouquet. After $N$ deformations, we obtain the quiver with one $T[\SU(N)]$ tail and $(N+1)$ $T_{[N-1,1]}[\SU(N)]$ tails:
\bes{ \label{defTNp1}
    \begin{tikzpicture}[baseline=10,font=\footnotesize]
        \node[gauge,label=right:$N$, fill=green] (t11)  {};
        \node[gauge,label=right:$N-1$] (t12) [above=4mm of t11] {};
        \node (t13) [above=3mm of t12] {$\vdots$};
        \node[gauge,label=right:$2$] (t14) [above=1.5mm of t13] {};
        \node[gauge,label=right:$1$] (t15) [above=4mm of t14] {};
        \node[gauge,label=below:$1$] (t11a) [below left=6mm of t11] {};
        \node[gauge,label=below:$1$] (t12a) [below right=6mm of t11] {};
         \node (t13a) [below=6mm of t11] {$\ldots$};
        \draw[thick] [decorate,decoration={mirror,brace,amplitude=5pt},xshift=0, yshift=0cm]
([yshift=-0.6cm]t11a.west) -- ([yshift=-0.6cm]t12a.east) node [black,midway,below,yshift=-2mm] 
{$N+1\,\U(1)$};
\draw[thick] (t11) -- (t11a) (t11) -- (t12a);
\draw[thick] (t11) -- (t12) -- (t13) -- (t14) -- (t15);
\end{tikzpicture}
}
It is worth pointing out that, in this case, the number of tails, which is $(N+2)$, is equal to the number of deformations plus 2. Note that the mirror theory of \eref{defTNp1} is the linear quiver studied in \cite{Hayashi:2013qwa, Aganagic:2014oia, Bergman:2014kza, Hayashi:2014hfa}:
\bes{
[\SU(N+1)]-\SU(N)-\SU(N-1)-\ldots-\SU(2)-\SU(1)
}

\subsubsection{First Generalization} 

We can generalize this analysis to the following quiver:
\begin{equation}\label{eq:genE6quiv-1}
	\begin{tikzpicture}[baseline=7,font=\footnotesize]
        \node[gauge,label=below:{$N+k$}] (Nk) {};
        \node[gauge,label=right:$N$] (t11) [above=4mm of Nk] {};
        \node (t12) [above=4mm of t11] {$T_{\rho_3}$};
        \node (t21) [right=6mm of Nk] {$Y_2'$};
        \node[gauge,label=below:$k$] (t22) [right=6mm of t21] {};
        \node  (t23) [right=3mm of t22] {$\cdots$};
        \node[gauge,label=below:$1$] (t24) [right=3mm of t23] {};
        \node (t31) [left=6mm of Nk] {$Y_1'$};
        \node[gauge,label=below:$k$] (t32) [left=6mm of t31] {};
        \node  (t33) [left=3mm of t32] {$\cdots$};
        \node[gauge,label=below:$1$] (t34) [left=3mm of t33] {};
        \draw[thick] (Nk) -- (t11) -- (t12);
        \draw[thick] (Nk) -- (t21) -- (t22) -- (t23) -- (t24);
        \draw[thick] (Nk) -- (t31) -- (t32) -- (t33) -- (t34);
	\end{tikzpicture}
\end{equation}
where we have defined $Y_i'$ as $T_{\rho_i}$ theories but with all the ranks shifted by $k$.\footnote{For example, if $\rho_i=\left[1^N\right]$, then the corresponding segment of the tail is $(1+k)-(2+k)-(3+k)-\cdots-[N+k]$, and the $[N+k]$ node is then gauged.} Upon turning on FI deformations with opposite sign on the leftmost $\U(1)$ node and rightmost $\U(1)$ node and reiterating, the resulting quiver is 
\bes{
\begin{tikzpicture}[baseline=0,font=\footnotesize]
        \node[gauge,label=below:{$N$}] (N1) {};
        \node[gauge,label=below:{$N$}] (N2) [right=10mm of N1] {};   \node (Trho2) [below left=10mm of N1] {$T_{\rho_2}$}; \node(Trho1) [above left=10mm of N1] {$T_{\rho_1}$};         \node (Trho3) [right=10mm of N2] {$T_{\rho_3}$};
        \node[gauge,label=above:{$1$}] (1a) [above left=4mm of N2] {};
        \node (1b) [above=5mm of N2] {$\ldots$};
        \node[gauge,label=above:{$1$}] (1c) [above right=4mm of N2] {};
        \draw[thick] (Trho1)--(N1)--(N2)--(Trho3);
        \draw[thick] (Trho2)--(N1);
        \draw[thick] (N2)--(1a);
        \draw[thick] (N2)--(1c);
\end{tikzpicture}
}
If we now further turn on the FI deformation (with opposite sign) at the two nodes of rank $N$, we obtain the final result:
\begin{equation} \label{ssqT1T2T3kU1}
    \begin{tikzpicture}[baseline=0,font=\footnotesize]
        \node[gauge,label=below:{$N$}] (Nk) {};
        \node[gauge,label=below:$1$] (t11) [below left=6mm of Nk] {};
        \node[gauge,label=below:$1$] (t12) [below right=6mm of Nk] {};
         \node (t13) [below=6mm of Nk] {$\cdots$};
        \node (trho3) [above=6mm of Nk] {$T_{\rho_3}$};
        \node (trho1) [left=9mm of Nk] {$T_{\rho_1}$};
        \node (trho2) [right=9mm of Nk] {$T_{\rho_2}$};
        \draw[thick] [decorate,decoration={mirror,brace,amplitude=5pt},xshift=0, yshift=0cm]
([yshift=-0.6cm]t11.west) -- ([yshift=-0.6cm]t12.east) node [black,midway,below,yshift=-2mm] 
{$k\,\U(1)$};
\draw[thick] (Nk) -- (t11) (Nk) -- (t12) (Nk) -- (trho1) (Nk) -- (trho2) (Nk) -- (trho3);
	\end{tikzpicture}
\end{equation}

\subsubsection{Further Generalization} 

More generally, we can consider the following $E_6$ quiver
\begin{equation}\label{eq:genE6quiv-2}
	\begin{tikzpicture}[baseline=7,font=\footnotesize]
		\node (T1) {$T_{\rho_1,\rho_2,\ldots,\rho_k}$};
		\node[gauge,label=below:{$kN$}] (Nk) [right=6mm of T1] {};
		\node (T2) [right=6mm of Nk]{$T_{\rho_1',\rho_2',\ldots,\rho_{k'}'}$};
		\node[gauge,label=left:{$N$}] (N) [above=4mm of Nk] {};
		\node (T3) [above=4mm of N] {$T_\rho$};
		\draw[thick] (T1.east) -- (Nk.west);
		\draw[thick] (Nk.east) -- (T2.west);
		\draw[thick] (Nk.north) -- (N.south);
		\draw[thick] (N.north) -- (T3.south);
	\end{tikzpicture}
\end{equation}
where we have defined $T_{\rho_1,\rho_2,\ldots,\rho_k}$ as a tail
\begin{equation}\label{eq:Trho123}
	\begin{tikzpicture}[baseline=0,font=\footnotesize]
		\node (T1) {$T_{\rho_1}$};
		\node[gauge,label=below:{$N$}] (N) [right=6mm of T1] {};
		\node (Y2) [right=6mm of N] {$Y_2$};
		\node[gauge,label=below:{$2N$}]  (N2) [right=6mm of Y2] {};
		\node (Y3) [right=6mm of N2] {$Y_3$};
		\node (dots) [right=6mm of Y3] {$\cdots$};
		\node[gauge,label=below:{$(k-1)N$}]  (Nk) [right=6mm of dots] {};
		\node (Yk) [right=6mm of Nk] {$Y_k$};
		\draw[thick] (T1) -- (N) -- (Y2) -- (N2) -- (Y3) -- (dots) -- (Nk) -- (Yk);
	\end{tikzpicture}
\end{equation}
where $Y_i$ are $T_{\rho_i}$ tails but with the rank nodes shifted by $(i-1)N$. For instance, by considering $\rho_i = \left[1^N\right]$, the quiver for $T_{\rho_i}$ is
\begin{equation}
	\begin{tikzpicture}[baseline=0,font=\footnotesize]
		\node[gauge,label=below:{$1$}] (u1) {};
		\node[gauge,label=below:{$2$}] (u2) [right=6mm of u1] {};
		\node (dots) [right=6mm of u2] {$\cdots$};
		\node[gauge,label=below:{$N-1$}] (uN1) [right=6mm of dots]{};
		\node[flavor,label=below:{$N$}] (fN) [right=9mm of uN1] {};
		\draw[thick] (u1) -- (u2) -- (dots) -- (uN1) -- (fN);
	\end{tikzpicture}
\end{equation}
the corresponding $Y_i$ is then
\begin{equation}
	\begin{tikzpicture}[baseline=0,font=\footnotesize]
		\node[gauge,label=below:{$1+(i-1)N$}] (u1) {};
		\node[gauge,label=below:{\qquad $2+(i-1)N$}] (u2) [right=18mm of u1] {};
		\node (dots) [right=15mm of u2]{$\cdots$};
		\node[gauge,label=below:{$iN-1$}] (uM1) [right=9mm of dots]{};
		\node[flavor,label=below:{$iN$}] (fM) [right=9mm of uM1] {};
		\draw[thick] (u1) -- (u2) -- (dots) -- (uM1) -- (fM);
	\end{tikzpicture}
\end{equation}
and the flavor node is gauged with the first node of $Y_{i+1}$. Note that nodes $N$, $2N$, \ldots, $kN$ can possibly be bad. Let us suppose for now that this is not the case. If we turn FI deformations in \eqref{eq:genE6quiv-2} at nodes $N$ and $N$, $2N$ and $2N$, and so on sequentially, we obtain a star-shaped quiver of the form
\begin{equation} \label{genericssq}
	\begin{tikzpicture}[baseline=0,font=\footnotesize]
		\node[gauge,label=left:{$N$}] (N) {};
		\node (T1) [above right=9mm of N] {$T_{\rho_k}$};
		\node (dots1) [above=7mm of N] {$\cdots$};
		\node (Tk) [above left=9mm of N] {$T_{\rho_1}$};
		\node (T1p) [below left=9mm of N] {$T_{\rho_1'}$};
		\node (dots2) [below=7mm of N] {$\cdots$};
		\node (Tkp) [below right=9mm of N] {$T_{\rho_{k'}'}$};
		\node (T3) [right=9mm of N] {$T_{\rho}$};
		\draw[thick] (N) -- (T1);
		\draw[thick] (N) -- (Tk);
		\draw[thick] (N) -- (T1p);
		\draw[thick] (N) -- (Tkp);
		\draw[thick] (N) -- (T3);
    \draw[thick] [decorate,decoration={brace,amplitude=5pt},xshift=0cm, yshift=0cm]
([yshift=5mm]Tk.west) -- ([yshift=5mm]T1.east) node [black,pos=0.5,yshift=0.4cm] 
{$k$ tails};
\draw[thick] [decorate,decoration={mirror,brace,amplitude=5pt},xshift=0cm, yshift=0cm]
([yshift=-5mm]T1p.west) -- ([yshift=-5mm]Tkp.east) node [black,pos=0.5,yshift=-0.5cm] 
{$k'$ tails};
	\end{tikzpicture}
\end{equation}
which is nothing but a generic star-shaped quiver with an arbitrary number of tails. 

The analysis we have carried out so far shows that a vast class of star-shaped quivers originates from $E_6$ quivers via FI deformation. Were it not for the fact that the quiver \eqref{eq:genE6quiv-2} can include bad nodes, what we have done would be enough to conclude that all star-shaped quivers have this property. In the next section, we will explain how to deal with this issue and see that it does not actually affect the main conclusion that all class $\mathcal{S}$ theories on the sphere originate from orbi-instanton theories via mass deformation.

\section{Inverse Algorithm for Determining a Parent Theory}
\label{sec:InvFormulas}
In this section, we discuss the so-called \textit{inverse algorithm}, which is a procedure that, given a descendant, allows us to determine a parent theory, including the corresponding $E_8$ orbi-instanton. The general idea is to choose a candidate parent theory, assuming that it comes from one of the mass deformations described in Section \ref{sec:MassDef}. In general, this candidate parent quiver may contain a gauge node that is underbalanced, rendering the quiver bad. We then deal with this by dualizing the underbalanced node using the prescription provided in \cite{Yaakov:2013fza}, also reviewed in Appendix \ref{sec:Itamarduality}. This dualization is always possible since there are FI parameters turned on precisely at the underbalanced nodes in question.  This process can be repeated until we obtain a good quiver, which is then a proper magnetic quiver of the parent theory. We emphasize that the parent theory determined by this method is by no means unique. Of course, there can be many other parent theories which, upon various deformations, flow to the same descendant. Our approach determines {\it a parent theory} out of many possibilities, given a descendant.  Moreover, it does not matter which candidate parent theory we pick, the end result after a series of dualizations of the underbalanced nodes will give a parent theory.

\subsection{Uplifting \texorpdfstring{$E_7$}{E7} Quivers to \texorpdfstring{$E_8$}{E8} Quivers}
\label{sec:InvFormulas-E7E8}
We discussed in Section \ref{sec:MassDefE8E7}, in particular in \cref{tab:E8toE7-singletail-body,tab:E8toE7-2tail-body}, how to obtain various $E_7$ descendants from an $E_8$ theory using FI deformations.  Let us now ask a reversed question: namely, given an $E_7$-shaped quiver, how we can determine a parent $E_8$-shaped quiver that, upon deformations, flows to the given $E_7$ quiver in question.

Let us first consider $E_7$ quivers with a single tail, i.e.
\begin{equation}
	\begin{tikzpicture}[baseline=7,font=\footnotesize]
		\node (dots) {$\cdots$};
		\node[gauge, label=below:{$k$}] (k) [right=6mm of dots] {};
		\node[gauge, bodyE7, label=below:{$A$}] (A) [right=6mm of k] {};
		\node[gauge, bodyE7, label=below:{$B$}] (B) [right=6mm of A] {};
		\node[gauge, bodyE7, label=below:{$C$}] (C) [right=6mm of B] {};
		\node[gauge, bodyE7, label=below:{$D$}] (D) [right=6mm of C] {};
		\node[gauge, bodyE7, label=below:{$E$}] (E) [right=6mm of D] {};
		\node[gauge, bodyE7, label=below:{$F$}] (F) [right=6mm of E] {};
		\node[gauge, bodyE7, label=below:{$G$}] (G) [right=6mm of F] {};
		\node[gauge, bodyE7, label=above:{$H$}] (FG) [above=4mm of D] {};
		\draw[thick] (dots.east) -- (k.west);
		\draw[thick] (k.east) -- (A.west);
		\draw[thick,bodyE7] (A.east) -- (B.west);
		\draw[thick,bodyE7] (B.east) -- (C.west);
		\draw[thick,bodyE7] (C.east) -- (D.west);
		\draw[thick,bodyE7] (D.east) -- (E.west);
		\draw[thick,bodyE7] (E.east) -- (F.west);
		\draw[thick,bodyE7] (F.east) -- (G.west);
		\draw[thick,bodyE7] (D.north) -- (FG.south);
	\end{tikzpicture}
	\label{eq:E71tail}
\end{equation}
A practical way to determine its $E_8$-shaped parent quiver is to choose one of the quivers listed in Table \ref{tab:E7toE8-singletail-body} to be a candidate. For definiteness, let us choose the one corresponding to the $n_6$ lift:
\begin{equation}
	\begin{tikzpicture}[baseline=7,font=\footnotesize]
		\node (dots) {$\cdots$};
		\node[gauge, label=below:{$k$}] (k) [right=6mm of dots] {};
		\node[gauge, bodyE8, label=below:{$A$}] (A) [right=6mm of k] {};
		\node[gauge, bodyE8, label=below:{$B$}] (B) [right=6mm of A] {};
		\node[gauge, bodyE8, label=below:{$C$}] (C) [right=6mm of B] {};
		\node[gauge, bodyE8, label=below:{$D$}] (D) [right=6mm of C] {};
		\node[gauge, bodyE8, label=above:{$E+H$}] (E) [right=6mm of D] {};
		\node[gauge, bodyE8, label=below:{$E+G+H$}] (F) [right=9mm of E] {};
		\node[gauge, bodyE8, label=above:{$E+G$}] (G) [right=9mm of F] {};
		\node[gauge, bodyE8, label=below:{$F$}] (H) [right=6mm of G] {};
		\node[gauge, bodyE8, label=above:{$G+H$}] (FEG) [above=4mm of F] {};
		\draw[thick] (dots.east) -- (k.west);
		\draw[thick] (k.east) -- (A.west);
		\draw[thick,bodyE8] (A.east) -- (B.west);
		\draw[thick,bodyE8] (B.east) -- (C.west);
		\draw[thick,bodyE8] (C.east) -- (D.west);
		\draw[thick,bodyE8] (D.east) -- (E.west);
		\draw[thick,bodyE8] (E.east) -- (F.west);
		\draw[thick,bodyE8] (F.east) -- (G.west);
		\draw[thick,bodyE8] (G.east) -- (H.west);
		\draw[thick,bodyE8] (F.north) -- (FEG.south);
	\end{tikzpicture}
\label{candidateE81} 
\end{equation} 
In order for \eref{candidateE81} to be a good quiver, the following conditions must be satisfied:
\bes{
\mathtt{e}_G\leq \mathtt{e}_H+\mathtt{e}_E~, \quad \text{for \eref{eq:generalE8quiv}}~.
}
If the above inequality is violated, then our candidate parent theory \eref{candidateE81} contains an underbalanced node and is not a good theory. We can then perform a series of dualizations until we obtain a good quiver. 

We list in \cref{tab:E7toE8-singletail-body} the parent $E_8$ theories obtained by inverting the quivers in \cref{tab:E8toE7-singletail-body}. The condition stated in the second column means that, if it is satisfied, there is no underbalanced node and no dualization is required. As emphasized above, in practice, we can use any uplift stated in \cref{tab:E7toE8-singletail-body} and, if there is an underbalanced node, one simply needs to dualize it. In other words, there is no preferred uplift; however, there exists a minimal path from which one can start, and by dualizing the quiver one obtains all the other possible uplifts. The path is 
\begin{equation}
	\begin{tikzpicture}[baseline=0,font=\footnotesize]
		\def\x{1.75}
		\node (n2p) at (-2,2) {Lift \ref{case:n2p}};
		\node (n4p) at (-1,1) {Lift \ref{case:n4p}};
		\node (n3p) at (-1,-1) {Lift \ref{case:n3p}};
		\node (n6) at (0,0) {Lift \ref{case:n6}};
		\node (n5) at (1.1*\x,0) {Lift \ref{case:n5}};
		\node (n4) at (2.2*\x,0) {Lift \ref{case:n4}};
		\node (n3) at (3.3*\x,0) {Lift \ref{case:n3}};
		\node (n2) at (4.4*\x,0) {Lift \ref{case:n2}};
		\node (n1) at (5.5*\x,0) {Lift \ref{case:n1}};
		\draw[-Triangle] (n2p) -- (n4p);
		\draw[-Triangle] (n4p) -- (n6);
		\draw[-Triangle] (n6) -- (n5);
		\draw[-Triangle] (n5) -- (n4);
		\draw[-Triangle] (n4) -- (n3);
		\draw[-Triangle] (n3) -- (n2);
		\draw[-Triangle] (n2) -- (n1);
		\draw[-Triangle] (n3p) -- (n6);
	\end{tikzpicture}
\end{equation}
Note that, if none of the conditions listed in Table \ref{tab:E7toE8-singletail-body} is satisfied, such as in the following quiver 
\begin{equation}
	\begin{tikzpicture}[baseline=7,font=\footnotesize]
		\node (dots) {$\cdots$};
		\node[gauge, label=below:{$k'$}] (kp) [right=6mm of dots] {};
		\node[gauge, label=below:{$k$}] (k) [right=6mm of kp] {};
		\node[gauge, bodyE7, label=below:{$A$}] (A) [right=6mm of k] {};
		\node[gauge, bodyE7, label=below:{$B$}] (B) [right=6mm of A] {};
		\node[gauge, bodyE7, label=below:{$C$}] (C) [right=6mm of B] {};
		\node[gauge, bodyE7, label=below:{$D$}] (D) [right=6mm of C] {};
		\node[gauge, bodyE7, label=below:{$E$}] (E) [right=6mm of D] {};
		\node[gauge, bodyE7, label=below:{$F$}] (F) [right=6mm of E] {};
		\node[gauge, bodyE7, label=below:{$G$}] (G) [right=6mm of F] {};
		\node[gauge, bodyE7, label=above:{$H$}] (FG) [above=4mm of D] {};
		\draw[thick] (dots.east) -- (kp.west);
		\draw[thick] (kp.east) -- (k.west);
		\draw[thick] (k.east) -- (A.west);
		\draw[thick,bodyE7] (A.east) -- (B.west);
		\draw[thick,bodyE7] (B.east) -- (C.west);
		\draw[thick,bodyE7] (C.east) -- (D.west);
		\draw[thick,bodyE7] (D.east) -- (E.west);
		\draw[thick,bodyE7] (E.east) -- (F.west);
		\draw[thick,bodyE7] (F.east) -- (G.west);
		\draw[thick,bodyE7] (D.north) -- (FG.south);
	\end{tikzpicture}
\end{equation} with  
\begin{equation}
		A>k+G \coma \qquad  k>G \text{ and } k'<k-G~,
\end{equation}
the dualization of any candidate $E_8$ parent quiver chosen from \cref{tab:E7toE8-singletail-body} will propagate to the tail, meaning that there is a tail deformation involved upon going from the $E_8$-shaped quiver to $E_7$-shaped quiver.

	\begin{landscape}
		\pagestyle{empty}
		\begin{center}
			\renewcommand{\arraystretch}{1.1}
			\setlength{\LTcapwidth}{1.2\textwidth}
			\begin{longtable}{c|c|c}
				\caption{Possible uplifts to $E_8$-shaped parent quivers of \eqref{eq:E71tail}. We list the nodes that are assumed to be balanced, where the FI deformations are turned on, the conditions for goodness of the $E_8$ quiver, and the corresponding $E_8$ quiver.} \label{tab:E7toE8-singletail-body}\\
                Lift & Condition for \eref{eq:generalE8quiv} & $E_8$ Quiver \\
				\hline
                \endfirsthead 
				Lift & Conditions & $E_8$ Quiver \\
				\hline
				\endhead
				\crtcrossreflabel{$n_{2'}$}[case:n2p] & $F\geq H$ & 
				\begin{tikzpicture}[baseline=0,font=\footnotesize]
					\node (dots) {$\cdots$};
					\node[gauge, label=below:{$k$}] (k) [right=6mm of dots] {};
					\node[gauge, bodyE8, label=below:{$A$}] (A) [right=6mm of k] {};
					\node[gauge, bodyE8, label=below:{$B$}] (B) [right=6mm of A] {};
					\node[gauge, bodyE8, label=below:{$C$}] (C) [right=6mm of B] {};
					\node[gauge, bodyE8, label=below:{$D$}] (D) [right=6mm of C] {};
					\node[gauge, bodyE8, label=above:{$E+H$}] (E) [right=6mm of D] {};
					\node[gauge, bodyE8, label=below:{$F+2H$}] (F) [right=9mm of E] {};
					\node[gauge, bodyE8, label=above:{$2H$}] (G) [right=9mm of F] {};
					\node[gauge, bodyE8, label=below:{$H$}] (H) [right=6mm of G] {};
					\node[gauge, bodyE8, label=above:{$G+H$}] (FEG) [above=4mm of F] {};
					\draw[thick] (dots.east) -- (k.west);
					\draw[thick] (k.east) -- (A.west);
					\draw[thick,bodyE8] (A.east) -- (B.west);
					\draw[thick,bodyE8] (B.east) -- (C.west);
					\draw[thick,bodyE8] (C.east) -- (D.west);
					\draw[thick,bodyE8] (D.east) -- (E.west);
					\draw[thick,bodyE8] (E.east) -- (F.west);
					\draw[thick,bodyE8] (F.east) -- (G.west);
					\draw[thick,bodyE8] (G.east) -- (H.west);
					\draw[thick,bodyE8] (F.north) -- (FEG.south);
				\end{tikzpicture} \\ \hline
				\crtcrossreflabel{$n_{3'}$}[case:n3p] & $\texttt{e}_G\leq \texttt{e}_E-\texttt{e}_H$ & 
				\begin{tikzpicture}[baseline=0,font=\footnotesize]
					\node (dots) {$\cdots$};
					\node[gauge, label=below:{$k$}] (k) [right=6mm of dots] {};
					\node[gauge, bodyE8, label=below:{$A$}] (A) [right=6mm of k] {};
					\node[gauge, bodyE8, label=below:{$B$}] (B) [right=6mm of A] {};
					\node[gauge, bodyE8, label=below:{$C$}] (C) [right=6mm of B] {};
					\node[gauge, bodyE8, label=below:{$D$}] (D) [right=6mm of C] {};
					\node[gauge, bodyE8, label=above:{$E+H$}] (E) [right=6mm of D] {};
					\node[gauge, bodyE8, label=below:{$2E$}] (F) [right=9mm of E] {};
					\node[gauge, bodyE8, label=above:{$G+E$}] (G) [right=9mm of F] {};
					\node[gauge, bodyE8, label=below:{$F$}] (H) [right=6mm of G] {};
					\node[gauge, bodyE8, label=above:{$E$}] (FEG) [above=4mm of F] {};
					\draw[thick] (dots.east) -- (k.west);
					\draw[thick] (k.east) -- (A.west);
					\draw[thick,bodyE8] (A.east) -- (B.west);
					\draw[thick,bodyE8] (B.east) -- (C.west);
					\draw[thick,bodyE8] (C.east) -- (D.west);
					\draw[thick,bodyE8] (D.east) -- (E.west);
					\draw[thick,bodyE8] (E.east) -- (F.west);
					\draw[thick,bodyE8] (F.east) -- (G.west);
					\draw[thick,bodyE8] (G.east) -- (H.west);
					\draw[thick,bodyE8] (F.north) -- (FEG.south);
				\end{tikzpicture} \\ \hline
				\crtcrossreflabel{$n_{4'}$}[case:n4p] & $\texttt{e}_G\leq \texttt{e}_H-\texttt{e}_E$ & 
				\begin{tikzpicture}[baseline=0,font=\footnotesize]
					\node (dots) {$\cdots$};
					\node[gauge, label=below:{$k$}] (k) [right=6mm of dots] {};
					\node[gauge, bodyE8, label=below:{$A$}] (A) [right=6mm of k] {};
					\node[gauge, bodyE8, label=below:{$B$}] (B) [right=6mm of A] {};
					\node[gauge, bodyE8, label=below:{$C$}] (C) [right=6mm of B] {};
					\node[gauge, bodyE8, label=below:{$D$}] (D) [right=6mm of C] {};
					\node[gauge, bodyE8, label=above:{$E+H$}] (E) [right=6mm of D] {};
					\node[gauge, bodyE8, label=below:{$F+2H$}] (F) [right=9mm of E] {};
					\node[gauge, bodyE8, label=above:{$F+H$}] (G) [right=9mm of F] {};
					\node[gauge, bodyE8, label=below:{$F$}] (H) [right=6mm of G] {};
					\node[gauge, bodyE8, label=above:{$G+H$}] (FEG) [above=4mm of F] {};
					\draw[thick] (dots.east) -- (k.west);
					\draw[thick] (k.east) -- (A.west);
					\draw[thick,bodyE8] (A.east) -- (B.west);
					\draw[thick,bodyE8] (B.east) -- (C.west);
					\draw[thick,bodyE8] (C.east) -- (D.west);
					\draw[thick,bodyE8] (D.east) -- (E.west);
					\draw[thick,bodyE8] (E.east) -- (F.west);
					\draw[thick,bodyE8] (F.east) -- (G.west);
					\draw[thick,bodyE8] (G.east) -- (H.west);
					\draw[thick,bodyE8] (F.north) -- (FEG.south);
				\end{tikzpicture} \\ \hline
				\crtcrossreflabel{$n_{6}$}[case:n6] & $\texttt{e}_G\leq \texttt{e}_H+\texttt{e}_E$ & 
				\begin{tikzpicture}[baseline=0,font=\footnotesize]
					\node (dots) {$\cdots$};
					\node[gauge, label=below:{$k$}] (k) [right=6mm of dots] {};
					\node[gauge, bodyE8, label=below:{$A$}] (A) [right=6mm of k] {};
					\node[gauge, bodyE8, label=below:{$B$}] (B) [right=6mm of A] {};
					\node[gauge, bodyE8, label=below:{$C$}] (C) [right=6mm of B] {};
					\node[gauge, bodyE8, label=below:{$D$}] (D) [right=6mm of C] {};
					\node[gauge, bodyE8, label=above:{$E+H$}] (E) [right=6mm of D] {};
					\node[gauge, bodyE8, label=below:{$E+G+H$}] (F) [right=9mm of E] {};
					\node[gauge, bodyE8, label=above:{$E+G$}] (G) [right=9mm of F] {};
					\node[gauge, bodyE8, label=below:{$F$}] (H) [right=6mm of G] {};
					\node[gauge, bodyE8, label=above:{$G+H$}] (FEG) [above=4mm of F] {};
					\draw[thick] (dots.east) -- (k.west);
					\draw[thick] (k.east) -- (A.west);
					\draw[thick,bodyE8] (A.east) -- (B.west);
					\draw[thick,bodyE8] (B.east) -- (C.west);
					\draw[thick,bodyE8] (C.east) -- (D.west);
					\draw[thick,bodyE8] (D.east) -- (E.west);
					\draw[thick,bodyE8] (E.east) -- (F.west);
					\draw[thick,bodyE8] (F.east) -- (G.west);
					\draw[thick,bodyE8] (G.east) -- (H.west);
					\draw[thick,bodyE8] (F.north) -- (FEG.south);
				\end{tikzpicture} \\ \hline
				\crtcrossreflabel{$n_{5}$}[case:n5] & $\texttt{e}_G\leq \texttt{e}_H+\texttt{e}_E+2\texttt{e}_D$ & 
				\begin{tikzpicture}[baseline=0,font=\footnotesize]
					\node (dots) {$\cdots$};
					\node[gauge, label=below:{$k$}] (k) [right=6mm of dots] {};
					\node[gauge, bodyE8, label=below:{$A$}] (A) [right=6mm of k] {};
					\node[gauge, bodyE8, label=below:{$B$}] (B) [right=6mm of A] {};
					\node[gauge, bodyE8, label=below:{$C$}] (C) [right=6mm of B] {};
					\node[gauge, bodyE8, label=below:{$D$}] (D) [right=6mm of C] {};
					\node[gauge, bodyE8, label=above:{$D+G$}] (E) [right=6mm of D] {};
					\node[gauge, bodyE8, label=below:{$D+2G$}] (F) [right=9mm of E] {};
					\node[gauge, bodyE8, label=above:{$E+G$}] (G) [right=9mm of F] {};
					\node[gauge, bodyE8, label=below:{$F$}] (H) [right=6mm of G] {};
					\node[gauge, bodyE8, label=above:{$G+H$}] (FEG) [above=4mm of F] {};
					\draw[thick] (dots.east) -- (k.west);
					\draw[thick] (k.east) -- (A.west);
					\draw[thick,bodyE8] (A.east) -- (B.west);
					\draw[thick,bodyE8] (B.east) -- (C.west);
					\draw[thick,bodyE8] (C.east) -- (D.west);
					\draw[thick,bodyE8] (D.east) -- (E.west);
					\draw[thick,bodyE8] (E.east) -- (F.west);
					\draw[thick,bodyE8] (F.east) -- (G.west);
					\draw[thick,bodyE8] (G.east) -- (H.west);
					\draw[thick,bodyE8] (F.north) -- (FEG.south);
				\end{tikzpicture} \\ \hline
				\crtcrossreflabel{$n_{4}$}[case:n4] & $\texttt{e}_G\leq \texttt{e}_H+\texttt{e}_E+2\texttt{e}_D+2\texttt{e}_C$ &
				\begin{tikzpicture}[baseline=0,font=\footnotesize]
					\node (dots) {$\cdots$};
					\node[gauge, label=below:{$k$}] (k) [right=6mm of dots] {};
					\node[gauge, bodyE8, label=below:{$A$}] (A) [right=6mm of k] {};
					\node[gauge, bodyE8, label=below:{$B$}] (B) [right=6mm of A] {};
					\node[gauge, bodyE8, label=below:{$C$}] (C) [right=6mm of B] {};
					\node[gauge, bodyE8, label=below:{$C+G$}] (D) [right=9mm of C] {};
					\node[gauge, bodyE8, label=above:{$C+2G$}] (E) [right=9mm of D] {};
					\node[gauge, bodyE8, label=below:{$D+2G$}] (F) [right=9mm of E] {};
					\node[gauge, bodyE8, label=above:{$E+G$}] (G) [right=9mm of F] {};
					\node[gauge, bodyE8, label=below:{$F$}] (H) [right=6mm of G] {};
					\node[gauge, bodyE8, label=above:{$G+H$}] (FEG) [above=4mm of F] {};
					\draw[thick] (dots.east) -- (k.west);
					\draw[thick] (k.east) -- (A.west);
					\draw[thick,bodyE8] (A.east) -- (B.west);
					\draw[thick,bodyE8] (B.east) -- (C.west);
					\draw[thick,bodyE8] (C.east) -- (D.west);
					\draw[thick,bodyE8] (D.east) -- (E.west);
					\draw[thick,bodyE8] (E.east) -- (F.west);
					\draw[thick,bodyE8] (F.east) -- (G.west);
					\draw[thick,bodyE8] (G.east) -- (H.west);
					\draw[thick,bodyE8] (F.north) -- (FEG.south);
				\end{tikzpicture} \\ \hline
				\crtcrossreflabel{$n_{3}$}[case:n3] & $\texttt{e}_G\leq \texttt{e}_H+\texttt{e}_E+2\texttt{e}_D+2\texttt{e}_C+2\texttt{e}_B$ & 
				\begin{tikzpicture}[baseline=0,font=\footnotesize]
					\node (dots) {$\cdots$};
					\node[gauge, label=below:{$k$}] (k) [right=6mm of dots] {};
					\node[gauge, bodyE8, label=below:{$A$}] (A) [right=6mm of k] {};
					\node[gauge, bodyE8, label=below:{$B$}] (B) [right=6mm of A] {};
					\node[gauge, bodyE8, label=above:{$B+G$}] (C) [right=9mm of B] {};
					\node[gauge, bodyE8, label=below:{$B+2G$}] (D) [right=9mm of C] {};
					\node[gauge, bodyE8, label=above:{$C+2G$}] (E) [right=9mm of D] {};
					\node[gauge, bodyE8, label=below:{$D+2G$}] (F) [right=9mm of E] {};
					\node[gauge, bodyE8, label=above:{$E+G$}] (G) [right=9mm of F] {};
					\node[gauge, bodyE8, label=below:{$F$}] (H) [right=6mm of G] {};
					\node[gauge, bodyE8, label=above:{$G+H$}] (FEG) [above=4mm of F] {};
					\draw[thick] (dots.east) -- (k.west);
					\draw[thick] (k.east) -- (A.west);
					\draw[thick,bodyE8] (A.east) -- (B.west);
					\draw[thick,bodyE8] (B.east) -- (C.west);
					\draw[thick,bodyE8] (C.east) -- (D.west);
					\draw[thick,bodyE8] (D.east) -- (E.west);
					\draw[thick,bodyE8] (E.east) -- (F.west);
					\draw[thick,bodyE8] (F.east) -- (G.west);
					\draw[thick,bodyE8] (G.east) -- (H.west);
					\draw[thick,bodyE8] (F.north) -- (FEG.south);
				\end{tikzpicture}\\ \hline
				\crtcrossreflabel{$n_{2}$}[case:n2] & $\texttt{e}_G\leq \texttt{e}_H+\texttt{e}_E+2\texttt{e}_D+2\texttt{e}_C+2\texttt{e}_B+2\texttt{e}_A$ & 
				\begin{tikzpicture}[baseline=0,font=\footnotesize]
					\node (dots) {$\cdots$};
					\node[gauge, label=below:{$k$}] (k) [right=6mm of dots] {};
					\node[gauge, bodyE8, label=below:{$A$}] (A) [right=6mm of k] {};
					\node[gauge, bodyE8, label=below:{$A+G$}] (B) [right=9mm of A] {};
					\node[gauge, bodyE8, label=above:{$A+2G$}] (C) [right=9mm of B] {};
					\node[gauge, bodyE8, label=below:{$B+2G$}] (D) [right=9mm of C] {};
					\node[gauge, bodyE8, label=above:{$C+2G$}] (E) [right=9mm of D] {};
					\node[gauge, bodyE8, label=below:{$D+2G$}] (F) [right=9mm of E] {};
					\node[gauge, bodyE8, label=above:{$E+G$}] (G) [right=9mm of F] {};
					\node[gauge, bodyE8, label=below:{$F$}] (H) [right=6mm of G] {};
					\node[gauge, bodyE8, label=above:{$G+H$}] (FEG) [above=4mm of F] {};
					\draw[thick] (dots.east) -- (k.west);
					\draw[thick] (k.east) -- (A.west);
					\draw[thick,bodyE8] (A.east) -- (B.west);
					\draw[thick,bodyE8] (B.east) -- (C.west);
					\draw[thick,bodyE8] (C.east) -- (D.west);
					\draw[thick,bodyE8] (D.east) -- (E.west);
					\draw[thick,bodyE8] (E.east) -- (F.west);
					\draw[thick,bodyE8] (F.east) -- (G.west);
					\draw[thick,bodyE8] (G.east) -- (H.west);
					\draw[thick,bodyE8] (F.north) -- (FEG.south);
				\end{tikzpicture} \\ \hline
				\crtcrossreflabel{$n_{1}$}[case:n1] & $A>k+G$ & 
				\begin{tikzpicture}[baseline=0,font=\footnotesize]
					\node (dots) {$\cdots$};
					\node[gauge, label=below:{$k$}] (k) [right=6mm of dots] {};
					\node[gauge, bodyE8, label=above:{$k+G$}] (A) [right=9mm of k] {};
					\node[gauge, bodyE8, label=below:{$k+2G$}] (B) [right=9mm of A] {};
					\node[gauge, bodyE8, label=above:{$A+2G$}] (C) [right=9mm of B] {};
					\node[gauge, bodyE8, label=below:{$B+2G$}] (D) [right=9mm of C] {};
					\node[gauge, bodyE8, label=above:{$C+2G$}] (E) [right=9mm of D] {};
					\node[gauge, bodyE8, label=below:{$D+2G$}] (F) [right=9mm of E] {};
					\node[gauge, bodyE8, label=above:{$E+G$}] (G) [right=9mm of F] {};
					\node[gauge, bodyE8, label=below:{$F$}] (H) [right=6mm of G] {};
					\node[gauge, bodyE8, label=above:{$G+H$}] (FEG) [above=4mm of F] {};
					\draw[thick] (dots.east) -- (k.west);
					\draw[thick] (k.east) -- (A.west);
					\draw[thick,bodyE8] (A.east) -- (B.west);
					\draw[thick,bodyE8] (B.east) -- (C.west);
					\draw[thick,bodyE8] (C.east) -- (D.west);
					\draw[thick,bodyE8] (D.east) -- (E.west);
					\draw[thick,bodyE8] (E.east) -- (F.west);
					\draw[thick,bodyE8] (F.east) -- (G.west);
					\draw[thick,bodyE8] (G.east) -- (H.west);
					\draw[thick,bodyE8] (F.north) -- (FEG.south);
				\end{tikzpicture}  
			\end{longtable}
		\end{center}
	\end{landscape}

\subsubsection{Obtaining 6d Parent Theories and Realizing Deformations} 
\label{sec:procedure}

Given an $E_7$ quiver, we can derive an $E_8$ parent quiver by using any uplift in Table \ref{tab:E7toE8-singletail-body}. If there are any underbalanced nodes, we dualize them until all gauge nodes are balanced or overbalanced. After this process, we can assume that the $E_8$ parent quiver takes the form \eref{magorbi1}. From the latter, we can then derive the F-theory quiver of the corresponding 6d theory at a generic point on the tensor branch using the following procedure.

From the $E_8$ parent quiver \eref{magorbi1}, we can derive the 6d theory at a generic point on the tensor branch by exploiting $E_8$ quiver subtractions \cite{Hanany:2018uhm, Cabrera:2018ann}, corresponding to the small $E_8$ instanton (or Kraft--Procesi) transition. The process is as follows. First, we subtract an $\fm \times E_8$ quiver, depicted in \eref{mtimesE8}, below from the $E_8$ parent quiver \eref{magorbi1}, where $\fm$ is defined around \eref{defri}:
\bes{ \label{mtimesE8}
\fm \vec{d} \,\, = \,\, \begin{array}{cccccccc}
		    &   &   &   &    & 3\fm &   &   \\
	        \fm & 2\fm & 3\fm & 4\fm & 5\fm & 6\fm & 4\fm & 2\fm  \\
	\end{array}
}
Explicitly, this subtraction is
\begin{equation}
\begin{tikzpicture}[baseline=-50,font=\footnotesize]
		\node (dots) {$\cdots$};
        \node[gauge, label=below:{$2s_l$}] (t2) [left=3mm of dots] {};
        \node[gauge, label=below:{$s_l$}] (t1) [left=6mm of t2] {};
        \node[gauge, label=below:{$k-s_1$}] (tkm1) [right=3mm of dots] {};	
		\node[gauge, label=below:{$k$}] (k) [right=12mm of dots] {};
		\node[gauge, bodyE8, label=below:{$N_1$}] (A) [right=6mm of k] {};
		\node[gauge, bodyE8, label=below:{$N_2$}] (B) [right=6mm of A] {};
		\node[gauge, bodyE8, label=below:{$N_3$}] (C) [right=6mm of B] {};
		\node[gauge, bodyE8, label=below:{$N_4$}] (D) [right=6mm of C] {};
		\node[gauge, bodyE8, label=below:{$N_5$}] (E) [right=6mm of D] {};
		\node[gauge, bodyE8, label=below:{$N_6$}] (F) [right=6mm of E] {};
		\node[gauge, bodyE8, label=below:{$N_{4'}$}] (G) [right=6mm of F] {};
		\node[gauge, bodyE8, label=below:{$N_{2'}$}] (H) [right=6mm of G] {};
		\node[gauge, bodyE8, label=above:{$N_{3'}$}] (L) [above=4mm of F] {};
        \draw[thick] (t1) -- (t2) -- (dots) -- (tkm1) -- (k) -- (A);
        \draw[thick,bodyE8] (A) -- (B) -- (C) -- (D) -- (E) -- (F) -- (G) -- (H);
        \draw[thick,bodyE8] (F) -- (L);
        
		\node[gauge, label=below:{$\fm$}] (I) [below=16mm of A] {};
		\node[gauge, label=below:{$2\fm$}] (J) [right=6mm of I] {};
		\node[gauge, label=below:{$3\fm$}] (K) [right=6mm of J] {};
		\node[gauge, label=below:{$4\fm$}] (L) [right=6mm of K] {};
		\node[gauge, label=below:{$5\fm$}] (M) [right=6mm of L] {};
		\node[gauge, label=below:{$6\fm$}] (N) [right=6mm of M] {};
		\node[gauge, label=below:{$4\fm$}] (O) [right=6mm of N] {};
		\node[gauge, label=below:{$2\fm$}] (P) [right=6mm of O] {};
		\node[gauge, label=above:{$3\fm$}] (Q) [above=4mm of N] {};
        \draw[thick] (I) -- (J) -- (K) -- (L) -- (M) -- (N) -- (O) -- (P);
        \draw[thick] (N) -- (Q);
        
		\node (dots2) [below=32mm of dots] {$\cdots$};
        \node[gauge, label=below:{$2s_l$}] (t3) [left=3mm of dots2] {};
        \node[gauge, label=below:{$s_l$}] (t4) [left=6mm of t3] {};
        \node[gauge, label=below:{$k-s_1$}] (tkm2) [right=3mm of dots2] {};	
		\node[gauge, label=below:{$k$}] (k2) [right=12mm of dots2] {};
		\node[gauge, bodyE8, label=below:{$\fr_1$}] (R) [right=6mm of k2] {};
		\node[gauge, bodyE8, label=below:{$\fr_2$}] (S) [right=6mm of R] {};
		\node[gauge, bodyE8, label=below:{$\fr_3$}] (T) [right=6mm of S] {};
		\node[gauge, bodyE8, label=below:{$\fr_4$}] (U) [right=6mm of T] {};
		\node[gauge, bodyE8, label=below:{$\fr_5$}] (V) [right=6mm of U] {};
		\node[gauge, bodyE8, label=below:{$\fr_6$}] (W) [right=6mm of V] {};
		\node[gauge, bodyE8, label=below:{$\fr_{4'}$}] (X) [right=6mm of W] {};
		\node[gauge, bodyE8, label=below:{$\fr_{2'}$}] (Y) [right=6mm of X] {};
		\node[gauge, bodyE8, label=above:{$\fr_{3'}$}] (Z) [above=4mm of W] {};
        \draw[thick] (t4) -- (t3) -- (dots2) -- (tkm2) -- (k2) -- (R);
        \draw[thick,bodyE8] (R) -- (S) -- (T) -- (U) -- (V) -- (W) -- (X) -- (Y);
        \draw[thick,bodyE8] (W) -- (Z);

        \node (equal) [left=3mm of t4] {$=$};
        \node (minus) [above=16mm of equal] {$-$};
    \end{tikzpicture}
\end{equation}
where $\fr_i$ are defined as in \eref{defri}.

Next, we rebalance the resulting quiver by adding $\fm$ $\U(1)$ gauge nodes to the $\U(k)$ node such that there is a hypermultiplet in the bifundamental representation between $\U(k)$ and each $\U(1)$ node. The resulting quiver after subtraction and rebalancing is
\begin{equation} \label{E8subreb}
    \begin{tikzpicture}[baseline=7,font=\footnotesize]
		\node (dots) {$\cdots$};
        \node[gauge, label=below:{$2s_l$}] (t2) [left=3mm of dots] {};
        \node[gauge, label=below:{$s_l$}] (t1) [left=6mm of t2] {};
        \node[gauge, label=below:{$k-s_1$}] (tkm1) [right=3mm of dots] {};	
		\node[gauge, label=below:{$k$}] (k) [right=12mm of dots] {};
		\node[gauge, bodyE8, label=below:{$\fr_1$}] (A) [right=6mm of k] {};
		\node[gauge, bodyE8, label=below:{$\fr_2$}] (B) [right=6mm of A] {};
		\node[gauge, bodyE8, label=below:{$\fr_3$}] (C) [right=6mm of B] {};
		\node[gauge, bodyE8, label=below:{$\fr_4$}] (D) [right=6mm of C] {};
		\node[gauge, bodyE8, label=below:{$\fr_5$}] (E) [right=6mm of D] {};
		\node[gauge, bodyE8, label=below:{$\fr_6$}] (F) [right=6mm of E] {};
		\node[gauge, bodyE8, label=below:{$\fr_{4'}$}] (G) [right=6mm of F] {};
		\node[gauge, bodyE8, label=below:{$\fr_{2'}$}] (H) [right=6mm of G] {};
		\node[gauge, bodyE8, label=above:{$\fr_{3'}$}] (L) [above=4mm of F] {};
            \node[gauge, label=above:{$1$}] (kk) [above left=5mm of k] {};
		\node[gauge, label=above:{$1$}] (kkk) [above right=5mm of k] {};
            \node (xxx) [above=2mm of k] {$\cdots$};
        \draw[thick] (t1) -- (t2) -- (dots) -- (tkm1) -- (k) -- (A);
        \draw[thick] (k)--(kk);
        \draw[thick] (k)--(kkk);
        \draw[thick,bodyE8] (A) -- (B) -- (C) -- (D) -- (E) -- (F) -- (G) -- (H);
        \draw[thick,bodyE8] (F) -- (L);
        \draw[thick] [decorate,decoration={brace,amplitude=5pt},xshift=0cm, yshift=0cm]
([yshift=7mm]kk.west) -- ([yshift=7mm]kkk.east) node [black,midway,yshift=0.4cm] 
{\footnotesize $\fm$};
	\end{tikzpicture}
\end{equation}
Note that this quiver has an overall $\U(1)$ that needs to be decoupled. 

Next, we compute the mirror theory of the quiver we obtained. This contains information about the gauge group and how hypermultiplets transform in the 6d tensor branch description. Finally, to complete the process, we insert the $(-1)$- and $(-2)$-curves in the 6d F-theory quiver in such a way that 
\bi
\item the $E_8$ global symmetry (or its subgroup) on the Coulomb branch of the original $E_8$ parent quiver \eref{magorbi1} is correctly present next to $(-1)$-curve, and
\item the gauge anomalies in the 6d theory are canceled on each curve. 
\ei

A convenient approach to see the mass deformation from the theory associated with the $E_8$ parent theory to that associated with the $E_7$ theory is by performing the $E_7$ quiver subtraction and rebalancing on the latter. The process is very similar to that discussed before, with the $\fm \times E_8$ quiver replaced by an $\fm \times E_7$ quiver:
\bes{
\begin{array}{ccccccc}
		     &   &    & 2\fm &   &  &   \\
	           \fm & 2\fm & 3\fm & 4\fm & 3\fm & 2\fm & \fm  \\
	\end{array}
}
The result will take the following form:
\begin{equation} \label{E7subreb}
    \begin{tikzpicture}[baseline=7,font=\footnotesize]
		\node (dots) {$\cdots$};
        \node[gauge, label=below:{$2s_l$}] (t2) [left=3mm of dots] {};
        \node[gauge, label=below:{$s_l$}] (t1) [left=6mm of t2] {};
        \node[gauge, label=below:{$k-s_1$}] (tkm1) [right=3mm of dots] {};	
		\node[gauge, label=below:{$k$}] (k) [right=12mm of dots] {};
		\node[gauge,bodyE7, label=below:{$\alpha_1$}] (A) [right=6mm of k] {};
		\node[gauge,bodyE7, label=below:{$\alpha_2$}] (B) [right=6mm of A] {};
		\node[gauge,bodyE7, label=below:{$\alpha_3$}] (C) [right=6mm of B] {};
		\node[gauge,bodyE7, label=below:{$\alpha_4$}] (D) [right=6mm of C] {};
		\node[gauge,bodyE7, label=below:{$\alpha_{3'}$}] (E) [right=6mm of D] {};
		\node[gauge,bodyE7, label=below:{$\alpha_{2'}$}] (F) [right=6mm of E] {};
		\node[gauge,bodyE7, label=below:{$\alpha_{1'}$}] (G) [right=6mm of F] {};
		\node[gauge,bodyE7, label=above:{$\alpha_{2''}$}] (FG) [above=4mm of D] {};
		\draw[thick,bodyE7] (A.east) -- (B.west);
		\draw[thick,bodyE7] (B.east) -- (C.west);
		\draw[thick,bodyE7] (C.east) -- (D.west);
		\draw[thick,bodyE7] (D.east) -- (E.west);
		\draw[thick,bodyE7] (E.east) -- (F.west);
		\draw[thick,bodyE7] (F.east) -- (G.west);
		\draw[thick,bodyE7] (D.north) -- (FG.south);
        \node[gauge, label=above:{$1$}] (kk) [above left=5mm of k] {};
		\node[gauge, label=above:{$1$}] (kkk) [above right=5mm of k] {};
        \node (xxx) [above=2mm of k] {$\cdots$};
        \draw[thick] (t1) -- (t2) -- (dots) -- (tkm1) -- (k) -- (A);
        \draw[thick] (k)--(kk);
        \draw[thick] (k)--(kkk);
        \draw[thick] [decorate,decoration={brace,amplitude=5pt},xshift=0cm, yshift=0cm]
([yshift=7mm]kk.west) -- ([yshift=7mm]kkk.east) node [black,midway,yshift=0.4cm] 
{\footnotesize $\fm$};
	\end{tikzpicture}
\end{equation}
After determining the mirror theory of \eref{E7subreb}, we see how the theory associated with the $E_8$ parent theory is mass deformed as follows:
\bi
\item If the mirror dual of \eref{E7subreb} is the same as that of \eref{E8subreb}, this means that, after taking into account the Kraft--Procesi transitions of the $E_7$ and $E_8$ small transitions in each theory, the resulting theories are the same. This also means that only the flavor symmetry attached to the $(-1)$-curve, but not those attached to the $(-2)$-curves, of the 6d parent theory gets mass-deformed. The change of the flavor symmetries upon deformation can be seen by contrasting the Coulomb branch symmetries of the $E_7$ quiver and the $E_8$ parent quiver. This can be read off from the sub-Dynkin diagrams after removing the overbalanced nodes, as instructed in \cite{Gaiotto:2008ak} (see also \cite{Mekareeya:2015bla, Mekareeya:2017jgc}). 
\item If the mirror dual of \eref{E7subreb} is not the same as that of \eref{E8subreb}, it is clear which hypermultiplets or flavor symmetries in the parent theory get mass-deformed.
\ei

\subsubsection{Examples: \texorpdfstring{$E_7$}{E7} Quivers with Four-dimensional Higgs Branch and One Tail} 
\label{sec:rank4onetail}

Let us list the $E_7$ quivers, containing one tail, with a four-dimensional Higgs branch, where the rank of the central node is less than or equal to $12$, below. 
\begin{longtable}{c|c|c|c}
	Number & $E_7$ Quiver & $E_8$ Parent Quiver & 6d Parent Theory   \\
	\hline
 \endhead
	 \hypertarget{rank4-th1}{} $1$ &
	$\begin{array}{ccc @{} >{\color{cE7}}c >{\color{cE7}}c >{\color{cE7}}c >{\color{cE7}}c >{\color{cE7}}c >{\color{cE7}}c >{\color{cE7}}c @{}}
		  &   &   &   &   &   & 6 &   &   &   \\
		1 & 2 &   & 3 & 6 & 9 & 12 & 9 & 6 & 3 \\
	\end{array}$
	& 
	$\begin{array}{ccc @{} >{\color{cE8}}c >{\color{cE8}}c >{\color{cE8}}c >{\color{cE8}}c >{\color{cE8}}c >{\color{cE8}}c >{\color{cE8}}c >{\color{cE8}}c @{}}
		  &   &   &   &   &   &   &   & 9 &   &   \\
		1 & 2 &   & 3 & 6 & 9 & 12 & 15 & 18 & 12 & 6 \\
	\end{array}$
	& $[\mathfrak{e}_8] \,\, 1\,\, \overset{\su(1)}{2} \,\, \overset{\su(2)}{2} \,\, [\su(3)]$  \hypertarget{rank4-th2}{}  \\
	\hline
 $2$ & 
	$\begin{array}{ccc @{} >{\color{cE7}}c >{\color{cE7}}c >{\color{cE7}}c >{\color{cE7}}c >{\color{cE7}}c >{\color{cE7}}c >{\color{cE7}}c @{}}
		  &   &   &   &   &   & 5 &   &   &   \\
		1 & 2 &   & 4 & 6 & 8 & 10 & 7 & 4 & 2 \\
	\end{array}$
	& 
	$\begin{array}{ccc @{} >{\color{cE8}}c >{\color{cE8}}c >{\color{cE8}}c >{\color{cE8}}c >{\color{cE8}}c >{\color{cE8}}c >{\color{cE8}}c >{\color{cE8}}c @{}}
		  &   &   &   &   &   &   &   & 7 &   &   \\
		1 & 2 &   & 4 & 6 & 8 & 10 & 12 & 14 & 9 & 4 \\
	\end{array}$
	& $[\so(16)] \,\, \overset{\usp(2)}{1} \,\, \overset{\su(2)}{2} \,\, [\su(2)]$ 	\hypertarget{rank4-th3}{}\\
	\hline
 $3$ & 
	$\begin{array}{cc @{} >{\color{cE7}}c >{\color{cE7}}c >{\color{cE7}}c >{\color{cE7}}c >{\color{cE7}}c >{\color{cE7}}c >{\color{cE7}}c @{}}
		  &   &   &   &   & 6 &   &   &   \\
		2 &   & 4 & 6 & 9 & 12 & 9 & 6 & 3 \\
	\end{array}$
	& 
	$\begin{array}{cc @{} >{\color{cE8}}c >{\color{cE8}}c >{\color{cE8}}c >{\color{cE8}}c >{\color{cE8}}c >{\color{cE8}}c >{\color{cE8}}c >{\color{cE8}}c @{}}
		  &   &   &   &   &   &   & 9 &   &   \\
		2 &   & 4 & 6 & 9 & 12 & 15 & 18 & 12 & 6 \\
	\end{array}$
	& $[\mathfrak{e}_8] \,\, 1 \,\, \overset{\su(1)}{2}\,\, \overset{\su(2)}{2}\,\, [\su(3)]$  	\hypertarget{rank4-th4}{}\\
	\hline
 $4$ &
	$\begin{array}{cccc @{} >{\color{cE7}}c >{\color{cE7}}c >{\color{cE7}}c >{\color{cE7}}c >{\color{cE7}}c >{\color{cE7}}c >{\color{cE7}}c @{}}
		  &   &   &   &   &   &   & 4 &   &   &   \\
		1 & 2 & 3 &   & 4 & 5 & 6 & 8 & 6 & 4 & 2 \\
	\end{array}$
	& 
	$\begin{array}{cccc @{} >{\color{cE8}}c >{\color{cE8}}c >{\color{cE8}}c >{\color{cE8}}c >{\color{cE8}}c >{\color{cE8}}c >{\color{cE8}}c >{\color{cE8}}c @{}}
		  &   &   &   &   &   &   &   &   & 6 &   &   \\
		1 & 2 & 3 &   & 4 & 5 & 6 & 8 & 10 & 12 & 8 & 4 \\
	\end{array}$
	&$[\mathfrak{e}_6]\,\, 1 \,\, \overset{\su(3)}{2}\,\, [\su(6)]$ 	\hypertarget{rank4-th5}{} \\
	\hline
 $5$ &
	$\begin{array}{ccccc @{} >{\color{cE7}}c >{\color{cE7}}c >{\color{cE7}}c >{\color{cE7}}c >{\color{cE7}}c >{\color{cE7}}c >{\color{cE7}}c @{}}
		  &   &   &   &   &   &   &   & 4 &   &   &   \\
		1 & 2 & 3 & 4 &   & 5 & 6 & 7 & 8 & 5 & 3 & 1 \\
	\end{array}$
	& 
	$\begin{array}{ccccc @{} >{\color{cE8}}c >{\color{cE8}}c >{\color{cE8}}c >{\color{cE8}}c >{\color{cE8}}c >{\color{cE8}}c >{\color{cE8}}c >{\color{cE8}}c @{}}
		  &   &   &   &   &   &   &   &   &   & 5 &   &   \\
		1 & 2 & 3 & 4 &   & 5 & 6 & 7 & 8 & 9 & 10 & 6 & 3 \\
	\end{array}$
	& $\underset{[\wedge^2]}{\overset{\su(4)}{1}}\,\, [\su(12)]$ \hypertarget{rank4-th6}{} \\
	\hline
	 $6$ & 
	$\begin{array}{ccccc @{} >{\color{cE7}}c >{\color{cE7}}c >{\color{cE7}}c >{\color{cE7}}c >{\color{cE7}}c >{\color{cE7}}c >{\color{cE7}}c @{}}
		  &   &   &   &   &   &   &   & 3 &   &   &   \\
		1 & 2 & 3 & 4 &   & 5 & 6 & 7 & 8 & 6 & 4 & 2 \\
	\end{array}$
	& 
	$\begin{array}{ccccc @{} >{\color{cE8}}c >{\color{cE8}}c >{\color{cE8}}c >{\color{cE8}}c >{\color{cE8}}c >{\color{cE8}}c >{\color{cE8}}c >{\color{cE8}}c @{}}
		  &   &   &   &   &   &   &   &   &   & 5 &   &   \\
		1 & 2 & 3 & 4 &   & 5 & 6 & 7 & 8 & 9 & 10 & 6 & 3 \\
	\end{array}$
	& $\underset{[\wedge^2]}{\overset{\su(4)}{1}}\,\, [\su(12)]$ 	\hypertarget{rank4-th7}{} \\
	\hline
$7$ & 
	$\begin{array}{ccc @{} >{\color{cE7}}c >{\color{cE7}}c >{\color{cE7}}c >{\color{cE7}}c >{\color{cE7}}c >{\color{cE7}}c >{\color{cE7}}c @{}}
		  &   &   &   &   &   & 6 &   &   &   \\
		2 & 4 &   & 6 & 8 & 10 & 12 & 8 & 4 & 1 \\
	\end{array}$
	& 
	$\begin{array}{ccc @{} >{\color{cE8}}c >{\color{cE8}}c >{\color{cE8}}c >{\color{cE8}}c >{\color{cE8}}c >{\color{cE8}}c >{\color{cE8}}c >{\color{cE8}}c @{}}
		  &   &   &   &   &   &   &   & 7 &   &   \\
		1 & 2 &   & 4 & 6 & 8 & 10 & 12 & 14 & 9 & 4 \\
	\end{array}$
	& $[\so(16)]\,\, \overset{\usp(2)}{1}\,\, \overset{\su(2)}{2} \, [\su(2)]$  \hypertarget{rank4-th8}{} \\
	\hline
	$8$ & 
	$\begin{array}{ccccccc @{} >{\color{cE7}}c >{\color{cE7}}c >{\color{cE7}}c >{\color{cE7}}c >{\color{cE7}}c >{\color{cE7}}c >{\color{cE7}}c @{}}
		  &   &   &   &   &   &   &   &   &   & 5 &   &   &   \\
		1 & 2 & 3 & 4 & 5 & 6 &   & 7 & 8 & 9 & 10 & 6 & 2 & 1 \\
	\end{array}$
	& 
	$\begin{array}{ccccccc @{} >{\color{cE8}}c >{\color{cE8}}c >{\color{cE8}}c >{\color{cE8}}c >{\color{cE8}}c >{\color{cE8}}c >{\color{cE8}}c >{\color{cE8}}c @{}}
		  &   &   &   &   &   &   &   &   &   &   &   & 6 &   &   \\
		1 & 2 & 3 & 4 & 5 & 6 &   & 7 & 8 & 9 & 10 & 11 & 12 & 7 & 2 \\
	\end{array}$
	& $\overset{\usp(6)}{1} \, [\so(28)]$ 
\end{longtable}
Let us demonstrate the procedure presented in \cref{sec:procedure} in many examples as follows.
\ben
\item \label{defso16} Let us consider Theory \hyperlink{rank4-th2}{2}. If we subtract the $2 \times E_8$ quiver out of the $E_8$ parent quiver and rebalance, we obtain \eref{E8subreb}, which is
\bes{
	\begin{tikzpicture}[baseline=0,font=\footnotesize]
		\node[gauge, label=below:{$1$}] (dots) {};
		\node[gauge, label=below:{$2$}] (k) [right=6mm of dots] {};
		\node[gauge, bodyE8, label=below:{$2$}] (A) [right=6mm of k] {};
		\node[gauge, bodyE8, label=below:{$2$}] (B) [right=6mm of A] {};
		\node[gauge, bodyE8, label=below:{$2$}] (C) [right=6mm of B] {};
		\node[gauge, bodyE8, label=below:{$2$}] (D) [right=6mm of C] {};
		\node[gauge, bodyE8, label=below:{$2$}] (E) [right=6mm of D] {};
		\node[gauge, bodyE8, label=below:{$2$}] (F) [right=6mm of E] {};
		\node[gauge, bodyE8, label=below:{$1$}] (G) [right=6mm of F] {};
		\node[gauge, bodyE8, label=below:{$0$}] (H) [right=6mm of G] {};
		\node[gauge, bodyE8, label=above:{$1$}] (FEG) [above=4mm of F] {};
		\node[gauge, label=above:{$1$}] (kk) [above left=5mm of k] {};
		\node[gauge, label=above:{$1$}] (kkk) [above right=5mm of k] {};
		\draw[thick] (dots.east) -- (k.west);
		\draw[thick] (k.east) -- (A.west);
		\draw[thick] (k) -- (kk);
		\draw[thick] (k) -- (kkk);
		\draw[thick] (k.east) -- (A.west);
		\draw[thick,bodyE8] (A.east) -- (B.west);
		\draw[thick,bodyE8] (B.east) -- (C.west);
		\draw[thick,bodyE8] (C.east) -- (D.west);
		\draw[thick,bodyE8] (D.east) -- (E.west);
		\draw[thick,bodyE8] (E.east) -- (F.west);
		\draw[thick,bodyE8] (F.east) -- (G.west);
		\draw[thick,bodyE8] (F.north) -- (FEG.south);
	\end{tikzpicture}
}
This was discussed in \cite[(3.22)]{Cabrera:2019izd}$_{n=2}$ as a mirror theory of the following quiver
\bes{ \label{mirrrank4a}
	[\su(2)]-\SU(2)-\USp(2)-[\so(16)]
}
In order for the gauge anomalies to be cancelled, we need to put the $\USp(2)$ gauge group on the $(-1)$-curve and put the $\SU(2)$ gauge group on the $(-2)$-curve. This explains the 6d parent quiver we depicted above in Theory \hyperlink{rank4-th2}{2}.  

On the other hand, we can also subtract the $2 \times E_7$ quiver out of the $E_7$ quiver and rebalance. The result is \eref{E7subreb} which is
\bes{ \label{mirrrank4b}
	\begin{tikzpicture}[baseline=0,font=\footnotesize]
		\node[gauge, label=below:{$1$}] (dots) {};
		\node[gauge, label=below:{$2$}] (k) [right=6mm of dots] {};
		\node[gauge, bodyE7, label=below:{$2$}] (A) [right=6mm of k] {};
		\node[gauge, bodyE7, label=below:{$2$}] (B) [right=6mm of A] {};
		\node[gauge, bodyE7, label=below:{$2$}] (C) [right=6mm of B] {};
		\node[gauge, bodyE7, label=below:{$2$}] (D) [right=6mm of C] {};
		\node[gauge, bodyE7, label=below:{$1$}] (F) [right=6mm of D] {};
		\node[gauge, bodyE7, label=below:{$0$}] (G) [right=6mm of F] {};
		\node[gauge, bodyE7, label=below:{$0$}] (H) [right=6mm of G] {};
		\node[gauge, bodyE7, label=above:{$1$}] (FEG) [above=4mm of D] {};
		\node[gauge, label=above:{$1$}] (kk) [above left=5mm of k] {};
		\node[gauge, label=above:{$1$}] (kkk) [above right=5mm of k] {};
		\draw[thick] (dots.east) -- (k.west);
		\draw[thick] (k.east) -- (A.west);
		\draw[thick] (k) -- (kk);
		\draw[thick] (k) -- (kkk);
		\draw[thick] (k.east) -- (A.west);
		\draw[thick,bodyE7] (A.east) -- (B.west);
		\draw[thick,bodyE7] (B.east) -- (C.west);
		\draw[thick,bodyE7] (C.east) -- (D.west);
		\draw[thick,bodyE7] (D.east) -- (F.west);
		\draw[thick,bodyE7] (D.north) -- (FEG.south);
	\end{tikzpicture}
}
This is a mirror theory of
\bes{
[\su(2)]-\SU(2)-\USp(2)-[\so(14)]
}
Comparing \eref{mirrrank4a} and \eref{mirrrank4b}, we see that a doublet of hypermultiplets in the fundamental representation of the $\USp(2)$ gauge group in the former becomes massive and is integrated out.
\item Let us consider Theory \hyperlink{rank4-th4}{4}. Upon subtracting the $2\times E_8$ quiver from the $E_8$ parent theory and rebalancing, we obtain
\bes{
	\begin{tikzpicture}[baseline=0,font=\footnotesize]
            \node[gauge, label=below:{$2$}] (dots) {};
            \node[gauge, label=below:{$1$}] (0) [left=6mm of dots] {};
		\node[gauge, label=below:{$3$}] (k) [right=6mm of dots] {};
		\node[gauge, bodyE8, label=below:{$2$}] (A) [right=6mm of k] {};
		\node[gauge, bodyE8, label=below:{$1$}] (B) [right=6mm of A] {};
		\node[gauge, bodyE8, label=below:{$0$}] (C) [right=6mm of B] {};
		\node[gauge, bodyE8, label=below:{$0$}] (D) [right=6mm of C] {};
		\node[gauge, bodyE8, label=below:{$0$}] (E) [right=6mm of D] {};
		\node[gauge, bodyE8, label=below:{$0$}] (F) [right=6mm of E] {};
		\node[gauge, bodyE8, label=below:{$0$}] (G) [right=6mm of F] {};
		\node[gauge, bodyE8, label=below:{$0$}] (H) [right=6mm of G] {};
		\node[gauge, bodyE8, label=above:{$0$}] (FEG) [above=4mm of F] {};
		\node[gauge, label=above:{$1$}] (kk) [above left=5mm of k] {};
		\node[gauge, label=above:{$1$}] (kkk) [above right=5mm of k] {};
		\draw[thick] (0)--(dots) -- (k);
		\draw[thick] (k.east) -- (A.west);
		\draw[thick] (k) -- (kk);
		\draw[thick] (k) -- (kkk);
		\draw[thick] (k.east) -- (A.west);
		\draw[thick,bodyE8] (A.east) -- (B.west);
	\end{tikzpicture}
}
This theory is mirror dual to the $\SU(3)$ gauge theory with $6$ flavors:
\bes{ \label{SU3w6flv}
\SU(3)-[\su(6)]
}
Here, we can put the gauge group $\SU(3)$ on the $(-2)$-curve. However, since node {\blue 6} is overbalanced in the original $E_8$ parent theory, it has the Coulomb branch symmetry $\mathfrak{e}_6 \times \su(6)$. We need to add a $(-1)$-curve and put $[\mathfrak{e}_6]$ next to it in order to take into account this $\mathfrak{e}_6$ flavor symmetry. We then arrive at the quiver description of the 6d theory as stated above.

Upon subtracting the $2 \times E_7$ quiver from the corresponding $E_7$ quiver and rebalancing, we also obtain the mirror dual of the $\SU(3)$ gauge theory with $6$ flavors. This means that, after considering the Kraft--Procesi transitions of the $E_8$ and $E_7$ small instantons in each theory, the end results of the two theories are the same.  Note, however, that since node {\red 6} in the $E_7$ quiver is overbalanced, the Coulomb branch symmetry of the latter is $\su(6) \times \su(6)$. This means that the $\mathfrak{e}_6$ flavor symmetry of the parent theory is mass-deformed to $\su(6)$.

\item Let us consider Theory  \hyperlink{rank4-th7}{7}. Although it has the same $E_8$ parent quiver and hence the same 6d parent theory as Theory  \hyperlink{rank4-th2}{2}, it comes from a different mass deformation. This can be seen as follows. We take the $E_7$ quiver in Theory \hyperlink{rank4-th7}{7} and decouple an overall $\U(1)$ from the rightmost $\U(1)$ gauge node. As a result, we obtain
\bes{
\begin{tikzpicture}[baseline=0,font=\footnotesize]
		\node[gauge, label=below:{$2$}] (tail1) {};
		\node[gauge, label=below:{$4$}] (k) [right=6mm of tail1] {};
		\node[gauge, bodyE7, label=below:{$6$}] (A) [right=6mm of k] {};
		\node[gauge, bodyE7, label=below:{$8$}] (B) [right=6mm of A] {};
		\node[gauge, bodyE7, label=below:{$10$}] (C) [right=6mm of B] {};
		\node[gauge, bodyE7, label=below:{$12$}] (D) [right=6mm of C] {};
		\node[gauge, bodyE7, label=below:{$8$}] (F) [right=6mm of D] {};
		\node[gauge, bodyE7, label=below:{$4$}] (G) [right=6mm of F] {};
		\node[flavor, bodyE7, label=below:{$1$}] (H) [right=6mm of G] {};
		\node[gauge, bodyE7, label=above:{$6$}] (I) [above=4mm of D] {};
		\draw[thick] (tail1) -- (k) -- (A);
        \draw[thick,bodyE7] (A) -- (B) -- (C) -- (D) -- (F) -- (G) -- (H);
        \draw[thick,bodyE7] (D) -- (I);
	\end{tikzpicture}
}
According to \cite{Mekareeya:2015bla}, the Coulomb branch of such a quiver realizes the moduli space of two $E_8$ instantons on $\BC^2/\BZ_2$ with the holonomy such that $E_8$ is broken to $\SO(16)$. Here, we see that the $\su(2)$ flavor symmetry of the corresponding parent 6d theory was mass-deformed.  This is opposite to the discussion in Point \ref{defso16}.
\item\label{antisymrank4} Let us consider Theories \hyperlink{rank4-th7}{5} and \hyperlink{rank4-th7}{6}. Upon subtracting the $1 \times E_8$ quiver out of the $E_8$ parent quiver and rebalancing, we obtain the theory
\bes{
\begin{tikzpicture}[baseline=0,font=\footnotesize]
		\node[gauge, label=below:{$1$}] (tail1) {};
        \node[gauge, label=below:{$2$}] (tail2) [right=6mm of tail1] {};
        \node[gauge, label=below:{$3$}] (tail3) [right=6mm of tail2] {};
		\node[gauge, label=below:{$4$}] (k) [right=6mm of tail3] {};
        \node[gauge, label=above:{$1$}] (tail4) [above=4mm of k] {};
		\node[gauge, bodyE8,label=below:{$4$}] (A) [right=6mm of k] {};
		\node[gauge, bodyE8,label=below:{$4$}] (B) [right=6mm of A] {};
		\node[gauge, bodyE8,label=below:{$4$}] (C) [right=6mm of B] {};
		\node[gauge, bodyE8,label=below:{$4$}] (D) [right=6mm of C] {};
		\node[gauge, bodyE8,label=below:{$4$}] (F) [right=6mm of D] {};
		\node[gauge, bodyE8,label=below:{$4$}] (G) [right=6mm of F] {};
		\node[gauge, bodyE8,label=below:{$2$}] (H) [right=6mm of G] {};
		\node[gauge, bodyE8,label=below:{$1$}] (I) [right=6mm of H] {};
        \node[gauge, bodyE8,label=above:{$2$}] (J) [above=4mm of G] {};
		\draw[thick] (tail1) -- (tail2) -- (tail3) -- (k) -- (A);
        \draw[thick,bodyE8] (A) -- (B) -- (C) -- (D) -- (F) -- (G) -- (H) -- (I);
        \draw[thick,bodyE8] (G) -- (J);
        \draw[thick] (k) -- (tail4);
	\end{tikzpicture}
}
This theory was mentioned explicitly in \cite[(3.106b)]{Cabrera:2019izd}$_{n=1}$ (see also \cite{Hanany:1999sj, Garozzo:2019hbf}). This is a mirror theory of the $\SU(4)$ gauge theory with one rank-two antisymmetric ($\wedge^2$) and $12$ fundamental hypermultiplets. To account for the small $E_8$ instanton transition, we introduce a $(-1)$-curve into the F-theory quiver, as detailed in the discussion following \eref{E8subreb}. This leads to the corresponding 6d parent theory depicted in the rightmost column, which is the expected result.

It is also instructive to compare this with the subtraction of $1 \times E_7$ quiver out of the $E_7$ quiver (and rebalancing appropriately). For Theory  \hyperlink{rank4-th5}{5}, we obtain
\bes{
\begin{tikzpicture}[baseline=0,font=\footnotesize]
		\node[gauge, label=below:{$1$}] (tail1) {};
        \node[gauge, label=below:{$2$}] (tail2) [right=6mm of tail1] {};
        \node[gauge, label=below:{$3$}] (tail3) [right=6mm of tail2] {};
		\node[gauge, label=below:{$4$}] (k) [right=6mm of tail3] {};
        \node[gauge, label=above:{$1$}] (tail4) [above=4mm of k] {};
		\node[gauge, bodyE7, label=below:{$4$}] (A) [right=6mm of k] {};
		\node[gauge, bodyE7, label=below:{$4$}] (B) [right=6mm of A] {};
		\node[gauge, bodyE7, label=below:{$4$}] (C) [right=6mm of B] {};
		\node[gauge, bodyE7, label=below:{$4$}] (D) [right=6mm of C] {};
		\node[gauge, bodyE7, label=below:{$2$}] (F) [right=6mm of D] {};
		\node[gauge, bodyE7, label=below:{$1$}] (G) [right=6mm of F] {};
		\node[gauge, bodyE7, label=below:{$0$}] (H) [right=6mm of G] {};
		\node[gauge, bodyE7, label=above:{$2$}] (I) [above=4mm of D] {};
		\draw[thick] (tail1) -- (tail2) -- (tail3) -- (k) -- (A);
        \draw[thick,bodyE7] (A) -- (B) -- (C) -- (D) -- (F) -- (G);
        \draw[thick,bodyE7] (I) -- (D);
        \draw[thick] (k) -- (tail4);
	\end{tikzpicture}
}
This is a mirror theory of the $\SU(4)$ gauge theory with one rank-two antisymmetric and $10$ fundamental hypermultiplets. We see that two fundamental hypermultiplets of the theories discussed in the previous paragraph become massive and are integrated out. On the other hand, for Theory  \hyperlink{rank4-th6}{6}, we obtain
\bes{
\begin{tikzpicture}[baseline=0,font=\footnotesize]
		\node[gauge, label=below:{$1$}] (tail1) {};
        \node[gauge, label=below:{$2$}] (tail2) [right=6mm of tail1] {};
        \node[gauge, label=below:{$3$}] (tail3) [right=6mm of tail2] {};
		\node[gauge, label=below:{$4$}] (k) [right=6mm of tail3] {};
        \node[gauge, label=above:{$1$}] (tail4) [above=4mm of k] {};
		\node[gauge, bodyE7, label=below:{$4$}] (A) [right=6mm of k] {};
		\node[gauge, bodyE7, label=below:{$4$}] (B) [right=6mm of A] {};
		\node[gauge, bodyE7, label=below:{$4$}] (C) [right=6mm of B] {};
		\node[gauge, bodyE7, label=below:{$4$}] (D) [right=6mm of C] {};
		\node[gauge, bodyE7, label=below:{$3$}] (F) [right=6mm of D] {};
		\node[gauge, bodyE7, label=below:{$2$}] (G) [right=6mm of F] {};
		\node[gauge, bodyE7, label=below:{$1$}] (H) [right=6mm of G] {};
		\node[gauge, bodyE7, label=above:{$1$}] (I) [above=4mm of D] {};
		\draw[thick] (tail1) -- (tail2) -- (tail3) -- (k) -- (A);
        \draw[thick,bodyE7] (A) -- (B) -- (C) -- (D) -- (F) -- (G) -- (H);
        \draw[thick,bodyE7] (I) -- (D);
        \draw[thick] (k) -- (tail4);
	\end{tikzpicture}
}
This is a mirror theory of the $\SU(4)$ gauge theory with $12$ fundamental hypermultiplets \cite[Figure 10]{Hanany:1996ie}. In this case, the adjoint hypermultiplet of the previously discussed theory becomes massive and is integrated out. 

\item Theories \hyperlink{rank4-th1}{1} and \hyperlink{rank4-th3}{3} have the same 6d parent theories even though the $E_8$ parent quivers are different. This can be seen by subtracting the $3 \times E_8$ quiver from both $E_8$ parent theories. After rebalancing, the resulting theory is
\bes{ \label{mirrorsu2with4flv}
	\begin{tikzpicture}[baseline=0,font=\footnotesize]
		\node[gauge, label=below:{$1$}] (dots) {};
		\node[gauge, label=below:{$2$}] (k) [right=6mm of dots] {};
		\node[gauge, label=below:{$1$}] (A) [right=6mm of k] {};
		\node[gauge, label=above:{$1$}] (kk) [above left=5mm of k] {};
		\node[gauge, label=above:{$1$}] (kkk) [above right=5mm of k] {};
		\draw[thick] (dots.east) -- (k.west);
		\draw[thick] (k.east) -- (A.west);
		\draw[thick] (k) -- (kk);
		\draw[thick] (k) -- (kkk);
		\draw[thick] (k.east) -- (A.west);
	\end{tikzpicture}
}
This is a mirror theory of the $\SU(2)$ gauge theory with 4 flavors. Due to the gauge anomaly cancellation condition, we know that this $\SU(2)$ has to be on a $(-2)$-curve. Since we have made a three times subtraction, the number of the tensor multiplet in the 6d theory must be three, and we need two more curves. Since the Coulomb branch symmetry of the $E_8$ parent theory is $\mathfrak{e}_8 \times \su(3)$, one of them has to be a $(-1)$-curve with $[\mathfrak{e}_8]$ attached, and the aforementioned $(-2)$-curve decorated with the $\su(2)$ gauge group must be attached to $[\su(3)]$. The remaining curve must be a $(-2)$-curve decorated with $\su(1)$. 

On the other hand, if we subtract the $3 \times E_7$ quiver from both $E_7$ quivers in Theories  \hyperlink{rank4-th1}{1} and  \hyperlink{rank4-th3}{3}, we also obtain \eref{mirrorsu2with4flv}. Upon contrasting the Coulomb branch symmetries of the $E_7$ quiver and $E_8$ parent quiver in each case, we see that in Theory \hyperlink{rank4-th1}{1}, the $\mathfrak{e}_8$ flavor symmetry of the parent theory is mass-deformed to $\mathfrak{e}_7$, whereas in Theory \hyperlink{rank4-th1}{3}, the $\mathfrak{e}_8$ flavor symmetry of the parent theory is mass-deformed to $\mathfrak{so}(12)$.
\item Let us consider Theory \hyperlink{rank4-th8}{8}. After subtracting the $1 \times E_8$ quiver from the $E_8$ parent quiver and rebalancing, we have
\bes{
\begin{tikzpicture}[baseline=7,font=\footnotesize]
		\node[gauge, label=below:{$1$}] (tail1) {};
        \node[gauge, label=below:{$2$}] (tail2) [right=6mm of tail1] {};
        \node (tail3) [right=6mm of tail2] {$\cdots$};
		\node[gauge, label=below:{$6$}] (k) [right=6mm of tail3] {};
        \node[gauge, label=above:{$1$}] (tail4) [above=4mm of k] {};
		\node[gauge, bodyE8,label=below:{$6$}] (A) [right=6mm of k] {};
		\node[gauge, bodyE8,label=below:{$6$}] (B) [right=6mm of A] {};
		\node[gauge, bodyE8,label=below:{$6$}] (C) [right=6mm of B] {};
		\node[gauge, bodyE8,label=below:{$6$}] (D) [right=6mm of C] {};
		\node[gauge, bodyE8,label=below:{$6$}] (F) [right=6mm of D] {};
		\node[gauge, bodyE8,label=below:{$6$}] (G) [right=6mm of F] {};
		\node[gauge, bodyE8,label=below:{$3$}] (H) [right=6mm of G] {};
		\node[gauge, bodyE8,label=below:{$0$}] (I) [right=6mm of H] {};
        \node[gauge, bodyE8,label=above:{$3$}] (J) [above=4mm of G] {};
		\draw[thick] (tail1) -- (tail2) -- (tail3) -- (k) -- (A);
        \draw[thick,bodyE8] (A) -- (B) -- (C) -- (D) -- (F) -- (G) -- (H);
        \draw[thick,bodyE8] (G) -- (J);
        \draw[thick] (k) -- (tail4);
	\end{tikzpicture}
}
From \cite[Figure 12]{Hanany:1999sj}, this is a mirror theory of the $\USp(6)$ gauge theory with $14$ flavors:
\bes{ \label{usp6w14}
\USp(6)-[\so(28)]
}
Putting the $\USp(6)$ gauge group on the $(-1)$-curve, we obtained the required F-theory quiver for the 6d parent theory.

We can also consider the deformation by subtracting the $1 \times E_7$ quiver from the corresponding $E_7$ quiver.  Upon rebalancing, we obtain
\bes{
\begin{tikzpicture}[baseline=7,font=\footnotesize]
		\node[gauge, label=below:{$1$}] (tail1) {};
        \node[gauge, label=below:{$2$}] (tail2) [right=6mm of tail1] {};
        \node (tail3) [right=6mm of tail2] {$\cdots$};
		\node[gauge, label=below:{$6$}] (k) [right=6mm of tail3] {};
        \node[gauge, label=above:{$1$}] (tail4) [above=4mm of k] {};
		\node[gauge, bodyE7, label=below:{$6$}] (A) [right=6mm of k] {};
		\node[gauge, bodyE7, label=below:{$6$}] (B) [right=6mm of A] {};
		\node[gauge, bodyE7, label=below:{$6$}] (C) [right=6mm of B] {};
		\node[gauge, bodyE7, label=below:{$6$}] (D) [right=6mm of C] {};
		\node[gauge, bodyE7, label=below:{$3$}] (F) [right=6mm of D] {};
		\node[gauge, bodyE7, label=below:{$0$}] (G) [right=6mm of F] {};
		\node[gauge, bodyE7, label=below:{$0$}] (H) [right=6mm of G] {};
		\node[gauge, bodyE7, label=above:{$3$}] (I) [above=4mm of D] {};
		\draw[thick] (tail1) -- (tail2) -- (tail3) -- (k) -- (A);
        \draw[thick,bodyE7] (A) -- (B) -- (C) -- (D) -- (F);
        \draw[thick,bodyE7] (I) -- (D);
        \draw[thick] (k) -- (tail4);
	\end{tikzpicture}
}
Similarly to the above, this is a mirror theory of the $\USp(6)$ gauge theory with $12$ flavors:
\bes{ \label{usp6w12}
\USp(6)-[\so(24)]
}
We see that, upon deformation, two flavors of the hypermultiplets become massive.
\een

\subsubsection{Examples: \texorpdfstring{$E_7$}{E7} Quivers with Six-dimensional Higgs Branch and One Tail} \label{sec:rank6onetail}
We list the $E_7$ quivers, containing one tail, with a six-dimensional Higgs branch, where the rank of the central node is less than or equal to $12$, below. The derivation of the following results is as described at the beginning of Section \ref{sec:rank4onetail}.
\begin{longtable}{c|c|c|c}
		Number & $E_7$ Quiver & $E_8$ Parent Quiver & 6d Parent Theory \\
		\hline
  \endhead
		\hypertarget{rank6-th1}{} 1 & 
		$\begin{array}{cccc @{} >{\color{cE7}}c >{\color{cE7}}c >{\color{cE7}}c >{\color{cE7}}c >{\color{cE7}}c >{\color{cE7}}c >{\color{cE7}}c @{}}
			  &  &  &   &   &   &   & 6 &   &   &   \\
			1 & 2 & 3  &   &4  & 6 & 9 & 12 & 9 & 6 & 3 \\
		\end{array}$
		& 
		$\begin{array}{cccc @{} >{\color{cE8}}c >{\color{cE8}}c >{\color{cE8}}c >{\color{cE8}}c >{\color{cE8}}c >{\color{cE8}}c >{\color{cE8}}c >{\color{cE8}}c @{}}
			  &  &   &   &   &   &   &   &   & 7 &   &   \\
			1 & 2 & 3 &   & 4 & 6 & 9 & 12 & 15 & 18 & 12 & 6 \\
		\end{array}$
		& $[\mathfrak{e}_7] \,\, \overset{}{1}  \,\, \underset{[N_f=1]}{\overset{\su(2)}{2}} \,\, \overset{\su(3)}{2} \,\, [\su(4)]$ \hypertarget{rank6-th2}{} \\
		\hline
		2 & 
		$\begin{array}{cccc @{} >{\color{cE7}}c >{\color{cE7}}c >{\color{cE7}}c >{\color{cE7}}c >{\color{cE7}}c >{\color{cE7}}c >{\color{cE7}}c @{}}
			  &  &  &   &   &   &   & 5 &   &   &   \\
			1 & 2 & 3  &   &5  & 7 & 9 & 11 & 8 & 5 & 2 \\
		\end{array}$
		& 
		$\begin{array}{cccc @{} >{\color{cE8}}c >{\color{cE8}}c >{\color{cE8}}c >{\color{cE8}}c >{\color{cE8}}c >{\color{cE8}}c >{\color{cE8}}c >{\color{cE8}}c @{}}
			  &  &   &   &   &   &   &   &   & 7 &   &   \\
			1 & 2 & 3 &   & 5 & 7 & 9 & 11 & 13 & 15 & 10 & 5 \\
		\end{array}$
		& $[\su(9)] \,\, \overset{\su(3)}{1}  \,\, \overset{\su(3)}{2}  \,\, [\su(3)]$ \hypertarget{rank6-th3}{}\\
		\hline
		3 & 
		$\begin{array}{ccc @{} >{\color{cE7}}c >{\color{cE7}}c >{\color{cE7}}c >{\color{cE7}}c >{\color{cE7}}c >{\color{cE7}}c >{\color{cE7}}c @{}}
			  &  &  &     &   &   &6 &   &   &   \\
			1 & 3 &     &5  & 7 & 9 & 12 & 9 & 6 & 3 \\
		\end{array}$
		& 
		$\begin{array}{ccc @{} >{\color{cE8}}c >{\color{cE8}}c >{\color{cE8}}c >{\color{cE8}}c >{\color{cE8}}c >{\color{cE8}}c >{\color{cE8}}c >{\color{cE8}}c @{}}
			  &  &     &   &   &   &   &   & 9 &   &   \\
			1 & 3 &    & 5 & 7 & 9 & 12 & 15 & 18 & 12 & 6 \\
		\end{array}$
		& $[\su(4)] \,\, \overset{\su(3)}{2}  \,\, \overset{\su(2)}{2}  \,\, [N_f=1]$ \hypertarget{rank6-th4}{}\\
		\hline			
		4 & 
		$\begin{array}{ccccc @{} >{\color{cE7}}c >{\color{cE7}}c >{\color{cE7}}c >{\color{cE7}}c >{\color{cE7}}c >{\color{cE7}}c >{\color{cE7}}c @{}}
			  &  &  &  & &   &   &   &5 &   &   &   \\
			1 & 2 & 3 & 4  &   &5  & 6 & 8 & 10 & 7 & 4 & 2 \\
		\end{array}$
		& 
		$\begin{array}{ccccc @{} >{\color{cE8}}c >{\color{cE8}}c >{\color{cE8}}c >{\color{cE8}}c >{\color{cE8}}c >{\color{cE8}}c >{\color{cE8}}c >{\color{cE8}}c @{}}
			  &  &   &  & &   &   &   &   & 7 &   &   &   \\
			1 & 2 & 3 & 4 &   & 5 & 8 & 10 & 12 & 14  & 9 & 4 \\
		\end{array}$
		& $[\so(12)] \,\, \overset{\usp(2)}{1}  \,\, \overset{\su(4)}{2}  \,\, [\su(6)]$ \hypertarget{rank6-th5}{}\\
		\hline			
		5 & 
		$\begin{array}{ccccc @{} >{\color{cE7}}c >{\color{cE7}}c >{\color{cE7}}c >{\color{cE7}}c >{\color{cE7}}c >{\color{cE7}}c >{\color{cE7}}c @{}}
			  &  &  &  & &   &   &   &6 &   &   &   \\
			1 & 2 & 3 & 4  &   &6  & 8& 10 & 12 & 8 & 4 & 1 \\
		\end{array}$
		& 
		$\begin{array}{ccccc @{} >{\color{cE8}}c >{\color{cE8}}c >{\color{cE8}}c >{\color{cE8}}c >{\color{cE8}}c >{\color{cE8}}c >{\color{cE8}}c >{\color{cE8}}c @{}}
			  &  &   & &  &   &   &   &   &   & 7 &   &   \\
			1 & 2 & 3 & 4 &   & 5 & 6 & 8 & 10 & 12 & 14  & 9 & 4 \\
		\end{array}$
		& $[\so(12)] \,\, \overset{\usp(2)}{1}  \,\, \overset{\su(4)}{2}  \,\, [\su(6)]$ \hypertarget{rank6-th6}{}\\
		\hline			
		6 & 
		$\begin{array}{cccc @{} >{\color{cE7}}c >{\color{cE7}}c >{\color{cE7}}c >{\color{cE7}}c >{\color{cE7}}c >{\color{cE7}}c >{\color{cE7}}c @{}}
			  &  &  &   &   &   &   &6 &   &   &   \\
			1 & 2 & 4  &   &6  & 8& 10 & 12 & 8 & 4 & 2 \\
		\end{array}$
		& 
		$\begin{array}{cccc @{} >{\color{cE8}}c >{\color{cE8}}c >{\color{cE8}}c >{\color{cE8}}c >{\color{cE8}}c >{\color{cE8}}c >{\color{cE8}}c >{\color{cE8}}c @{}}
			  &  &   &   &   &   &   &   &   & 8 &   &   \\
			1 & 2 & 4 &   & 6 & 8 & 10 & 12 & 14 & 16  & 10 & 4 \\
		\end{array}$
		& $[\so(18)] \,\, \overset{\usp(4)}{1}  \,\, \overset{\su(3)}{2}  \,\, [\su(2)]$ \hypertarget{rank6-th7}{}\\			
		\hline			
		7 & 
		$\begin{array}{ccc @{} >{\color{cE7}}c >{\color{cE7}}c >{\color{cE7}}c >{\color{cE7}}c >{\color{cE7}}c >{\color{cE7}}c >{\color{cE7}}c @{}}
			  &  &    &   &   &   &6 &   &   &   \\
			2 & 4   &   &6  & 8& 10 & 12 & 8 & 5 & 2 \\
		\end{array}$
		& 
		$\begin{array}{ccc @{} >{\color{cE8}}c >{\color{cE8}}c >{\color{cE8}}c >{\color{cE8}}c >{\color{cE8}}c >{\color{cE8}}c >{\color{cE8}}c >{\color{cE8}}c @{}}
			  &  &     &   &   &   &   &   & 8 &   &   \\
			2 & 4 &    & 6 & 8 & 10 & 12 & 14 & 16  & 10 & 5 \\
		\end{array}$
		& $[\su(10)] \,\, \underset{[\wedge^2]}{\overset{\su(4)}{1}}  \,\, \overset{\su(2)}{2}$  \hypertarget{rank6-th8}{}\\
		\hline			
		8 & 
		$\begin{array}{ccc @{} >{\color{cE7}}c >{\color{cE7}}c >{\color{cE7}}c >{\color{cE7}}c >{\color{cE7}}c >{\color{cE7}}c >{\color{cE7}}c @{}}
			  &  &    &   &   &   &5 &   &   &   \\
			2 & 4   &   &6  & 8& 10 & 12 & 9 & 6 & 3 \\
		\end{array}$
		& 
		$\begin{array}{ccc @{} >{\color{cE8}}c >{\color{cE8}}c >{\color{cE8}}c >{\color{cE8}}c >{\color{cE8}}c >{\color{cE8}}c >{\color{cE8}}c >{\color{cE8}}c @{}}
			  &  &      &   &   &   &   &   & 8 &   &   \\
			2 & 4 &    & 6 & 8 & 10 & 12 & 14 & 16  & 10 & 5 \\
		\end{array}$
		& $[\su(10)] \,\, \underset{[\wedge^2]}{\overset{\su(4)}{1}}  \,\, \overset{\su(2)}{2}$  \hypertarget{rank6-th9}{}\\
		\hline			
		9 & 
		$\begin{array}{cccccc @{} >{\color{cE7}}c >{\color{cE7}}c >{\color{cE7}}c >{\color{cE7}}c >{\color{cE7}}c >{\color{cE7}}c >{\color{cE7}}c @{}}
			  &  &  &  & &  &   &   &   &4 &   &   &   \\
			1 & 2 & 3 & 4 & 5  &   &6  & 7& 8 & 9 & 6 & 4 & 2 \\
		\end{array}$
		& 
		$\begin{array}{cccccc @{} >{\color{cE8}}c >{\color{cE8}}c >{\color{cE8}}c >{\color{cE8}}c >{\color{cE8}}c >{\color{cE8}}c >{\color{cE8}}c >{\color{cE8}}c @{}}
			  &  &   & & &  &   &   &   &   &   & 6 &   &   \\
			1 & 2 & 3 & 4 & 5  &   & 6 & 7 & 8 & 9 & 10 & 12  & 8 & 4 \\
		\end{array}$
		& $[\su(5)] \,\, {1}  \,\, \overset{\su(5)}{2}\,\, [\su(10)]$ \hypertarget{rank6-th10}{} \\
		\hline			
		10 & 
		$\begin{array}{cccccc @{} >{\color{cE7}}c >{\color{cE7}}c >{\color{cE7}}c >{\color{cE7}}c >{\color{cE7}}c >{\color{cE7}}c >{\color{cE7}}c @{}}
			  &  &  &  & & &   &   &   &5 &   &   &   \\
			1 & 2 & 3 & 4 & 5  &   &6  & 7& 8 & 10 & 7 & 4 & 1 \\
		\end{array}$
		& 
		$\begin{array}{cccccc @{} >{\color{cE8}}c >{\color{cE8}}c >{\color{cE8}}c >{\color{cE8}}c >{\color{cE8}}c >{\color{cE8}}c >{\color{cE8}}c >{\color{cE8}}c @{}}
			  &  &   &  &  &   &   &   &   &   & 6 &   &   \\
			1 & 2 & 3 & 4 & 5  &   & 6 & 7 & 8 & 9 & 10 & 12  & 8 & 4 \\
		\end{array}$
		& $[\su(5)] \,\, {1}  \,\, \overset{\su(5)}{2}\,\, [\su(10)]$ \hypertarget{rank6-th11}{} \\
		\hline				
		11 & 
		$\begin{array}{ccccccc @{} >{\color{cE7}}c >{\color{cE7}}c >{\color{cE7}}c >{\color{cE7}}c >{\color{cE7}}c >{\color{cE7}}c >{\color{cE7}}c @{}}
			  &  &  & & & &  &   &   &   &5 &   &   &   \\
			1 & 2  & 3 & 4 &5 & 6  &   &7  & 8& 9 & 10 & 6 & 3 & 1 \\
		\end{array}$
		& 
		$\begin{array}{ccccccc @{} >{\color{cE8}}c >{\color{cE8}}c >{\color{cE8}}c >{\color{cE8}}c >{\color{cE8}}c >{\color{cE8}}c >{\color{cE8}}c >{\color{cE8}}c @{}}
			  &  &   & & & &  &   &   &   &   &   & 6 &   &   \\
			1 & 2  &3 &  4 & 5 & 6  &   & 7 & 8 & 9 & 10 & 11 & 12  & 7 & 3 \\
		\end{array}$
		& $[\su(14)] \,\, \underset{[\wedge^2]}{\overset{\su(6)}{1}}$  \hypertarget{rank6-th12}{} \\
		\hline			
		12 & 
		$\begin{array}{ccccccc @{} >{\color{cE7}}c >{\color{cE7}}c >{\color{cE7}}c >{\color{cE7}}c >{\color{cE7}}c >{\color{cE7}}c >{\color{cE7}}c @{}}
			  &  &  & & & &  &   &   &   &4 &   &   &   \\
			1 & 2 & 3 &  4 &5 & 6  &   &7  & 8& 9 & 10 & 7 & 4 & 1 \\
		\end{array}$
		& 
		$\begin{array}{ccccccc @{} >{\color{cE8}}c >{\color{cE8}}c >{\color{cE8}}c >{\color{cE8}}c >{\color{cE8}}c >{\color{cE8}}c >{\color{cE8}}c >{\color{cE8}}c @{}}
			  &  &   & & & &  &   &   &   &   &   & 5 &   &   \\
			1 & 2 & 3 & 4 & 5 & 6  &   & 7 & 8 & 9 & 10 & 11 & 12  & 8 & 4 \\
		\end{array}$
		& $[\su(15)] \,\, \underset{[\frac{1}{2}\wedge^3]}{\overset{\su(6)}{1}}$  					
\end{longtable}

We comment on some of the above theories as follows:
\ben
\item Let us consider Theories \hyperlink{rank6-th7}{7} and \hyperlink{rank6-th8}{8} that have the same 6d parent theory.\footnote{This can be obtained by appropriately modify \cite[(3.104) and (3.106b)]{Cabrera:2019izd}.} However, they correspond to different mass deformations. This can be seen by subtracting the $2 \times E_7$ quiver from each case and rebalancing. As a result, we obtain, respectively
\bes{
	&	\begin{tikzpicture}[baseline=0,font=\footnotesize]
		\node[gauge, label=below:{$2$}] (dots) {};
		\node[gauge, label=below:{$4$}] (k) [right=6mm of dots] {};
		\node[gauge, bodyE7, label=below:{$4$}] (A) [right=6mm of k] {};
		\node[gauge, bodyE7, label=below:{$4$}] (B) [right=6mm of A] {};
		\node[gauge, bodyE7, label=below:{$4$}] (C) [right=6mm of B] {};
		\node[gauge, bodyE7, label=below:{$4$}] (D) [right=6mm of C] {};
		\node[gauge, bodyE7, label=below:{$2$}] (F) [right=6mm of D] {};
		\node[gauge, bodyE7, label=below:{$1$}] (G) [right=6mm of F] {};
		\node[gauge, bodyE7, label=below:{$0$}] (H) [right=6mm of G] {};
		\node[gauge, bodyE7, label=above:{$2$}] (FEG) [above=4mm of D] {};
		\node[gauge, label=above:{$1$}] (kk) [above left=5mm of k] {};
		\node[gauge, label=above:{$1$}] (kkk) [above right=5mm of k] {};
		\draw[thick] (dots.east) -- (k.west);
		\draw[thick] (k.east) -- (A.west);
		\draw[thick] (k) -- (kk);
		\draw[thick] (k) -- (kkk);
		\draw[thick] (k.east) -- (A.west);
		\draw[thick,bodyE7] (A.east) -- (B.west);
		\draw[thick,bodyE7] (B.east) -- (C.west);
		\draw[thick,bodyE7] (C.east) -- (D.west);
		\draw[thick,bodyE7] (D.east) -- (F.west);
		\draw[thick,bodyE7] (F.east) -- (G.west);
		\draw[thick,bodyE7] (D.north) -- (FEG.south);
	\end{tikzpicture}
\\
&
	\begin{tikzpicture}[baseline=0,font=\footnotesize]
	\node[gauge, label=below:{$2$}] (dots) {};
	\node[gauge, label=below:{$4$}] (k) [right=6mm of dots] {};
	\node[gauge, bodyE7, label=below:{$4$}] (A) [right=6mm of k] {};
	\node[gauge, bodyE7, label=below:{$4$}] (B) [right=6mm of A] {};
	\node[gauge, bodyE7, label=below:{$4$}] (C) [right=6mm of B] {};
	\node[gauge, bodyE7, label=below:{$4$}] (D) [right=6mm of C] {};
	\node[gauge, bodyE7, label=below:{$3$}] (F) [right=6mm of D] {};
	\node[gauge, bodyE7, label=below:{$2$}] (G) [right=6mm of F] {};
	\node[gauge, bodyE7, label=below:{$1$}] (H) [right=6mm of G] {};
	\node[gauge, bodyE7, label=above:{$1$}] (FEG) [above=4mm of D] {};
	\node[gauge, label=above:{$1$}] (kk) [above left=5mm of k] {};
	\node[gauge, label=above:{$1$}] (kkk) [above right=5mm of k] {};
	\draw[thick] (dots.east) -- (k.west);
	\draw[thick] (k.east) -- (A.west);
	\draw[thick] (k) -- (kk);
	\draw[thick] (k) -- (kkk);
	\draw[thick] (k.east) -- (A.west);
	\draw[thick,bodyE7] (A.east) -- (B.west);
	\draw[thick,bodyE7] (B.east) -- (C.west);
	\draw[thick,bodyE7] (C.east) -- (D.west);
	\draw[thick,bodyE7] (D.east) -- (F.west);
	\draw[thick,bodyE7] (F.east) -- (G.west);
	\draw[thick,bodyE7] (G.east) -- (H.west);
	\draw[thick,bodyE7] (D.north) -- (FEG.south);
\end{tikzpicture}
}
These are the mirror theories of the following theories, respectively
\bes{
[\su(8)]-\underset{[\wedge^2]}{\SU(4)}-\SU(2)~, \qquad \quad [\su(10)]-\SU(4)-\SU(2)
}
Comparing these with the corresponding 6d parent theories, we see that, in the former, two fundamental hypermultiplets of $\su(4)$ become massive and integrated out, whereas, in the latter, the antisymmetric hypermultiplet becomes massive and integrated out.  This is very similar to the discussion in Point \ref{antisymrank4} of the rank-4 case.
\item Similarly, Theories \hyperlink{rank6-th9}{9} and \hyperlink{rank6-th10}{10} have the same 6d parent theory, but they correspond to different mass deformations. This can be seen by subtracting the $2 \times E_7$ quiver from Theory \hyperlink{rank6-th9}{9} and rebalancing, and by subtracting the $1 \times E_7$ quiver from Theory \hyperlink{rank6-th10}{10} and rebalancing. We obtain the following quivers, respectively
\begin{equation}
		\begin{tikzpicture}[baseline=-20,font=\footnotesize]
		\node[gauge, label=below:{$1$}] (tail1) {};
		\node[gauge, label=below:{$2$}] (tail2) [right=6mm of tail1] {};
		\node[gauge, label=below:{$3$}] (tail3) [right=6mm of tail2] {};
		\node[gauge, label=below:{$4$}] (k) [right=6mm of tail3] {};
        \node[gauge, label=above:{$1$}] (tail4) [above left=5mm of k] {};
		\node[gauge, label=above:{$1$}] (tail5) [above right=5mm of k] {};
		\node[gauge, bodyE7, label=below:{$3$}] (A) [right=6mm of k] {};
		\node[gauge, bodyE7, label=below:{$2$}] (B) [right=6mm of A] {};
		\node[gauge, bodyE7, label=below:{$2$}] (C) [right=6mm of B] {};
		\node[gauge, bodyE7, label=below:{$2$}] (D) [right=6mm of C] {};
		\node[gauge, bodyE7, label=below:{$1$}] (F) [right=6mm of D] {};
		\node[gauge, bodyE7, label=below:{$0$}] (G) [right=6mm of F] {};
		\node[gauge, bodyE7, label=below:{$0$}] (H) [right=6mm of G] {};
		\node[gauge, bodyE7, label=above:{$1$}] (I) [above=4mm of D] {};
		\draw[thick] (tail1) -- (tail2) -- (tail3) -- (k) -- (A);
        \draw[thick,bodyE7] (A) -- (B) -- (C) -- (D) -- (F);
        \draw[thick,bodyE7] (I) -- (D);
        \draw[thick] (k) -- (tail4);
        \draw[thick] (k) -- (tail5);

        \node[gauge, bodyE7, label=below:{$5$}] (J) [below=16mm of A] {};
        \node[gauge, label=below:{$5$}] (k2) [left=6mm of J] {};
        \node[gauge, label=below:{$4$}] (tail9) [left=6mm of k2] {};
        \node[gauge, label=below:{$3$}] (tail8) [left=6mm of tail9] {};
        \node[gauge, label=below:{$2$}] (tail7) [left=6mm of tail8] {};
        \node[gauge, label=below:{$1$}] (tail6) [left=6mm of tail7]{};
        \node[gauge, label=above:{$1$}] (tail10) [above=4mm of k2] {};
		
		\node[gauge, bodyE7, label=below:{$5$}] (K) [right=6mm of J] {};
		\node[gauge, bodyE7, label=below:{$5$}] (L) [right=6mm of K] {};
		\node[gauge, bodyE7, label=below:{$6$}] (M) [right=6mm of L] {};
		\node[gauge, bodyE7, label=below:{$4$}] (N) [right=6mm of M] {};
		\node[gauge, bodyE7, label=below:{$2$}] (O) [right=6mm of N] {};
		\node[gauge, bodyE7, label=below:{$0$}] (P) [right=6mm of O] {};
		\node[gauge, bodyE7, label=above:{$3$}] (Q) [above=4mm of M] {};
		\draw[thick] (tail6) -- (tail7) -- (tail8) -- (tail9) -- (k2) -- (J);
        \draw[thick,bodyE7] (J) -- (K) -- (L) -- (M) -- (N) -- (O);
        \draw[thick,bodyE7] (M) -- (Q);
        \draw[thick] (k2) -- (tail10);

	\end{tikzpicture}
\end{equation}
The mirror theories of these quivers are respectively
\bes{
\SU(5)-[\su(10)]~, \qquad \quad [\su(5)]- \begin{tikzpicture}[baseline, font=\footnotesize, yshift=0.1cm] \draw[gray!=20,fill=gray!20] (0,0) circle (1.5ex); \node at (0,0) {$\mathfrak{e}_8$}; \end{tikzpicture}-\SU(5)-[\su(8)]
}
where, in the second diagram, the gray circle with the label $\mathfrak{e}_8$ denotes the rank-1 $E_8$ SCFT such that one $\su(5)$ in the maximal subalgebra $\su(5) \oplus \su(5)$ of $\mathfrak{e}_8$ is gauged and coupled to $8$ hypermultiplets in fundamental representation. Comparing with the 6d parent theory, we see that, in the latter case, two fundamental hypermultiplets in the $\su(5)$ gauge group are massive and integrated out, whereas, in the former case, the $\su(5)$ flavor symmetry associated with the $(-1)$-curve is completely higgsed.
\item For Theory \hyperlink{rank6-th11}{11}, upon subtracting the $1\times E_7$ quiver from the $E_7$ quiver, we obtain
\bes{
\begin{tikzpicture}[baseline=0,font=\footnotesize]
		\node[gauge, label=below:{$1$}] (tail1) {};
        \node[gauge, label=below:{$2$}] (tail2) [right=6mm of tail1] {};
        \node[gauge, label=below:{$3$}] (tail3) [right=6mm of tail2] {};
        \node[gauge, label=below:{$4$}] (tail4) [right=6mm of tail3] {};
        \node[gauge, label=below:{$5$}] (tail5) [right=6mm of tail4] {};
		\node[gauge, label=below:{$6$}] (k) [right=6mm of tail5] {};
        \node[gauge, label=above:{$1$}] (tail6) [above=4mm of k] {};
		\node[gauge, bodyE7, label=below:{$6$}] (A) [right=6mm of k] {};
		\node[gauge, bodyE7, label=below:{$6$}] (B) [right=6mm of A] {};
		\node[gauge, bodyE7, label=below:{$6$}] (C) [right=6mm of B] {};
		\node[gauge, bodyE7, label=below:{$6$}] (D) [right=6mm of C] {};
		\node[gauge, bodyE7, label=below:{$6$}] (F) [right=6mm of D] {};
		\node[gauge, bodyE7, label=below:{$1$}] (G) [right=6mm of F] {};
		\node[gauge, bodyE7, label=below:{$0$}] (H) [right=6mm of G] {};
		\node[gauge, bodyE7, label=above:{$3$}] (I) [above=4mm of D] {};
		\draw[thick] (tail1) -- (tail2) -- (tail3) -- (tail4) -- (tail5) -- (k) -- (A);
        \draw[thick,bodyE7] (A) -- (B) -- (C) -- (D) -- (F) -- (G);
        \draw[thick,bodyE7] (I) -- (D);
        \draw[thick] (k) -- (tail6);
	\end{tikzpicture}
}
This is a mirror theory of the $\su(6)$ gauge theory with 12 fundamental and 1 antisymmetric hypermultiplets. Comparing with the corresponding 6d parent theory, we see that two hypermultiplets are massive and integrated out.	

Similarly, for Theory \hyperlink{rank6-th12}{12}, upon subtracting the $1\times E_7$ quiver from the $E_7$ quiver, we obtain
\bes{
\begin{tikzpicture}[baseline=0,font=\footnotesize]
		\node[gauge, label=below:{$1$}] (tail1) {};
        \node[gauge, label=below:{$2$}] (tail2) [right=6mm of tail1] {};
        \node[gauge, label=below:{$3$}] (tail3) [right=6mm of tail2] {};
        \node[gauge, label=below:{$4$}] (tail4) [right=6mm of tail3] {};
        \node[gauge, label=below:{$5$}] (tail5) [right=6mm of tail4] {};
		\node[gauge, label=below:{$6$}] (k) [right=6mm of tail5] {};
        \node[gauge, label=above:{$1$}] (tail6) [above=4mm of k] {};
		\node[gauge, bodyE7, label=below:{$6$}] (A) [right=6mm of k] {};
		\node[gauge, bodyE7, label=below:{$6$}] (B) [right=6mm of A] {};
		\node[gauge, bodyE7, label=below:{$6$}] (C) [right=6mm of B] {};
		\node[gauge, bodyE7, label=below:{$6$}] (D) [right=6mm of C] {};
		\node[gauge, bodyE7, label=below:{$4$}] (F) [right=6mm of D] {};
		\node[gauge, bodyE7, label=below:{$2$}] (G) [right=6mm of F] {};
		\node[gauge, bodyE7, label=below:{$0$}] (H) [right=6mm of G] {};
		\node[gauge, bodyE7, label=above:{$2$}] (I) [above=4mm of D] {};
		\draw[thick] (tail1) -- (tail2) -- (tail3) -- (tail4) -- (tail5) -- (k) -- (A);
        \draw[thick,bodyE7] (A) -- (B) -- (C) -- (D) -- (F) -- (G);
        \draw[thick,bodyE7] (I) -- (D);
        \draw[thick] (k) -- (tail6);
	\end{tikzpicture}
}
This is a mirror theory of the $\su(6)$ gauge theory with 13 fundamental and 1 rank-three antisymmetric half-hypermultiplets. Comparing with the corresponding 6d parent theory, we see that two hypermultiplets are massive and integrated out.	
\een

\subsubsection{Generalizations: \texorpdfstring{$E_7$}{E7} Quiver with Two Tails}\label{sec:InvFormulasE72tailstoE8}

One can generalize the inverse algorithm described above in the case in which the $E_7$ theory has two tails, i.e.
\begin{equation}
	\begin{tikzpicture}[baseline=7,font=\footnotesize]
		\node (dots1) {$\cdots$};
		\node[gauge, label=below:{$k_1$}] (k1) [right=6mm of dots1] {};
		\node[gauge, bodyE7, label=below:{$A$}] (A) [right=6mm of k1] {};
		\node[gauge, bodyE7, label=below:{$B$}] (B) [right=6mm of A] {};
		\node[gauge, bodyE7, label=below:{$C$}] (C) [right=6mm of B] {};
		\node[gauge, bodyE7, label=below:{$D$}] (D) [right=6mm of C] {};
		\node[gauge, bodyE7, label=below:{$E$}] (E) [right=6mm of D] {};
		\node[gauge, bodyE7, label=below:{$F$}] (F) [right=6mm of E] {};
		\node[gauge, bodyE7, label=below:{$G$}] (G) [right=6mm of F] {};
		\node[gauge, bodyE7, label=above:{$H$}] (FG) [above=4mm of D] {};
		\node[gauge, label=below:{$k_2$}] (k2) [right=6mm of G] {};
		\node (dots2) [right=6mm of k2] {$\cdots$};
		\draw[thick] (dots1.east) -- (k1.west);
		\draw[thick] (k1.east) -- (A.west);
		\draw[thick,bodyE7] (A.east) -- (B.west);
		\draw[thick,bodyE7] (B.east) -- (C.west);
		\draw[thick,bodyE7] (C.east) -- (D.west);
		\draw[thick,bodyE7] (D.east) -- (E.west);
		\draw[thick,bodyE7] (E.east) -- (F.west);
		\draw[thick,bodyE7] (F.east) -- (G.west);
		\draw[thick,bodyE7] (D.north) -- (FG.south);
		\draw[thick] (G.east) -- (k2.west);
		\draw[thick] (k2.east) -- (dots2.west);
	\end{tikzpicture}
	\label{eq:E72tail}
\end{equation}

Without losing generality, we can assume $k_1\geq k_2$,\footnote{If $k_1<k_2$, it is sufficient to redefine the nodes to mirror the quiver with respect to the vertical line.} and we call $k = k_1+k_2$. The general resulting quiver will be
\begin{equation}
	\begin{tikzpicture}[baseline=7,font=\footnotesize]
		\node (dots) {$\cdots$};
		\node[gauge, label=below:{$k$}] (k) [right=6mm of dots] {};
		\node[gauge, bodyE8] (A) [right=6mm of k] {};
		\node[gauge, bodyE8] (B) [right=6mm of A] {};
		\node[gauge, bodyE8] (C) [right=6mm of B] {};
		\node[gauge, bodyE8] (D) [right=6mm of C] {};
		\node[gauge, bodyE8] (E) [right=6mm of D] {};
		\node[gauge, bodyE8] (F) [right=6mm of E] {};
		\node[gauge, bodyE8] (G) [right=6mm of F] {};
		\node[gauge, bodyE8] (H) [right=6mm of G] {};
		\node[gauge, bodyE8] (FEG) [above=4mm of F] {};
		\draw[thick] (dots.east) -- (k.west);
		\draw[thick] (k.east) -- (A.west);
		\draw[thick,bodyE8] (A.east) -- (B.west);
		\draw[thick,bodyE8] (B.east) -- (C.west);
		\draw[thick,bodyE8] (C.east) -- (D.west);
		\draw[thick,bodyE8] (D.east) -- (E.west);
		\draw[thick,bodyE8] (E.east) -- (F.west);
		\draw[thick,bodyE8] (F.east) -- (G.west);
		\draw[thick,bodyE8] (G.east) -- (H.west);
		\draw[thick,bodyE8] (F.north) -- (FEG.south);
	\end{tikzpicture}
\end{equation} 
where the ranks of the $E_8$ type quiver depend on the kind of deformation that is used in order to get the $E_7$ theory, up to dualizations \cite{Yaakov:2013fza}. Let us, for instance, consider a deformation involving nodes $3$ and $3'$  of the $E_8$ quiver, then \eqref{eq:E72tail} can be obtained by starting from an $E_8$ quiver of the form
\begin{equation}\label{eq:33plift}
	\begin{tikzpicture}[baseline=7,font=\footnotesize]
		\node (dots) {$\cdots$};
		\node[gauge, label=below:{$k$}] (k) [right=6mm of dots] {};
		\node[gauge, bodyE8, label=above:{$A+k_2$}] (A) [right=9mm of k] {};
		\node[gauge, bodyE8, label=below:{$B+k_2$}] (B) [right=9mm of A] {};
		\node[gauge, bodyE8, label=above:{$C+k_2$}] (C) [right=9mm of B] {};
		\node[gauge, bodyE8, label=below:{$G+C$}] (D) [right=6mm of C] {};
		\node[gauge, bodyE8, label=above:{$F+C$}] (E) [right=6mm of D] {};
		\node[gauge, bodyE8, label=below:{$E+C$}] (F) [right=9mm of E] {};
		\node[gauge, bodyE8, label=above:{$D$}] (G) [right=9mm of F] {};
		\node[gauge, bodyE8, label=below:{$H$}] (H) [right=9mm of G] {};
		\node[gauge, bodyE8, label=above:{$C$}] (FEG) [above=4mm of F] {};
		\draw[thick] (dots.east) -- (k.west);
		\draw[thick] (k.east) -- (A.west);
		\draw[thick,bodyE8] (A.east) -- (B.west);
		\draw[thick,bodyE8] (B.east) -- (C.west);
		\draw[thick,bodyE8] (C.east) -- (D.west);
		\draw[thick,bodyE8] (D.east) -- (E.west);
		\draw[thick,bodyE8] (E.east) -- (F.west);
		\draw[thick,bodyE8] (F.east) -- (G.west);
		\draw[thick,bodyE8] (G.east) -- (H.west);
		\draw[thick,bodyE8] (F.north) -- (FEG.south);
	\end{tikzpicture}
\end{equation}
It is easy to check that all the excess numbers of the $E_8$ quiver for the nodes that are not involved in the deformation are related to the excess number of the $E_7$ quiver, namely
\begin{equation}
	n_1 = \texttt{e}_A\coma n_2 = \texttt{e}_B\coma n_4 = \texttt{e}_G\coma n_5=\texttt{e}_F\coma n_6=\texttt{e}_E\coma n_{4'}=\texttt{e}_D\text{ and } n_{2'} = \texttt{e}_H\fstop
\end{equation}
On the other hand, the excess numbers $n_3$ and $n_{3'}$ cannot be obtained from the $E_7$ quiver, and we must check whether they are positive or not. If the nodes are bad, we show in Appendix \ref{sec:MixDef2tail} that a set of dualizations starting from those nodes will always lead to a good $E_8$  quiver. We collect all the inversion of the formulas of Table \ref{tab:E8toE7-2tail-body} (up to reflection of the $E_7$ quiver tails with respect to the central nodes) in Table \ref{tab:E7toE8-2tail}. However, we stress that all these uplifts are equivalent up to dualizations, as explained in Appendix \ref{sec:MixDef2tail}.

\begin{table}[!htp]
	\begin{center}
	\begin{tabular}{c|c}
		Nodes & $E_8$ Quiver \\
		\hline
		$2$ / $2'$ & \begin{tikzpicture}[baseline=0,font=\footnotesize]
		\node (dots) {$\cdots$};
		\node[gauge, label=below:{$k$}] (k) [right=6mm of dots] {};
		\node[gauge, bodyE8, label=above:{$A+k_2$}] (A) [right=9mm of k] {};
		\node[gauge, bodyE8, label=below:{$B+k_2$},fill=cyan] (B) [right=9mm of A] {};
		\node[gauge, bodyE8, label=above:{$G+B$}] (C) [right=9mm of B] {};
		\node[gauge, bodyE8, label=below:{$F+B$}] (D) [right=6mm of C] {};
		\node[gauge, bodyE8, label=above:{$E+B$}] (E) [right=6mm of D] {};
		\node[gauge, bodyE8, label=below:{$D+B$}] (F) [right=9mm of E] {};
		\node[gauge, bodyE8, label=above:{$H+B$}] (G) [right=9mm of F] {};
		\node[gauge, bodyE8, label=below:{$B$},fill=cyan] (H) [right=9mm of G] {};
		\node[gauge, bodyE8, label=above:{$C$}] (FEG) [above=4mm of F] {};
		\draw[thick] (dots.east) -- (k.west);
		\draw[thick] (k.east) -- (A.west);
		\draw[thick,bodyE8] (A.east) -- (B.west);
		\draw[thick,bodyE8] (B.east) -- (C.west);
		\draw[thick,bodyE8] (C.east) -- (D.west);
		\draw[thick,bodyE8] (D.east) -- (E.west);
		\draw[thick,bodyE8] (E.east) -- (F.west);
		\draw[thick,bodyE8] (F.east) -- (G.west);
		\draw[thick,bodyE8] (G.east) -- (H.west);
		\draw[thick,bodyE8] (F.north) -- (FEG.south);
	\end{tikzpicture} \\ \hline
		$3$ / $3'$ & \begin{tikzpicture}[baseline=0,font=\footnotesize]
		\node (dots) {$\cdots$};
		\node[gauge, label=below:{$k$}] (k) [right=6mm of dots] {};
		\node[gauge, bodyE8, label=above:{$A+k_2$}] (A) [right=9mm of k] {};
		\node[gauge, bodyE8, label=below:{$B+k_2$}] (B) [right=9mm of A] {};
		\node[gauge, bodyE8, label=above:{$C+k_2$},fill=cyan] (C) [right=9mm of B] {};
		\node[gauge, bodyE8, label=below:{$G+C$}] (D) [right=6mm of C] {};
		\node[gauge, bodyE8, label=above:{$F+C$}] (E) [right=6mm of D] {};
		\node[gauge, bodyE8, label=below:{$E+C$}] (F) [right=9mm of E] {};
		\node[gauge, bodyE8, label=above:{$D$}] (G) [right=9mm of F] {};
		\node[gauge, bodyE8, label=below:{$H$}] (H) [right=9mm of G] {};
		\node[gauge, bodyE8, label=above:{$C$},fill=cyan] (FEG) [above=4mm of F] {};
		\draw[thick] (dots.east) -- (k.west);
		\draw[thick] (k.east) -- (A.west);
		\draw[thick,bodyE8] (A.east) -- (B.west);
		\draw[thick,bodyE8] (B.east) -- (C.west);
		\draw[thick,bodyE8] (C.east) -- (D.west);
		\draw[thick,bodyE8] (D.east) -- (E.west);
		\draw[thick,bodyE8] (E.east) -- (F.west);
		\draw[thick,bodyE8] (F.east) -- (G.west);
		\draw[thick,bodyE8] (G.east) -- (H.west);
		\draw[thick,bodyE8] (F.north) -- (FEG.south);
	\end{tikzpicture} \\ \hline
		$4$ / $4'$ & \begin{tikzpicture}[baseline=0,font=\footnotesize]
		\node (dots) {$\cdots$};
		\node[gauge, label=below:{$k$}] (k) [right=6mm of dots] {};
		\node[gauge, bodyE8, label=above:{$A+k_2$}] (A) [right=9mm of k] {};
		\node[gauge, bodyE8, label=below:{$B+k_2$}] (B) [right=9mm of A] {};
		\node[gauge, bodyE8, label=above:{$C+k_2$}] (C) [right=9mm of B] {};
		\node[gauge, bodyE8, label=below:{$D+k_2$},fill=cyan] (D) [right=6mm of C] {};
		\node[gauge, bodyE8, label=above:{$G+D$}] (E) [right=6mm of D] {};
		\node[gauge, bodyE8, label=below:{$F+D$}] (F) [right=9mm of E] {};
		\node[gauge, bodyE8, label=above:{$D$},fill=cyan] (G) [right=9mm of F] {};
		\node[gauge, bodyE8, label=below:{$H$}] (H) [right=9mm of G] {};
		\node[gauge, bodyE8, label=above:{$E$}] (FEG) [above=4mm of F] {};
		\draw[thick] (dots.east) -- (k.west);
		\draw[thick] (k.east) -- (A.west);
		\draw[thick,bodyE8] (A.east) -- (B.west);
		\draw[thick,bodyE8] (B.east) -- (C.west);
		\draw[thick,bodyE8] (C.east) -- (D.west);
		\draw[thick,bodyE8] (D.east) -- (E.west);
		\draw[thick,bodyE8] (E.east) -- (F.west);
		\draw[thick,bodyE8] (F.east) -- (G.west);
		\draw[thick,bodyE8] (G.east) -- (H.west);
		\draw[thick,bodyE8] (F.north) -- (FEG.south);
	\end{tikzpicture} \\ \hline
		$5$ / $3'+2'$ & \begin{tikzpicture}[baseline=0,font=\footnotesize]
		\node (dots) {$\cdots$};
		\node[gauge, label=below:{$k$}] (k) [right=6mm of dots] {};
		\node[gauge, bodyE8, label=above:{$A+k_2$}] (A) [right=9mm of k] {};
		\node[gauge, bodyE8, label=below:{$B+k_2$}] (B) [right=9mm of A] {};
		\node[gauge, bodyE8, label=above:{$C+k_2$}] (C) [right=9mm of B] {};
		\node[gauge, bodyE8, label=below:{$D+k_2$}] (D) [right=6mm of C] {};
		\node[gauge, bodyE8, label=above:{$E+H+k_2$},fill=cyan] (E) [right=6mm of D] {};
		\node[gauge, bodyE8, label=below:{$G+E+H$}] (F) [right=9mm of E] {};
		\node[gauge, bodyE8, label=above:{$F+H$}] (G) [right=9mm of F] {};
		\node[gauge, bodyE8, label=below:{$H$},fill=cyan] (H) [right=9mm of G] {};
		\node[gauge, bodyE8, label=above:{$E$},fill=cyan] (FEG) [above=4mm of F] {};
		\draw[thick] (dots.east) -- (k.west);
		\draw[thick] (k.east) -- (A.west);
		\draw[thick,bodyE8] (A.east) -- (B.west);
		\draw[thick,bodyE8] (B.east) -- (C.west);
		\draw[thick,bodyE8] (C.east) -- (D.west);
		\draw[thick,bodyE8] (D.east) -- (E.west);
		\draw[thick,bodyE8] (E.east) -- (F.west);
		\draw[thick,bodyE8] (F.east) -- (G.west);
		\draw[thick,bodyE8] (G.east) -- (H.west);
		\draw[thick,bodyE8] (F.north) -- (FEG.south);
	\end{tikzpicture} \\ \hline
		$6$ / $3'+2'+1$ & \begin{tikzpicture}[baseline=0,font=\footnotesize]
		\node (dots) {$\cdots$};
		\node[gauge, label=below:{$k$}] (k) [right=6mm of dots] {};
		\node[gauge, bodyE8, label=above:{$A+k_2$},fill=cyan] (A) [right=9mm of k] {};
		\node[gauge, bodyE8, label=below:{$G+A$}] (B) [right=9mm of A] {};
		\node[gauge, bodyE8, label=above:{$F+A$}] (C) [right=9mm of B] {};
		\node[gauge, bodyE8, label=below:{$E+A$}] (D) [right=6mm of C] {};
		\node[gauge, bodyE8, label=above:{$D+A$}] (E) [right=6mm of D] {};
		\node[gauge, bodyE8, label=below:{$A+C+H$},fill=cyan] (F) [right=9mm of E] {};
		\node[gauge, bodyE8, label=above:{$B+H$}] (G) [right=9mm of F] {};
		\node[gauge, bodyE8, label=below:{$H$},fill=cyan] (H) [right=9mm of G] {};
		\node[gauge, bodyE8, label=above:{$C$},fill=cyan] (FEG) [above=4mm of F] {};
		\draw[thick] (dots.east) -- (k.west);
		\draw[thick] (k.east) -- (A.west);
		\draw[thick,bodyE8] (A.east) -- (B.west);
		\draw[thick,bodyE8] (B.east) -- (C.west);
		\draw[thick,bodyE8] (C.east) -- (D.west);
		\draw[thick,bodyE8] (D.east) -- (E.west);
		\draw[thick,bodyE8] (E.east) -- (F.west);
		\draw[thick,bodyE8] (F.east) -- (G.west);
		\draw[thick,bodyE8] (G.east) -- (H.west);
		\draw[thick,bodyE8] (F.north) -- (FEG.south);
	\end{tikzpicture}
	\end{tabular}
\end{center}
	\caption{Some uplifts of $E_7$ quivers to their parent $E_8$ orbi-instanton theories obtained by inverting Table \ref{tab:E8toE7-2tail-body}, up to reflection of the $E_7$ quiver tails with respect to central node. We colored in {\color{cyan} cyan} the nodes whose excess numbers cannot be determined directly from the $E_7$ quiver. In case they are negative, the parent theory is obtained with a series of dualizations \cite{Yaakov:2013fza}  as we explain in Appendix \ref{sec:MixDef2tail}.}
	\label{tab:E7toE8-2tail}
\end{table}

\subsubsection{Obtaining 6d Parent Theories and Realizing Deformations}
\label{sec:obtainE8fromE7}

Let us suppose that the good $E_8$ parent quiver is of the form \eref{magorbi1}. The procedure for obtaining the F-theory quiver associated with the 6d parent theory is exactly as described in \cref{sec:procedure}, namely finding a mirror theory of \eref{E8subreb} and inserting appropriate $(-1)$ and $(-2)$-curves.

We emphasize, however, that, upon subtracting the $\fm \times E_7$ quiver out of a given $E_7$ quiver in the two-tail case, the rebalancing procedure is different from the one-tail case, which was illustrated in \eref{E7subreb}.  In particular, after subtracting the $\fm \times E_7$ quiver, the rebalancing procedure yields instead the following quiver: 
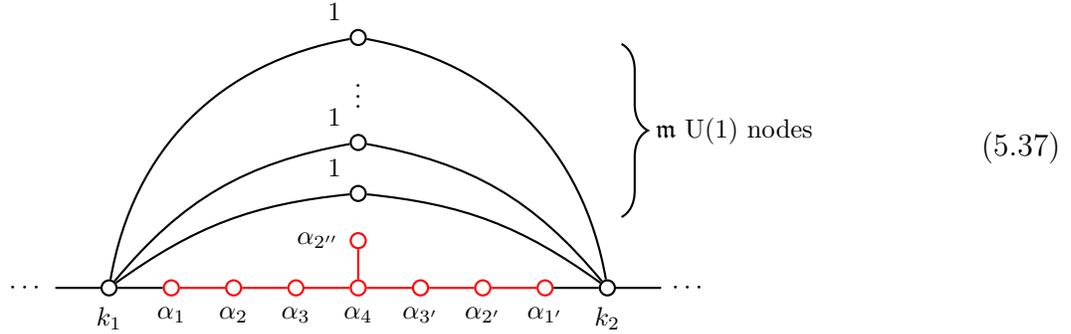
\begin{equation} \label{E7subreb2tails}
	\begin{tikzpicture}[baseline=50,font=\footnotesize]
		\node (dots1) {$\cdots$};
		\node[gauge, label=below:{$k_1$}] (k1) [right=6mm of dots1] {};
		\node[gauge, bodyE7, label=below:{$\alpha_1$}] (A) [right=6mm of k1] {};
		\node[gauge, bodyE7, label=below:{$\alpha_2$}] (B) [right=6mm of A] {};
		\node[gauge, bodyE7, label=below:{$\alpha_3$}] (C) [right=6mm of B] {};
		\node[gauge, bodyE7, label=below:{$\alpha_4$}] (D) [right=6mm of C] {};
		\node[gauge, bodyE7, label=below:{$\alpha_{3'}$}] (E) [right=6mm of D] {};
		\node[gauge, bodyE7, label=below:{$\alpha_{2'}$}] (F) [right=6mm of E] {};
		\node[gauge, bodyE7, label=below:{$\alpha_{1'}$}] (G) [right=6mm of F] {};
		\node[gauge, bodyE7, label=left:{$\alpha_{2''}$}] (FG) [above=4mm of D] {};
		\node[gauge, label=below:{$k_2$}] (k2) [right=6mm of G] {};
		\node (dots2) [right=6mm of k2] {$\cdots$};
            \node[gauge, label=above left:{$1$}] (1a) [above=4mm of FG] {};
            \node[gauge, label=above left:{$1$}] (1a1) [above=4.5mm of 1a] {};
            \node (1b) [above=2mm of 1a1] {$\vdots$};
            \node[gauge, label=above left:{$1$}] (1c) [above=1.5mm of 1b] {};
		\draw[thick] (dots1.east) -- (k1.west);
		\draw[thick] (k1.east) -- (A.west);
		\draw[thick,bodyE7] (A.east) -- (B.west);
		\draw[thick,bodyE7] (B.east) -- (C.west);
		\draw[thick,bodyE7] (C.east) -- (D.west);
		\draw[thick,bodyE7] (D.east) -- (E.west);
		\draw[thick,bodyE7] (E.east) -- (F.west);
		\draw[thick,bodyE7] (F.east) -- (G.west);
		\draw[thick,bodyE7] (D.north) -- (FG.south);
		\draw[thick] (G.east) -- (k2.west);
		\draw[thick] (k2.east) -- (dots2.west);
            \draw[thick] (k1) to[bend left=15] (1a) (1a) to[bend left=15] (k2);
            \draw[thick] (k1) to[bend left=20] (1a1) (1a1) to[bend left=20](k2);
            \draw[thick] (k1) to[bend left=35](1c) (1c)to[bend left=35] (k2);
            \draw[thick] [decorate,decoration={brace,amplitude=10pt, mirror},xshift=0cm,yshift=0cm]
([xshift=3.5cm,yshift=-2mm]1a.south) -- ([xshift=3.5cm,yshift=-2mm]1c.north) node [black,midway,xshift=1.5cm] 
{$\fm$ $\U(1)$ nodes};
	\end{tikzpicture}
\end{equation}
where an overall $\U(1)$ must be modded out. Upon finding the mirror dual of this quiver, we can determine the deformation as described in \cref{sec:procedure}.

\subsubsection{Examples: \texorpdfstring{$E_7$}{E7} Quivers with Four-dimensional Higgs Branch and Two Tails} 
\label{sec:rank4twotails}

Let us list below the $E_7$ quivers containing two tails with a four-dimensional Higgs branch, where the rank of the central node less than or equal to $10$.
\begin{longtable}{c|c|c|c}
		Number & $E_7$ Quiver & $E_8$ Parent Quiver & 6d Parent Theory\\
		\hline
  \endhead
		\hypertarget{rank4-2t-th1}{} 1 & 
		$\begin{array}{cc @{} >{\color{cE7}}c >{\color{cE7}}c >{\color{cE7}}c >{\color{cE7}}c >{\color{cE7}}c >{\color{cE7}}c >{\color{cE7}}cc @{}}
			  &   &   &   &   & 4 &   &   &   &     \\
			1 &   & 3 & 5 & 7 & 9 & 7 & 5 & 3 &   1 \\
		\end{array}$
		& 
		$\begin{array}{ccc @{} >{\color{cE8}}c >{\color{cE8}}c >{\color{cE8}}c >{\color{cE8}}c >{\color{cE8}}c >{\color{cE8}}c >{\color{cE8}}c >{\color{cE8}}c @{}}
			  &   &   &   &   &   &   &   & 7 &   &   \\
			1 & 2 &   & 4 & 6 & 8 & 10 & 12 & 14 & 9 & 4 \\
		\end{array}$
		& $[\so(16)]\,\, \overset{\usp(2)}{1}\,\, \overset{\su(2)}{2} \, [\so(4)]$  \hypertarget{rank4-2t-th2}{}\\
		\hline
		2 & 
		$\begin{array}{cc @{} >{\color{cE7}}c >{\color{cE7}}c >{\color{cE7}}c >{\color{cE7}}c >{\color{cE7}}c >{\color{cE7}}c >{\color{cE7}}cc @{}}
			  &   &   &   &   & 5 &   &   &   &      \\
			2 &   & 4 & 6 & 8 & 10 & 7 & 4 & 2    & 1 \\
		\end{array}$
		& 
		$\begin{array}{ccc @{} >{\color{cE8}}c >{\color{cE8}}c >{\color{cE8}}c >{\color{cE8}}c >{\color{cE8}}c >{\color{cE8}}c >{\color{cE8}}c >{\color{cE8}}c @{}}
			  &   &   &   &   &   &   &   & 7 &   &   \\
			1 & 2 &   & 4 & 6 & 8 & 10 & 12 & 14 & 9 & 4 \\
		\end{array}$
		& $[\so(16)] \,\, \overset{\usp(2)}{1} \,\, \overset{\su(2)}{2} \,\, [\so(4)]$  \hypertarget{rank4-2t-th3}{}\\
		\hline
		3 &
		$\begin{array}{ccc@{} >{\color{cE7}}c >{\color{cE7}}c >{\color{cE7}}c >{\color{cE7}}c >{\color{cE7}}c >{\color{cE7}}c >{\color{cE7}}cc @{}}
			  &   &   &   &   &   & 4 &   &   &   &     \\
			1 & 2 &   & 3 & 4 & 6 & 8 & 6 & 4 & 2    & 1 \\
		\end{array}$
		& 
		$\begin{array}{cccc @{} >{\color{cE8}}c >{\color{cE8}}c >{\color{cE8}}c >{\color{cE8}}c >{\color{cE8}}c >{\color{cE8}}c >{\color{cE8}}c >{\color{cE8}}c @{}}
			  &   &   &   &   &   &   &   &   & 6 &   &   \\
			1 & 2 & 3 &   & 4 & 5 & 6 & 8 & 10 & 12 & 8 & 4 \\
		\end{array}$
		& $[\mathfrak{e}_6]\,\, 1 \,\, \overset{\su(3)}{2}\,\, [\su(6)]$ \hypertarget{rank4-2t-th4}{}\\
		\hline
		4 &
		$\begin{array}{cccc@{} >{\color{cE7}}c >{\color{cE7}}c >{\color{cE7}}c >{\color{cE7}}c >{\color{cE7}}c >{\color{cE7}}c >{\color{cE7}}cc @{}}
			  &   &   &   &   &   &   & 3 &   &   &   &      \\
			1 & 2 & 3 &   & 4 & 5 & 6 & 7 & 5 & 3 & 2    & 1 \\
		\end{array}$
		& 
		$\begin{array}{ccccc @{} >{\color{cE8}}c >{\color{cE8}}c >{\color{cE8}}c >{\color{cE8}}c >{\color{cE8}}c >{\color{cE8}}c >{\color{cE8}}c >{\color{cE8}}c @{}}
			  &   &   &   &   &   &   &   &   &   & 5 &   &   \\
			1 & 2 & 3 & 4 &   & 5 & 6 & 7 & 8 & 9 & 10 & 6 & 3 \\
		\end{array}$
		& $\underset{[\wedge^2]}{\overset{\su(4)}{1}}\,\, [\su(12)]$  \hypertarget{rank4-2t-th5}{}\\
		\hline
		5 & 
		$\begin{array}{ccc@{} >{\color{cE7}}c >{\color{cE7}}c >{\color{cE7}}c >{\color{cE7}}c >{\color{cE7}}c >{\color{cE7}}c >{\color{cE7}}ccc @{}}
			  &   &   &   &   &   & 3 &   &   &   &      &   \\
			1 & 2 &   & 3 & 4 & 5 & 6 & 5 & 4 & 3    & 2 & 1 \\
		\end{array}$
		& 
		$\begin{array}{ccccc @{} >{\color{cE8}}c >{\color{cE8}}c >{\color{cE8}}c >{\color{cE8}}c >{\color{cE8}}c >{\color{cE8}}c >{\color{cE8}}c >{\color{cE8}}c @{}}
			  &   &   &   &   &   &   &   &   &   & 5 &   &   \\
			1 & 2 & 3 & 4 &   & 5 & 6 & 7 & 8 & 9 & 10 & 6 & 3 \\
		\end{array}$
		& $\underset{[\wedge^2]}{\overset{\su(4)}{1}}\,\, [\su(12)]$ \hypertarget{rank4-2t-th6}{}\\
		\hline
		6 & 
		$\begin{array}{cccc@{} >{\color{cE7}}c >{\color{cE7}}c >{\color{cE7}}c >{\color{cE7}}c >{\color{cE7}}c >{\color{cE7}}c >{\color{cE7}}cccc @{}}
			  &   &   &   &   &   &   & 2 &   &   &   &   &   &      \\
			1 & 2 & 3 &   & 4 & 5 & 6 & 7 & 6 & 5 & 4 &    3 & 2 & 1 \\
		\end{array}$
		& 
		$\begin{array}{ccccccc @{} >{\color{cE8}}c >{\color{cE8}}c >{\color{cE8}}c >{\color{cE8}}c >{\color{cE8}}c >{\color{cE8}}c >{\color{cE8}}c >{\color{cE8}}c @{}}
			  &   &   &   &   &   &   &   &   &   &   &   & 6 &   &   \\
			1 & 2 & 3 & 4 & 5 & 6 &   & 7 & 8 & 9 & 10 & 11 & 12 & 7 & 2 \\
		\end{array}$
		& $\overset{\usp(6)}{1} \, [\so(28)]$ 
\end{longtable}	

\newpage
Let us comment on the above results as follows. 
\ben
\item Subtracting the $E_7$ quiver in Theory \hyperlink{rank4-2t-th1}{1} by the $2 \times E_7$ quiver and rebalancing, we arrive at \eref{E7subreb2tails} where, upon decoupling an overall $\U(1)$, we obtain  
\begin{equation}
\begin{tikzpicture}[baseline=7,font=\footnotesize]
\node at (0,0) (dots) {$\cdots$};
\node[gauge, label=below:{$1$}] (n1) [left=6mm of dots] {};
\node[gauge, label=below:{$1$}] (n0) [left=6mm of n1] {};
\node[gauge, label=below:{$1$}] (n2) [right=6mm of dots] {};
\node[gauge, label=below:{$1$}] (n3) [right=6mm of n2] {};
\node[gauge, label=above:{$1$}] (nt) [above=6mm of dots] {};
\node[flavor, label=below:{$1$}] (fL) [left=6mm of n0] {};
\node[flavor, label=below:{$1$}] (fR) [right=6mm of n3] {};
\draw[thick] (fL)--(n0)--(n1)--(dots)--(n2)--(n3)--(fR);
\draw[thick] (n0) to[bend left=15] (nt) (nt) to[bend left=15](n3);
\draw[thick] [decorate,decoration={brace,amplitude=10pt, mirror},xshift=0cm, yshift=-0.7cm]
([yshift=-0.75cm]n0.west) -- ([yshift=-0.75cm]n3.east) node [black,midway,yshift=-0.75cm] 
{\footnotesize $10$ $\U(1)$ nodes in total};
\end{tikzpicture}
\end{equation}
The Coulomb branch of this theory describes the moduli space of $\su(10)$ instantons on $\BC^2/\BZ_2$ with the holonomy such that $\su(10)$ is broken to $\su(8) \oplus \su(2) \oplus \mathfrak{u}(1)$ \cite{Mekareeya:2015bla}. The mirror theory of this is
\bes{ \label{theory1subE7}
\begin{tikzpicture}[baseline=5,font=\footnotesize]
\node[gauge, label=below:{$1$}] (nL)  {};
\node[gauge, label=below:{$1$}] (nR) [right=9mm of nL] {};
\node[flavor, label=below:{$8$}] (fL) [left=6mm of nL] {};
\node[flavor, label=below:{$2$}] (fR) [right=6mm of nR] {};
\draw[thick] (nL) to [bend left=20] (nR);
\draw[thick] (nL) to [bend right=20] (nR);
\draw[thick] (nL)--(fL);
\draw[thick] (nR)--(fR);
\end{tikzpicture}
}
Comparing this with the corresponding 6d parent theory, we see that the $\so(16) \oplus \su(2)^2$ flavor symmetry of the latter\footnote{See the discussion in \cite[Section 5.1]{Mekareeya:2017jgc} and \cite[Section 3]{Hanany:2018vph} regarding the global symmetry of the 6d parent theory of Theories \hyperlink{rank4-2t-th1}{1} and \hyperlink{rank4-2t-th2}{2}.} is broken to $\su(8) \oplus \su(2)^2 \oplus \u(1)$,\footnote{Note that $\su(8) \oplus \su(2) \oplus \mathfrak{u}(1)$ can be seen from the holonomy of the instantons. The extra $\su(2)$ factor corresponds to the isometry of $\BC^2/\BZ_2$. This was pointed out in \cite{Mekareeya:2015bla}.} as manifest in \eref{theory1subE7}, upon mass deformation. Note that the gauge groups also get modified. From the 4d perspective, this means that the deformation in question involves not only the Higgs branch, but also the Coulomb branch. 
\item Upon subtracting the $E_7$ quiver in Theory \hyperlink{rank4-2t-th2}{2} by the $2 \times E_7$ quiver and rebalancing, we arrive at \eref{E7subreb2tails}, which can be rewritten as
\bes{
\begin{tikzpicture}[baseline=7,font=\footnotesize]
\node[gauge, label=below:{$2$}] (n1)  {};
\node[gauge, label=above:{$1$}] (nLU) [above left=6mm of n1] {};
\node[gauge, label=below:{$1$}] (nLD) [below left=6mm of n1] {};
\node[gauge, label=below:{$2$}] (n2) [right=6mm of n1] {};
\node[gauge, label=below:{$2$}] (n3) [right=6mm of n2] {};
\node[gauge, label=below:{$2$}] (n4) [right=6mm of n3] {};
\node[gauge, label=below:{$2$}] (n5) [right=6mm of n4] {};
\node[gauge, label=above:{$1$}] (nRU) [above right=6mm of n5] {};
\node[gauge, label=below:{$1$}] (nRD) [below right=6mm of n5] {};
\node[gauge, label=below:{$1$}] (n6) [below right=6mm of nRU] {};
\draw[thick] (nLU)--(n1)--(n2)--(n3)--(n4)--(n5)--(nRU)--(n6)--(nRD)--(n5);
\draw[thick] (nLD)--(n1);
\end{tikzpicture}
}
Decoupling an overall $\U(1)$ from the rightmost node, we obtain
\bes{
\begin{tikzpicture}[baseline=0,font=\footnotesize]
\node[gauge, label=below:{$2$}] (n1)  {};
\node[gauge, label=above:{$1$}] (nLU) [above left=6mm of n1] {};
\node[gauge, label=below:{$1$}] (nLD) [below left=6mm of n1] {};
\node[gauge, label=below:{$2$}] (n2) [right=6mm of n1] {};
\node[gauge, label=below:{$2$}] (n3) [right=6mm of n2] {};
\node[gauge, label=below:{$2$}] (n4) [right=6mm of n3] {};
\node[gauge, label=below:{$2$}] (n5) [right=6mm of n4] {};
\node[gauge, label=above:{$1$}] (nRU) [above right=6mm of n5] {};
\node[gauge, label=below:{$1$}] (nRD) [below right=6mm of n5] {};
\node[flavor, label=above:{$1$}] (f1) [right=6mm of nRU] {};
\node[flavor, label=below:{$1$}] (f2) [right=6mm of nRD] {};
\draw[thick] (nLU)--(n1)--(n2)--(n3)--(n4)--(n5)--(nRU)--(f1);
\draw[thick] (nLD)--(n1);
\draw[thick] (n5)--(nRD)--(f2);
\end{tikzpicture}
}
whose Coulomb branch describes the moduli space of one $\so(16)$ instanton on $\BC^2/\BZ_2$ with the holonomy such that $\so(16)$ is broken to $\so(14) \oplus \so(2)$ \cite{Mekareeya:2015bla}. The mirror theory is
\bes{ \label{theory2subE7}
[\so(14)] - \USp(2) - \USp(2) - [\so(2)]
}
Comparing this with the corresponding 6d parent theory, we see that the $\so(16) \oplus \su(2)^2$ flavor symmetry of the latter is broken to $\so(14)  \oplus \su(2) \oplus \u(1)  $, as manifest in \eref{theory2subE7}, upon mass deformation. In this case, the gauge groups do not get modified.
\item Similarly to the previous case, for Theory \hyperlink{rank4-2t-th2}{3}, after subtracting the $2 \times E_7$ quiver and rebalancing, we obtain
\bes{
\begin{tikzpicture}[baseline=7,font=\footnotesize]
\node[gauge, label=below:{$2$}] (n1)  {};
\node[gauge, label=below:{$1$}] (nLU) [above left=6mm of n1] {};
\node[gauge, label=below:{$1$}] (nLD) [below left=6mm of n1] {};
\node[gauge, label=below:{$1$}] (nRU) [above right=6mm of n1] {};
\node[gauge, label=below:{$1$}] (nRD) [below right=6mm of n1] {};
\node[flavor, label=below:{$1$}] (f1) [right=6mm of nRU] {};
\node[flavor, label=below:{$1$}] (f2) [right=6mm of nRD] {};
\draw[thick] (nLU)--(n1)--(nRU)--(f1);
\draw[thick] (nLD)--(n1)--(nRD)--(f2);
\end{tikzpicture}
}
Its Coulomb branch describes the moduli space of one $\so(8)$ instanton on $\BC^2/\BZ_2$ with the holonomy such that $\so(8)$ is broken to $\so(6) \oplus \so(2) \cong \su(4) \oplus \u(1)$ (see \cite{Mekareeya:2015bla}). The mirror theory is described by
\bes{ \label{theory2subE7-2}
[\so(6)] - \USp(2) - \USp(2) - [\so(2)]
}
This theory has the flavor symmetry $\so(6) \oplus \su(2) \oplus \so(2) \cong \su(4) \oplus \su(2) \oplus \u(1)$. Note, however, that, in addition to this, the Coulomb branch symmetry of the $E_7$ quiver also contains $\so(10)$.
Hence, we see that the $\mathfrak{e}_6 \oplus \su(6) \oplus \u(1)$ flavor symmetry of the 6d parent theory is broken to $\so(10)  \oplus \su(4) \oplus \su(2) \oplus \u(1)$ upon deformation.
\item For Theory \hyperlink{rank4-2t-th2}{4}, after subtracting the $1 \times E_7$ quiver and rebalancing, we obtain
\bes{
\begin{tikzpicture}[baseline=0,font=\footnotesize]
		\node[gauge, label=below:{$1$}] (tail1) {};
        \node[gauge, label=below:{$2$}] (tail2) [right=6mm of tail1] {};
		\node[gauge, label=below:{$3$}] (k) [right=6mm of tail2] {};
        \node[flavor, label=above:{$1$}] (tail3) [above=4mm of k] {};
		\node[gauge, bodyE7, label=below:{$3$}] (A) [right=6mm of k] {};
		\node[gauge, bodyE7, label=below:{$3$}] (B) [right=6mm of A] {};
		\node[gauge, bodyE7, label=below:{$3$}] (C) [right=6mm of B] {};
		\node[gauge, bodyE7, label=below:{$3$}] (D) [right=6mm of C] {};
		\node[gauge, bodyE7, label=below:{$2$}] (F) [right=6mm of D] {};
		\node[gauge, bodyE7, label=below:{$1$}] (G) [right=6mm of F] {};
		\node[gauge, bodyE7, label=below:{$1$}] (H) [right=6mm of G] {};
		\node[gauge, bodyE7, label=above:{$1$}] (I) [above=4mm of D] {};
        \node[gauge, label=below:{$1$}] (k2) [right=6mm of H] {};
        \node[flavor, label=above:{$1$}] (tail4) [above=4mm of k2] {};
		\draw[thick] (tail1) -- (tail2) -- (k) -- (A);
        \draw[thick,bodyE7] (A) -- (B) -- (C) -- (D) -- (F) -- (G) -- (H);
        \draw[thick,bodyE7] (I) -- (D);
        \draw[thick] (k) -- (tail3);
        \draw[thick] (H) -- (k2) -- (tail4);
	\end{tikzpicture}
}
The mirror theory of this is	
\bes{ \label{theory2subE7-3}
[\su(9)] - \su(3) - \u(1) - [\su(3)]
}	
In this case, the $\su(12) \oplus \su(2)$ flavor symmetry of the 6d parent theory is broken to $\su(9)  \oplus \su(3) \oplus \u(1)$.
\item For Theory \hyperlink{rank4-2t-th2}{5}, after subtracting the $1 \times E_7$ quiver and rebalancing, we obtain
\bes{
\begin{tikzpicture}[baseline=0,font=\footnotesize]
		\node[gauge, label=below:{$1$}] (tail1) {};
		\node[gauge, label=below:{$2$}] (k) [right=6mm of tail1] {};
        \node[flavor, label=above:{$1$}] (tail2) [above=4mm of k] {};
		\node[gauge, bodyE7, label=below:{$2$}] (A) [right=6mm of k] {};
		\node[gauge, bodyE7, label=below:{$2$}] (B) [right=6mm of A] {};
		\node[gauge, bodyE7, label=below:{$2$}] (C) [right=6mm of B] {};
		\node[gauge, bodyE7, label=below:{$2$}] (D) [right=6mm of C] {};
		\node[gauge, bodyE7, label=below:{$2$}] (F) [right=6mm of D] {};
		\node[gauge, bodyE7, label=below:{$2$}] (G) [right=6mm of F] {};
		\node[gauge, bodyE7, label=below:{$2$}] (H) [right=6mm of G] {};
		\node[gauge, bodyE7, label=above:{$1$}] (I) [above=4mm of D] {};
        \node[gauge, label=below:{$2$}] (k2) [right=6mm of H] {};
        \node[gauge, label=below:{$1$}] (tail3) [right=6mm of k2] {};
        \node[flavor, label=above:{$1$}] (tail4) [above=4mm of k2] {};
		\draw[thick] (tail1) -- (k) -- (A);
        \draw[thick,bodyE7] (A) -- (B) -- (C) -- (D) -- (F) -- (G) -- (H);
        \draw[thick,bodyE7] (I) -- (D);
        \draw[thick] (k) -- (tail2);
        \draw[thick] (H) -- (k2) -- (tail3);
        \draw[thick] (k2) -- (tail4);
	\end{tikzpicture}
}
The mirror theory of this is	
\bes{ \label{theory2subE7-4}
[\su(6)] - \SU(2) - \SU(2) - [\su(6)] \qquad /\U(1)_B
}	
where $/\U(1)_B$ denotes gauging of the common baryonic symmetry for the $\SU(2)$ gauge theory with 6 flavors on each side.
In this case, the $\su(12) \oplus \su(2)$ flavor symmetry of the 6d parent theory is broken to $\su(6)  \oplus \su(6) \oplus \su(2) \oplus 1$.
\item For Theory \hyperlink{rank4-2t-th2}{6}, after subtracting the $1 \times E_7$ quiver and rebalancing, we obtain
\bes{
\begin{tikzpicture}[baseline=0,font=\footnotesize]
		\node[gauge, label=below:{$1$}] (tail1) {};
        \node[gauge, label=below:{$2$}] (tail2) [right=6mm of tail1] {};
		\node[gauge, label=below:{$3$}] (k) [right=6mm of tail2] {};
        \node[flavor, label=above:{$1$}] (tail3) [above=4mm of k] {};
		\node[gauge, bodyE7, label=below:{$3$}] (A) [right=6mm of k] {};
		\node[gauge, bodyE7, label=below:{$3$}] (B) [right=6mm of A] {};
		\node[gauge, bodyE7, label=below:{$3$}] (C) [right=6mm of B] {};
		\node[gauge, bodyE7, label=below:{$3$}] (D) [right=6mm of C] {};
		\node[gauge, bodyE7, label=below:{$3$}] (F) [right=6mm of D] {};
		\node[gauge, bodyE7, label=below:{$3$}] (G) [right=6mm of F] {};
		\node[gauge, bodyE7, label=below:{$3$}] (H) [right=6mm of G] {};
		\node[gauge, bodyE7, label=above:{$0$}] (I) [above=4mm of D] {};
        \node[gauge, label=below:{$3$}] (k2) [right=6mm of H] {};
        \node[gauge, label=below:{$2$}] (tail4) [right=6mm of k2] {};
        \node[gauge, label=below:{$1$}] (tail5) [right=6mm of tail4] {};
        \node[flavor, label=above:{$1$}] (tail6) [above=4mm of k2] {};
		\draw[thick] (tail1) -- (tail2) -- (k) -- (A);
        \draw[thick,bodyE7] (A) -- (B) -- (C) -- (D) -- (F) -- (G) -- (H);
        \draw[thick,bodyE7] (I) -- (D);
        \draw[thick] (k) -- (tail3);
        \draw[thick] (H) -- (k2) -- (tail4) -- (tail5);
        \draw[thick] (k2) -- (tail6);
	\end{tikzpicture}
}
The mirror theory of this is	
\bes{ \label{theory2subE7-5}
\mathrm{U}(3) - [\su(14)]
}	
In this case, the $\so(28)$ flavor symmetry of the 6d parent theory is broken to $\su(14)$.
	
\een

\subsection{Uplifting \texorpdfstring{$E_6$}{E6} Quivers to \texorpdfstring{$E_7$}{E7} Quivers}
\label{sec:InvFormulas-E6E7}

In Section \ref{sec:MassDefE7E6}, we discussed how to obtain various $E_6$ descendants from an $E_7$ theory. In Table \ref{tab:E7toE6-singletail-body}, we collected various possibilities, and, in the following section, we have shown how it is possible to generalize the prescription, generating $E_6$ quivers with three tails. In this section, instead, we will show how to invert those transformations, in order to predict the parent $E_7$ theory. In general, as we have seen in Section \ref{sec:InvFormulas-E7E8}, the resulting $E_7$-shaped quiver will be bad, with some nodes underbalanced. Once again, we can dualize this quiver at the bad nodes using the duality explained in Appendix \ref{sec:Itamarduality}, since these bad nodes have always FI deformations turned on. The resulting parent theory will be a good $E_7$-shaped quiver, which will be the proper parent quiver of the $E_6$ theory. 

It is, then, sufficient to consider a generic $E_6$-shaped quiver with three tails, i.e.
\begin{equation}
    \begin{tikzpicture}[baseline=30,font=\footnotesize]
			\node (dots) {$\cdots$};
			\node[gauge, label=below:{$k_1$}] (k1) [right=3mm of dots] {};
			\node[gauge, bodyE6, label=below:{$a$}] (A) [right=6mm of k1] {};
			\node[gauge, bodyE6, label=below:{$b$}] (B) [right=6mm of A] {};
			\node[gauge, bodyE6, label=below:{$c$}] (C) [right=6mm of B] {};
			\node[gauge, bodyE6, label=below:{$d$}] (D) [right=6mm of C] {};
			\node[gauge, bodyE6, label=below:{$e$}] (E) [right=6mm of D] {};
			\node[gauge, bodyE6, label=left:{$f$}] (F) [above=4mm of C] {};
			\node[gauge, bodyE6, label=left:{$g$}] (G) [above=4mm of F] {};
			\node[gauge, label=left:{$k_3$}] (k2) [above=4mm of G] {};
            \node (dots2) [above=3mm of k2] {$\vdots$};
            \node[gauge, label=below:{$k_2$}] (tk) [right=6mm of E] {};
            \node (dots3) [right=3mm of tk] {$\cdots$};
			\draw[thick] (dots.east) -- (k1.west);
			\draw[thick] (k1.east) -- (A.west);
			\draw[thick,bodyE6] (A.east) -- (B.west);
			\draw[thick,bodyE6] (B.east) -- (C.west);
			\draw[thick,bodyE6] (C.east) -- (D.west);
			\draw[thick,bodyE6] (D.east) -- (E.west);
			\draw[thick,bodyE6] (C.north) -- (F.south);
			\draw[thick,bodyE6] (F.north) -- (G.south);
			\draw[thick] (E.east) -- (tk.west);
            \draw[thick] (tk.east) -- (dots3.west);
            \draw[thick] (G.north) -- (k2.south);
            \draw[thick] (k2.north) -- (dots2.south);
		\end{tikzpicture}
  \label{eq:generalE6quiv-3tail}
\end{equation}
and inverting the conditions that lead, e.g., to \eqref{eq:BHdef-E71tailtoE63tail}, we have the following parent $E_7$ theory:
\begin{equation}
\label{eq:guessedE7fromE6}
    \begin{tikzpicture}[baseline=7,font=\footnotesize]
			\node (dots) {$\cdots$};
			\node[gauge, label=below:{$k_1+k_2$}] (k1) [right=3mm of dots] {};
			\node[gauge, bodyE7, label=above:{$a+k_2$}] (A) [right=9mm of k1] {};
			\node[gauge, bodyE7, label=below:{$b+k_2$}] (B) [right=9mm of A] {};
			\node[gauge, bodyE7, label=above:{$b+e$}] (C) [right=9mm of B] {};
			\node[gauge, bodyE7, label=below:{$b+d$}] (D) [right=9mm of C] {};
			\node[gauge, bodyE7, label=above:{$c$}] (E) [right=6mm of D] {};
			\node[gauge, bodyE7, label=below:{$f$}] (F) [right=6mm of E] {};
			\node[gauge, bodyE7, label=below:{$g$}] (G) [right=6mm of F] {};
			\node[gauge, bodyE7, label=above:{$b$}] (H) [above=4mm of D] {};
			\node[gauge, label=below:{$k_3$}] (k2) [right=6mm of G] {};
            \node (dots2) [right=3mm of k2] {$\cdots$};
			\draw[thick] (dots.east) -- (k1.west);
			\draw[thick] (k1.east) -- (A.west);
			\draw[thick,bodyE7] (A.east) -- (B.west);
			\draw[thick,bodyE7] (B.east) -- (C.west);
			\draw[thick,bodyE7] (C.east) -- (D.west);
			\draw[thick,bodyE7] (D.east) -- (E.west);
			\draw[thick,bodyE7] (E.east) -- (F.west);
			\draw[thick,bodyE7] (F.east) -- (G.west);
			\draw[thick,bodyE7] (D.north) -- (H.south);
            \draw[thick] (G.east) -- (k2.west);
            \draw[thick] (k2.east) -- (dots2.west);
		\end{tikzpicture}
\end{equation}
In this way, one can also obtain the parent $E_7$ quivers with a single tail considering $k_3 =0$. The nodes with ranks $b+k_2$ and $b$ are not manifestly good, and there can be cases in which their excess numbers are negative. However, by dualizing the quiver, the resulting parent theory will be good. 

\subsubsection{Example: the Trinion Theory}
\label{sec:TN}

Once again, we first consider the $T_{N+1}$ theory, whose mirror theory is illustrated in \eqref{eq:mirrorTN+1}, namely
\bes{
    \begin{tikzpicture}[baseline=0,font=\footnotesize]
        \node[gauge,bodyE6,label=below:$N+1$] (C) {};
        \node[gauge,bodyE6,label=right:$N$] (t11) [above=4mm of C] {};
        \node[gauge,bodyE6,label=right:$N-1$] (t12) [above=4mm of t11] {};
        \node (t13) [above=3mm of t12] {$\vdots$};
        \node[gauge,label=right:$1$] (t14) [above=1.5mm of t13] {};
        \node[gauge,bodyE6,label=below:$N$] (t21) [right=9mm of C] {};
        \node[gauge,bodyE6,label=below:$N-1$] (t22) [right=9mm of t21] {};
        \node  (t23) [right=3mm of t22] {$\cdots$};
        \node[gauge,label=below:$1$] (t24) [right=3mm of t23] {};
        \node[gauge,bodyE6,label=below:$N$] (t31) [left=9mm of C] {};
        \node[gauge,bodyE6,label=below:$N-1$] (t32) [left=9mm of t31] {};
        \node  (t33) [left=3mm of t32] {$\cdots$};
        \node[gauge,label=below:$1$] (t34) [left=3mm of t33] {};
        \draw[thick] (t12)--(t13) -- (t14);
        \draw[thick] (t22)--(t23) -- 
        (t24);
        \draw[thick] (t32)--(t33) -- (t34);
        \draw[thick,bodyE6] (C) -- (t11) -- (t12);
        \draw[thick,bodyE6] (C) -- (t21) -- (t22);
        \draw[thick,bodyE6] (C) -- (t31) -- (t32);
    \end{tikzpicture}
}
By applying \eqref{eq:guessedE7fromE6}, one possible resulting parent theory is
\begin{equation}
\label{eq:E7fromTN+1}
    \begin{tikzpicture}[baseline=7,font=\footnotesize]
			\node[gauge,label=below:{$1$}] (11) {};
            \node (dots) [right=3mm of 11] {$\cdots$};
			\node[gauge, label=below:{$2N-4$}] (k1) [right=3mm of dots] {};
			\node[gauge, bodyE7, label=above:{$2N-3$}] (A) [right=9mm of k1] {};
			\node[gauge, bodyE7, label=below:{$2N-2$}] (B) [right=9mm of A] {};
			\node[gauge, bodyE7, label=above:{$2N-1$}] (C) [right=9mm of B] {};
			\node[gauge, bodyE7, label=below:{$2N$}] (D) [right=9mm of C] {};
			\node[gauge, bodyE7, label=above:{$N+1$}] (E) [right=6mm of D] {};
			\node[gauge, bodyE7, label=below:{$N$}] (F) [right=6mm of E] {};
			\node[gauge, bodyE7, label=above:{$N-1$}] (G) [right=6mm of F] {};
			\node[gauge, bodyE7, label=above:{$N$}] (H) [above=4mm of D] {};
			\node[gauge, label=below:{$N-2$}] (k2) [right=6mm of G] {};
            \node (dots2) [right=3mm of k2] {$\cdots$};
            \node[gauge,label=below:{$1$}] (12) [right=3mm of dots2] {};
			\draw[thick] (11.east) -- (dots.west) (dots.east) -- (k1.west);
			\draw[thick] (k1.east) -- (A.west);
			\draw[thick,bodyE7] (A.east) -- (B.west);
			\draw[thick,bodyE7] (B.east) -- (C.west);
			\draw[thick,bodyE7] (C.east) -- (D.west);
			\draw[thick,bodyE7] (D.east) -- (E.west);
			\draw[thick,bodyE7] (E.east) -- (F.west);
			\draw[thick,bodyE7] (F.east) -- (G.west);
			\draw[thick,bodyE7] (D.north) -- (H.south);
            \draw[thick] (G.east) -- (k2.west);
            \draw[thick] (k2.east) -- (dots2.west) (12.west) -- (dots2.east);
		\end{tikzpicture}
\end{equation}
In this case, nodes $b+k_2=2N-2$ and $b=N$ are good by construction, so it is not necessary to perform any dualization.  In fact, this quiver was also presented in \cite[Figure 27(b)]{Eckhard:2020jyr}, and our method allows us to derive this result very quickly. Note also that this is a magnetic quiver for the 5d theory described in \cite[Figure 21]{Zafrir:2015rga}. 

At this point, it is also possible to find the $E_8$ parent theory, by inverting one of the proposals in Table \ref{tab:E8toE7-2tail-body}, as listed in Table \ref{tab:E7toE8-2tail}. Some inversions may lead to underbalanced nodes, and the quiver will become good only after performing a series of dualities. However, it is possible to show explicitly that, without any dualization, \eqref{eq:E7fromTN+1} can be obtained from an $E_8$ quiver given by
    \begin{equation}
	\begin{tikzpicture}[baseline=7,font=\footnotesize]
		\node[gauge,label=below:$1$] (11) {};
        \node (dots) [right=3mm of 11] {$\cdots$};
		\node[gauge, label=below:{$3N-6$}] (k) [right=3mm of dots] {};
		\node[gauge, bodyE8, label=above:{$3N-5$}] (A) [right=9mm of k] {};
		\node[gauge, bodyE8, label=below:{$3N-4$}] (B) [right=9mm of A] {};
		\node[gauge, bodyE8, label=above:{$3N-3$}] (C) [right=9mm of B] {};
		\node[gauge, bodyE8, label=below:{$3N-2$}] (D) [right=6mm of C] {};
		\node[gauge, bodyE8, label=above:{$3N-1$}] (E) [right=9mm of D] {};
		\node[gauge, bodyE8, label=below:{$3N$}] (F) [right=9mm of E] {};
		\node[gauge, bodyE8, label=above:{$2N$}] (G) [right=9mm of F] {};
		\node[gauge, bodyE8, label=below:{$N$}] (H) [right=9mm of G] {};
		\node[gauge, bodyE8, label=above:{$N+1$}] (FEG) [above=4mm of F] {};
		\draw[thick] (11.east) -- (dots.west) (dots.east) -- (k.west);
		\draw[thick] (k.east) -- (A.west);
		\draw[thick,bodyE8] (A.east) -- (B.west);
		\draw[thick,bodyE8] (B.east) -- (C.west);
		\draw[thick,bodyE8] (C.east) -- (D.west);
		\draw[thick,bodyE8] (D.east) -- (E.west);
		\draw[thick,bodyE8] (E.east) -- (F.west);
		\draw[thick,bodyE8] (F.east) -- (G.west);
		\draw[thick,bodyE8] (G.east) -- (H.west);
		\draw[thick,bodyE8] (F.north) -- (FEG.south);
	\end{tikzpicture}
\end{equation}
via an FI deformation involving nodes $4$ / $4'$, as in Table \ref{tab:E7toE8-2tail}. Note that this quiver was also presented in \cite[Figure 27(c)]{Eckhard:2020jyr}. It is a magnetic quiver for the 6d theory described in \cite[Figure 22]{Zafrir:2015rga}.

\subsubsection{Examples: \texorpdfstring{$E_6$}{E6} Quivers with Four-dimensional Higgs Branch}
\label{sec:E64dimHB-exam}

Let us list in Table \ref{listE6} below some $E_6$ quivers with a four-dimensional Higgs branch and such that the rank of the central node is less than or equal to $7$.  An uplift to the $E_7$ and $E_8$ parent quivers can be obtained by applying \eref{eq:guessedE7fromE6} and \eref{eq:33plift}, as well as dualizing all nodes that are underbalanced.  

In order to realize the deformations using the $\fm \times E_6$ quiver subtraction, the rebalancing procedure has to be generalized from that presented in \eref{E7subreb2tails} in the following way. We add $\fm$ $\U(1)$ nodes as in \eref{E7subreb2tails}, but we have to connect each of them by a hypermultiplet to the nodes $k_1$, $k_2$ and $k_3$ in \eref{eq:generalE6quiv-3tail}. Let us illustrate this for the case of $\fm=1$ as follows:
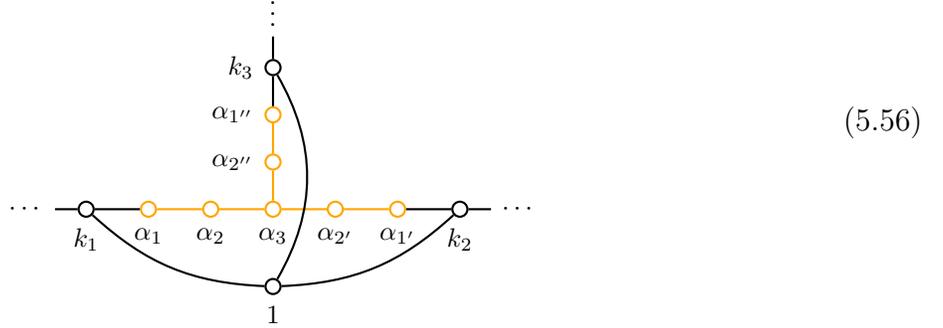
\begin{equation}
    \begin{tikzpicture}[baseline=30,font=\footnotesize]
			\node (dots) {$\cdots$};
			\node[gauge, label=below:{$k_1$}] (k1) [right=3mm of dots] {};
			\node[gauge, bodyE6, label=below:{$\alpha_1$}] (A) [right=6mm of k1] {};
			\node[gauge, bodyE6, label=below:{$\alpha_2$}] (B) [right=6mm of A] {};
			\node[gauge, bodyE6, label=below:{$\alpha_3$}] (C) [right=6mm of B] {};
   \node[gauge, label=below:{$1$}] (1a) [below=8mm of C] {};
			\node[gauge, bodyE6, label=below:{$\alpha_{2'}$}] (D) [right=6mm of C] {};
			\node[gauge, bodyE6, label=below:{$\alpha_{1'}$}] (E) [right=6mm of D] {};
			\node[gauge, bodyE6, label=left:{$\alpha_{2''}$}] (F) [above=4mm of C] {};
			\node[gauge, bodyE6, label=left:{$\alpha_{1''}$}] (G) [above=4mm of F] {};
			\node[gauge, label=left:{$k_3$}] (k2) [above=4mm of G] {};
            \node (dots2) [above=3mm of k2] {$\vdots$};
            \node[gauge, label=below:{$k_2$}] (tk) [right=6mm of E] {};
            \node (dots3) [right=3mm of tk] {$\cdots$};
			\draw[thick] (dots.east) -- (k1.west);
			\draw[thick] (k1.east) -- (A.west);
			\draw[thick,bodyE6] (A.east) -- (B.west);
			\draw[thick,bodyE6] (B.east) -- (C.west);
			\draw[thick,bodyE6] (C.east) -- (D.west);
			\draw[thick,bodyE6] (D.east) -- (E.west);
			\draw[thick,bodyE6] (C.north) -- (F.south);
			\draw[thick,bodyE6] (F.north) -- (G.south);
			\draw[thick] (E.east) -- (tk.west);
            \draw[thick] (tk.east) -- (dots3.west);
            \draw[thick] (G.north) -- (k2.south);
            \draw[thick] (k2.north) -- (dots2.south);
            \draw[thick] (k1) to [bend right=20] (1a);
            \draw[thick] (1a) to [bend right=20] (tk);
            \draw[thick] (1a) to [bend right=30] (k2);
		\end{tikzpicture}
  \label{eq:generalE6quiv-3tail-2}
\end{equation}
where an overall $\U(1)$ needs to be ungauged.  Let us demonstrate this in some examples presented in Table \ref{listE6}.
\ben
\item Let us consider Theory \hyperlink{rank4-E6-th3}{3}. Upon subtraction the $1 \times E_6$ quiver from the corresponding $E_6$ quiver, we obtain
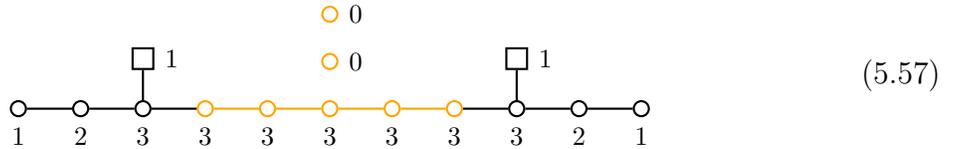
\bes{
\begin{tikzpicture}[baseline=20, font=\footnotesize]
\node[gauge, label=below:{$1$}] (1) {};
\node[gauge, label=below:{$2$}] (2) [right=6mm of 1] {};
\node[gauge, label=below:{$3$}] (3) [right=6mm of 2] {};
\node[gauge, bodyE6, label=below:{$3$}] (31) [right=6mm of 3] {};
\node[gauge, bodyE6, label=below:{$3$}] (32) [right=6mm of 31] {};
\node[gauge, bodyE6, label=below:{$3$}] (33) [right=6mm of 32] {};
\node[gauge, bodyE6, label=below:{$3$}] (34) [right=6mm of 33] {};
\node[gauge, bodyE6, label=below:{$3$}] (35) [right=6mm of 34] {};
\node[gauge, label=below:{$3$}] (3a) [right=6mm of 35] {};
\node[gauge, label=below:{$2$}] (2a) [right=6mm of 3a] {};
\node[gauge, label=below:{$1$}] (1a) [right=6mm of 2a] {};
\node[gauge, bodyE6, label=right:{$0$}] (0a) [above=4mm of 33] {};
\node[gauge, bodyE6, label=right:{$0$}] (0b) [above=4mm of 0a] {};
\node[flavor, label=right:{$1$}] (f1) [above=4mm of 3] {};
\node[flavor, label=right:{$1$}] (f2) [above=4mm of 3a] {};
\draw[thick] (1)--(2)--(3)--(31);
\draw[thick] (1a)--(2a)--(3a)--(35);
\draw[thick, bodyE6] (31)--(32)--(33)--(34)--(35);
\draw[thick] (3)--(f1);
\draw[thick] (3a)--(f2);
\end{tikzpicture}
}
This is the mirror theory of the $\U(3)$ gauge theory with 12 flavors, whose flavor symmetry is $\su(12)$. We see that the FI deformations from the $E_8$ to $E_7$ to $E_6$ quivers correspond to the following mass deformations, after taking into account the small $E_{8,7,6}$ instanton transitions:
\bes{
\USp(6)-[\so(28)] \quad \longrightarrow \quad  \USp(6)-[\so(24)] \quad \longrightarrow \quad
\U(3)-[\su(12)]
}
where the first two theories were discussed in \eref{usp6w14} and \eref{usp6w12}.
\item Let us consider Theory \hyperlink{rank4-E6-th8}{8}. The $1 \times E_6$ quiver subtraction from the corresponding $E_6$ quiver and rebalancing gives
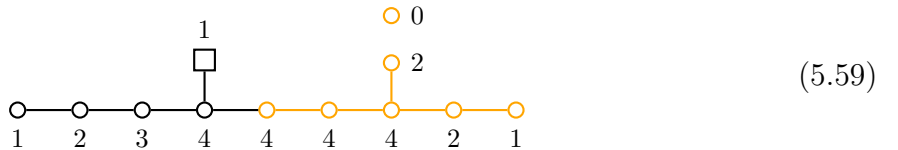
\bes{
\begin{tikzpicture}[baseline=20, font=\footnotesize]
\node[gauge, label=below:{$1$}] (1) {};
\node[gauge, label=below:{$2$}] (2)[right=6mm of 1] {};
\node[gauge, label=below:{$3$}] (2i)[right=6mm of 2] {};
\node[gauge, label=below:{$4$}] (3)
[right=6mm of 2i] {};
\node[gauge, bodyE6, label=below:{$4$}] (31)
[right=6mm of 3] {};
\node[gauge, bodyE6, label=below:{$4$}] (32)
[right=6mm of 31] {};
\node[gauge, bodyE6, label=below:{$4$}] (33)
[right=6mm of 32] {};
\node[gauge, bodyE6, label=below:{$2$}] (34)
[right=6mm of 33] {};
\node[gauge, bodyE6, label=below:{$1$}] (35)
[right=6mm of 34] {};
\node[gauge, bodyE6, label=right:{$2$}] (0a)
[above=4mm of 33] {};
\node[gauge, bodyE6, label=right:{$0$}] (0b)
[above=4mm of 0a] {};
\node[flavor, label=above:{$1$}] (f1)
[above=4mm of 3] {};
\draw[thick] (1)--(2)--(2i)--(3)--(31);
\draw[thick] (f1)--(3);
\draw[thick, bodyE6] (31)--(32)--(33)--(34)--(35);
\draw[thick, bodyE6] (33)--(0a);
\end{tikzpicture}
}
This is mirror dual to the $\SU(4)$ gauge theory with $9$ hypermultiplets in the fundamental representation and $1$ hypermultiplet in the rank-two antisymmetric representation (cf. \cite[(3.104), (3.106b)]{Cabrera:2019izd}):
\bes{
\underset{[\wedge^2]}{\SU(4)}-[\su(9)]
}
We see that the FI deformations from the $E_8$ to $E_7$ to $E_6$ quivers correspond to the following mass deformations, after taking into account the small $E_{8,7,6}$ instanton transitions:
\bes{
\underset{[\wedge^2]}{\SU(4)}-[\su(12)] \quad \longrightarrow \quad  \underset{[\wedge^2]}{\SU(4)}-[\su(10)] \quad \longrightarrow \quad
\underset{[\wedge^2]}{\SU(4)}-[\su(9)]
}
where the first two theories were discussed in Point \ref{antisymrank4} in Section \ref{sec:rank4onetail}.
\item Let us now consider Theory \hyperlink{rank4-E6-th7}{7}. After subtracting the $2 \times E_6$ quiver from the $E_6$ quiver and rebalancing, we obtain
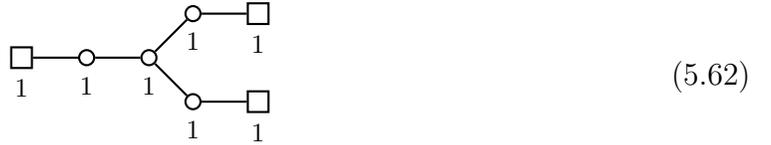
\bes{
\begin{tikzpicture}[baseline=7, font=\footnotesize]
\node[gauge, label=below:{$1$}] (1c) {}; 
\node[gauge, label=below:{$1$}] (1a)
[left=6mm of 1c] {};
\node[gauge, label=below:{$1$}] (1b)
[above right=6mm of 1c] {};
\node[gauge, label=below:{$1$}] (1d)
[below right=6mm of 1c] {};
\node[flavor, label=below:{$1$}] (fa)
[left=6mm of 1a] {};
\node[flavor, label=below:{$1$}] (fb)
[right=6mm of 1b] {};
\node[flavor, label=below:{$1$}] (fd)
[right=6mm of 1d] {};
\draw[thick] (fa)--(1a)--(1c)--(1b)--(fb);
\draw[thick] (1c)--(1d)--(fd);
\end{tikzpicture}
}
The mirror dual of this theory is 
\bes{ \label{su2U1USp2so4}
[\su(2)]-\U(1)-\USp(2)-[\so(4)]
}
We see that the FI deformations from the $E_8$ to $E_7$ to $E_6$ quivers correspond to the following mass deformations, after taking into account the small $E_{8,7,6}$ instanton transitions:
\bes{
\eref{SU3w6flv} \quad \longrightarrow \quad \eref{theory2subE7-2} \quad \longrightarrow \quad 
\eref{su2U1USp2so4}\fstop
}
\een

	\begin{landscape}
		\pagestyle{empty}
		\begin{center}
			\renewcommand{\arraystretch}{1.1}
			\setlength{\LTcapwidth}{1.2\textwidth}
			\begin{longtable}{c|c|c|c|c}
 \caption{Examples of rank-4 $E_6$ theories and the corresponding parent theories.} \label{listE6}  \\
  Number & $E_6 $ Quiver& $E_7$ Parent Quiver & $E_8$ Parent Quiver & 6d Parent Theory\\ 
		\hline
  \endfirsthead
		Number & $E_6 $ Quiver& $E_7$ Parent Quiver & $E_8$ Parent Quiver & 6d Parent Theory\\
		\hline
  \endhead
		\hypertarget{rank4-E6-th1}{} 1 &$
\begin{array}{ccc@{} >{\color{cE6}}c >{\color{cE6}}c >{\color{cE6}}c >{\color{cE6}}c >{\color{cE6}}c @{}ccc}
  &  &  &  &  & 1 &  &  &  &  &  \\
  &  &  &  &  & 3 &  &  &  &  &  \\
 1 & 2 &  & 3 & 4 & 5 & 4 & 3 &  & 2 & 1 \\
\end{array}
$ & $
\begin{array}{ccccc@{} >{\color{cE7}}c >{\color{cE7}}c >{\color{cE7}}c >{\color{cE7}}c >{\color{cE7}}c >{\color{cE7}}c >{\color{cE7}}c @{}}
  &  &  &  &  &  &  &  & 4 &  &  &  \\
 1 & 2 & 3 & 4 &  & 5 & 6 & 7 & 8 & 5 & 3 & 1 \\
\end{array}$ & $
\begin{array}{ccccc@{} >{\color{cE8}}c >{\color{cE8}}c >{\color{cE8}}c >{\color{cE8}}c >{\color{cE8}}c >{\color{cE8}}c >{\color{cE8}}c >{\color{cE8}}c @{}}
  &  &  &  &  &  &  &  &  &  & 5 &  &  \\
 1 & 2 & 3 & 4 &  & 5 & 6 & 7 & 8 & 9 & 10 & 6 & 3 \\
\end{array}
$ & $\overset{\su(4)}{\underset{[\Lambda^2]}{1}}\,[\su(12)]$\hypertarget{rank4-E6-th2}{}\\
\hline
2 &$
\begin{array}{ccc@{} >{\color{cE6}}c >{\color{cE6}}c >{\color{cE6}}c >{\color{cE6}}c >{\color{cE6}}c @{}cc}
  &  &  &  &  & \color{black}{1} &  &  &  &  \\
  &  &  &  &  & 2 &  &  &  &  \\
  &  &  &  &  & 3 &  &  &  &  \\
 1 & 2 &  & 3 & 4 & 5 & 3 & 2 &  & 1 \\
\end{array}
$ & $
\begin{array}{cccc@{} >{\color{cE7}}c >{\color{cE7}}c >{\color{cE7}}c >{\color{cE7}}c >{\color{cE7}}c >{\color{cE7}}c >{\color{cE7}}c @{}cc}
  &  &  &  &  &  &  & 3 &  &  &  &  &  \\
 1 & 2 & 3 &  & 4 & 5 & 6 & 7 & 5 & 3 & 2 &  & 1 \\
\end{array}
$ & $
\begin{array}{ccccc@{} >{\color{cE8}}c >{\color{cE8}}c >{\color{cE8}}c >{\color{cE8}}c >{\color{cE8}}c >{\color{cE8}}c >{\color{cE8}}c >{\color{cE8}}c @{}}
  &  &  &  &  &  &  &  &  &  & 5 &  &  \\
 1 & 2 & 3 & 4 &  & 5 & 6 & 7 & 8 & 9 & 10 & 6 & 3 \\
\end{array}
$ & $\overset{\su(4)}{\underset{[\Lambda^2]}{1}}\,[\su(12)]$\hypertarget{rank4-E6-th3}{}\\
\hline
3 &$
\begin{array}{cccc@{} >{\color{cE6}}c >{\color{cE6}}c >{\color{cE6}}c >{\color{cE6}}c >{\color{cE6}}c @{}cccc}
  &  &  &  &  &  & 1 &  &  &  &  &  &  \\
  &  &  &  &  &  & 2 &  &  &  &  &  &  \\
 1 & 2 & 3 &  & 4 & 5 & 6 & 5 & 4 &  & 3 & 2 & 1 \\
\end{array}
$ & $
\begin{array}{ccccccc@{} >{\color{cE7}}c >{\color{cE7}}c >{\color{cE7}}c >{\color{cE7}}c >{\color{cE7}}c >{\color{cE7}}c >{\color{cE7}}c @{}}
  &  &  &  &  &  &  &  &  &  & 5 &  &  &  \\
 1 & 2 & 3 & 4 & 5 & 6 &  & 7 & 8 & 9 & 10 & 6 & 2 & 1 \\
\end{array}
$ & $
\begin{array}{ccccccc@{} >{\color{cE8}}c >{\color{cE8}}c >{\color{cE8}}c >{\color{cE8}}c >{\color{cE8}}c >{\color{cE8}}c >{\color{cE8}}c >{\color{cE8}}c @{}}
  &  &  &  &  &  &  &  &  &  &  &  & 6 &  &  \\
 1 & 2 & 3 & 4 & 5 & 6 &  & 7 & 8 & 9 & 10 & 11 & 12 & 7 & 2 \\
\end{array}
$ & $\overset{\usp(6)}{1}\,[\so(28)]$\hypertarget{rank4-E6-th4}{}\\
\hline
4 &$
\begin{array}{cccc@{} >{\color{cE6}}c >{\color{cE6}}c >{\color{cE6}}c >{\color{cE6}}c >{\color{cE6}}c @{}cc}
  &  &  &  &  &  & 1 &  &  &  &  \\
  &  &  &  &  &  & 3 &  &  &  &  \\
 1 & 2 & 3 &  & 4 & 5 & 6 & 4 & 2 &  & 1 \\
\end{array}
$ & $
\begin{array}{ccccc@{} >{\color{cE7}}c >{\color{cE7}}c >{\color{cE7}}c >{\color{cE7}}c >{\color{cE7}}c >{\color{cE7}}c >{\color{cE7}}c @{}}
  &  &  &  &  &  &  &  & 4 &  &  &  \\
 1 & 2 & 3 & 4 &  & 5 & 6 & 7 & 8 & 5 & 3 & 1 \\
\end{array}
$ & $
\begin{array}{ccccc@{} >{\color{cE8}}c >{\color{cE8}}c >{\color{cE8}}c >{\color{cE8}}c >{\color{cE8}}c >{\color{cE8}}c >{\color{cE8}}c >{\color{cE8}}c @{}}
  &  &  &  &  &  &  &  &  &  & 5 &  &  \\
 1 & 2 & 3 & 4 &  & 5 & 6 & 7 & 8 & 9 & 10 & 6 & 3 \\
\end{array}
$ & $\overset{\su(4)}{\underset{[\Lambda^2]}{1}}\,[\su(12)]$\hypertarget{rank4-E6-th5}{}\\
\hline
5 &$
\begin{array}{cccc@{} >{\color{cE6}}c >{\color{cE6}}c >{\color{cE6}}c >{\color{cE6}}c >{\color{cE6}}c @{}}
  &  &  &  &  &  & 2 &  &  \\
  &  &  &  &  &  & 4 &  &  \\
 1 & 2 & 3 &  & 4 & 5 & 6 & 4 & 2 \\
\end{array}
$ & $
\begin{array}{cccc@{} >{\color{cE7}}c >{\color{cE7}}c >{\color{cE7}}c >{\color{cE7}}c >{\color{cE7}}c >{\color{cE7}}c >{\color{cE7}}c @{}}
  &  &  &  &  &  &  & 4 &  &  &  \\
 1 & 2 & 3 &  & 4 & 5 & 6 & 8 & 6 & 4 & 2 \\
\end{array}
$ & $
\begin{array}{cccc@{} >{\color{cE8}}c >{\color{cE8}}c >{\color{cE8}}c >{\color{cE8}}c >{\color{cE8}}c >{\color{cE8}}c >{\color{cE8}}c >{\color{cE8}}c @{}}
  &  &  &  &  &  &  &  &  & 6 &  &  \\
 1 & 2 & 3 &  & 4 & 5 & 6 & 8 & 10 & 12 & 8 & 4 \\
\end{array}
$ & $[\mathfrak{e}_6] \, 1\,\overset{\su(3)}{2}[\su(6)]$\hypertarget{rank4-E6-th6}{}\\
\hline
6 &$
\begin{array}{ccc@{} >{\color{cE6}}c >{\color{cE6}}c >{\color{cE6}}c >{\color{cE6}}c >{\color{cE6}}c @{}cc}
  &  &  &  &  & 2 &  &  &  &  \\
  &  &  &  &  & 4 &  &  &  &  \\
 1 & 2 &  & 3 & 4 & 6 & 4 & 2 &  & 1 \\
\end{array}
$ & $
\begin{array}{cccc@{} >{\color{cE7}}c >{\color{cE7}}c >{\color{cE7}}c >{\color{cE7}}c >{\color{cE7}}c >{\color{cE7}}c >{\color{cE7}}c @{}}
  &  &  &  &  &  &  & 4 &  &  &  \\
 1 & 2 & 3 &  & 4 & 5 & 6 & 8 & 6 & 4 & 2 \\
\end{array}
$ & $
\begin{array}{cccc@{} >{\color{cE8}}c >{\color{cE8}}c >{\color{cE8}}c >{\color{cE8}}c >{\color{cE8}}c >{\color{cE8}}c >{\color{cE8}}c >{\color{cE8}}c @{}}
  &  &  &  &  &  &  &  &  & 6 &  &  \\
 1 & 2 & 3 &  & 4 & 5 & 6 & 8 & 10 & 12 & 8 & 4 \\
\end{array}
$ & $[\mathfrak{e}_6] \, 1\,\overset{\su(3)}{2}[\su(6)]$\hypertarget{rank4-E6-th7}{}\\
\hline
7 &$
\begin{array}{cc@{} >{\color{cE6}}c >{\color{cE6}}c >{\color{cE6}}c >{\color{cE6}}c >{\color{cE6}}c @{}cc}
  &  &  &  & \color{black}{1} &  &  &  &  \\
  &  &  &  & 2 &  &  &  &  \\
  &  &  &  & 4 &  &  &  &  \\
 1 &  & 2 & 4 & 6 & 4 & 2 &  & 1 \\
\end{array}
$ & $
\begin{array}{ccc@{} >{\color{cE7}}c >{\color{cE7}}c >{\color{cE7}}c >{\color{cE7}}c >{\color{cE7}}c >{\color{cE7}}c >{\color{cE7}}c @{}cc}
  &  &  &  &  &  & 4 &  &  &  &  &  \\
 1 & 2 &  & 3 & 4 & 6 & 8 & 6 & 4 & 2 &  & 1 \\
\end{array}
$ & $
\begin{array}{cccc@{} >{\color{cE8}}c >{\color{cE8}}c >{\color{cE8}}c >{\color{cE8}}c >{\color{cE8}}c >{\color{cE8}}c >{\color{cE8}}c >{\color{cE8}}c @{}}
  &  &  &  &  &  &  &  &  & 6 &  &  \\
 1 & 2 & 3 &  & 4 & 5 & 6 & 8 & 10 & 12 & 8 & 4 \\
\end{array}
$ & $[\mathfrak{e}_6] \, 1\,\overset{\su(3)}{2}[\su(6)]$\hypertarget{rank4-E6-th8}{}\\
\hline
8 &$
\begin{array}{ccccc@{} >{\color{cE6}}c >{\color{cE6}}c >{\color{cE6}}c >{\color{cE6}}c >{\color{cE6}}c @{}}
  &  &  &  &  &  &  & 1 &  &  \\
  &  &  &  &  &  &  & 4 &  &  \\
 1 & 2 & 3 & 4 &  & 5 & 6 & 7 & 4 & 2 \\
\end{array}
$ & $
\begin{array}{ccccc@{} >{\color{cE7}}c >{\color{cE7}}c >{\color{cE7}}c >{\color{cE7}}c >{\color{cE7}}c >{\color{cE7}}c >{\color{cE7}}c @{}}
  &  &  &  &  &  &  &  & 4 &  &  &  \\
 1 & 2 & 3 & 4 &  & 5 & 6 & 7 & 8 & 5 & 3 & 1 \\
\end{array}
$ & $
\begin{array}{ccccc@{} >{\color{cE8}}c >{\color{cE8}}c >{\color{cE8}}c >{\color{cE8}}c >{\color{cE8}}c >{\color{cE8}}c >{\color{cE8}}c >{\color{cE8}}c @{}}
  &  &  &  &  &  &  &  &  &  & 5 &  &  \\
 1 & 2 & 3 & 4 &  & 5 & 6 & 7 & 8 & 9 & 10 & 6 & 3 \\
\end{array}
$ & $\overset{\su(4)}{\underset{[\Lambda^2]}{1}}\,[\su(12)]$\hypertarget{rank4-E6-th9}{}\\
\hline
9 &$
\begin{array}{ccc@{} >{\color{cE6}}c >{\color{cE6}}c >{\color{cE6}}c >{\color{cE6}}c >{\color{cE6}}c @{}cc}
  &  &  &  &  & 1 &  &  &  &  \\
  &  &  &  &  & 4 &  &  &  &  \\
 1 & 2 &  & 3 & 5 & 7 & 5 & 3 &  & 1 \\
\end{array}
$ & $
\begin{array}{cccc@{} >{\color{cE7}}c >{\color{cE7}}c >{\color{cE7}}c >{\color{cE7}}c >{\color{cE7}}c >{\color{cE7}}c >{\color{cE7}}c @{}}
  &  &  &  &  &  &  & 5 &  &  &  \\
 1 & 2 & 3 &  & 4 & 6 & 8 & 10 & 7 & 4 & 1 \\
\end{array}
$ & $
\begin{array}{cccc@{} >{\color{cE8}}c >{\color{cE8}}c >{\color{cE8}}c >{\color{cE8}}c >{\color{cE8}}c >{\color{cE8}}c >{\color{cE8}}c >{\color{cE8}}c @{}}
  &  &  &  &  &  &  &  &  & 6 &  &  \\
 1 & 2 & 3 &  & 4 & 5 & 6 & 8 & 10 & 12 & 8 & 4 \\
\end{array}
$ & $[\mathfrak{e}_6] \, 1\,\overset{\su(3)}{2}[\su(6)]$\hypertarget{rank4-E6-th10}{}\\
\hline
10 &$
\begin{array}{cc@{} >{\color{cE6}}c >{\color{cE6}}c >{\color{cE6}}c >{\color{cE6}}c >{\color{cE6}}c @{}cc}
  &  &  &  & 2 &  &  &  &  \\
  &  &  &  & 4 &  &  &  &  \\
 1 &  & 3 & 5 & 7 & 5 & 3 &  & 1 \\
\end{array}
$ & $
\begin{array}{ccc@{} >{\color{cE7}}c >{\color{cE7}}c >{\color{cE7}}c >{\color{cE7}}c >{\color{cE7}}c >{\color{cE7}}c >{\color{cE7}}c @{}}
  &  &  &  &  &  & 5 &  &  &  \\
 1 & 2 &  & 4 & 6 & 8 & 10 & 7 & 4 & 2 \\
\end{array}
$ & $
\begin{array}{ccc@{} >{\color{cE8}}c >{\color{cE8}}c >{\color{cE8}}c >{\color{cE8}}c >{\color{cE8}}c >{\color{cE8}}c >{\color{cE8}}c >{\color{cE8}}c @{}}
  &  &  &  &  &  &  &  & 7 &  &  \\
 1 & 2 &  & 4 & 6 & 8 & 10 & 12 & 14 & 9 & 4 \\
\end{array}
$ & $[\su(8)] \, \overset{\su(2)}{1}\,\overset{\su(2)}{2}[\su(2)]$
\end{longtable}	
\end{center}
	\end{landscape}

\subsection{Uplifting Generic Star-shaped Quivers to \texorpdfstring{$E_6$}{E6} Quivers}
\label{sec:InvFormulas-StarShapedE6}

Let us consider a generic star-shaped quiver as in Section \ref{sec:E6toStarShaped}, i.e.
\begin{equation}\label{eq:genStarShaped}
	\begin{tikzpicture}[baseline=0,font=\footnotesize]
		\node[gauge,label=left:{$N$}] (N) {};
		\node (T1) [above right=9mm of N] {$T_{\rho_k}$};
		\node (dots1) [above=7mm of N] {$\cdots$};
		\node (Tk) [above left=9mm of N] {$T_{\rho_1}$};
		\node (T1p) [below left=9mm of N] {$T_{\rho_1'}$};
		\node (dots2) [below=7mm of N] {$\cdots$};
		\node (Tkp) [below right=9mm of N] {$T_{\rho_{k'}'}$};
		\node (T3) [right=9mm of N] {$T_{\rho}$};
		\draw[thick] (N) -- (T1);
		\draw[thick] (N) -- (Tk);
		\draw[thick] (N) -- (T1p);
		\draw[thick] (N) -- (Tkp);
		\draw[thick] (N) -- (T3);
    \draw[thick] [decorate,decoration={brace,amplitude=5pt},xshift=0cm, yshift=0cm]
([yshift=5mm]Tk.west) -- ([yshift=5mm]T1.east) node [black,pos=0.5,yshift=0.4cm] 
{$k$ tails};
\draw[thick] [decorate,decoration={mirror,brace,amplitude=5pt},xshift=0cm, yshift=0cm]
([yshift=-5mm]T1p.west) -- ([yshift=-5mm]Tkp.east) node [black,pos=0.5,yshift=-0.5cm] 
{$k'$ tails};
	\end{tikzpicture}
\end{equation}
The strategy is to come up with a candidate $E_6$ parent theory \eqref{eq:genE6quiv-2}, with the tails defined as in \eqref{eq:Trho123}.\footnote{If the number of tails in \eqref{eq:genStarShaped} is even, it is sufficient to see the quiver as having an extra tail with trivial partition.} It is possible that this candidate $E_6$ theory contains underbalanced nodes, rendering the whole quiver bad. As before, we can repeatedly dualize the underbalanced nodes until the quiver becomes good. The result is then the $E_6$ parent theory from which the star-shaped quiver in question originates. We will provide a number of examples to illustrate this point in the following sections. By a similar procedure as described in \cref{sec:procedure,sec:obtainE8fromE7,sec:E64dimHB-exam}, one can uplift generic star-shaped quivers to their parent theories, and we list some examples of star-shaped quivers with four and five tails of four-dimensional Higgs branch with their parent theories in Table \ref{listStarShaped}.

\subsubsection{Example: Two Full Tails and Three Minimal Tails}

Let us consider the following star-shaped quiver with two $T\left[\SU(N)\right]$ tails and three $T_{[N-1,1]}\left[\SU(N)\right]$ ones:
\bes{ \label{ssq2full3min}
    \begin{tikzpicture}[baseline=0,font=\footnotesize]
        \node[gauge,label=below:$N$] (t11)  {};
        \node[gauge,label=below:$N-1$] (t12) [left=9mm of t11] {};
        \node (t13) [left=6mm of t12] {$\cdots$};
        \node[gauge,label=below:$2$] (t14) [left=6mm of t13] {};
        \node[gauge,label=below:$1$] (t15) [left=9mm of t14] {};
        \node[gauge,label=below:$N-1$] (t12l) [right=9mm of t11] {};
        \node (t13l) [right=6mm of t12l] {$\cdots$};
        \node[gauge,label=below:$2$] (t14l) [right =6mm of t13l] {};
        \node[gauge,label=below:$1$] (t15l) [right=9mm of t14l] {};
        \node[gauge,label=above:$1$] (t11a) [above left=6mm of t11] {};
        \node[gauge,label=above:$1$] (t12a) [above right=6mm of t11] {};
        \node[gauge,label=above:$1$] (t13a) [above=6mm of t11] {};
\draw[thick] (t11) -- (t11a) (t11) -- (t12a) (t11) -- (t13a);
\draw[thick] (t11) -- (t12) -- (t13) -- (t14) -- (t15);
\draw[thick] (t11) -- (t12l) -- (t13l) -- (t14l) -- (t15l);
\end{tikzpicture}
}
Since this corresponds to \eref{genericssq} with $k=2$, $\rho_1 = \rho'_1 = \left[1^N\right]$ and $\rho_2 = \rho'_2 = \rho = [N-1,1]$, a parent $E_6$ quiver is given by \eref{eq:genE6quiv-2}, which, in this case, yields the following candidate $E_6$ quiver.
\bes{
    \begin{tikzpicture}[baseline=7,font=\footnotesize]
        \node[gauge,bodyE6,label=below:$2 N$, label={[xshift=2mm, yshift=-1.5mm] \purple{\scriptsize $\lambda$}}] (C) {};
        \node[gauge,bodyE6,label=right:$N$] (t11) [above=6mm of C] {};
        \node[gauge,bodyE6,label=right:$1$] (t12) [above=6mm of t11] {};
        \node[gauge,bodyE6,label=above:$N+1$, label=below:{\purple{\scriptsize $-\lambda - \eta$}}] (t21) [right=9mm of C] {};
        \node[gauge,bodyE6,label=below:$N$] (t22) [right=9mm of t21] {};
        \node  (t23) [right=6mm of t22] {$\cdots$};
        \node[gauge,label=below:$1$, label=above:{\purple{\scriptsize $\lambda$}}] (t24) [right=6mm of t23] {};
        \node[gauge,bodyE6,label=above:$N+1$, label=below:{\purple{\scriptsize $-\lambda + \eta$}}] (t31) [left=9mm of C] {};
        \node[gauge,bodyE6,label=below:$N$] (t32) [left=9mm of t31] {};
        \node  (t33) [left=6mm of t32] {$\cdots$};
        \node[gauge,label=below:$1$, label=above:{\purple{\scriptsize $\lambda$}}] (t34) [left=6mm of t33] {};
        \draw[thick] (t22)--(t23) -- 
        (t24);
        \draw[thick] (t32)--(t33) -- (t34);
        \draw[thick,bodyE6] (C) -- (t11) -- (t12);
        \draw[thick,bodyE6] (C) -- (t21) -- (t22);
        \draw[thick,bodyE6] (C) -- (t31) -- (t32);
    \end{tikzpicture}
}
Let us now turn on some FI parameters $\lambda$ and $\eta$, highlighted in {\purple purple}, at different nodes. Let us suppose that $4 N > 3 N +2$, \ie $N > 2$, then the middle node is bad. We can dualize at such node, obtaining the following quiver, whose middle node becomes good:
\bes{ \label{ssqNp2N}
    \begin{tikzpicture}[baseline=20,font=\footnotesize]
        \node[gauge,bodyE6,label=below:$N+2$, label={[xshift=2.3mm, yshift=-1.5mm] \purple{\scriptsize $-\lambda$}}] (C) {};
        \node[gauge,bodyE6,label=right:$N$, label=left:{\purple{\scriptsize $\lambda$}}] (t11) [above=6mm of C] {};
        \node[gauge,bodyE6,label=right:$1$] (t12) [above=6mm of t11] {};
        \node[gauge,bodyE6,label=above:$N+1$, label=below:{\purple{\scriptsize $- \eta$}}] (t21) [right=9mm of C] {};
        \node[gauge,bodyE6,label=below:$N$] (t22) [right=9mm of t21] {};
        \node  (t23) [right=6mm of t22] {$\cdots$};
        \node[gauge,label=below:$1$, label=above:{\purple{\scriptsize $\lambda$}}] (t24) [right=6mm of t23] {};
        \node[gauge,bodyE6,label=above:$N+1$, label=below:{\purple{\scriptsize $\eta$}}] (t31) [left=9mm of C] {};
        \node[gauge,bodyE6,label=below:$N$] (t32) [left=9mm of t31] {};
        \node  (t33) [left=6mm of t32] {$\cdots$};
        \node[gauge,label=below:$1$, label=above:{\purple{\scriptsize $\lambda$}}] (t34) [left=6mm of t33] {};
        \draw[thick] (t22)--(t23) -- 
        (t24);
        \draw[thick] (t32)--(t33) -- (t34);
        \draw[thick,bodyE6] (C) -- (t11) -- (t12);
        \draw[thick,bodyE6] (C) -- (t21) -- (t22);
        \draw[thick,bodyE6] (C) -- (t31) -- (t32);
    \end{tikzpicture}
}
Observe that such quiver is equivalent to \eref{eq:genE6quiv-1} with $k=2$, $\rho_1 = \rho_2 = \left[1^N\right]$ and $\rho_3 = \left[N-1,1\right]$. Thus, upon turning on the FI parameters appropriately, this can be deformed to quiver \eref{ssq2full3min}.

\begin{figure}[!htp]
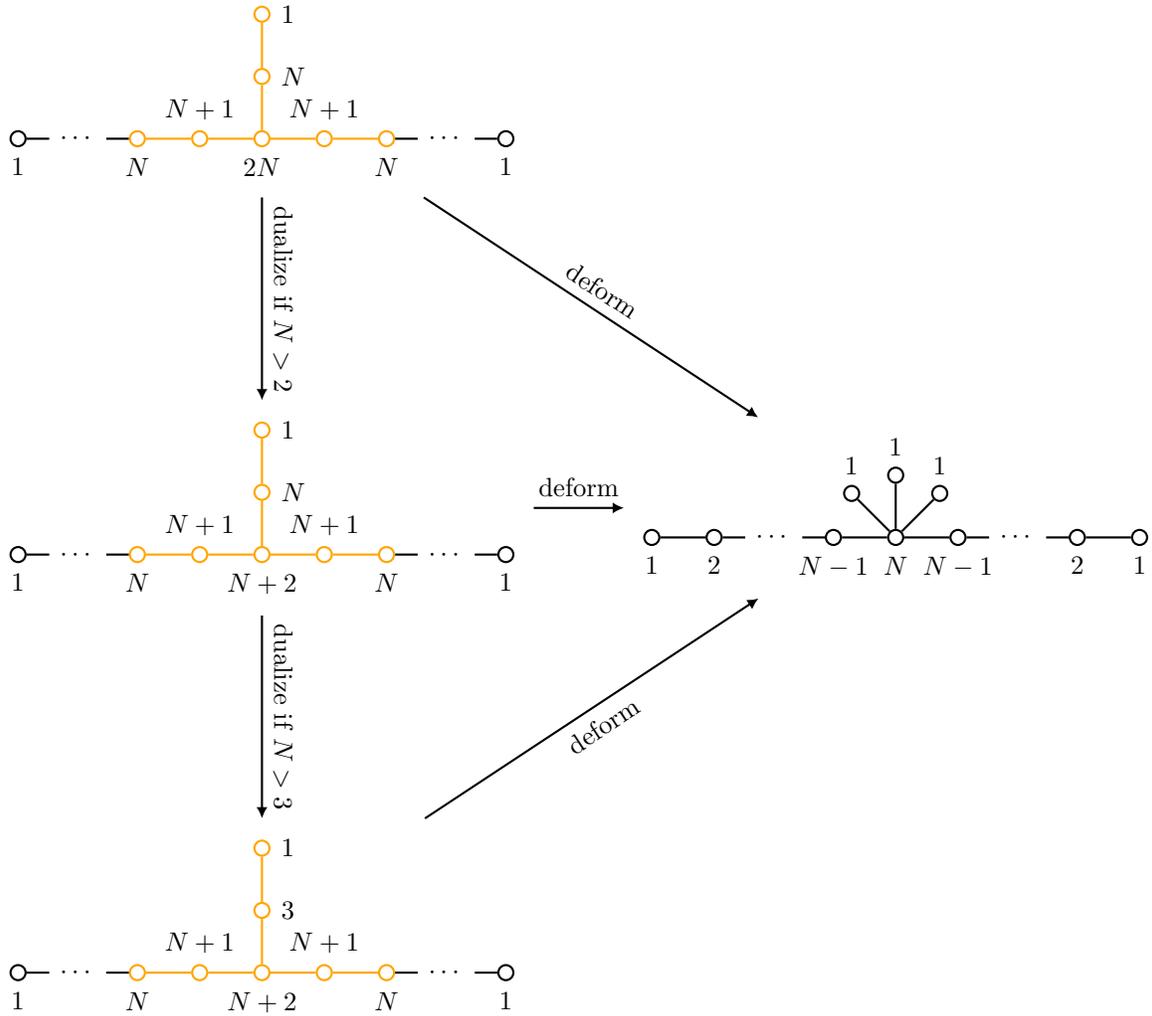

    \centering
    %
 
    \caption{Summary of the deformations and the dualizations that lead to \eqref{ssq2full3min}.}
    \label{fig:ssq2full3min}
\end{figure}

However, the node labelled by $N$ above the one labelled by $N+2$ in \eref{ssqNp2N} is bad if $2 N >  N + 3$, \ie $N > 3$. Upon dualizing at such node, we finally obtain the following good quiver:
\bes{ \label{ssqNp23}
    \begin{tikzpicture}[baseline=20,font=\footnotesize]
        \node[gauge,bodyE6,label=below:$N+2$,] (C) {};
        \node[gauge,bodyE6,label=right:$3$, label=left:{\purple{\scriptsize $-\lambda$}}] (t11) [above=6mm of C] {};
        \node[gauge,bodyE6,label=right:$1$, label=left:{\purple{\scriptsize $\lambda$}}] (t12) [above=6mm of t11] {};
        \node[gauge,bodyE6,label=above:$N+1$, label=below:{\purple{\scriptsize $- \eta$}}] (t21) [right=9mm of C] {};
        \node[gauge,bodyE6,label=below:$N$] (t22) [right=9mm of t21] {};
        \node  (t23) [right=3mm of t22] {$\cdots$};
        \node[gauge,label=below:$1$, label=above:{\purple{\scriptsize $\lambda$}}] (t24) [right=3mm of t23] {};
        \node[gauge,bodyE6,label=above:$N+1$, label=below:{\purple{\scriptsize $\eta$}}] (t31) [left=9mm of C] {};
        \node[gauge,bodyE6,label=below:$N$] (t32) [left=9mm of t31] {};
        \node  (t33) [left=3mm of t32] {$\cdots$};
        \node[gauge,label=below:$1$, label=above:{\purple{\scriptsize $\lambda$}}] (t34) [left=3mm of t33] {};
        \draw[thick] (t22)--(t23) -- 
        (t24);
        \draw[thick] (t32)--(t33) -- (t34);
        \draw[thick,bodyE6] (C) -- (t11) -- (t12);
        \draw[thick,bodyE6] (C) -- (t21) -- (t22);
        \draw[thick,bodyE6] (C) -- (t31) -- (t32);
    \end{tikzpicture}
}
Due to the presence of two independent FI parameters turned on in quiver \eref{ssqNp23}, the corresponding deformation can be analyzed via the following sequence of subtractions. First, we subtract two (non-affine) $A_{N+3}$ abelian quivers, where each of them is aligned with the subquivers connecting one of the two $\U(1)$ nodes, one on the left and one on the right of the middle node, with the $\U(3)$ node on top of the middle node. Below, we color in {\red red} the nodes with the FI parameter $\lambda$ turned on and in {\blue blue} the $\U(1)$ nodes arising to rebalance after subtraction.
\bes{
\scalebox{0.95}{$
 
$}
}
Next, we subtract a (non-affine) $A_2$ abelian quiver which is aligned with the two $\U(1)$ nodes with the FI parameter $\lambda$ turned on and, again, we rebalance with a $\U(1)$ node, colored in {\brown brown}.
\bes{
%
 
}
Finally, we subtract a (non-affine) $A_3$ non-abelian quiver with $\U(N)$ nodes such that its leftmost and rightmost nodes are aligned with the two $\U(N)$ nodes in {\color{magenta} magenta} with the FI parameter $\eta$ turned on. Subsequently, we rebalance with a $\U(N)$ node, colored in {\color{teal} teal}.
\bes{
%
 
}
Note that the final outcome of the deformation we have implemented coincides with quiver \eref{ssq2full3min}, that is, at each step of dualization, the $E_6$-shaped quiver in question can be deformed to the same parent star-shaped quiver, given by \eref{ssq2full3min}. We summarize the deformations and the dualizations that lead to \eqref{ssq2full3min} in Figure \ref{fig:ssq2full3min}.

\begin{landscape}
		\pagestyle{empty}
		\begin{center}
			\renewcommand{\arraystretch}{1.1}
			\setlength{\LTcapwidth}{1.2\textwidth}
			\begin{longtable}{c|c|c|c|c|c}
        \caption{Examples of rank-4 star-shaped theories and the corresponding parent theories.} \label{listStarShaped}\\
        Number & Star-shaped Quiver & $E_6 $ Quiver& $E_7$ Parent Quiver & $E_8$ Parent Quiver & 6d Parent Theory\\
		\hline
        \endfirsthead
		Number & Star-shaped Quiver & $E_6 $ Quiver& $E_7$ Parent Quiver & $E_8$ Parent Quiver & 6d Parent Theory\\
		\hline
        \endhead
		\hypertarget{rank4-Star-th1}{} 1 &
   \begin{tikzpicture}[baseline=0,font=\footnotesize]
		\node[gauge,label=below:$3$] (C) {};
        \node[gauge,label=above:$1$] (t11) [above=4mm of C] {};
        \node[gauge,label=above:$1$] (t21) [above left=4mm of C] {};
        \node[gauge,label=above:$1$] (t31) [above right=4mm of C] {};
        \node[gauge,label=below:$2$] (t41) [below left=4mm of C] {};
        \node[gauge,label=below:$1$] (t42) [below left=4mm of t41] {};
        \node[gauge,label=below:$2$] (t51) [below right=4mm of C] {};
        \node[gauge,label=below:$1$] (t52) [below right=4mm of t51] {};
        \draw[thick] (C) -- (t11);
        \draw[thick] (C) -- (t21);
        \draw[thick] (C) -- (t31);
        \draw[thick] (C) -- (t41) -- (t42);
        \draw[thick] (C) -- (t51) -- (t52);
		 \end{tikzpicture} 
   &$
\begin{array}{ccc@{} >{\color{cE6}}c >{\color{cE6}}c >{\color{cE6}}c >{\color{cE6}}c >{\color{cE6}}c @{}ccc}
  &  &  &  &  & 1 &  &  &  &  &  \\
  &  &  &  &  & 3 &  &  &  &  &  \\
 1 & 2 &  & 3 & 4 & 5 & 4 & 3 &  & 2 & 1 \\
\end{array}
$ & $
\begin{array}{ccccc@{} >{\color{cE7}}c >{\color{cE7}}c >{\color{cE7}}c >{\color{cE7}}c >{\color{cE7}}c >{\color{cE7}}c >{\color{cE7}}c @{}}
  &  &  &  &  &  &  &  & 4 &  &  &  \\
 1 & 2 & 3 & 4 &  & 5 & 6 & 7 & 8 & 5 & 3 & 1 \\
\end{array}$ & $
\begin{array}{ccccc@{} >{\color{cE8}}c >{\color{cE8}}c >{\color{cE8}}c >{\color{cE8}}c >{\color{cE8}}c >{\color{cE8}}c >{\color{cE8}}c >{\color{cE8}}c @{}}
  &  &  &  &  &  &  &  &  &  & 5 &  &  \\
 1 & 2 & 3 & 4 &  & 5 & 6 & 7 & 8 & 9 & 10 & 6 & 3 \\
\end{array}
$ & $\overset{\su(4)}{\underset{[\Lambda^2]}{1}}\,[\su(12)]$ \hypertarget{rank4-Star-th2}{}\\
\end{longtable}	
\end{center}
	\end{landscape}

\section{Comments on Higher Genus Theories}
\label{sec:HigherGenusTheories}

So far, we have discussed only class $\mathcal{S}$ theories on the sphere. In this section, we will explain how to get class $\mathcal{S}$ theories on a genus-$g$ Riemann surface starting from theories on the sphere. Again, we will work at the level of the magnetic quivers and implement the RG flow by activating FI deformations. We will proceed by predicting the relevant star-shaped quiver and deformation leading to unitary quivers with $g$ adjoint hypermultiplets for the central node, according to \cite{Benini:2010uu}. Let us start with the following star-shaped quiver
\begin{equation}\label{eq:examgtailsquiver}
    \begin{tikzpicture}[baseline=0,font=\footnotesize]
		\node[gauge,label=below:{$2N$}] (N) {};
		\node[gauge,label={below:$N$}] (T11) [above left=8mm of N] {};
        \node (T12) [above left=4mm of T11] {$T_{\rho_1}$};
		\node (dots1) [above=8mm of N] {$\cdots$};
		\node[gauge,label={below:$N$}] (T21) [above right=8mm of N] {};
        \node (T22) [above right=4mm of T21] {$T_{\rho_{g+1}}$};
		\node[gauge,label={below:$N$},label={left:{\purple \scriptsize $\lambda$}}] (T1p) [below left=8mm of N] {};
		\node[gauge,label={below:$N$},label={right:{\purple \scriptsize $-\lambda$}}] (Tkp) [below right=8mm of N] {};
		\draw[thick] (N) -- (T11) -- (T12);
		\draw[thick] (N) -- (T21) -- (T22);
		\draw[thick] (N) -- (T1p);
		\draw[thick] (N) -- (Tkp);
         \draw[thick] [decorate,decoration={brace,amplitude=5pt},xshift=0cm, yshift=0cm]
([yshift=5mm]T12.west) -- ([yshift=5mm]T22.east) node [black,pos=0.5,yshift=0.4cm] 
{$g+1$ tails};
	\end{tikzpicture}
\end{equation}
where $\lambda$ is the FI parameter turned on at $\U(N)$ nodes. The effect of the FI deformation is to turn \eqref{eq:examgtailsquiver} into
\begin{equation}
    \begin{tikzpicture}[baseline=0,font=\footnotesize]
		\node[gauge,label=below:{$N$},label=left:{\purple \scriptsize $\eta_1$}] (N) {};
		\node[gauge,label={above:$N$}] (T11) [above left=8mm of N] {};
        \node (T12) [above left=4mm of T11] {$T_{\rho_1}$};
		\node (dots1) [above=8mm of N] {$\cdots$};
		\node[gauge,label={above:$N$},label=right:{\purple \scriptsize $-\eta_1$}] (T21) [above right=8mm of N] {};
        \node (T22) [above right=4mm of T21] {$T_{\rho_{g+1}}$};
        \node[gauge,label=below:{$N$},fill=cyan] (N2) [below=8mm of N] {};
		\draw[thick] (N) -- (T11) -- (T12);
		\draw[thick] (N) -- (T21) -- (T22);
        \draw[thick] (T11) to[bend right=30] (N2);
        \draw[thick] (N2) to[bend right=30] (T21);
         \draw[thick] [decorate,decoration={brace,amplitude=5pt},xshift=0cm, yshift=0cm]
([yshift=5mm]T12.west) -- ([yshift=5mm]T22.east) node [black,pos=0.5,yshift=0.4cm] 
{$g+1$ tails};
	\end{tikzpicture}
\end{equation}
where we have now turned on another FI deformation with parameter $\eta_1$, and the {\color{cyan}{cyan}} node is the rebalancing node. This second FI deformation leads to 
\begin{equation}
   \begin{tikzpicture}[baseline=0,font=\footnotesize]
		\node[gauge,label=below right:{$N$},label=left:{\purple \scriptsize $\eta_2$}] (N) {};
		\node[gauge,label={above:$N$}] (T11) [above left=8mm of N] {};
        \node (T12) [above left=4mm of T11] {$T_{\rho_1}$};
		\node (dots1) [above=9mm of N] {$\cdots$};
		\node[gauge,label={above:$N$},label=right:{\purple \scriptsize $-\eta_2$}] (T21) [above right=8mm of N] {};
        \node (T22) [above right=4mm of T21] {$T_{\rho_{g}}$};
        \node[gauge,label=below:{$N$}] (N2) [below=8mm of N] {};
        \node (T3) [right=8mm of N] {$T_{\rho_{g+1}}$};
		\draw[thick] (N) -- (T11) -- (T12);
		\draw[thick] (N) -- (T21) -- (T22);
        \draw[thick] (T11) to[bend right=30] (N2);
        \draw[thick] (N2) to[bend right=30] (T21);
        \draw[thick] (N) -- (T3);
        \draw[thick] (N2) -- (N);
         \draw[thick] [decorate,decoration={brace,amplitude=5pt},xshift=0cm, yshift=0cm]
([yshift=5mm]T12.west) -- ([yshift=5mm]T22.east) node [black,pos=0.5,yshift=0.4cm] 
{$g$ tails};
	\end{tikzpicture}
\end{equation}
Reiterating this process $g+1$ times, with FI parameters $\eta_i$ for $i=1,\dots,g+1$, yields the quiver 
\begin{equation}
    \begin{tikzpicture}[baseline=0,font=\footnotesize]
		\node[gauge,label=below:{$N$}] (N) {};
		\node (T1) [above left=9mm of N] {$T_{\rho_1}$};
		\node (dots1) [above=6mm of N] {$\cdots$};
		\node (Tk) [above right=9mm of N] {$T_{\rho_{g+1}}$};
		\node[gauge,label=below:{$N$}] (N2) [right=15mm of N] {};
		\draw[thick] (N) -- (T1);
        \draw[thick] (N) -- (Tk);
        \draw[thick] [decorate,decoration={brace,amplitude=5pt},xshift=0cm, yshift=0cm]
([yshift=5mm]T1.west) -- ([yshift=5mm]Tk.east) node [black,pos=0.5,yshift=0.4cm] 
{$g+1$ tails};
    \draw[thick] (N) --node[below,midway] {$\begin{array}{c}g+1\\ \text{edges}\end{array}$} (N2);
	\end{tikzpicture}
\end{equation}
Finally, with a further deformation involving the two $\U(N)$ nodes, we find precisely the quiver we are looking for:
\begin{equation}\label{loopy}
    \begin{tikzpicture}[baseline=0,font=\footnotesize]
		\node[gauge,label=right:{$N$}] (N) {};
		\node (T1) [above left=9mm of N] {$T_{\rho_1}$};
		\node (dots1) [above=6mm of N] {$\cdots$};
		\node (Tk) [above right=9mm of N] {$T_{\rho_{g+1}}$};
		\draw[thick] (N) -- (T1);
        \draw[thick] (N) -- (Tk);
        \draw[thick] [decorate,decoration={brace,amplitude=5pt},xshift=0cm, yshift=0cm]
([yshift=5mm]T1.west) -- ([yshift=5mm]Tk.east) node [black,pos=0.5,yshift=0.4cm] 
{$g+1$ tails};
    \draw[thick] (N) to[out=225, in=315, looseness=12] node[below,pos=0.5] {$g$ loops} (N);
	\end{tikzpicture}
\end{equation}
This describes a higher genus class $\mathcal{S}$ theory. However, in the process, we have also produced $g$ free hypermultiplets (the trace components of the adjoint hypermultiplets), which, on the mirror side, correspond to free vector multiplets. We therefore conclude that the FI deformations of a star-shaped quiver with $g+3$ tails lead to a genus $g$ theory with $g$ free vector multiplets. This is a new feature we have not encountered so far and deserves a detailed discussion.

In order to elucidate the physical meaning of the above finding, let us consider the simplest nontrivial example with $N=2$ and $g=1$. If we further set $T_{\rho_2}$ to be trivial, the quiver \eqref{eq:examgtailsquiver} reduces to 
\begin{equation}\label{uspquiv}
    \begin{tikzpicture}[baseline=0,font=\footnotesize]
		\node[gauge,label=below:{$4$}] (N) {};
		\node[gauge,label={below:$2$}] (T11) [above left=8mm of N] {};
            \node[gauge,label={below:$1$}] (T12) [above left=8mm of T11] {};
		\node[gauge,label={below:$2$}] (T21) [above right=8mm of N] {};
		\node[gauge,label={below:$2$}] (T1p) [below left=8mm of N] {};
		\node[gauge,label={below:$2$}] (Tkp) [below right=8mm of N] {};
		\draw[thick] (N) -- (T11) -- (T12);
		\draw[thick] (N) -- (T21);
		\draw[thick] (N) -- (T1p);
		\draw[thick] (N) -- (Tkp);
	\end{tikzpicture}
\end{equation}
which is known to be the magnetic quiver of the $\USp(4)$ gauge theory with $1$ antisymmetric and 4 fundamental hypermultiplets. After four FI deformations, we find \eqref{loopy}, which, in the present case, has one adjoint and one tail:
\begin{equation}\label{loopy1}
    \begin{tikzpicture}[baseline=0,font=\footnotesize]
		\node[gauge,label=below:{$2$}] (N) {};
		\node[gauge,label=below:{$1$}] (1) [right=6mm of N] {};
		\draw[thick] (N) -- (1);
    \draw[thick] (N) to[out=135, in=225, looseness=12] (N);
	\end{tikzpicture}
\end{equation}
This is known to be the magnetic quiver for 4d $\mathcal{N}=4$ SYM with gauge group $\SU(2)$.\footnote{Theory \eref{uspquiv} and its mirror dual, which is the $\USp(4)$ gauge theory with 1 antisymmetric and 4 fundamental hypermultiplets, are in fact ugly theories: the former contains a free twisted hypermultiplet due to the presence of the monopole operator with dimension $\frac{1}{2}$, and the latter contains a free hypermultiplet due to the fact that the rank-two antisymmetric representation $\mathbf{6}$ of $\USp(4)$ is reducible to $\mathbf{5} \oplus \mathbf{1}$, where $\mathbf{1}$ is the trace. The Coulomb branch of the former and the Higgs branch of the latter describe the moduli space of two $\SO(8)$ instantons on $\mathbb{C}^2$, and the free sector corresponds to the center of the instantons; see \cite{Cremonesi:2014xha}. Note also that theory \eref{loopy1} has a free twisted hypermultiplet. We therefore conclude that, at each deformation step from theory \eref{uspquiv} all the way to theory \eref{loopy1}, the theory is ugly. The free hypermultiplet and twisted hypermultiplet mentioned so far in this footnote do not play any role in the discussion in the main text. This should not be confused with the free multiplet mentioned in the main text, which corresponds to the trace part of theory \eref{loopy1}. Note that this has no analogue in theory \eref{uspquiv}.} The latter theory can also be considered as 4d $\cN=2$ $\SO(3)$ SQCD with one flavor in the vector representation. 

The effect of the four FI deformations on \eqref{uspquiv} is to break the $\SO(8)$ symmetry on the Coulomb branch, and therefore it is natural to interpret them in 4d as turning on deformations which give a mass to the $4$ flavors one-by-one. As a result, we end up with a $\USp(4)$ gauge theory with $1$ hypermultiplet in the antisymmetric representation, which we can also think of as $\SO(5)$ SQCD with $1$ flavor. The Coulomb branch of the latter theory has been studied in detail in \cite{Giacomelli:2013qz},\footnote{See also \cite{Argyres:1996hc, Carlino:2001ya, Giacomelli:2012ea} for previous studies on the moduli space of $\SO(N)$ SQCD.} where it was shown that there is a one-dimensional locus in the Coulomb branch hosting $\SU(2)$ $\mathcal{N}=4$ SYM plus a free vector multiplet as low-energy theory. The statement therefore is that the sequence of FI deformations displayed above amounts to decoupling the flavors of the $\USp(4)$ gauge theory and zooming around one point inside the special locus in the Coulomb branch we have just mentioned. This is similar to the procedure proposed originally by Argyres and Douglas in \cite{Argyres:1995xn}. 

In conclusion, the relevant deformation we have activated in the quiver corresponds, at the level of the 4d theory, to turning on mass deformations which lead to an asymptotically free theory, whose Coulomb branch is not a cone with a well-defined notion of origin, and tuning to a specific singular point on the Coulomb branch. This motion in the moduli space of the theory is a feature of the flow to higher genus theories and does not arise for the other flows we have studied in this paper, relating class $\mathcal{S}$ theories on the sphere.

\subsection{Generalization}

In the previous section, we have obtained a genus $g$ class $\mathcal{S}$ theory starting from a class $\mathcal{S}$ theory on a sphere with $g+3$ punctures. The problem with \eqref{loopy} is that there is a correlation between the number of loops and the number of tails in the quiver. However, no such constraint appears in the class $\mathcal{S}$ context, where the genus of the Riemann surface and the number of punctures are two completely independent parameters.

In this section, we show that it is possible to relax the constraint and obtain from genus zero models any genus $g$ class $\mathcal{S}$ theory with an arbitrary number of punctures. As before, the physical interpretation of the RG flow is exactly as before: we start from a genus zero theory, turn on mass deformations which make the theory asymptotically free and zoom around a singular point in the Coulomb branch.

The starting point is the following star-shaped quiver:
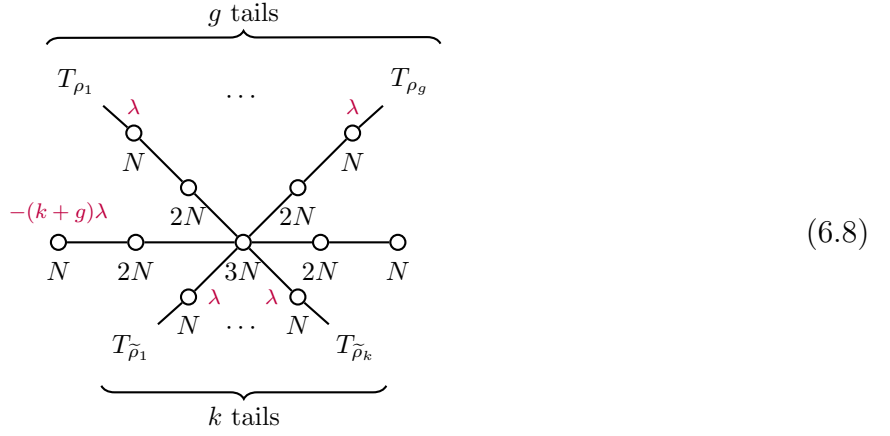
\begin{equation}\label{eq:genstarshapedwithgenus}
    \begin{tikzpicture}[baseline=0,font=\footnotesize]
		\node[gauge,label=below:{$3N$}] (N) {};
		\node[gauge,label={below:$2N$}] (T11) [above left=8mm of N] {};
        \node[gauge,label={below:$N$},label=above:{\purple\scriptsize$\lambda$}] (T12) [above left=8mm of T11] {};
        \node (T13) [above left=4mm of T12] {$T_{\rho_1}$};
		\node (dots1) [above=16mm of N] {$\cdots$};
		\node[gauge,label={below:$2N$}] (T21) [above right=8mm of N] {};
        \node[gauge,label={below:$N$},label=above:{\purple\scriptsize$\lambda$}] (T22) [above right=8mm of T21] {};
        \node (T23) [above right=4mm of T22] {$T_{\rho_g}$};
		\node[gauge,label={below:$2N$}] (T31) [left=12mm of N] {};
        \node[gauge,label={below:$N$},label=above:{\purple\scriptsize$-(k+g)\lambda$}] (T32) [left=8mm of T31] {};
        \node[gauge,label={below:$2N$}] (T41) [right=8mm of N] {};
        \node[gauge,label={below:$N$}] (T42) [right=8mm of T41] {};
        \node[gauge,label={below:$N$},label=right:{\purple\scriptsize$\lambda$}] (T51) [below left=8mm of N] {};
        \node (T52) [below left=4mm of T51] {$T_{\tilde{\rho}_1}$};
		\node (dots2) [below=8mm of N] {$\cdots$};
		\node[gauge,label={below:$N$},label=left:{\purple\scriptsize$\lambda$}] (T61) [below right=8mm of N] {};
        \node (T62) [below right=4mm of T61] {$T_{\tilde{\rho}_k}$};
		\draw[thick] (N) -- (T11) -- (T12) -- (T13);
		\draw[thick] (N) -- (T21) -- (T22) -- (T23);
		\draw[thick] (N) -- (T31) -- (T32);
		\draw[thick] (N) -- (T41) -- (T42);
        \draw[thick] (N) -- (T51) -- (T52);
		\draw[thick] (N) -- (T61) -- (T62);
         \draw[thick] [decorate,decoration={brace,amplitude=5pt},xshift=0cm, yshift=0cm]
([yshift=5mm]T13.west) -- ([yshift=5mm]T23.east) node [black,pos=0.5,yshift=0.4cm] 
{$g$ tails};
\draw[thick] [decorate,decoration={mirror,brace,amplitude=5pt},xshift=0cm, yshift=0cm]
([yshift=-5mm]T52.west) -- ([yshift=-5mm]T62.east) node [black,pos=0.5,yshift=-0.4cm] 
{$k$ tails};
	\end{tikzpicture}
\end{equation}
Note that, by construction, none of the nodes in this quiver are underbalanced. The effect of the FI deformation is to lead to the following quiver:
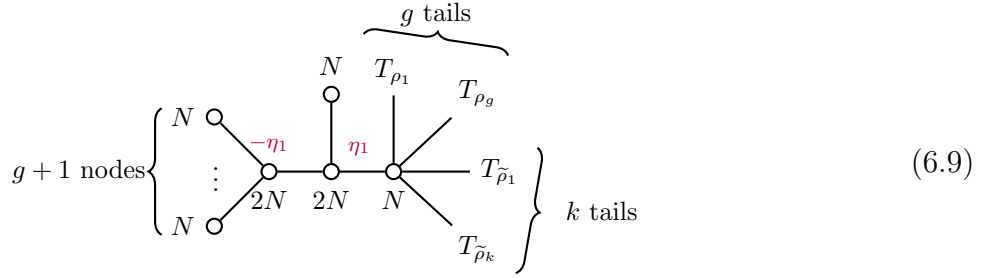
\begin{equation}
      \begin{tikzpicture}[baseline=0,font=\footnotesize]
		\node[gauge,label=below:{$N$}] (N) {};
        \node (T13) [above=9mm of N] {$T_{\rho_1}$};
        \node (T23) [above right=9mm of N] {$T_{\rho_g}$};
        \node[gauge,label={below:$2N$},label={above right:{\purple \scriptsize $\eta_1$}}] (T30) [left=6mm of N] {};
		\node[gauge,label={below:$2N$},label={above:{\purple \scriptsize $-\eta_1$}}] (T31) [left=6mm of T30] {};
        \node (T32) [left=4mm of T31] {$\vdots$};
        \node[gauge,label={left:$N$}] (T33) [above left=8mm of T31] {};
        \node[gauge,label={left:$N$}] (T34) [below left=8mm of T31] {};
        \node[gauge,label={above:$N$}] (T41) [above=8mm of T30] {};
        \node (T52) [right=9mm of N] {$T_{\tilde{\rho}_1}$};
        \node (T62) [below right=9mm of N] {$T_{\tilde{\rho}_k}$};
		\draw[thick] (N)  -- (T13);
        \draw[thick] (N)  -- (T23);
        \draw[thick] (N)  -- (T52);
        \draw[thick] (N)  -- (T62);
		\draw[thick] (N) -- (T30) -- (T31);
		\draw[thick] (T31) -- (T33);
        \draw[thick] (T31) -- (T34);
		\draw[thick] (T30) -- (T41);
         \draw[thick] [decorate,decoration={brace,amplitude=5pt},xshift=0cm, yshift=0cm]
([yshift=5mm]T13.west) -- ([yshift=5mm]T23.east) node [black,pos=0.5,yshift=0.4cm] 
{$g$ tails};
\draw[thick] [decorate,decoration={brace,amplitude=5pt},xshift=0cm, yshift=0cm]
([xshift=5mm]T52.north) -- ([xshift=5mm]T62.south) node [black,pos=0.5,xshift=1cm] 
{$k$ tails};
\draw[thick] [decorate,decoration={mirror,brace,amplitude=5pt},xshift=0cm, yshift=0cm]
([xshift=-7mm]T33.north) -- ([xshift=-7mm]T34.south) node [black,pos=0.5,xshift=-1.1cm] 
{$g+1$ nodes};
	\end{tikzpicture}
\end{equation}
The deformation parametrized by $\eta_1$ in turn leads to 
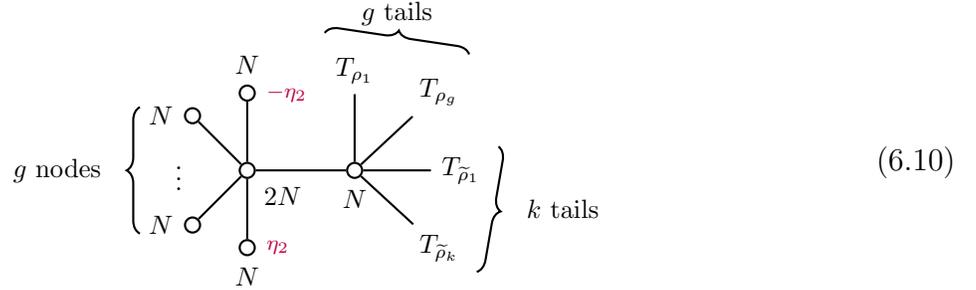
\begin{equation}
      \begin{tikzpicture}[baseline=0,font=\footnotesize]
		\node[gauge,label=below:{$N$}] (N) {};
        \node (T13) [above=9mm of N] {$T_{\rho_1}$};
        \node (T23) [above right=9mm of N] {$T_{\rho_g}$};
		\node[gauge,label={below right:$2N$}] (T31) [left=12mm of N] {};
        \node (T32) [left=6mm of T31] {$\vdots$};
        \node[gauge,label={left:$N$}] (T33) [above left=8mm of T31] {};
        \node[gauge,label={left:$N$}] (T34) [below left=8mm of T31] {};
        \node[gauge,label={below:$N$},label={right:{\purple \scriptsize $\eta_2$}}] (T41) [below=8mm of T31] {};
        \node[gauge,label={above:$N$},label={right:{\purple \scriptsize $-\eta_2$}}] (T71) [above=8mm of T31] {};
        \node (T52) [right=9mm of N] {$T_{\tilde{\rho}_1}$};
        \node (T62) [below right=9mm of N] {$T_{\tilde{\rho}_k}$};
		\draw[thick] (N)  -- (T13);
        \draw[thick] (N)  -- (T23);
        \draw[thick] (N)  -- (T52);
        \draw[thick] (N)  -- (T62);
		\draw[thick] (N) -- (T31);
		\draw[thick] (T31) -- (T33);
        \draw[thick] (T31) -- (T34);
		\draw[thick] (T31) -- (T41);
        \draw[thick] (T31) -- (T71);
         \draw[thick] [decorate,decoration={brace,amplitude=5pt},xshift=0cm, yshift=0cm]
([yshift=5mm]T13.west) -- ([yshift=5mm]T23.east) node [black,pos=0.5,yshift=0.4cm] 
{$g$ tails};
\draw[thick] [decorate,decoration={brace,amplitude=5pt},xshift=0cm, yshift=0cm]
([xshift=5mm]T52.north) -- ([xshift=5mm]T62.south) node [black,pos=0.5,xshift=1cm] 
{$k$ tails};
\draw[thick] [decorate,decoration={mirror,brace,amplitude=5pt},xshift=0cm, yshift=0cm]
([xshift=-7mm]T33.north) -- ([xshift=-7mm]T34.south) node [black,pos=0.5,xshift=-1.1cm] 
{$g$ nodes};
	\end{tikzpicture}
\end{equation}
Notice that we have isolated two $\U(N)$ nodes attached to the $\U(2N)$ node, and the left side of the quiver morally looks like \eqref{eq:examgtailsquiver}. If, in fact, we continue with the deformation parametrized by $\eta_2$, we reach
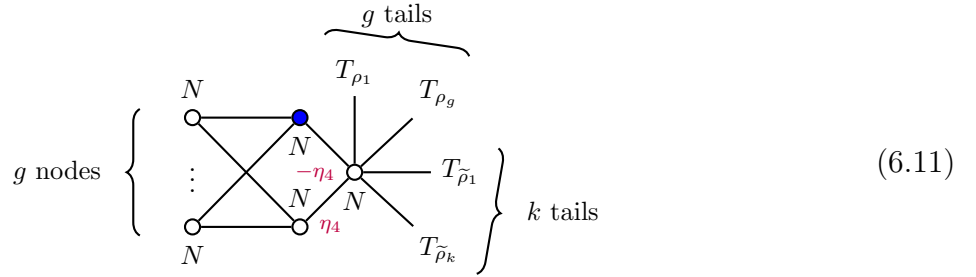
\begin{equation}
      \begin{tikzpicture}[baseline=0,font=\footnotesize]
		\node[gauge,label=below:{$N$},label={left:{\purple \scriptsize $-\eta_4$}}] (N) {};
        \node (T13) [above=9mm of N] {$T_{\rho_1}$};
        \node (T23) [above right=9mm of N] {$T_{\rho_g}$};
        \node[gauge,label={below:$N$},fill=blue] (T41) [above left=8mm of N] {};
        \node[gauge,label={above:$N$},label={right:{\purple \scriptsize $\eta_4$}}] (T51) [below left=8mm of N] {};
        \node[gauge,label={above:$N$}] (T31) [left=12mm of T41] {};
        \node[gauge,label={below:$N$}] (T33) [left=12mm of T51] {};
        \node (T52) [right=9mm of N] {$T_{\tilde{\rho}_1}$};
        \node (T62) [below right=9mm of N] {$T_{\tilde{\rho}_k}$};
        \node (T32) [below=2mm of T31] {$\vdots$};
		\draw[thick] (N)  -- (T13);
        \draw[thick] (N)  -- (T23);
        \draw[thick] (N)  -- (T41);
        \draw[thick] (N)  -- (T51);
        \draw[thick] (N)  -- (T52);
        \draw[thick] (N)  -- (T62);
		\draw[thick] (T41) -- (T31);
        \draw[thick] (T41) -- (T33);
        \draw[thick] (T51) -- (T31);
        \draw[thick] (T51) -- (T33);
         \draw[thick] [decorate,decoration={brace,amplitude=5pt},xshift=0cm, yshift=0cm]
([yshift=5mm]T13.west) -- ([yshift=5mm]T23.east) node [black,pos=0.5,yshift=0.4cm] 
{$g$ tails};
\draw[thick] [decorate,decoration={brace,amplitude=5pt},xshift=0cm, yshift=0cm]
([xshift=5mm]T52.north) -- ([xshift=5mm]T62.south) node [black,pos=0.5,xshift=1cm] 
{$k$ tails};
\draw[thick] [decorate,decoration={mirror,brace,amplitude=5pt},xshift=0cm, yshift=0cm]
([xshift=-7mm]T31.north) -- ([xshift=-7mm]T33.south) node [black,pos=0.5,xshift=-1.1cm] 
{$g$ nodes};
	\end{tikzpicture}
\end{equation}
where the {\color{blue}{blue}} node has been added for rebalancing. The deformations now proceed as in the previous section and, after two deformations, we obtain
\begin{equation}
      \begin{tikzpicture}[baseline=0,font=\footnotesize]
		\node[gauge,label=below:{$N$},label=left:{\purple \scriptsize$-g\chi$}] (N) {};
        \node (T13) [above=9mm of N] {$T_{\rho_1}$};
        \node (T23) [above right=9mm of N] {$T_{\rho_g}$};
        \node[gauge,label={above:$N$},label=left:{\purple \scriptsize$\chi$}] (T31) [above left=10mm and 16mm of N] {};
        \node[gauge,label={below:$N$},label=left:{\purple \scriptsize$\chi$}] (T33) [below left=10mm and 16mm of N] {};
        \node (T52) [right=9mm of N] {$T_{\tilde{\rho}_1}$};
        \node (T62) [below right=9mm of N] {$T_{\tilde{\rho}_k}$};
        \node (T32) [below=4mm of T31] {$\vdots$};
		\draw[thick] (N)  -- (T13);
        \draw[thick] (N)  -- (T23);
        \draw[thick] (N)  -- (T52);
        \draw[thick] (N)  -- (T62);
		\draw[thick] (N) -- node[above,midway,sloped] {$2$ edges}  (T31);
        \draw[thick] (N) -- node[below,midway,sloped] {$2$ edges} (T33);
         \draw[thick] [decorate,decoration={brace,amplitude=5pt},xshift=0cm, yshift=0cm]
([yshift=5mm]T13.west) -- ([yshift=5mm]T23.east) node [black,pos=0.5,yshift=0.4cm] 
{$g$ tails};
\draw[thick] [decorate,decoration={brace,amplitude=5pt},xshift=0cm, yshift=0cm]
([xshift=5mm]T52.north) -- ([xshift=5mm]T62.south) node [black,pos=0.5,xshift=1cm] 
{$k$ tails};
\draw[thick] [decorate,decoration={mirror,brace,amplitude=5pt},xshift=0cm, yshift=0cm]
([xshift=-7mm]T31.north) -- ([xshift=-7mm]T33.south) node [black,pos=0.5,xshift=-1.1cm] 
{$g$ nodes};
	\end{tikzpicture}
\end{equation}
Finally, performing $g$ deformations, we obtain
\begin{equation}
      \begin{tikzpicture}[baseline=0,font=\footnotesize]
		\node[gauge,label=below:{$N$}] (N) {};
        \node (T13) [above=9mm of N] {$T_{\rho_1}$};
        \node (T23) [above right=9mm of N] {$T_{\rho_g}$};
        \node (T52) [right=9mm of N] {$T_{\tilde{\rho}_1}$};
        \node (T62) [below right=9mm of N] {$T_{\tilde{\rho}_k}$};
		\draw[thick] (N)  -- (T13);
        \draw[thick] (N)  -- (T23);
        \draw[thick] (N)  -- (T52);
        \draw[thick] (N)  -- (T62);
        \draw[thick] (N) to[out=135, in=225, looseness=12] node[left,pos=0.5] {$g$ loops} (N);
         \draw[thick] [decorate,decoration={brace,amplitude=5pt},xshift=0cm, yshift=0cm]
([yshift=5mm]T13.west) -- ([yshift=5mm]T23.east) node [black,pos=0.5,yshift=0.4cm] 
{$g$ tails};
\draw[thick] [decorate,decoration={brace,amplitude=5pt},xshift=0cm, yshift=0cm]
([xshift=5mm]T52.north) -- ([xshift=5mm]T62.south) node [black,pos=0.5,xshift=1cm] 
{$k$ tails};
	\end{tikzpicture}
 \label{eq:genusgquiver}
\end{equation}
We have then obtained a generic class $\mathcal{S}$ theory on a genus $g$ Riemann surface with $k+g$ punctures. However, since the number of punctures $g$ and $k$ are arbitrary, we can obtain any kind of theory, by choosing some punctures to be trivial if needed.

Interestingly, because \eqref{eq:genstarshapedwithgenus} is by construction a good quiver, the inversion algorithm is straightforward, and the candidate parent star-shaped quiver of \eqref{eq:genusgquiver} is directly encoded in \eqref{eq:genstarshapedwithgenus}. However, differently from what we have discussed in the previous sections, such deformations are not preserving the Higgs branch dimensions of the quiver theory, if we focus on the interacting sector of the infrared theory. This will have a Higgs branch which has lower dimension than that of the original parent theory, i.e. \eqref{eq:genstarshapedwithgenus}, since the deformation generates $g$ free hypermultiplets along the way. This is again a manifestation of the specific nature of the RG flows connecting genus zero and higher genus theories, which are not standard mass deformations as those we have discussed in the previous sections.

\acknowledgments 

The authors thank Florent Baume, Federico Carta, Julius Grimminger, Craig Lawrie, Lorenzo Mansi and Raffaele Savelli for useful discussions. The work of S. G. is supported by the INFN grant ``Per attività di formazione per sostenere progetti di ricerca'' (GRANT 73/STRONGQFT). N. M. and W. H. are partially supported by the MUR-PRIN grant No. 2022NY2MXY. A. M. is supported in part by the DOE grant DE-SC0017647. Until September 2024, A. M. was supported in part by Deutsche Forschungsgemeinschaft under Germany's Excellence Strategy EXC 2121 Quantum Universe 390833306, Deutsche Forschungsgemeinschaft through a German-Israeli Project Cooperation (DIP) grant ``Holography and the Swampland'' and Deutsche Forschungsgemeinschaft through the Collaborative Research Center SFB1624 ``Higher Structures, Moduli Spaces, and Integrability''. N. M. gratefully acknowledges support from the Simons Center for Geometry and Physics, Stony Brook University, at which part of this research project was conducted during the Simons Physics Summer Workshop 2024.

\appendix

\section{Review of Dualities for Bad Theories}
\label{sec:Itamarduality}

In \cite{Yaakov:2013fza}, it was proposed that a bad SQCD theory described by the $\U(N)$ gauge theory with $N\leq N_f\leq 2N-2$ flavors is Seiberg dual to a good SQCD theory described by the $\U(N_f-N)$ gauge theory and $N_f$ flavors plus $2N-N_f$ free twisted hypermultiplets. As emphasized in \cite{Assel:2017jgo, Giacomelli:2023zkk}, this statement is really a duality if and only if a non-zero FI parameter is turned on at the $\U(N)$ gauge group of the bad theory. In particular, it was shown using the three-sphere partition function that, if $\eta \neq 0$ is the real FI parameter for the latter, then, in the dual theory, the FI parameter for the $\U(N_f-N)$ gauge group is $-\eta$ and the real mass for the twisted free hypermultiplets is $\eta$. 

Such dualization plays an important role in our paper. In particular, upon applying the inverse algorithm presented in Section \ref{sec:InvFormulas}, we find the parent theory in terms of a mirror dual star-shaped quiver that possibly contains underbalanced nodes. The latter implies that the candidate parent theory is bad. As we will demonstrate explicitly in Appendix \ref{sec:MixDef2tail} for lifting the $E_7$ quivers to the $E_8$ ones, the FI parameters corresponding to the mass deformations in question are {\it always turned on at the underbalanced nodes}. We conjecture that this statement holds generally, and can be tested in all examples that we know of. This implies that we can dualize such a node as described above and obtain a new quiver that is indeed dual to the original one. Let us suppose that the underbalanced node of our interest has FI parameter $\lambda$. Upon dualizing it, it now possesses a new FI parameter $-\lambda$ as discussed above, and each of the nodes adjacent to it now acquires the FI parameter $\lambda$; see \cite[(3.34)]{Yaakov:2013fza} and \cite{Giacomelli:2024laq}. Reiterating this procedure, we arrive at a good star-shaped quiver, which can now be interpreted as a magnetic quiver of the parent theory. The free twisted hypermultiplets, which are a by-product of dualization, are guaranteed to be massive and can be integrated out upon flowing to the infrared.

\section{Good Quivers via Inversion Algorithm and Dualization}
	\label{sec:MixDef2tail}
 
	In this section, we show that, given a generic $E_7$ quiver with two tails, it is always possible to determine its lift to a good parent $E_8$ quiver using the mixed deformation inverse algorithm, presented in Section \ref{sec:InvFormulasE72tailstoE8}, and performing a sequence of dualities \cite{Yaakov:2013fza}. Let us consider an $E_7$ quiver with two tails of the form
	\begin{equation} \label{E7quivk1k2}

	\end{equation}
	where we take $E \ge C$ and we write down the tail on the left-hand side as
	\begin{equation} 
		%
	\end{equation}
	Quiver \eqref{E7quivk1k2} is good provided that the following conditions hold:
	\begin{equation} \label{GoodnesscondE7}
		\begin{cases}
			D-C \ge C-B \ge B-A \ge A-k_1 \ge k_1-\hat{k} \geq \ldots \\
			D-E \ge E-F \ge F-G \ge G-k_2 \geq \ldots \\
			2 H \le D \\
			2 D \le C+H+E \\
			D-E < D-C
		\end{cases}~.
	\end{equation}
	We can lift such quiver to a parent $E_8$ quiver related by a deformation involving the nodes $3$ and $3'$ in the latter. Using \eqref{eq:33plift}, the $E_8$ quiver is then realized as
	\begin{equation}
		%
	\end{equation}
	where $k = k_1 + k_2$ and we highlight in {\purple purple} the various FI parameters we turn on at different nodes. 
	\paragraph*{First Sequence of Dualizations} 
	If $C - B > G - k_2$, we can dualize at node $3$ and the resulting quiver is
	\begin{equation} \label{E8dualizenode3}
		%
	\end{equation}
	Note that, upon dualizing, the sign of the FI parameter at node $3$ is flipped, and we have to turn on FI parameters at adjacent nodes, \ie nodes $2$ and $4$. Now, we can dualize both on the left-hand side and on the right-hand side of node $3$, leading to the sequence of dualizations described in Figure \ref{figE8dual1}.
	\begin{figure} 
		\centering
		\scalebox{0.92}{$
			%
			$}
		\caption{Sequence of dualities obtained by starting from quiver \eqref{E8dualizenode3} and dualizing at each node both on the left-hand side and on the right-hand side of node $3$. Observe that we do not dualize at node $3'$ since $2 C \le B + D$ from \eqref{GoodnesscondE7}.}
		\label{figE8dual1}
	\end{figure}
	Let us summarize the inequalities we have to solve in order to perform the complete sequence of dualizations reported in Figure \ref{figE8dual1}.
	\bes{ \label{Dualizeseq1}
		%
	}
	Note that the inequalities at the bottom in \eqref{Dualizeseq1}	are stronger than the inequalities at the top, \ie if the inequality at the bottom holds, then also the inequality at the top is verified, whereas the opposite is not necessarily true.

\begin{figure}[!htp]
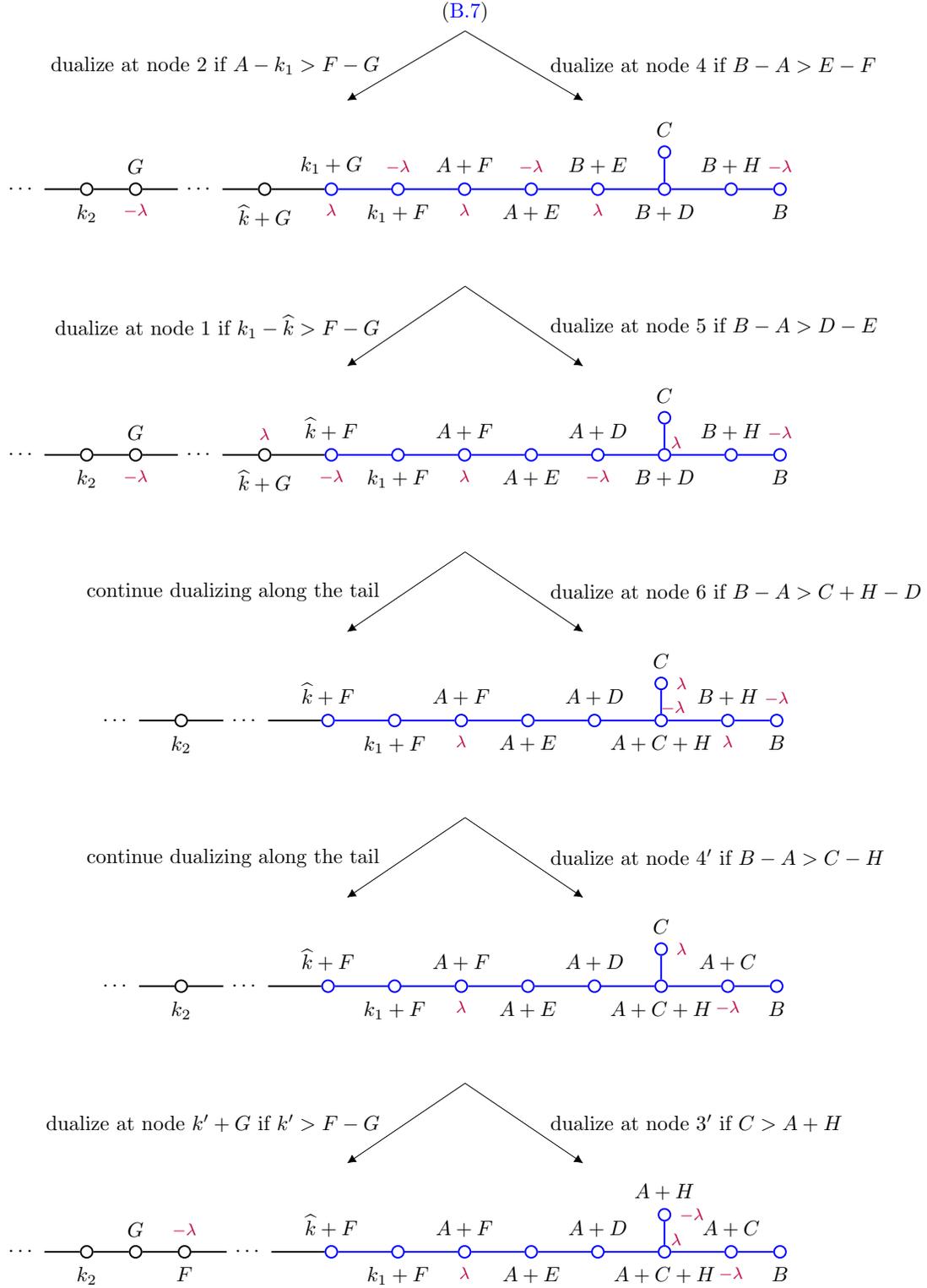

		\centering
		\scalebox{0.93}{$
			%
			$}
		\caption{Sequence of dualities obtained by starting from quiver \eqref{E8dualizenode3seq2} and dualizing at each node both on the left-hand side and on the right-hand side of node $3$. Note that, even if we cannot dualize at node $2'$ since there is no FI parameter turned on, such node is good given that $2 B \le A + C$ from \eqref{GoodnesscondE7}.}
		\label{figE8dual2}
	\end{figure}
 
	\paragraph*{Second Sequence of Dualizations} 
 
 We have to dualize again at node $3$ if $B-A > F-G$, and we obtain the following quiver:
	\begin{equation} \label{E8dualizenode3seq2}
		%
	\end{equation}
	Again, we can start dualizing both on the left-hand side and on the right-hand side of node $3$ in quiver \eqref{E8dualizenode3seq2}. This second sequence of dualizations, which is summarized in Figure \ref{figE8dual2}, is accompanied by the following inequalities:
	\bes{ \label{Dualizeseq2}
		%
	}

	\paragraph*{Third Sequence of Dualizations} 
 
	If $A-k_1 > E - F$, node $3$ of the quiver at the bottom of Figure \ref{figE8dual2} is still bad. Nevertheless, since there is an FI parameter turned on at such node, we can dualize and obtain the following quiver:
	\begin{equation} \label{E8dualizenode3seq3}
		%
	\end{equation}
	Note that we have to dualize at node $2$ of the quiver above if $k_1 - \hat{k} > E-F$ and at node $4$ if $A-k_1 > D-E$. This results in
	\begin{equation} \label{E8dualizenode24seq3}
		%
	\end{equation}
	Node $5$ of quiver \eqref{E8dualizenode24seq3} is bad if $A-k_1 > C+H-D$ and, in such a case, we have to dualize at this node. Furthermore, we might have to continue dualizing on the left-hand side of node $2$. Let us suppose that we dualize at all nodes along the tail, we finally reach, if $k' > E - F$, the following quiver:
	\begin{equation} \label{E8dualizenodetail5seq3}
		\scalebox{0.89}{$
			%
			$}
	\end{equation}
	Let us make a few comments regarding quiver \eqref{E8dualizenodetail5seq3}.
	\bi
	\item The goodness condition at node $6$ is $k_1 \ge 0$, which is always satisfied given that our starting point is the $E_7$ quiver with two tails \eqref{E7quivk1k2}.
	\item Node $4'$ is bad if $B-A < C-H$. However, since, during the second sequence of dualizations summarized in Figure \ref{figE8dual2}, we had to dualize at node $4'$ if $B-A > C-H$, it follows that such node cannot be bad at this step.
	\item The goodness condition at node $2'$ reads $B-A \le C-B$, which is always satisfied thanks to \eqref{GoodnesscondE7}.
	\item Node $3'$ is bad if $A+H > C$. This cannot happen, since, during the second sequence of dualizations implemented in Figure \ref{figE8dual2}, we had to dualize at such node if $C > A+H$.
	\ei
	It follows that we do not have to dualize at nodes $6$, $4'$, $2'$ and $3'$ of quiver \eqref{E8dualizenodetail5seq3}, which is reached if the following inequalities are satisfied:
	\bes{ \label{Dualizeseq3}
		\begin{tikzpicture}
			\node[draw=white] (n3) at (0,3) {$A-k_1 > E-F$}; 
			\node[draw=white] (LHS) at (-3,0) {$\begin{array}{rcl}
					k_1-\hat{k} &>& E-F  \\ &\vdots& \\ k' &>& E-F
			\end{array}$}; 
			\node[draw=white] (RHS) at (3.27,0.33) {$\begin{array}{rcl}
					A-k_1 &>& D-E \\ A-k_1 &>& C+H-D
			\end{array}$};  
			\draw[->] (n3.south) to node[midway, left=0.3] {dualize on the left} (LHS.north east);
			\draw[->] (n3.south) to node[midway, right=0.3] {dualize on the right} (RHS.north west);
			\node (t) at (-6,4) {};
			\node (b) at (-6,-1) {};
			\draw[<-] (b) to node[midway, above,sloped] {stronger inequality} (t);
		\end{tikzpicture}
	}
 
	\paragraph*{Fourth Sequence of Dualizations} 
 
	Node $3$ of quiver \eqref{E8dualizenodetail5seq3} is still bad if $k_1 - \hat{k} > D - E$. We can make it good by dualizing at such a node, which yields
	\begin{equation} \label{E8dualizenode3seq4}
		\scalebox{0.89}{$
			\begin{tikzpicture}[baseline=7,font=\footnotesize]
				\node (dotsv) {$\cdots$};
				\node[gauge, label=below:{$k_2$}] (k2v) [right=6mm of dotsv] {};
				\node[gauge, label=above:{$G$}] (kpv) [right=6mm of k2v] {};
				\node[gauge, label=below:{$F$}] (kv) [right=6mm of kpv] {};
				\node[gauge, label=above:{$E$}, label=below:{\purple{\scriptsize $-\lambda$}}] (kv2) [right=6mm of kv] {};
				\node[gauge, label=below:{$k'+E$}] (kv3) [right=9mm of kv2] {};
				\node (dots2v) [right=6mm of kv3] {$\cdots$};
				\node[gauge, bodyE8, label=above:{$\tilde{k}+E$}] (Av) [right=9mm of dots2v] {};
				\node[gauge, bodyE8, label=below:{$\hat{k}+E$}, label=above:{\purple{\scriptsize $\lambda$}}] (Bv) [right=12mm of Av] {};
				\node[gauge, bodyE8, label=above:{$\hat{k}+D$}, label=below:{\purple{\scriptsize $-\lambda$}}] (Cv) [right=12mm of Bv] {};
				\node[gauge, bodyE8, label=below:{$k_1+D$}, label=above:{\purple{\scriptsize $\lambda$}}] (Dv) [right=12mm of Cv] {};
				\node[gauge, bodyE8, label=above:{$k_1+C+H$}, label=below:{\purple{\scriptsize $-\lambda$}}] (Ev) [right=12mm of Dv] {};
				\node[gauge, bodyE8, label=below:{$A+C+H$}, label={[xshift=2.5mm, yshift=-1mm] \purple{\scriptsize $2 \lambda$}}] (Fv) [right=12mm of Ev] {};
				\node[gauge, bodyE8, label=above:{$A+C$}, label=below:{\purple{\scriptsize $-\lambda$}}] (Gv) [right=12mm of Fv] {};
				\node[gauge, bodyE8, label=below:{$B$}] (Hv) [right=9mm of Gv] {};
				\node[gauge, bodyE8, label=above:{$A+H$}, label=right:{\purple{\scriptsize $-\lambda$}}] (Lv) [above=6mm of Fv] {};
				\draw[thick] (dotsv.east) -- (k2v.west);
				\draw[thick] (k2v.east) -- (kpv.west);
				\draw[thick] (kpv.east) -- (kv.west);
				\draw[thick] (kv.east) -- (kv2.west);
				\draw[thick] (kv2.east) -- (kv3.west);
				\draw[thick] (kv3.east) -- (dots2v.west);
				\draw[thick] (dots2v.east) -- (Av.west);
				\draw[thick,bodyE8] (Av.east) -- (Bv.west);
				\draw[thick,bodyE8] (Bv.east) -- (Cv.west);
				\draw[thick,bodyE8] (Cv.east) -- (Dv.west);
				\draw[thick,bodyE8] (Dv.east) -- (Ev.west);
				\draw[thick,bodyE8] (Ev.east) -- (Fv.west);
				\draw[thick,bodyE8] (Fv.east) -- (Gv.west);
				\draw[thick,bodyE8] (Gv.east) -- (Hv.west);
				\draw[thick,bodyE8] (Fv.north) -- (Lv.south);
			\end{tikzpicture}
			$}
	\end{equation}
	If $k_1 - \hat{k} > C + H - D$, we have to dualize at node $4$. Moreover, if there are bad nodes on the left-hand side of node $3$, we keep dualizing, until we finally reach the following quiver:
	\begin{equation} \label{E8dualizenodetail4seq4}
		\scalebox{0.91}{$
			\begin{tikzpicture}[baseline=0,font=\footnotesize]
				\node (dotsv) {$\cdots$};
				\node[gauge, label=below:{$k_2$}] (k2v) [right=6mm of dotsv] {};
				\node[gauge, label=above:{$G$}] (kpv) [right=6mm of k2v] {};
				\node[gauge, label=below:{$F$}] (kv) [right=6mm of kpv] {};
				\node[gauge, label=above:{$E$}] (kv2) [right=6mm of kv] {};
				\node[gauge, label=below:{$D$}, label=above:{\purple{\scriptsize $-\lambda$}}] (kv3) [right=6mm of kv2] {};
				\node (dots2v) [right=6mm of kv3] {$\cdots$};
				\node[gauge, bodyE8, label=above:{$\bar{k}+D$}] (Av) [right=9mm of dots2v] {};
				\node[gauge, bodyE8, label=below:{$\tilde{k}+D$}] (Bv) [right=12mm of Av] {};
				\node[gauge, bodyE8, label=above:{$\hat{k}+D$}, label=below:{\purple{\scriptsize $\lambda$}}] (Cv) [right=12mm of Bv] {};
				\node[gauge, bodyE8, label=below:{$\hat{k}+C+H$}, label=above:{\purple{\scriptsize $-\lambda$}}] (Dv) [right=12mm of Cv] {};
				\node[gauge, bodyE8, label=above:{$k_1+C+H$}] (Ev) [right=12mm of Dv] {};
				\node[gauge, bodyE8, label=below:{$A+C+H$}, label={[xshift=2.5mm, yshift=-1mm] \purple{\scriptsize $2 \lambda$}}] (Fv) [right=12mm of Ev] {};
				\node[gauge, bodyE8, label=above:{$A+C$}, label=below:{\purple{\scriptsize $-\lambda$}}] (Gv) [right=12mm of Fv] {};
				\node[gauge, bodyE8, label=below:{$B$}] (Hv) [right=9mm of Gv] {};
				\node[gauge, bodyE8, label=above:{$A+H$}, label=right:{\purple{\scriptsize $-\lambda$}}] (Lv) [above=6mm of Fv] {};
				\draw[thick] (dotsv.east) -- (k2v.west);
				\draw[thick] (k2v.east) -- (kpv.west);
				\draw[thick] (kpv.east) -- (kv.west);
				\draw[thick] (kv.east) -- (kv2.west);
				\draw[thick] (kv2.east) -- (kv3.west);
				\draw[thick] (kv3.east) -- (dots2v.west);
				\draw[thick] (dots2v.east) -- (Av.west);
				\draw[thick,bodyE8] (Av.east) -- (Bv.west);
				\draw[thick,bodyE8] (Bv.east) -- (Cv.west);
				\draw[thick,bodyE8] (Cv.east) -- (Dv.west);
				\draw[thick,bodyE8] (Dv.east) -- (Ev.west);
				\draw[thick,bodyE8] (Ev.east) -- (Fv.west);
				\draw[thick,bodyE8] (Fv.east) -- (Gv.west);
				\draw[thick,bodyE8] (Gv.east) -- (Hv.west);
				\draw[thick,bodyE8] (Fv.north) -- (Lv.south);
			\end{tikzpicture}
			$}
	\end{equation}
	Note that we cannot dualize at node $5$ of quiver \eqref{E8dualizenodetail4seq4} since there is no FI parameter turned on. This is not a problem, given that node $5$ is good if $k_1 - \hat{k} \le A - k_1$, which is always satisfied from \eqref{GoodnesscondE7}. The remaining part of the quiver above, namely the nodes on the right-hand side of node $5$, is identical to quiver \eqref{E8dualizenodetail5seq3} and we do not have to dualize it. In summary, in order to reach quiver \eqref{E8dualizenodetail4seq4} starting from \eqref{E8dualizenodetail5seq3}, the inequalities we have to satisfy are the following:
	\bes{ \label{Dualizeseq4}
		\begin{tikzpicture}
			\node[draw=white] (n3) at (0,3) {$k_1 - \hat{k} > D-E$}; 
			\node[draw=white] (LHS) at (-3,0) {$\begin{array}{rcl}
					\hat{k}-\tilde{k} &>& D-E  \\ &\vdots& \\ k' &>& D-E
			\end{array}$}; 
			\node[draw=white] (RHS) at (3.27,0.60) {$\begin{array}{rcl}
					k_1 - \hat{k} &>& C+H-D
			\end{array}$};  
			\draw[->] (n3.south) to node[midway, left=0.3] {dualize on the left} (LHS.north east);
			\draw[->] (n3.south) to node[midway, right=0.3] {dualize on the right} (RHS.north west);
			\node (t) at (-6,4) {};
			\node (b) at (-6,-1) {};
            \draw[<-] (b) to node[midway, above,sloped] {stronger inequality} (t);
		\end{tikzpicture}
	}
 
	\paragraph*{Fifth Sequence of Dualizations} 
 
	Node $3$ of quiver \eqref{E8dualizenodetail4seq4} is bad if $\hat{k}-\tilde{k} > C+H-D$. Upon dualizing at such node, we obtain
	\begin{equation} \label{E8dualizenode3seq5}
		\scalebox{0.91}{$
			\begin{tikzpicture}[baseline=7,font=\footnotesize]
				\node (dotsv) {$\cdots$};
				\node[gauge, label=below:{$k_2$}] (k2v) [right=6mm of dotsv] {};
				\node[gauge, label=above:{$G$}] (kpv) [right=6mm of k2v] {};
				\node[gauge, label=below:{$F$}] (kv) [right=6mm of kpv] {};
				\node[gauge, label=above:{$E$}] (kv2) [right=6mm of kv] {};
				\node[gauge, label=below:{$D$}, label=above:{\purple{\scriptsize $-\lambda$}}] (kv3) [right=6mm of kv2] {};
				\node (dots2v) [right=6mm of kv3] {$\cdots$};
				\node[gauge, bodyE8, label=above:{$\bar{k}+D$}] (Av) [right=9mm of dots2v] {};
				\node[gauge, bodyE8, label=below:{$\tilde{k}+D$}, label=above:{\purple{\scriptsize $\lambda$}}] (Bv) [right=12mm of Av] {};
				\node[gauge, bodyE8, label=above:{$\tilde{k}+C+H$}, label=below:{\purple{\scriptsize $-\lambda$}}] (Cv) [right=12mm of Bv] {};
				\node[gauge, bodyE8, label=below:{$\hat{k}+C+H$}] (Dv) [right=12mm of Cv] {};
				\node[gauge, bodyE8, label=above:{$k_1+C+H$}] (Ev) [right=12mm of Dv] {};
				\node[gauge, bodyE8, label=below:{$A+C+H$}, label={[xshift=2.5mm, yshift=-1mm] \purple{\scriptsize $2 \lambda$}}] (Fv) [right=12mm of Ev] {};
				\node[gauge, bodyE8, label=above:{$A+C$}, label=below:{\purple{\scriptsize $-\lambda$}}] (Gv) [right=12mm of Fv] {};
				\node[gauge, bodyE8, label=below:{$B$}] (Hv) [right=9mm of Gv] {};
				\node[gauge, bodyE8, label=above:{$A+H$}, label=right:{\purple{\scriptsize $-\lambda$}}] (Lv) [above=6mm of Fv] {};
				\draw[thick] (dotsv.east) -- (k2v.west);
				\draw[thick] (k2v.east) -- (kpv.west);
				\draw[thick] (kpv.east) -- (kv.west);
				\draw[thick] (kv.east) -- (kv2.west);
				\draw[thick] (kv2.east) -- (kv3.west);
				\draw[thick] (kv3.east) -- (dots2v.west);
				\draw[thick] (dots2v.east) -- (Av.west);
				\draw[thick,bodyE8] (Av.east) -- (Bv.west);
				\draw[thick,bodyE8] (Bv.east) -- (Cv.west);
				\draw[thick,bodyE8] (Cv.east) -- (Dv.west);
				\draw[thick,bodyE8] (Dv.east) -- (Ev.west);
				\draw[thick,bodyE8] (Ev.east) -- (Fv.west);
				\draw[thick,bodyE8] (Fv.east) -- (Gv.west);
				\draw[thick,bodyE8] (Gv.east) -- (Hv.west);
				\draw[thick,bodyE8] (Fv.north) -- (Lv.south);
			\end{tikzpicture}
			$}
	\end{equation}
	We cannot dualize at node $4$, since there is no FI parameter turned on at such node. Fortunately, node $4$ is good if $\hat{k}-\tilde{k} \le k_1 - \hat{k}$, which is always satisfied thanks to \eqref{GoodnesscondE7}. Upon dualizing on the left-hand side of node $3$ of quiver \eqref{E8dualizenode3seq5}, if the following inequalities are satisfied
	\bes{ \label{Dualizeseq5}
		\begin{tikzpicture}
			\node[draw=white] (n3) at (0,3) {$\hat{k}-\tilde{k} > C+H-D$}; 
			\node[draw=white] (LHS) at (0,0) {$\begin{array}{rcl}
					\tilde{k}-\bar{k} &>& C+H-D \\ &\vdots& \\ k' &>& C+H-D
			\end{array}$};  
			\draw[->] (n3) to node[midway, right=0.3] {dualize on the left} (LHS);
			\node (t) at (-3,4) {};
			\node (b) at (-3,-2) {};
			\draw[<-] (b) to node[midway, above,sloped] {stronger inequality} (t);
		\end{tikzpicture}
	}
	we finally reach the following quiver:
	\begin{equation} \label{E8dualizetailseq5}
		\scalebox{0.86}{$
			\begin{tikzpicture}[baseline=7,font=\footnotesize]
				\node (dotsv) {$\cdots$};
				\node[gauge, label=below:{$k_2$}] (k2v) [right=3mm of dotsv] {};
				\node[gauge, label=above:{$G$}] (kpv) [right=6mm of k2v] {};
				\node[gauge, label=below:{$F$}] (kv) [right=6mm of kpv] {};
				\node[gauge, label=above:{$E$}] (kv2) [right=6mm of kv] {};
				\node[gauge, label=below:{$D$}] (kv3) [right=6mm of kv2] {};
				\node[gauge, label=above:{$C+H$}, label=below:{\purple{\scriptsize $-\lambda$}}] (kv4) [right=9mm of kv3] {};
				\node (dots2v) [right=6mm of kv4] {$\cdots$};
				\node[gauge, bodyE8, label=above:{$k^{\star}+C+H$}] (Av) [right=9mm of dots2v] {};
				\node[gauge, bodyE8, label=below:{$\bar{k}+C+H$}] (Bv) [right=12mm of Av] {};
				\node[gauge, bodyE8, label=above:{$\tilde{k}+C+H$}] (Cv) [right=12mm of Bv] {};
				\node[gauge, bodyE8, label=below:{$\hat{k}+C+H$}] (Dv) [right=12mm of Cv] {};
				\node[gauge, bodyE8, label=above:{$k_1+C+H$}] (Ev) [right=12mm of Dv] {};
				\node[gauge, bodyE8, label=below:{$A+C+H$}, label={[xshift=2.5mm, yshift=-1mm] \purple{\scriptsize $2 \lambda$}}] (Fv) [right=12mm of Ev] {};
				\node[gauge, bodyE8, label=above:{$A+C$}, label=below:{\purple{\scriptsize $-\lambda$}}] (Gv) [right=12mm of Fv] {};
				\node[gauge, bodyE8, label=below:{$B$}] (Hv) [right=9mm of Gv] {};
				\node[gauge, bodyE8, label=above:{$A+H$}, label=right:{\purple{\scriptsize $-\lambda$}}] (Lv) [above=6mm of Fv] {};
				\draw[thick] (dotsv.east) -- (k2v.west);
				\draw[thick] (k2v.east) -- (kpv.west);
				\draw[thick] (kpv.east) -- (kv.west);
				\draw[thick] (kv.east) -- (kv2.west);
				\draw[thick] (kv2.east) -- (kv3.west);
				\draw[thick] (kv3.east) -- (kv4.west);
				\draw[thick] (kv4.east) -- (dots2v.west);
				\draw[thick] (dots2v.east) -- (Av.west);
				\draw[thick,bodyE8] (Av.east) -- (Bv.west);
				\draw[thick,bodyE8] (Bv.east) -- (Cv.west);
				\draw[thick,bodyE8] (Cv.east) -- (Dv.west);
				\draw[thick,bodyE8] (Dv.east) -- (Ev.west);
				\draw[thick,bodyE8] (Ev.east) -- (Fv.west);
				\draw[thick,bodyE8] (Fv.east) -- (Gv.west);
				\draw[thick,bodyE8] (Gv.east) -- (Hv.west);
				\draw[thick,bodyE8] (Fv.north) -- (Lv.south);
			\end{tikzpicture}
			$}
	\end{equation}
	Let us look at the tail of such quiver:
	\begin{equation} \label{tailseq5}
		\begin{tikzpicture}[baseline=0,font=\footnotesize]
			\node (dotsv) {$\cdots$};
			\node[gauge, label=below:{$k_2$}] (k2v) [right=3mm of dotsv] {};
			\node[gauge, label=above:{$G$}] (kpv) [right=9mm of k2v] {};
			\node[gauge, label=below:{$F$}] (kv) [right=9mm of kpv] {};
			\node[gauge, label=above:{$E$}] (kv2) [right=9mm of kv] {};
			\node[gauge, label=below:{$D$}] (kv3) [right=9mm of kv2] {};
			\node[gauge, label=above:{$C+H$}, label=below:{\purple{\scriptsize $-\lambda$}}] (kv4) [right=12mm of kv3] {};
			\node[gauge, label=below:{$k'+C+H$}] (kv5) [right=12mm of kv4] {};
			\node[gauge, label=above:{$k''+C+H$}] (kv6) [right=12mm of kv5] {};
			\node (dots2v) [right=3mm of kv6] {$\cdots$};
			\draw[thick] (dotsv.east) -- (k2v.west);
			\draw[thick] (k2v.east) -- (kpv.west);
			\draw[thick] (kpv.east) -- (kv.west);
			\draw[thick] (kv.east) -- (kv2.west);
			\draw[thick] (kv2.east) -- (kv3.west);
			\draw[thick] (kv3.east) -- (kv4.west);
			\draw[thick] (kv4.east) -- (kv5.west);
			\draw[thick] (kv5.east) -- (kv6.west);
			\draw[thick] (kv6.east) -- (dots2v.west);
		\end{tikzpicture}
	\end{equation}
	We point out that every node is good, including the one with the FI parameter turned on. In fact, the goodness condition at this node is $k' \ge C+H-D$, which is always true since, to reach the quiver \eqref{E8dualizetailseq5} starting from the quiver \eqref{E8dualizenodetail4seq4}, we had to satisfy the inequalities \eqref{Dualizeseq5}. This implies that the sequence of dualizations ends here, and we reach the good quiver \eqref{E8dualizetailseq5}. This shows that, starting from a generic good $E_7$ quiver with two tails, it is always possible to find a good parent $E_8$ quiver.
	
	For reference, let us report an explicit example of a good $E_7$ quiver, for which we have to go through all five sequences of dualizations described in this section in order to determine a good parent $E_8$ quiver. We look for an $E_7$ quiver satisfying \cref{GoodnesscondE7,Dualizeseq1,Dualizeseq2,Dualizeseq3,Dualizeseq4,Dualizeseq5}, which is given for instance by
	\begin{equation}
		\begin{tikzpicture}[baseline=7,font=\footnotesize]
			\node[gauge, label=below:{$13$}] (kp) {};
			\node[gauge, label=below:{$28$}] (hk) [right=9mm of kp] {};
			\node[gauge, label=below:{$45$}] (hk1) [right=9mm of hk] {};
			\node[gauge, label=below:{$64$}] (hk2) [right=9mm of hk1] {};
			\node[gauge, label=below:{$85$}] (k1) [right=9mm of hk2] {};
			\node[gauge, bodyE7, label=below:{$108$}] (A) [right=9mm of k1] {};
			\node[gauge, bodyE7, label=below:{$222$}] (B) [right=9mm of A] {};
			\node[gauge, bodyE7, label=below:{$338$}] (C) [right=9mm of B] {};
			\node[gauge, bodyE7, label=below:{$555$}] (D) [right=9mm of C] {};
			\node[gauge, bodyE7, label=below:{$547$}] (E) [right=9mm of D] {};
			\node[gauge, bodyE7, label=below:{$541$}] (F) [right=9mm of E] {};
			\node[gauge, bodyE7, label=below:{$537$}] (G) [right=9mm of F] {};
			\node[gauge, bodyE7, label=above:{$227$}] (H) [above=4mm of D] {};
			\node[gauge, label=below:{$535$}] (k2) [right=9mm of G] {};
			\node (dots2) [right=3mm of k2] {$\cdots$};
			\draw[thick] (kp.east) -- (hk.west);
			\draw[thick] (hk.east) -- (hk1.west);
			\draw[thick] (hk1.east) -- (hk2.west);
			\draw[thick] (hk2.east) -- (k1.west);
			\draw[thick] (k1.east) -- (A.west);
			\draw[thick,bodyE7] (A.east) -- (B.west);
			\draw[thick,bodyE7] (B.east) -- (C.west);
			\draw[thick,bodyE7] (C.east) -- (D.west);
			\draw[thick,bodyE7] (D.east) -- (E.west);
			\draw[thick,bodyE7] (E.east) -- (F.west);
			\draw[thick,bodyE7] (F.east) -- (G.west);
			\draw[thick,bodyE7] (D.north) -- (H.south);
			\draw[thick] (G.east) -- (k2.west);
			\draw[thick] (k2.east) -- (dots2.west);
		\end{tikzpicture}
	\end{equation}
	From \eqref{E8dualizetailseq5}, a corresponding good parent $E_8$ quiver is
	\begin{equation}
 \scalebox{0.95}{
		\begin{tikzpicture}[baseline=7,font=\footnotesize]
			\node (dotsv) {$\cdots$};
			\node[gauge, label=below:{$535$}] (k2v) [right=3mm of dotsv] {};
			\node[gauge, label=below:{$537$}] (kpv) [right=9mm of k2v] {};
			\node[gauge, label=below:{$541$}] (kv) [right=9mm of kpv] {};
			\node[gauge, label=below:{$547$}] (kv2) [right=9mm of kv] {};
			\node[gauge, label=below:{$555$}] (kv3) [right=9mm of kv2] {};
			\node[gauge, label=below:{$565$}, label=above:{\purple{\scriptsize $-\lambda$}}] (kv4) [right=9mm of kv3] {};
			\node[gauge, bodyE8, label=below:{$578$}] (Av) [right=9mm of kv4] {};
			\node[gauge, bodyE8, label=below:{$593$}] (Bv) [right=9mm of Av] {};
			\node[gauge, bodyE8, label=below:{$610$}] (Cv) [right=9mm of Bv] {};
			\node[gauge, bodyE8, label=below:{$629$}] (Dv) [right=9mm of Cv] {};
			\node[gauge, bodyE8, label=below:{$650$}] (Ev) [right=9mm of Dv] {};
			\node[gauge, bodyE8, label=below:{$673$}, label={[xshift=2.5mm, yshift=-1mm] \purple{\scriptsize $2 \lambda$}}] (Fv) [right=9mm of Ev] {};
			\node[gauge, bodyE8, label=below:{$446$}, label=above:{\purple{\scriptsize $-\lambda$}}] (Gv) [right=9mm of Fv] {};
			\node[gauge, bodyE8, label=below:{$222$}] (Hv) [right=9mm of Gv] {};
			\node[gauge, bodyE8, label=above:{$335$}, label=right:{\purple{\scriptsize $-\lambda$}}] (Lv) [above=6mm of Fv] {};
			\draw[thick] (dotsv.east) -- (k2v.west);
			\draw[thick] (k2v.east) -- (kpv.west);
			\draw[thick] (kpv.east) -- (kv.west);
			\draw[thick] (kv.east) -- (kv2.west);
			\draw[thick] (kv2.east) -- (kv3.west);
			\draw[thick] (kv3.east) -- (kv4.west);
			\draw[thick] (kv4.east) -- (Av.west);
			\draw[thick,bodyE8] (Av.east) -- (Bv.west);
			\draw[thick,bodyE8] (Bv.east) -- (Cv.west);
			\draw[thick,bodyE8] (Cv.east) -- (Dv.west);
			\draw[thick,bodyE8] (Dv.east) -- (Ev.west);
			\draw[thick,bodyE8] (Ev.east) -- (Fv.west);
			\draw[thick,bodyE8] (Fv.east) -- (Gv.west);
			\draw[thick,bodyE8] (Gv.east) -- (Hv.west);
			\draw[thick,bodyE8] (Fv.north) -- (Lv.south);
		\end{tikzpicture}
  }
	\end{equation}

\newpage

\bibliographystyle{JHEP}
\bibliography{mybib}

\end{document}